\documentclass[
  aps,
  prx,
  twocolumn,
  superscriptaddress,
  nofootinbib
]{revtex4-2}

\usepackage[T1]{fontenc}
\usepackage{lmodern}
\usepackage{microtype}
\usepackage{amsmath,amssymb,amsthm,mathtools}
\usepackage{bm}
\usepackage[normalem]{ulem}
\usepackage{enumitem}
\usepackage{booktabs}
\usepackage{graphicx}
\usepackage{xcolor}
\usepackage{tikz}
\usetikzlibrary{arrows.meta}
\usepackage{hyperref}
\hypersetup{colorlinks=true,linkcolor=blue,citecolor=blue,urlcolor=red}

\allowdisplaybreaks
\setlist{itemsep=0.35em,topsep=0.5em}

\newtheorem{theorem}{Theorem}%

\theoremstyle{definition}

\theoremstyle{remark}

\newcommand{\Z}{\mathbb{Z}}
\newcommand{\T}{\mathbb{T}}

\newcommand{\supp}{\operatorname{supp}}

\newcommand{\diam}{\operatorname{diam}}

\newcommand{\inc}{\mathrm{inc}}

\begin{document}
\raggedbottom

\title{Stability of energy gap of macroscopic quantum matter}

\author{Kangle Li}
\affiliation{Department of Physics, National University of Singapore, 117551, Singapore}

\author{Liujun Zou}
\affiliation{Department of Physics, National University of Singapore, 117551, Singapore}

\begin{abstract}

In quantum many-body physics and quantum field theory, it is often assumed that gapped systems remain gapped under sufficiently weak local perturbations. However, to what extent this assumption holds is elusive, except for some special systems. In this work, we examine the stability of energy gap of macroscopic quantum matter that is only subject to two physically natural conditions, locality and translation invariance. We first show that in any spatial dimension, there are gapped local Hamiltonians with a unique ground state and an unstable gap, i.e., they become gapless under certain arbitrarily weak local perturbation in the thermodynamic limit. We then prove that for both spin and fermionic systems in one dimension, such Hamiltonians with an unstable gap must not only break translation invariance but also be fine-tuned. We further show that in any dimension greater than one, there are fine-tuned local translation invariant Hamiltonians with a unique ground state and an unstable gap. These results greatly sharpen our understanding of the fundamental problem of the stability of quantum phases of matter. We also discuss implications of our results on, for example, anyon superconductivity and higher-form symmetries.

\end{abstract}

\maketitle

An energy gap, i.e., a finite energy difference between the ground states and excited states, has profound implications on macroscopic quantum systems. For example, gapped systems often have short-range correlations \cite{Hastings2003, Hastings2004, NachtergaeleSims2006, Hastings-Koma_2006}, satisfy entanglement area laws \cite{Hastings, Plenio2004, Cramer2005, Brandao2012, Arad2013, VanAcoleyen2013, Cho2017, Kuwahara2019, Liu2024, Ukai2024}, and can serve as platforms for topological phases and fault-tolerant quantum computing \cite{Kitaev1997, Freedman2001, Nayak2007}. Moreover, many influential open problems concern the presence of a gap, such as the Yang-Mills gap problem \cite{JaffeWitten2006}. Therefore, it is of fundamental importance to understand when a system is gapped.

Empirically supported by massive numerical and experimental studies, it is often assumed that gapped systems remain gapped under sufficiently weak local perturbations, which provides a practical way to argue that a system is gapped if its Hamiltonian is close to another Hamiltonian that is known to be gapped. This stability of gap can be proved for few-body systems using elementary linear algebra, but its validity in interacting many-body systems remains elusive, as the perturbations there can be extensive. Important earlier results include the proofs of a stable gap in some classical and frustration-free systems \cite{Yarotsky2006, BravyiHastingsMichalakis2010, BravyiHastings2011, MichalakisZwolak2013}. However, these systems are rather special, and rigorous results on more general systems that are only subject to physically natural conditions, such as locality and translation invariance, are rare.

In this work, we address the stability of gap of general macroscopic quantum matter with local Hamiltonians. We say that a gapped system has a {\it stable gap} if it remains gapped under {\it all} sufficiently weak local perturbations, and it has an {\it unstable gap} if it can become gapless under certain {\it arbitrarily} weak local perturbation. Our main results, which will be elaborated, are as follows.

\begin{enumerate}
    \item In any spatial dimension, there are gapped local translation non-invariant Hamiltonians with a unique ground state and an unstable gap.

    \item All gapped local one dimensional (1D) spin and fermionic systems with a unique ground state and an unstable gap are fine-tuned, i.e., under certain arbitrarily weak local perturbation they remain gapped but their gap becomes stable.

    \item All gapped local translation invariant 1D systems with a unique ground state have a stable gap.

    \item In all dimensions greater than 1, there are fine-tuned local translation invariant Hamiltonians with a unique ground state and an unstable gap.
\end{enumerate}

We get the first and fourth results from concrete examples. These examples host states where the numbers of particle-like excitations grow linearly with the system's volume, while their energies only grow sub-linearly, so an arbitrarily weak but extensive perturbation can close the gap by reducing the energy of such excitations, if the system is sufficiently large. For the second and third results, we first prove a general theorem that rules out intrinsic topological order in 1D, which allows us to understand the stability of gap by studying some very special Hamiltonians. For all cases under our consideration that have a stable gap, we prove that the gap is a continuous function of the strength of any perturbation. These results generalize the previous results to physically natural setups and greatly sharpen our understanding of the stability of quantum phases of matter.

{\bfseries Setup.} We consider a sequence of lattice systems, each with linear sizes $\vec L=(L_1, L_2, \cdots, L_d)$ along the $d$ spatial dimensions, where $L_i$ never decreases in this sequence for any $i\in\{1, 2, \cdots, d\}$. We discuss both qudit and fermionic systems, and each site of the lattice hosts a qudit with a finite dimensional Hilbert space or a finite number of fermionic modes. We demand that the lattice geometries and local degrees of freedom across different systems match, in a way that all systems share the same origin of coordinates and smaller systems can be viewed as subsystems of larger systems.

Denote by $H_{\vec {L}}$ the Hamiltonian of the system with linear sizes $\vec L$. The Hamiltonians are required to be local, which has two meanings. First, $H_{\vec L}=\sum_{\vec x}h_{\vec x, \vec L}$ includes only finite-range interactions, i.e., $\diam(\supp(h_{\vec x, \vec L}))\leqslant R$, with $\supp(h_{\vec x, \vec L})$ the support of $h_{\vec x, \vec L}$, $\diam(X)$ the linear size of a region $X$, and $R$ a constant independent of $\vec x$ or $\vec L$. In fermionic systems, each $h_{\vec x, \vec L}$ should also be even under fermion parity. Second, local dynamics is determined truly locally and compatible across different sizes, i.e., for any disk-like region containing the origin, when $\vec L$ is large enough{\footnote{In this paper, taking $\vec L$ to be large or $\vec L\rightarrow\infty$ means that $L_i\rightarrow\infty$ with (approximately) fixed aspect ratios $L_i/L_j$ for all $i, j\in\{1, 2, \cdots, d\}$.}}, $[H_{\vec L}, O]$ is independent of $\vec L$ for all operators $O$ supported in this region{\footnote{These commutators are operators in systems with different sizes, and we view them as operators in a common subsystem of all these systems, which is always possible in our setup and makes the comparison of these commutators meaningful.}}. This second requirement is motivated by the physical picture that local interactions are between nearby degrees of freedom and should not depend on the macroscopic length scale $\vec L$. It rules out interactions like $h_{\vec x, \vec L}=h_{\vec x}/|\vec L|$ with some $\vec L$-independent $h_{\vec x}$, and implies that the Hamiltonians contain only finite-range interactions if there is translation invariance \cite{supp}. Moreover, it provides a meaning of the thermodynamic limit of the Hamiltonians with increasing system sizes. Most physically relevant Hamiltonians, such as the Hubbard Hamiltonian and Heisenberg Hamiltonian, satisfy both requirements.

\begin{figure}[t]
\centering
\includegraphics[width=0.66\columnwidth]{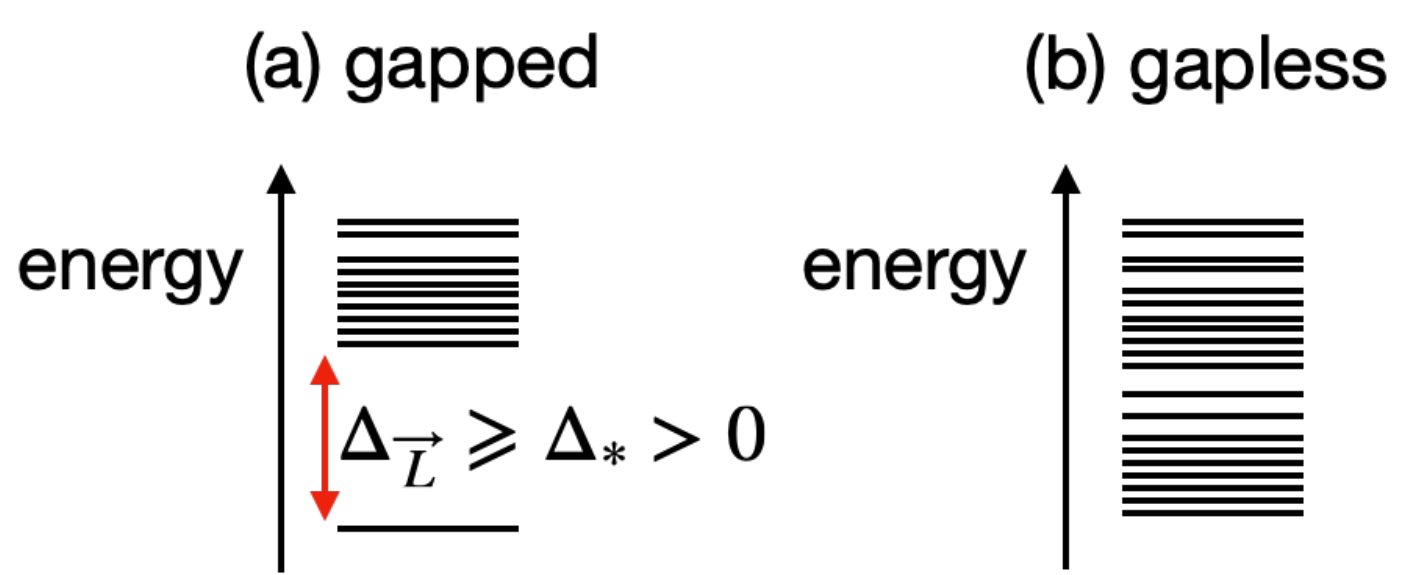}
\caption{Schematic gapped and gapless spectra.}
\label{fig:gapped-gapless-spectra}
\end{figure}

We call a {\it sequence} of Hamiltonians $\{H_{\vec L}\}$ gapped if there are an $\vec L$-independent scale $\delta>0$, a non-negative decreasing sequence $\{\epsilon_{\vec L}\}$ satisfying $\lim_{\vec L\rightarrow\infty}\epsilon_{\vec L}=0$, and linear sizes $\vec L_0$, such that if $L_i\geqslant L_{0i}$ for all $i$, then
\begin{equation}
\operatorname{spec}(H_{\vec L})
\subseteq [E_{0, \vec L},E_{0, \vec L}+\epsilon_{\vec L}]\cup[E_{0, \vec L}+\delta,\infty),
\end{equation}
and the dimension of the eigenspace with eigenvalues in $[E_{0, \vec L},E_{0, \vec L}+\epsilon_{\vec L}]$ is $\vec L$-independent, where $E_{0,\vec L}$ is the ground state energy of $H_{\vec L}$ (see Fig.~\ref{fig:gapped-gapless-spectra}). Physically, in gapped systems the ground states and excited states are separated by a gap $\Delta_*(\{H_{\vec L}\})=\sup\delta$, and the ground states can have an $\vec L$-independent (approximate) degeneracy with energy splitting $\epsilon_{\vec L}$. In contrast, $\{H_{\vec L}\}$ is gapless if the spectrum near $E_{0, \vec L}$ is quasi-continuous, i.e., there exists an $\vec L$-independent $W>0$, such that for any $E_1$ and $E_2$ satisfying $0\leqslant E_1<E_2\leqslant W$, there is an eigenenergy in $(E_{0,\vec L}+E_1, E_{0,\vec L}+E_2)$ for all sufficiently large $\vec L$. 

Gapped and gapless systems defined above are indeed identified as gapped and gapless, respectively, in most of the literature. For example, under our definitions, both the ferromagnetic and paramagnetic regimes of the transverse field Ising chain are gapped, while the critical point is gapless, which agrees with the usual definition. However, our definitions do not cover all systems. For example, systems with $H_{\vec L}=0$ are neither gapped nor gapless in our definitions. In this work, it suffices to consider gapped and gapless systems under our definitions, as the stability of gap is often assumed under these definitions.

{\bfseries Unstable gap in translation non-invariant systems.} Now we show that the generic assumption of the stability of gap is false in all spatial dimensions, by constructing a 1D example of gapped local Hamiltonians of qudits that have unique ground states and unstable gaps. Stacking many copies of such systems can lead to higher dimensional examples. These examples can also be viewed as fermionic examples, since qudits can be obtained from fermions with strong Hubbard-type repulsions.

Inspired by Ref. \cite{gap_undecidability}, we consider a chain of qutrits with a periodic boundary condition and length $L$, denote the basis of the qutrit at site $x$ by $\{|0_x\rangle, |\uparrow_x\rangle, |\downarrow_x\rangle\}$, and view $|0_x\rangle$ and $\{|\uparrow_x\rangle, |\downarrow_x\rangle\}$ as an on-site vacuum and a spin-1/2 defect, respectively. Let $P_x=|0_x\rangle\langle 0_x|$. The Hamiltonian of our example with an unstable gap is
\begin{equation} \label{eq: 1d unperturbed}
H_L^{(1)}=J\sum_x(P_x-P_{x+1})^2-h_0P_0,
\end{equation}
with $0<h_0<2J$. The first term favors states where {\it all} sites host either the vacuum or defect, and the second term, which breaks translation invariance, selects $\otimes_x|0_x\rangle$ as the unique ground state and makes all states $\otimes_x|\psi_x\rangle$ with $\langle 0_x|\psi_x\rangle=0$ the first excited states with a gap $h_0$.

The above already shows a peculiar feature of our Hamiltonian, i.e., the number of defects in the state $\otimes_i|\psi_i\rangle$ is $L$, but its energy $h_0$ is independent of $L$. Eq. \eqref{eq: 1d unperturbed} is inspired by an analogous gapped Hamiltonian in Ref. \cite{BravyiHastingsMichalakis2010}. Ref. \cite{BravyiHastingsMichalakis2010} gave an arbitrarily weak local perturbation that changes the ground state for large enough systems, but this perturbation does not make the systems gapless. However, we claim that the following local translation invariant perturbation makes our system gapless:
\begin{equation} \label{eq: 1d perturbation}
V_L^{(1)}=\varepsilon\sum_x(P_x-J_s\vec\sigma_x\cdot\vec\sigma_{x+1}),
\end{equation}
where $\varepsilon$ and $J_s$ are {\it any} positive constants, and $\vec\sigma_x=(\sigma_x^1, \sigma_x^2, \sigma_x^3)$ with $\sigma_x^1=|\uparrow_x\rangle\langle\downarrow_x|+{\rm h.c.}$, $\sigma_x^2=-i|\uparrow_x\rangle\langle\downarrow_x|+{\rm h.c.}$, and $\sigma_x^3=|\uparrow_x\rangle\langle\uparrow_x|-|\downarrow_x\rangle\langle\downarrow_x|$.

To understand the spectrum of $H^{(1)}_L+V^{(1)}_L$, we first consider the simpler case with $\varepsilon>0$ and $J_s=0$. Then $\otimes_x|0_x\rangle$ and $\otimes_x|\psi_x\rangle$ are still energy eigenstates, with energies $\varepsilon L$ and $h_0$, respectively. As long as $L>h_0/\varepsilon$, $\{\otimes_x|\psi_x\rangle\}$ spans a $2^L$-dimensional ground subspace of $H^{(1)}_L+V^{(1)}_L$, which can be viewed as a chain of decoupled qubits. Now suppose $\varepsilon>0$ and $J_s>0$. These qubits are coupled by the Heisenberg interaction and develop a gapless spectrum. Therefore, as long as $\varepsilon>0$ and $J_s>0$, the system is gapless, proving the instability of the gap of $H_L^{(1)}$.

Note that the ground state changes under the perturbation in this example. In the supplemental material (SM), we present another example with an unstable gap, where the unique ground states of the unperturbed Hamiltonians remain the unique ground states under the perturbations that make the system gapless \cite{supp}.

{\bfseries Stability of gap in 1D systems.} The above example may be surprising, as it contradicts the common numerical and experimental observation that the gap of a system is stable and varies continuously with the perturbation strength. However, we can prove that in 1D such examples are always fine-tuned and we conjecture that this is true in any dimension, which explains why such an unstable gap is not observed commonly.

To intuitively see why the unstable gap of $H_L^{(1)}$ requires fine tuning, consider adding another perturbation $V'^{(1)}_L=-\varepsilon'\sum_xP_x$ to $H_L^{(1)}$ with any $\varepsilon'>0$, which increases the excitation energy of $\otimes_x|\psi_x\rangle$ by $\varepsilon'L$. Now $V_L^{(1)}$ cannot close the gap of the new Hamiltonian $H_L^{(1)}+V_L'^{(1)}$ unless $\varepsilon\gtrsim\varepsilon'$, suggesting a stable gap of $H_L^{(1)}+V_L'^{(1)}$.

Such fine-tuned nature can be proved for {\it all} 1D local Hamiltonians with a unique ground state and unstable gap. Moreover, all 1D gapped local translation invariant Hamiltonians with a unique ground state have a stable gap. Here we explain the essential meaning of these statements and the intuition behind, while leaving the full statements and details of the proof to SM \cite{supp}.

Suppose the system with length $L$ obeys the periodic boundary condition, i.e., the lattice is on $\T_L:=\Z/(L\Z)$. The unperturbed Hamiltonian $H_L=\sum_{x\in\T_L}h_{x, L}$ and the perturbation $V_L=\sum_{x\in\T_L}v_{x, L}$. The strengths of $H_L$ and $V_L$ can be measured by $\|h_L\|_{\inc}:=\max_{y\in\T_L}\sum_{\supp(h_{x, L})\ni y}\|h_{x, L}\|$ and $\|v_L\|_\inc=\max_{y\in\T_L}\sum_{\supp(v_{x, L})\ni y}\|v_{x, L}\|$, respectively, where $\|\cdot\|$ is the operator norm. Throughout this section,  for any Hermitian operator $M$, we denote its gap by ${\rm gap}(M):=\lambda_2(M)-\lambda_1(M)$, with $\lambda_1$ and $\lambda_2$ the smallest and second smallest eigenvalues of $M$, and we only consider the case where the unperturbed Hamiltonians have finite strengths and satisfy ${\rm gap}(H_L)>\gamma_0$ for all $L>L_0$, with $\gamma_0>0$ and $L_0>0$ some constants. Note that $\gamma_0>0$ implies that $H_L$ has a unique ground state.

\begin{theorem} \label{thm: generic stable gap}
    Denote the ground state of $H_L$ by $|\psi_L\rangle$. If $\{|\psi_L\rangle\}$ has a thermodynamic limit, then there is an arbitrarily weak local perturbation $V'_L=\sum_{x\in\T_L}v'_{x, L}$, such that for any $R>0$, there are constants $L_R, c_R, D_R>0$, such that for all $L>L_R$ and all local perturbations $V_L=\sum_{x\in\T_L}v_{x, L}$ with $\|v_L\|_\inc<c_R\|v'_L\|_\inc$ and $\diam(\supp(v_{x, L}))\leqslant R$ for all $x$ and $L$,
    \begin{equation} \label{eq: continuity generic}
    \begin{split}
        &\inf_{L>L_R}{\rm gap}(H_L+V_L')>0,\\
        &|{\rm gap
        }(H_L+V_L'+V_L)-{\rm gap}(H_L+V'_L)|\leqslant D_R\|v_L\|_{\rm inc}.
    \end{split}
    \end{equation}
    Here $c_R, D_R>0$ are $L$-independent.
\end{theorem}

Theorem \ref{thm: generic stable gap} means that Hamiltonians with an unstable gap are fine-tuned. Moreover, Eq. \eqref{eq: continuity generic} means that the gap of the new Hamiltonians $H_L+V'_L$ is a continuous function of the strength of any additional perturbation $V_L$. This theorem is proved thanks to the following theorem \cite{supp}.

\begin{theorem} \label{thm: invertibility}
Denote by $\theta$ the complex conjugation operation with respect to the tensor product of some on-site basis and define $|\bar\psi_L\rangle=\theta|\psi_L\rangle$, so $|\psi_L\rangle\otimes|\bar\psi_L\rangle$ is a doubled state of $|\psi_L\rangle$. Then for sufficiently large $L$, there is a unitary operator $\widehat T_L$ in the doubled Hilbert space such that $\widehat T_L\left(|\psi_L\rangle\otimes|\bar\psi_L\rangle\right)$ is a product state, where $\widehat T_L$ is the time evolution generated by a quasi-local Hamiltonian over a finite duration.
\end{theorem}

Here a quasi-local Hamiltonian means a Hamiltonian whose interaction terms have finite magnitudes that decay exponentially as their ranges increase. Theorem \ref{thm: invertibility} means that $|\psi_L\rangle$ is invertible and thus has no intrinsic topological order, i.e., upon stacking with another system it can evolve into a product state under a finite-duration dynamics generated by a quasi-local Hamiltonian. This result has been anticipated for a long time \cite{Chen2010, Fidkowski2010, Schuch2010}, and recently it was proved for infinite spin chains in Ref. \cite{deroeck2025}, which did not even double the system and actually proved a stronger result that such states are short-range entangled. Our result concerns finite-size systems, which are more relevant to real systems, and we cover both spin and fermionic systems.

The existence of the above $\widehat T_L$ allows us to construct $\widehat K_L=\widehat T_L(\widehat H_L-2E_{0,L}\mathbf{1}) \widehat{T}_L^\dag$, where $\widehat H_L=H_L\otimes\mathbf{1}+\mathbf{1}\otimes(\theta H_L\theta)$. $\widehat K_L$ is a very special Hamiltonian, as it has an unentangled ground state, and it can be decomposed into a sum of quasi-local interactions due to the locality of $\widehat T_L$. In SM \cite{supp}, we show that the stability of the gap of the original Hamiltonians $\{H_L\}$ can be inferred from the stability of the gap of $\{\widehat K_L\}$. Denote the ground state of $\widehat K_L$ by $\otimes_x|\widehat 0_x\rangle$ where $|\widehat 0_x\rangle$ is a state at position $x$ of the doubled system, and call $\widehat{\mathsf{N}}_L=\sum_x(1-|\widehat 0_x\rangle\langle \widehat 0_x|)$ the defect number. Then by adding $\widehat V_L'=\varepsilon'\widehat{\mathsf{N}}_L$ to $\widehat K_L$ where $\varepsilon'>0$, intuitively, the excitation energy of any state is proportional to the defect number, and the gap of $\widehat K_L+\widehat V'_L$ cannot be closed unless the perturbation strength exceeds $\varepsilon'$. This also suggests that $\widehat T_L^\dag (\widehat K_L+\widehat V'_L)\widehat{T}_L=\widehat H_L-2E_{0,L}\mathbf{1} +\widehat T_L^\dag \widehat V'_L \widehat T_L$ has a stable gap. The locality of $\widehat T_L$ means that $\widehat T_L^\dag \widehat V'_L\widehat T_L$ is approximately local, and $V_L'$ in Theorem \ref{thm: generic stable gap} can be obtained by restricting $\widehat T_L^\dagger\widehat V_L' \widehat T_L$ to the undoubled system and truncating the tails of its interaction terms.

With the additional assumption of translation invariance of $H_L$, we can show that $\widehat K_L\geqslant\kappa\widehat{\mathsf{N}}_L$ with $\kappa>0$ an $L$-independent constant. So even without adding any $\widehat V'_L$, the excitation energy of any state for $\widehat K_L$ is already at least proportional to the defect number. This leads to the following useful result.

\begin{theorem} \label{thm: continuous gap translation invariant}
    Suppose that $H_L$ is translation invariant. Then for each $R\geqslant 0$, there are constants $L_R, c_R, D_R>0$, such that for all $L>L_R$ and all local (potentially translation non-invariant) perturbations $V_L=\sum_{x\in\T_L}v_{x, L}$ with $\|v_L\|_\inc<c_R$ and $\diam(\supp(v_{x, L}))\leqslant R$ for all $x$ and $L$, $H_L+V_L$ has a unique ground state, and
    \begin{equation}
        |{\rm gap}(H_L+V_L)-{\rm gap}(H_L)|\leqslant D_R\|v_L\|_{\rm inc}.
    \end{equation}
    Here $c_R, D_R>0$ are $L$-independent.
\end{theorem}

Theorem \ref{thm: continuous gap translation invariant} shows that 1D gapped local translation invariant Hamiltonians with a unique ground state always have a stable gap, and the gap varies continuously with the perturbation strength.

{\bf Unstable gap in higher dimensional translation invariant systems.} Theorems \ref{thm: generic stable gap} and \ref{thm: continuous gap translation invariant} are fundamental results about 1D quantum many-body systems. We expect that Theorem \ref{thm: generic stable gap} can be generalized to any dimension, although we cannot prove it. However, now we show that the generalization of Theorem \ref{thm: continuous gap translation invariant} is false in all dimensions greater than 1. Again, it suffices to give a counterexample of qudits in two dimensions (2D).

This example consists of qudits on a 2D lattice with periodic boundary conditions, i.e., the lattice is $\T^2_{\vec L}=\prod_{i=1,2}\Z/(L_i\Z)$. The site $\vec x$ hosts a 23-dimensional qudit with local basis $\{|0_{\vec x}\rangle, |a_{\vec x}\rangle\otimes |s_{\vec x}\rangle\}$, where $a_{\vec x}\in\mathcal{T}:=\{1,2,\cdots, 11\}$ and $s_{\vec x}\in\{\uparrow, \downarrow\}$. We view $|0_{\vec x}\rangle$ as an on-site vacuum, and $|a_{\vec x}\rangle\otimes|s_{\vec x}\rangle$ as 11 types of spin-1/2 defects.

\begin{figure}[tb]
\centering
\includegraphics[width=0.6\linewidth]{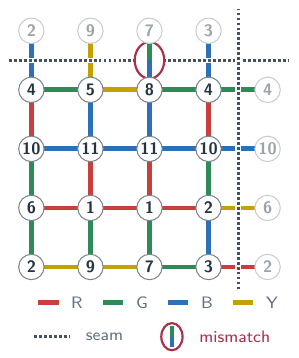}
\caption{
A $4\times4$ torus with one mismatch marked by a red circle. The number at each site indicates a type of tile, and the dashed lines are seams that wrap the torus.
}
\label{fig:gap-instability-examples}
\end{figure}

The Hamiltonian is constructed using some classic results about tiling \cite{Wang1961, JeandelRao2021, supp}. Tiling can be viewed as assigning colors to each link of the lattice (see Fig. \ref{fig:gap-instability-examples}). The 11 values of $a_{\vec x}$ can be viewed as 11 types of tiles, with each type being a site where its four links are colored as in Table \ref{tab:wang-tiles}. A valid tiling pattern satisfies $E(a_{\vec x})=W(a_{\vec x+\vec e_1})$ and $N(a_{\vec x})=S(a_{\vec x+\vec e_2})$ for all $\vec x$, where $\vec e_1=(1, 0)$ and $\vec e_2=(0, 1)$, i.e., both tiles connected by each link give consistent colors. It is known that Table \ref{tab:wang-tiles} gives the simplest tiles that can have valid tiling patterns on the infinite plane $\Z^2$, but {\it all} such patterns are aperiodic. So these tiles cannot lead to a valid tiling pattern on $\T^2_{\vec L}$ for any $\vec L$, otherwise the tiling pattern on $\Z^2$ can be periodic with period $\vec L$. In passing, we note that the following construction is inapplicable in 1D because all tiles with valid tiling patterns in $\Z$ allow periodic patterns \cite{LindMarcus1995}.

\begin{table}[h]
\centering
\caption{Link colors of the 11 tiles. For the tile $a$, the columns $W(a)$, $E(a)$, $S(a)$, and $N(a)$ give the west, east, south, and north link a color, respectively.}
\label{tab:wang-tiles}
\begin{tabular}{c rrrr @{\qquad} c rrrr}
\toprule
$a$ & $W(a)$ & $E(a)$ & $S(a)$ & $N(a)$ &
$a$ & $W(a)$ & $E(a)$ & $S(a)$ & $N(a)$ \\
\midrule
1 & R & R & G & R &  7 & Y & G & G & G \\
2 & R & Y & B & G &  8 & Y & G & B & B \\
3 & G & R & B & B &  9 & Y & Y & Y & G \\
4 & G & G & R & B & 10 & B & B & G & R \\
5 & G & Y & B & Y & 11 & B & B & R & B \\
6 & Y & R & G & G &    &   &   &   &   \\
\bottomrule
\end{tabular}
\end{table}

Based on this fact about the tiles, we let the unperturbed local translation invariant Hamiltonian be
\begin{equation} \label{eq: 2d unperturbed}
    H^{(2)}_{\vec L}
    =
    J
    \left[
    \sum_{\langle\vec x, \vec y\rangle}
    (P_{\vec x}-P_{\vec y})^2+\sum_{\vec x}(C^{(1)}_{\vec x}+C^{(2)}_{\vec x})\right],
\end{equation}
where $J>0$, $P_{\vec x}=|0_{\vec x}\rangle\langle 0_{\vec x}|$, and
\begin{equation}
\begin{split}
    C_{\vec x}^{(1)} &= \sum_{\substack{a_{\vec x},b_{\vec x+\vec e_1}\in\mathcal T\\ E(a_{\vec x}) \neq W(b_{\vec x+\vec e_1}) }} \Pi_{a_{\vec x}} \Pi_{b_{\vec x+\vec e_1}}, \\
    C_{\vec x}^{(2)} &=\sum_{\substack{a_{\vec x},b_{\vec x+\vec e_2}\in\mathcal T \\ N(a_{\vec x})\neq S(b_{\vec x+\vec e_2})}} \Pi_{a_{\vec x}} \Pi_{b_{\vec x+\vec e_2}},
    \end{split}
\end{equation}
with $\Pi_{a_{\vec x}}=|a_{\vec x}\rangle\langle a_{\vec x}|\otimes(|\uparrow\rangle\langle\uparrow|+|\downarrow\rangle\langle\downarrow|)_{\vec x}$.

To minimize the first term of $H^{(2)}_{\vec L}$, {\it all} sites must host either the vacuum or a defect. In the latter case, the defects form a tiling pattern, which, as discussed above, is always invalid on any torus. The second term in Eq. \eqref{eq: 2d unperturbed} penalizes invalid tilings. So $H_{\vec L}$ is gapped and $\otimes_{\vec x}|0_{\vec x}\rangle$ is its unique ground state. Moreover, wrapping a portion of a valid tiling pattern on $\Z^2$ onto $\T^2_{\vec L}$ gives a specific tiling pattern on $\T^2_{\vec L}$, where the number of links with inconsistent colors is at most $L_1+L_2$. Hence, there are excited states where {\it all} sites host defects but their energies are at most $J(L_1+L_2)$, which grows sub-linearly with the system's volume.

The perturbation that makes the system gapless is still
\begin{equation}
    V_{\vec L}^{(2)}=\varepsilon\left[\sum_{
    \vec x
    }P_{\vec x}-J_s\sum_{\langle\vec x, \vec y\rangle}\vec\sigma_{\vec x}\cdot\vec\sigma_{\vec y}\right],
\end{equation}
where $\varepsilon$ and $J_s$ are {\it any} positive constants, and $\vec\sigma_{\vec x}=(\sigma_{\vec x}^1, \sigma_{\vec x}^2, \sigma_{\vec x}^3)$ with $\sigma_{\vec x}^1=(\sum_{a_{\vec x}\in\mathcal{T}}|a_{\vec x}\rangle\langle a_{\vec x}|)\otimes|\uparrow_{\vec x}\rangle\langle\downarrow_{\vec x}|+{\rm h.c.}$, $\sigma_{\vec x}^2=(\sum_{a_{\vec x}\in\mathcal{T}}|a_{\vec x}\rangle\langle a_{\vec x}|)\otimes(-i|\uparrow_{\vec x}\rangle\langle\downarrow_{\vec x}|)+{\rm h.c.}$, and $\sigma_{\vec x}^3=(\sum_{a_{\vec x}\in\mathcal{T}}|a_{\vec x}\rangle\langle a_{\vec x}|)\otimes(|\uparrow_{\vec x}\rangle\langle\uparrow_{\vec x}|-|\downarrow_{\vec x}\rangle\langle\downarrow_{\vec x}|)$. The effect of $V_{\vec L}^{(2)}$ on $H^{(2)}_{\vec L}$ is similar to the effect of $V_L^{(1)}$ on $H_L^{(1)}$. Concretely, as long as $L_{1,2}$ satisfy $\varepsilon L_1L_2>J(L_1+L_2)$, the first term of $V_{\vec L}^{(2)}$ lets certain states where all sites host defects become the degenerate ground states \cite{supp}, and the second term makes the spectrum gapless. 

So $\{H_{\vec L}^{(2)}\}$ has an unstable gap. This unstable gap also requires fine-tuning. For example, by adding $V'^{(2)}_{\vec L}=-\varepsilon'\sum_{\vec x}P_{\vec x}$ to $H_{\vec L}^{(2)}$ with $\varepsilon'>0$, $V_{\vec L}^{(2)}$ cannot close the gap of $H_{\vec L}^{(2)}+V'^{(2)}_{\vec L}$ unless $\varepsilon\gtrsim \varepsilon'$. In fact, Theorem \ref{thm: generic stable gap} can be generalized to any dimension if we further assume that the ground state of each $H_{\vec L}$ is a product state \cite{supp}.

{\bf Discussion.} Our results greatly sharpen our understanding of the stability of quantum phases of matter, and they provide an explanation for the empirical observation that the gap of a macroscopic system varies continuously as the perturbation strength changes.

Now we discuss additional implications of our results. First, from our analysis, it is clear that even if a gapped local Hamiltonian has a ground state with zero correlation length, under certain arbitrarily weak local perturbations the new ground state can have an infinite correlation length. Recently, some mixed-state version of this phenomenon was also discovered \cite{Chen2026, Newsom2026, Negari2026}.

The ground states of our examples with an unstable gap, such as $H_{\vec L}^{(2)}$, are just $\otimes_{\vec x}|0_{\vec x}\rangle$. Another parent Hamiltonian of this state is $\tilde H_{\vec L}=-J\sum_{\vec x}|0_{\vec x}\rangle\langle 0_{\vec x}|$, which has a stable gap \cite{BravyiHastingsMichalakis2010} and is adiabatically connected to $H_{\vec L}^{(2)}$ via the path $(1-s)\tilde H_{\vec L}+sH_{\vec L}^{(2)}$ with $s\in[0, 1]$. So if we view these two Hamiltonians as in the same quantum phase, then whether the gap is stable is {\it not} a universal property of a gapped quantum phase.

In our examples with an unstable gap, there is also a $U(1)$ symmetry generated by $\sum_{\vec x}\sigma^3_{\vec x}$. Excitations carrying nonzero charge under this symmetry are gapped in the unperturbed Hamiltonians, but they become gapless when the arbitrarily weak local perturbation is added to close the gap. So our examples also show that the {\it charge gap} can be unstable.

This observation has multiple consequences. For example, Refs. \cite{Divic2024} and \cite{Kuhlenkamp2025} numerically studied a 2D Hamiltonian with anyon superconductivity, where the superconductivity is obtained by doping a quantum critical point where spinful excitations (e.g., single fermions) are gapped but spinless excitations (e.g., Cooper pairs) are gapless. An interesting question is whether the gap of the spinful excitations directly implies that the doped system cannot be a Fermi liquid, so it is likely a superconductor. This would be true if the charge gap were always stable. As we have demonstrated that the charge gap can be unstable, the answer to this question is no. Even if one can prove that in all dimensions unstable gaps only arise in fine-tuned Hamiltonians, one needs to show that the Hamiltonian in Refs. \cite{Divic2024} and \cite{Kuhlenkamp2025} is not one of those fine-tuned ones, otherwise numerical calculation is still needed to conclude that the doped system cannot be a Fermi liquid.

Another application of our results is to higher-form symmetries. Originally developed in quantum field theories \cite{Gaiotto2014}, higher-form symmetries also found many applications in condensed matter systems \cite{McGreevy2022, Luo2023}. However, Ref. \cite{Liu2026} pointed out that the higher-form symmetries often discussed in condensed matter systems may behave very differently from the topological higher-form symmetries in field theories, and a necessary condition for the former to mimic the latter is that excitations with a charge under the higher-form symmetries have a stable gap. Our results allow us to construct an interesting Hamiltonian where the gap of such excitations is unstable, and this example exhibits various properties beyond those in Ref. \cite{Liu2026} (see SM \cite{supp} for details).

Finally, there are many important open problems. First, it is important to understand whether Theorem \ref{thm: generic stable gap} can be generalized to higher dimensions. Second, it is valuable to extend our results to systems with long-range interactions and/or multiple ground states. Third, it is interesting to understand the geometry of the fine-tuned Hamiltonians with unstable gaps in the space of all local Hamiltonians, such as their (co-)dimension.

{\bf Acknowledgement.} ChatGPT is used to get a large part of the results. We have verified these results and take the full responsibility for the content. This work is supported by the National University of Singapore start-up
grants A-0009991-00-00 and A-0009991-01-00, and by the Ministry of Education, Singapore, under its Academic Research Fund Tier 2 Program (MOE-T2EP50225-0005). Any opinions, findings and conclusions or recommendations expressed in this material are those of the authors and do not reflect the views of National University of Singapore or Ministry of Education, Singapore.

\nocite{NSYstability2022,BachmannMichalakisNachtergaeleSims2012,Hastings2005,Kitaev2006,Verstraete_2008,DalzellBrandao2019,BrandaoHorodecki2015,Sopenko2023,Ranard_2022,Grimm_2002,Campostrini_2015,NSYspin,NSYfermion,Kato1995,Uhlmann1976,FuchsVanDeGraaf1999,Watrous,Junge_2018,choi1974schwarz}
\bibliography{reference}

\end{document}


\title{ Supplementary Material for ``Stability of energy gap of macroscopic quantum matter"}

\author{Kangle Li}
\affiliation{Department of Physics, National University of Singapore, 117551, Singapore}

\author{Liujun Zou}
\affiliation{Department of Physics, National University of Singapore, 117551, Singapore}

\maketitle

\tableofcontents

This supplemental material is written in a largely self-contained manner, which includes the setup, examples that falsify the general assumption of the stability of gap for gapped local Hamiltonians 1) in any spatial dimensions if there is no translation invariance and 2) in all spatial dimensions larger than 1 even if there is translation invariance, and the proof that establishes the stability and continuity of the gap in one dimensional local translation invariant Hamiltonians with unique ground states. We also provide a concrete example of lattice systems where the lattice version of 1-form symmetry is unstable to arbitrarily weak local perturbations. Specifically, Section~\ref{sec: setup and summary} introduces the setup, derives the consequences of locality of Hamiltonians, and summarizes this supplemental material. Section~\ref{sec: another 1D example}
presents another one dimensional example with an unstable gap, whose unique ground
state remains as the ground state even if its gap is closed by a weak local perturbation. The proof of the stability and continuity of gap of one dimensional local translation invariant Hamiltonians with unique ground states is organized in Section~\ref{subsec:intermediate_step_summary}, with the details of the arguments for the spin part given in Sections~\ref{sec:across_sizes}-\ref{sec:coercivity} and the fermionic
extensions given in Section~\ref{sec:fermions}. Section~\ref{sec:2d-instability}
provides more details of the two dimensional example, which shows that the gap can be unstable even if the local Hamiltonians with unique ground states are translation invariant, as long as the spatial dimension is larger than 1. This two dimensional example is then generalized in Sec. \ref{sec: 1-form}, which gives a gapped local translation invariant Hamiltonian with a lattice version of 1-form symmetry that is unstable to certain arbitrarily weak local perturbation. In Sec. \ref{sec: stabilization}, we discuss the phenomenon of the stabilization of energy gap, i.e., even if the Hamiltonian has an unstable gap, after adding certain arbitrarily weak local perturbation, the new Hamiltonian may still be gapped and have a stable gap. Section~\ref{app:technical} collects some general lemmas that do not rely on our setup and are used in our proofs.

\section{Setup and summary} \label{sec: setup and summary}

In this section, we first introduce the setup of this supplemental material in a self-contained manner. In particular, we will introduce the geometry of our lattices, the operator contents of our systems of qudits and fermions, and the notion of local Hamiltonians under our consideration. We will explicitly describe how systems with different sizes are compared, so that a thermodynamic limit of Hamiltonians in systems with increasing sizes makes sense. After introducing the setup, we discuss the consequences of the locality of Hamiltonians in Sec. \ref{app: locality} and summarize the main results of this supplemental material in Sec. \ref{subsec:summary-main}.

In the main text and this supplemental material, we mainly consider lattice systems with periodic boundary conditions{\footnote{However, we will discuss the infinite-size versions of some of our examples in Sec. \ref{subsec:1d-infinite-ground-states} and Sec. \ref{subsec:2d-infinite-ground-states}, and the formalism and conventions in the infinite-size settings are briefly introduced in Sec. \ref{subsubsec: review of infinite lattice}.}}. Let $d\ge1$ be the spatial dimension of our system. Each of our systems lives in a $d$-dimensional periodic lattice with linear sizes $\vec L=(L_1,\ldots,L_d)$, i.e.,
\[
 \T_{\vec L}=\prod_{i=1}^d\big(\Z/(L_i\Z)\big).
\]

At each site of the lattice, we may either have spins (i.e., qudits) or fermions. In general, the dimensions of the Hilbert spaces of the spins or the numbers of fermionic modes at different sites can vary from site to site, but in this work it suffices to focus on the case where these are all the same for all sites. So for spin systems, each site hosts a Hilbert space $\C^D$, with a fixed $D\ge2$. For any subset of sites $X\subseteq\mathbb{T}_{\vec L}$, its Hilbert space $\mathcal{H}_X$ and operator algebra $\mathcal{A}_X$ are, respectively,
\[
 \mathcal H_X=(\C^D)^{\otimes |X|},\qquad
 \A_X=\mathcal B(\mathcal H_X),
\]
where $\mathcal B(\mathcal H_X)$ is the algebra of bounded operators in $\mathcal H_X$. As usual, operators can be embedded in larger systems by tensoring with the
identity operator. For fermionic systems, each site hosts $\nu\ge1$ modes with a fixed $\nu$, so the on-site Hilbert space has a dimension $D=2^\nu$, and
$\A_X$ is the algebra generated by the creation and annihilation
operators of these fermionic modes, which obey the standard canonical anti-commutation relations (CAR). Denote by $\mathsf n_X$ the
fermion number in $X$, and by $\Gamma_X=(-1)^{\mathsf n_X}$ the fermion parity in $X$. An operator $A\in\A_X$ is even if $\Gamma_XA\Gamma_X=A$ and odd if
$\Gamma_XA\Gamma_X=-A$. The total fermion parity is $\Gamma_{\mathbb{T}_{\vec L}}$. More details of the fermionic setups can be found in Sec. \ref{sec:fermions}.

After specifying the lattice geometry and the degrees of freedom as above, now we can specify the Hamiltonian and interaction of each of our systems. An interaction $\Phi_{\vec L}$ assigns a Hermitian operator
$\Phi_{\vec L}(X)\in\A_X$ to each non-empty subset $X$ of the lattice, and the Hamiltonian (which can be the unperturbed Hamiltonian, perturbation, or the perturbed Hamiltonian) is
\begin{equation}\label{eq:general-interaction}
 H_{\vec L}=
 \sum_{\varnothing\ne X\subseteq\T_{\vec L}}\Phi_{\vec L}(X).
\end{equation}
We define the \emph{incidence norm} of the interaction by
\begin{equation}\label{eq:general-incidence}
 \|\Phi_{\vec L}\|_{\inc}
 =\sup_{x\in\T_{\vec L}}\sum_{X\ni x}\|\Phi_{\vec L}(X)\|,
\end{equation}
where $\|\cdot\|$ is the operator norm. Then the strength of the Hamiltonian $H_{\vec L}$ is measured by the supremum of $\|\Phi_{\vec L}\|_{\inc}$ over system sizes. So even if the strength of the Hamiltonian $H_{\vec L}$ is small, its operator norm can still grow without a bound as the system size increases. Throughout this work, we assume that $\sup_L\|\Phi_{\vec L}\|_{\rm inc}<\infty$, so all (unperturbed and perturbed) Hamiltonians have a finite strength.

A special class of Hamiltonians under our consideration are translation invariant Hamiltonians. Denote by $\tau_{\vec L, x_0}$ the unitary translation operation by an element $x_0$ on $\mathbb{T}_{\vec L}$, which transforms $x\in\mathbb{T}_{\vec L}$ as $x_i\rightarrow x_i+x_{0, i}\ {\rm (mod\ }L_i)$. Then the action of $\tau_{\vec L, x}$ on any operator $A$ is denoted by $\tau_{\vec L, x}(A):=\tau_{\vec L, x}^{-1}A\tau_{\vec L, x}$, and translation invariance of the Hamiltonian $H_{\vec L}$ means that $\tau_{\vec L, x}\left(H_{\vec L}\right)=\tau_{\vec L, x}^{-1}H_{\vec L}\tau_{\vec L, x}=H_{\vec L}$ for any $x\in\mathbb{T}_{\vec L}$.

For a given finite-dimensional Hermitian operator $H$, let
$\lambda_1(H)\le\lambda_2(H)\le\cdots$ be its eigenvalues that may be degenerate. In this supplemental material, we call $\gap(H)=\lambda_2(H)-\lambda_1(H)$ the gap of $H$. Note that this notion of gap is different from the one introduced in the main text, which is the gap of a {\it sequence} of Hamiltonians in systems with increasing sizes. For the purpose of this supplemental material, this notion of gap of an {\it individual} Hamiltonian $H$ is sufficient, and we can deduce the properties of the gap in the main text using this notion. 

We demand that our Hamiltonians should be local, which requires us to have a notion of distance on the periodic lattices and a way to compare systems with different sizes. To measure the distance between regions on such a lattice, we use the graph distance on the torus. Namely, for any integer representatives
of $x,y\in\T_{\vec L}$, their distance is
\[
 d_{\vec L}(x,y)
 =\sum_{i=1}^d\min_{n_i\in\Z}|x_i-y_i+n_iL_i|.
\]
For a non-empty subset $X\subseteq\T_{\vec L}$, we use $\diam(X)=\max_{x,y\in X}d_{\vec L}(x,y)$ to represent its ``diameter'', and
$d_{\vec L}(y,X)=\min_{x\in X}d_{\vec L}(y,x)$ to represent the distance between $y$ and $X$. Later, we also use $B_{\vec L}(x,r)=\{y:d_{\vec L}(x,y)\le r\}$
to represent a disk-like or ball-like region centered around $x$ with a ``radius'' $r$.

As emphasized in the main text and above, just having a notion of distance is not enough to define our locality condition of the Hamiltonians, and it is also important to clarify how systems with different sizes are compared. The precise relation between these systems also gives a notion of thermodynamic limit of the Hamiltonians in these systems. To compare systems with different sizes, for the system with linear sizes $\vec L$, we
consider its centered representation around the origin of the coordinates:
\begin{equation}\label{eq:origin_coordinate}
 \Lambda_{\vec L}
 =\prod_{i=1}^d\{-\lfloor L_i/2\rfloor,\ldots,\lceil L_i/2\rceil-1\}
 \subset\Z^d.
\end{equation}
We momentarily view $\Lambda_{\vec L}$ as a finite lattice with open boundary conditions. Then each $\Lambda_{\vec L_1}$, including its lattice sites and the associated operator algebras, can be identified as a subsystem of $\Lambda_{\vec L_2}$, if $L_{1i}\leqslant L_{2i}$ for all $i\in\{1, 2, \cdots, d\}$. Finally, we impose periodic boundary conditions on $\Lambda_{\vec L}$. Physically speaking, increasing $\vec L$ (as in the main text, increasing $\vec L$
means increasing all $L_i$ with approximately fixed aspect ratios) fixes the bulk of $\Lambda_{\vec L}$ but pushes its boundaries further and further to infinity. So from the perspective of an observer who only has access to observables with certain spatial extent, as long as the system size is larger than this spatial extent, the physics is expected to be approximately the same as the physics in the thermodynamic limit.

With the above notions, our assumption of locality on the Hamiltonians consists
of two requirements:
\begin{enumerate}[label=(\arabic*),ref=\arabic*]
 \item \textbf{Finite range.}\label{ass:locality-range}
 The decomposition Eq. \eqref{eq:general-interaction} can be chosen so that
 $\Phi_{\vec L}(X)=0$ whenever $\diam(X)>R_0$, with a common bound
 $R_0$ independent of $\vec L$ and $X$. For fermionic systems, each $\Phi_{\vec L}(X)$ should further be even under total fermion parity{\footnote{This condition is included in the ``finite-range'' property of interactions in fermionic systems because violating it leads to a violation of Lieb-Robinson bound, just like having long-range interactions.}}.

 \item \textbf{System-size independent local dynamics.}\label{ass:locality-dynamics}
 For every fixed $r$,
 there is a size threshold $\vec L_{0}(r)$, such that
 \begin{equation}\label{eq:local-dynamics-compatible}
  [H_{\vec L'},A]=[H_{\vec L},A],
  \quad
  \forall A\in\A_{B_{\vec L}(0, r)},
 \end{equation}
 as long as $L'_i\geqslant L_i\geqslant L_{0i}(r)$ for all $i\in\{1, 2, \cdots, d\}$. In the above, $0$ stands for the origin of the coordinates. Eq. \eqref{eq:local-dynamics-compatible} uses the
 identifications of local operator algebras described above, with the operators on both sides embedded in a common larger
 region whenever necessary. For fermionic systems, the operator $A$ above includes both even and odd operators under total fermion parity.
\end{enumerate}
The first requirement, i.e., (\ref{ass:locality-range}), bounds the spatial extents of the interactions.
The second requirement, i.e., (\ref{ass:locality-dynamics}), ensures that enlarging the system eventually leaves the generator of
Heisenberg dynamics of every fixed local operator unchanged, so that the family of Hamiltonians of systems with increasing sizes has a well defined thermodynamic limit, which is precisely characterized by the commutators in Eq. \eqref{eq:local-dynamics-compatible} for sufficiently large $\vec L$.

\subsection{Consequences of locality of Hamiltonians}\label{app: locality}\label{subsec:locality}

We now show that the requirement~(\ref{ass:locality-dynamics}) allows the
interaction terms in Eq. \eqref{eq:general-interaction} in any fixed disk-like region around the origin to be chosen in a way that is independent of the
system size, up to an additive constant, as long as the system is sufficiently large. If translation
invariance is further assumed, the
requirement~(\ref{ass:locality-dynamics}) alone also implies that Eq. \eqref{eq:general-interaction} can be written in more restrictive form, as indicated in Eq. \eqref{eq: decomposing translation invariant Hamiltonian} below. Having these consequences in mind greatly simplifies the following discussion.

To start, we first note that, up to an additive constant operator, each $\Phi_{\vec L}(X)$ in Eq. \eqref{eq:general-interaction} can be decomposed as $\Phi_{\vec L}(X)=\Phi_{\vec L, 1}(X_1)+\Phi_{\vec L, 2}(X_2)$, such that $\Phi_{\vec L, 1}(X_1)$ is supported on $X_1\subseteq B_{\vec L}(0, r)$ while $\Phi_{\vec L, 2}(X_2)$ is on $X_2\nsubseteq B_{\vec L}(0,r)$, and both $\Phi_{\vec L, 1}(X_1)$ and $\Phi_{\vec L, 2}(X_2)$ are finite-range operators (in fermionic systems, they are also even under total fermion parity). Specifically, to do this decomposition for a spin system, we first take an orthonormal basis for operators at each site that is made of identity operator and traceless operators, and then we tensor product these on-site basis operators to get a set of many-body basis operators. For example, if all spins are qubits, then these basis operators are Pauli strings. To do this decomposition for a fermionic system, we similarly take this many-body basis as the product of Majorana operators of each fermionic mode. Then Eq. \eqref{eq:general-interaction} can be written as $H_{\vec L}=\sum_{\varnothing\neq X_1\subseteq B_{\vec L}(0, r)}\Phi_{\vec L, 1}(X_1)+\sum_{\varnothing\neq X_2\nsubseteq B_{\vec L}(0, r)}\Phi_{\vec L, 2}(X_2)$, up to an additive constant operator that has no effect on our discussion. Note that every $\Phi_{\vec L,2}(X_2)$ has a traceless factor outside $B_{\vec L}(0,r)$.

Next, we note that the requirement (\ref{ass:locality-dynamics}) means that $[H_{\vec L'}-H_{\vec L}, A]=[\sum_{\varnothing\neq X_1\subseteq B_{\vec L}(0, r)}(\Phi_{\vec L', 1}(X_1)-\Phi_{\vec L, 1}(X_1)), A]+[\sum_{\varnothing\neq X_2\nsubseteq B_{\vec L}(0, r)}(\Phi_{\vec L', 2}(X_2)-\Phi_{\vec L, 2}(X_2)), A]=0$ for any $A\in\mathcal{A}_{B_{\vec L}(0, r)}$, as long as $L'_i\geqslant L_i\geqslant L_{0i}(r)$ for all $i$. So when $L'_i\geqslant L_i\geqslant L_{0i}(r)$ for all $i$, $[\sum_{\varnothing\neq X_1\subseteq B_{\vec L}(0, r)}(\Phi_{\vec L', 1}(X_1)-\Phi_{\vec L, 1}(X_1)), A]=[\sum_{\varnothing\neq X_2\nsubseteq B_{\vec L}(0, r)}(\Phi_{\vec L', 2}(X_2)-\Phi_{\vec L, 2}(X_2)), A]=0$ for all $A\in\mathcal{A}_{B_{\vec L}(0, r)}$, which implies that $\sum_{\varnothing\neq X_1\subseteq B_{\vec L}(0, r)}(\Phi_{\vec L', 1}(X_1)-\Phi_{\vec L, 1}(X_1))$ can only be a constant.

Therefore, the interaction terms in Eq. \eqref{eq:general-interaction} in any fixed disk-like region with radius $r$ around the origin can be chosen in a way that is independent of the system size $\vec L$, up to an additive constant, as long as the system is sufficiently large, i.e., $L_i\geqslant L_{0i}(r)$ for all $i\in\{1, 2, \cdots, d\}$.

In general, even under the requirement (\ref{ass:locality-dynamics}), the range of the interaction terms in Eq. \eqref{eq:general-interaction} can increase as $r$ or $\vec L$ increases. However, if there is translation symmetry, all interaction terms in Eq. \eqref{eq:general-interaction} can always be taken to be some finite-range interactions that are independent of the system size, as long as the system size is large enough. Namely, there is a Hermitian operator $\Phi_0$ and a linear size $\vec L_0$, both of which are independent of the system size $\vec L$, such that
\begin{equation} \label{eq: decomposing translation invariant Hamiltonian}
    H_{\vec L}=c_{\vec L}+\sum_{x\in\mathbb{T}_{\vec L}}\tau_{\vec L, x}\left(\Phi_0\right),
\end{equation}
as long as $L_i\geqslant L_{0 i}$ for all $i\in\{1, 2, \cdots, d\}$, where $c_{\vec L}$ is a c-number. Here $\Phi_0$ can be viewed as the Hamiltonian density, and translating it leads to the total Hamiltonian.

To derive the above result, consider the case with $r=0$ in the above discussion, so that $B_{\vec L}(0, r)$ contains only the origin. Then by previous analysis, $\sum_{ X_2: \{\vec 0\}\subsetneq X_2} [(\Phi_{\vec L',2}(X_2) - \Phi_{\vec L,2}(X_2) ), A_0] = 0$ for $A_0\in \A_{\{\vec 0\} }$. 
Denote by $\Phi_0' = \Phi_{\vec L,1}(\{\vec 0\}) + \sum_{X_2: \{\vec 0\}\subsetneq X_2} \Phi_{\vec L, 2}(X_2)$. The above discussion actually shows that, up to an additive constant, $\Phi_0'$ is independent of the system size \footnote{Here we use the fact that the decomposition into the orthonormal operator basis in last paragraph is unique.}, as long as $L_i\geqslant L_{0i}(0)$. So the range of $\Phi_0'$ is bounded by $\sum_{i}(L_{0i}(0)-1)$, which is independent of the system size $\vec L$. Translation symmetry implies that this same upper bound also applies to the range of the interactions in the Hamiltonian that have a support which contains {\it any} site. Therefore, all interaction terms in the Hamiltonian have a range with this upper bound, and the Hamiltonian satisfies the finite-range requirement in (\ref{ass:locality-range}).

To further see that the Hamiltonian can be written as Eq. \eqref{eq: decomposing translation invariant Hamiltonian}, we need to construct $\Phi_0$ from $\Phi_0'$. Since the support of $\Phi_0'$ is independent of the system size when $L_i\geqslant L_{0i}(0)$, for all operators supported in this region we can find a fixed complete  orthonormal basis of operators with respect to the Hilbert-Schmidt inner product. As discussed before, for example, for qubits this basis can be taken to be the Pauli strings, and for fermions this basis can be taken to be the products of Majorana operators. Then $\Phi_0'$ can be decomposed in this basis as $\Phi_0'=\sum_\alpha c_\alpha O_\alpha$, where $\alpha$ indexes the operator basis, $O_\alpha$ is a basis operator that acts nontrivially on the site $\vec 0$ and $c_\alpha$ is its expansion coefficient. Taking $\Phi_0=\sum_\alpha\frac{c_\alpha O_\alpha}{|{\rm supp}(O_\alpha)|}$, one can check that Eq. \eqref{eq: decomposing translation invariant Hamiltonian} holds.

In passing, we note that for the consequences discussed above to hold in fermionic systems, it is important that the operators $A$ in Eq. \eqref{eq:local-dynamics-compatible} include not only fermion-parity-even but also fermion-parity-odd operators. If only fermion-parity-even operators are included there, even if there is translation symmetry, the Hamiltonian could contain a term proportional to the total fermion parity and still satisfy the requirement (\ref{ass:locality-dynamics}). But such a Hamiltonian does not satisfy the finite-range condition in (\ref{ass:locality-range}).

\subsection{Summary of main results}\label{subsec:summary-main}

In this subsection, we summarize the main results in this supplemental material. This subsection also serves as the outline of the rest of this supplemental material.

First, in Sec. \ref{sec: another 1D example} we present an example of gapped local translation non-invariant Hamiltonians with a unique ground state, which can become gapless under certain arbitrarily weak local perturbation. In particular, in this example the Hamiltonians 1) have a unique ground state in both the finite-size and infinite-size settings, and 2) become gapless under certain arbitrarily weak local translation invariant perturbation that does not change the ground state (i.e., the unperturbed ground state is still the unique ground state under the perturbation, in both the finite-size and infinite-size settings).

Next, from Sec. \ref{sec: proof of main theorem} to Sec. \ref{sec:fermions}, we prove the theorem of the stability of gap of one dimensional local translation invariant Hamiltonians, which is Theorem 3 in the main text and will be reviewed now.

We restrict the setup to $d=1$ and impose translation
invariance on the unperturbed Hamiltonian. Our lattice now is a ring $\T_L=\Z/(L\Z)$, and it is convenient
to label the sites by $0,\cdots,L-1$. We call the bond between sites $L-1$ and 0 the {\it closing seam}.

By Sec. \ref{app: locality}, a local translation invariant Hamiltonian can be written as in Eq. \eqref{eq: decomposing translation invariant Hamiltonian}. Suppose that ${\rm diam}({\rm supp}(\Phi_0))=R_0$, in the following we write the Hamiltonian as
\begin{equation} \label{eq: unperturbed Hamiltonian family}
    H_L=\sum_{x\in\mathbb{T}_L}h_{x, L},
\end{equation}
where $L>2R_0$, $h_{x, L}=\tau_{x, L}(\Phi_0)=\tau_{x, L}^{-1}\Phi_0\tau_{x, L}$, and the constant $c_L$ has been dropped because it has no effect on our discussion. These Hamiltonians will be our unperturbed Hamiltonians.

We assume that the unperturbed Hamiltonians satisfy the following {\it uniform gap condition with a unique ground state}. Namely, there are $L_0\ge2R_0$ and $\gamma_0>0$ such that, for every
$L>L_0$, the ground space of $H_L$ is one dimensional and
\begin{equation}\label{eq:assumption}
 H_L-E_L\one\ge\gamma_0(\one-P_L),
\end{equation}
where $E_L=\lambda_1(H_L)$, $P_L=\ket{\psi_L}\bra{\psi_L}$ with $\ket{\psi_L}$ being the normalized ground state of $H_L$. 

We then consider perturbing the system with a local perturbation
\begin{equation}
    V_L=\sum_{\varnothing\neq X\subseteq\mathbb{T}_L}\Xi_L(X),
\end{equation}
which also satisfies the requirements (\ref{ass:locality-range}) and (\ref{ass:locality-dynamics}). However, the perturbation may not be translation invariant. We define the range of interaction $\Xi_L$ to be the minimal integer $R$ such that $\Xi_L(X)=0$ for any $X$ with $\diam(X)>R$, denoted as $\operatorname{range}(\Xi_L)$. For instance, a nearest-neighbor interaction has range one.

Within this setup, we have the following theorem on the stability of gap.

\begin{theorem}[Stability of gap]\label{thm:main}
Let $\{H_L\}$ be the unperturbed Hamiltonians in Eq. \eqref{eq: unperturbed Hamiltonian family} that satisfy
the uniform gap condition with a unique ground state as in Eq. \eqref{eq:assumption}.
For every fixed perturbation range $R$, there are
$L_*\ge L_0$ and $\varepsilon_R,\gamma_R>0$ such that, for every $L$
and every perturbing interaction $\Xi_L$ satisfying
\[
 L>L_*,
 \quad\operatorname{range}(\Xi_L)\leq R,
 \quad \|\Xi_L\|_{\inc}<\varepsilon_R,
\]
the perturbed Hamiltonian $H_L+V_L$ still has a unique ground state and obeys
\[
 \gap(H_L+V_L)\ge\gamma_R.
\]
The constants $L_*$, $\varepsilon_R$ and $\gamma_R$ can depend on $L_0$, $\gamma_0$, the on-site Hilbert space dimension $D$, the perturbation range $R$ and the unperturbed Hamiltonian, but do not depend on the system size $L$.
\end{theorem}

In fact, we can prove the following stronger corollary that shows that the gap of the perturbed Hamiltonian is a continuous function of the strength of the perturbation.

\begin{corollary}[Continuity of gap] \label{corollary: continuity of gap}
Under the assumptions of Theorem~\ref{thm:main}, for any fixed perturbation range $R$, there are constants $L_*,\eta_R,c_R>0$, such that for every $L>L_*$ and every interaction $\Xi_L$ satisfying $\operatorname{range}(\Xi_L)\le R$ and
$\|\Xi_L\|_{\inc}<\eta_R$,
\begin{equation}
 \bigl|\gap(H_L+V_L)-\gap(H_L)\bigr|
 \le c_R\|\Xi_L\|_{\inc},
 \label{eq:gap-continuity}
\end{equation}
with $V_L=\sum_X\Xi_L(X)$. The constants $L_*$, $\eta_R$ and $c_R$ are independent of $L$.
\end{corollary}

Theorem \ref{thm:main} and Corollary \ref{corollary: continuity of gap} together correspond to Theorem 3 in the main text, while the statements here are more complete and precise. To get the proof of these two results, we will establish an intermediate result, which is the more precise version of Theorem 2 in the main text, about the invertibility of the unique ground state of gapped local Hamiltonians in one dimension. This intermediate result is summarized in Sec. \ref{subsec:prep-lga}.

After proving Theorem \ref{thm:main} and Corollary \ref{corollary: continuity of gap}, in Sec. \ref{sec:2d-instability} we discuss more details on the 2D example in the main text, which shows that the direct generalization of Theorem \ref{thm:main} is false in any dimension larger than 1. This 2D example is then further generalized in Sec. \ref{sec: 1-form} to give an example of lattice system with a lattice version of 1-form symmetry that does not behave as a topological 1-form symmetry in quantum field theories.

In Sec. \ref{sec: stabilization}, we prove the results for the stabilization of the energy gap. Specifically, we have Theorem \ref{thm:gap-stabilization}, Corollary \ref{cor:stab-compatible} and Theorem \ref{thm:stabilization-product-higher-dim-MZ}. The full statements of these results are relatively complicated, so they will only be presented in Sec. \ref{sec: stabilization}. Theorem 1 in the main text corresponds to a relatively informal version of the combination of Theorem \ref{thm:gap-stabilization} and Corollary \ref{cor:stab-compatible}, while Theorem \ref{thm:stabilization-product-higher-dim-MZ} can be used to show that the 2D example discussed in Sec. \ref{sec:2d-instability} requires fine tuning.

Finally, we collect some general lemmas in Sec. \ref{app:technical} that are not restricted to our setup and are used in our proof.

\section{Another 1D example with an unstable gap} \label{sec: another 1D example}

In the main text, we present a 1D example of gapped local translation non-invariant Hamiltonians with a unique ground state, which can become gapless under certain arbitrarily weak local perturbation. In that example, the ground state changes under the perturbation. Moreover, the infinite-size version of that example, as we will discuss, has more than one ground state. In this section, we present another 1D example of gapped local translation non-invariant Hamiltonians of qutrits that 1) have a unique ground state in both the finite-size and infinite-size settings, and 2) become gapless under certain arbitrarily weak local translation invariant perturbation that does not change the ground state (i.e., the unperturbed ground state is still the unique ground state under the perturbation, in both the finite-size and infinite-size settings). Specifically, we introduce the setup, the unperturbed Hamiltonians and the perturbations in Sec. \ref{subsec:1d-block-example}, discuss the unperturbed and perturbed spectra in Sec. \ref{subapp: spectra of 1D example}, and extend our discussion to the infinite-size versions of this example in Sec. \ref{subsec:1d-infinite-ground-states}.

\subsection{Setup and Hamiltonians}\label{subsec:1d-block-example}

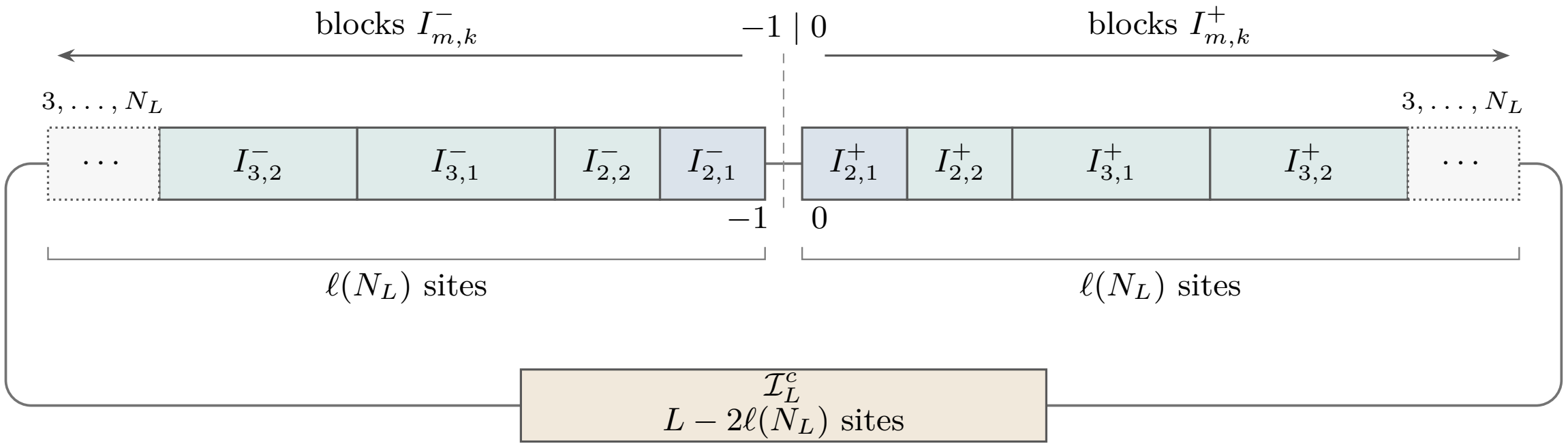
\begin{figure}[htbp]
\centering
\definecolor{blockStageOne}{RGB}{54,98,146}
\definecolor{blockStageTwo}{RGB}{42,125,116}
\definecolor{blockRemainder}{RGB}{171,123,49}
\begin{tikzpicture}[x=1cm,y=1cm,font=\small,
  block/.style={draw=black!65,line width=.6pt},
  geometry/.style={draw=black!55,line width=.7pt},
  direction/.style={-{Stealth[length=1.7mm]},draw=black!65,line width=.6pt},
  every node/.style={inner sep=2pt}]
  \draw[geometry] (-.18,.35)--(.18,.35);
  \draw[densely dashed,black!45] (0,-.08)--(0,1.42);
  \node at (0,1.66) {$-1\mid0$};
  \node[below] at (-.35,0) {$-1$};
  \node[below] at (.35,0) {$0$};

  \foreach \side/\sign in {-1/-,1/+}{
    \begin{scope}[xscale=\side]
      \draw[block,fill=blockStageOne!18] (.18,0) rectangle (1.20,.70);
      \node at (.69,.35) {$I_{2,1}^{\sign}$};
      \draw[block,fill=blockStageTwo!15] (1.20,0) rectangle (2.22,.70);
      \node at (1.71,.35) {$I_{2,2}^{\sign}$};
      \draw[block,fill=blockStageTwo!15] (2.22,0) rectangle (4.14,.70);
      \node at (3.18,.35) {$I_{3,1}^{\sign}$};
      \draw[block,fill=blockStageTwo!15] (4.14,0) rectangle (6.06,.70);
      \node at (5.10,.35) {$I_{3,2}^{\sign}$};
      \draw[block,densely dotted,fill=black!3] (6.06,0) rectangle (7.14,.70);
      \node at (6.60,.35) {$\cdots$};
      \node[font=\scriptsize] at (6.60,.94) {$3,\ldots,N_L$};
      \draw[direction] (.40,1.40)--(7.05,1.40);
      \node at (3.75,1.68) {blocks $I_{m,k}^{\sign}$};
      \draw[black!50,line width=.45pt]
        (.18,-.46)--(.18,-.58)--(7.14,-.58)--(7.14,-.46);
      \node at (3.66,-.85) {$\ell(N_L)$ sites};
      \draw[geometry,rounded corners=7pt]
        (7.14,.35)--(7.55,.35)--(7.55,-2.00)--(2.55,-2.00);
    \end{scope}
  }
  \draw[block,fill=blockRemainder!17] (-2.55,-2.35) rectangle (2.55,-1.65);
  \node[align=center] at (0,-2.00)
    {$\mathcal I_L^c$\\[-1pt]$L-2\ell(N_L)$ sites};
\end{tikzpicture}
\caption{Arrangement of the blocks. Arrows point away from the bond between sites 0 and $-1$. Dotted boxes collect remaining blocks $I^{\pm}_{m, k}$ with $2\leqslant m\leqslant N_L+1$ and $1\leqslant k\leqslant N_L$. The remaining
sites form $\mathcal I_L^c$ on the opposite arc of the ring, which is empty when
$L=2\ell(N_L)$.}
\label{fig:mirrored-block-distribution}
\end{figure}

For our setup, we again consider qutrits on a ring with sites $0,\ldots,L-1$, with all coordinates defined modulo $L$ (see Fig.~\ref{fig:mirrored-block-distribution}). The orthonormal basis states of the qutrit at site $x$ are still denoted by $\{\ket{0_x}, \ket{\uparrow_x}, \ket{\downarrow_x}\}$, and $\ket{0_x}$ and $\{\ket{\uparrow_x}, \ket{\downarrow_x}\}$ are still viewed as an on-site vacuum and spin-1/2 defect, respectively.

We will define local Hamiltonians on this system, which has a well defined thermodynamic limit to be discussed in Sec. \ref{subsec:1d-infinite-ground-states}. In order to get this thermodynamic limit, we first partition the system as in Fig. \ref{fig:mirrored-block-distribution}, with the following rules of partition.

We denote by $I_{m, k}^\pm$ a block made up of $m^2$ contiguous sites, where $m\geqslant 2$ and $k\geqslant 1$. Starting from site 0 and going to the right direction, we order the blocks as the following sequence and let the sites belong to one of these blocks:
\begin{equation} \label{eq: partition using blocks}
 \begin{aligned}
  &{\color{blue}I_{2,1}^+};\
  {\color{red}I_{2,2}^+},\ {\color{blue}I_{3,1}^+,\ I_{3,2}^+};\ 
  {\color{red}I_{2,3}^+,\ I_{3,3}^+},\ {\color{blue}I_{4,1}^+,\ I_{4,2}^+,\ I_{4,3}^+};
  \ \cdots
  \ 
  {\color{red} I_{2,k}^+,\ I_{3,k}^+,\ \cdots,\ I_{k,k}^+},\ {\color{blue} I_{k+1,1}^+,\ I_{k+1,2}^+,\ \cdots,\ I_{k+1,k}^+;} \ \cdots
 \end{aligned}
\end{equation}
In the above, starting from $k=2$, the blocks with red color are of the form $I_{2,k}^+,\ I_{3,k}^+,\ \cdots,\ I_{k,k}^+$, blocks with blue color are of the form $I_{k+1,1}^+,\ I_{k+1,2}^+,\ \cdots,\ I_{k+1,k}^+$, $k$ keeps increasing, and these two types of blocks appear in alternating order. For example, concretely,
\[
 I_{2,1}^+=\{0,\cdots,3\},\quad I_{2,2}^+=\{4,\cdots,7\},\quad
 I_{3,1}^+=\{8,\cdots,16\},\quad I_{3,2}^+=\{17,\cdots,25\},\quad
 I^+_{2,3}=\{26, \cdots, 29\}.
\]

On the left side of site 0, blocks are arranged so that the left and right sides of the bond between sites 0 and $-1$ are mirror symmetric. Concretely, if $I_{m, k}^+=\{x_{m, k, 0}+1, x_{m, k, 0}+2, \cdots, x_{m, k, 0}+m^2\}$ for some $x_{m, k, 0}\in\mathbb{Z}$, then $I_{m, k}^-=\{-1-(x_{m, k, 0}+m^2), -1-(x_{m, k, 0}+m^2)+1, \cdots, -1-x_{m, k, 0}-1\}$.

Our length-$L$ ring may not accommodate exactly these blocks, and there can be some remaining sites. We decide which sites should be included in the blocks and which sites remain as follows. For each side of the bond between sites -1 and 0, suppose that blocks with $2\le m\le N+1$ and $1\le k\le N$ are accommodated in the ring, then the number of sites covered by these blocks in each side is
\begin{equation}
 \ell(N)=N\sum_{m=2}^{N+1}m^2
 =\frac{N^2(2N^2+9N+13)}6,
\end{equation}
and the total number of sites included in the blocks on both sides is $2\ell(N)$. For sufficiently large $L$, we define
\begin{equation} \label{eq: number of blocks}
 N_L=\max\{N\ge2:2\ell(N)\le L\},
\end{equation}
and demand that blocks with $2\le m\le N_L+1$ and $1\le k\le N_L$ are accommodated on each side. In coordinates $0,\cdots,L-1$, the union of these blocks is
$\{0,\ldots,\ell(N_L)-1\}\cup\{L-\ell(N_L),\ldots,L-1\}$.
The remaining sites form
$\mathcal I_L^c=\{\ell(N_L),\ldots,L-\ell(N_L)-1\}$, which is empty if $L=2\ell(N_L)$. 

The above partition of the system into blocks and remaining sites ensures that the neighborhood of site 0 has a fixed pattern of blocks for sufficiently large $L$, based on which we can now define the local Hamiltonians that satisfy Eq. \eqref{eq:local-dynamics-compatible}, which gives a thermodynamic limit.

Our unperturbed Hamiltonian, written in a dimensionless form, is
\begin{equation} \label{eq: 1d unperturbed}
    H_L=\sum_{m=2}^{N_L+1}\sum_{k=1}^{N_L}(H_{m, k}^++H_{m, k}^-)+6\sum_{x\in\mathcal{I}_L^c}P_x,
\end{equation}
where $P_x=1-\ket{0_x}\bra{0_x}$, and $H_{m, k}^+$ and $H_{m, k}^-$ are supported on $I^+_{m, k}$ and $I^-_{m, k}$, respectively, and they are related to each other by reflection with respect to the bond between sites -1 and 0. More specifically, suppose that $I_{m, k}^+=\{x_{m, k, 0}+1, x_{m, k, 0}+2, \cdots, x_{m, k, 0}+m^2\}$ for some $x_{m, k, 0}\in\mathbb{Z}$, then
\begin{equation} \label{eq: 1d block Hamiltonian}
H_{m, k}^+
=
P_{x_{m, k, 0}+1} + P_{x_{m, k, 0}+m^2}+\sum_{x=x_{m, k, 0}+1}^{x_{m, k, 0}+m^2-1}(P_x-P_{x+1})^2+\sum_{x=x_{m, k, 0}+1}^{x_{m, k, 0}+m^2}\left(4\hat P_x+\frac{P_x}{km^4}\right),
\end{equation}
with
\begin{equation}\label{eq:preferred_projection}
 \hat P_x=\begin{cases}
 \ket{\downarrow_x}\bra{\downarrow_x},&x\in A_{m, k}^+,\\
 \ket{\uparrow_x}\bra{\uparrow_x},&x\notin A_{m, k}^+,
 \end{cases}
\end{equation}
where $A_{m, k}^+=\{x_{m, k, 0}+1+u(m+1): u\in\mathbb Z,0\leqslant u\leqslant m-1\}$. The sites in $A_{m, k}^+$ are colored in red in Fig. \ref{fig: segments of block}.

The perturbation, which is translation invariant and on-site, is
\begin{equation} \label{eq: 1d perturbation}
V_{L, n}=\sum_xv_{x, n},
\quad
v_{x, n}=\frac{4}{n+1}\left(-\ket{\uparrow_x}\bra{\uparrow_x}+\frac{1}{2n}\ket{\downarrow_x}\bra{\downarrow_x}\right),
\end{equation}
where $n$ is an arbitrary integer that is at least 2. The strength of this perturbation, which is determined by the incidence norm in Eq. \eqref{eq:general-incidence}, is $4/(n+1)$ and can be arbitrarily small by making $n$ arbitrarily large.

\subsection{Spectra} \label{subapp: spectra of 1D example}

Now we analyze the spectra of the unperturbed Hamiltonian $H_L$ in Eq. \eqref{eq: 1d unperturbed} and the perturbed Hamiltonian $H_L+V_{L, n}$, with $V_{L, n}$ given by Eq. \eqref{eq: 1d perturbation}. For $\{H_L\}$, we will show that 1) it is gapped and has a unique ground state $\otimes_x\ket{0_x}$, and 2) its gap $\Delta\geqslant 2$. For $\{H_L+V_{L, n}\}$, we will show that it is gapless and $\otimes_x\ket{0_x}$ is still its unique ground state for sufficiently large $L$. This analysis can be carried out largely because both $H_L$ and $H_L+V_{L, n}$ are classical Hamiltonians, and all eigenstates can just be specified by the configuration of vacuum and defect at each site.

We start with $H_L$. 

Because all blocks and also $\mathcal{I}^c_L$ are decoupled from each other, it suffices to consider the blocks and $\mathcal{I}^c_L$ separately. The Hamiltonian for $\mathcal{I}^c_L$ (if nonempty) can be easily solved, and it has a unique ground state $\otimes_{x\in\mathcal{I}^c_L}\ket{0_x}$ and a gap 6. For the blocks, without loss of generality, we consider the Hamiltonian $H_{m, k}^+$ in Eq. \eqref{eq: 1d block Hamiltonian}, which consists of non-negative operators and clearly $\otimes_{x\in I_{m, k}^+}\ket{0_x}$ is a unique ground state and the Hamiltonian is gapped. This discussion establishes that $\otimes_x\ket{0_x}$ is the unique ground state of $H_L$.

\begin{figure}[h!]
    \centering
    \includegraphics[width=0.4\linewidth]{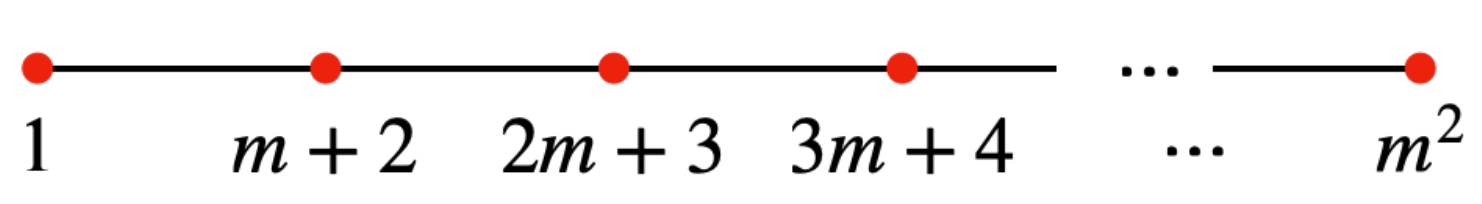}
    \caption{In the block $I_{m, k}^+$, $A_{m, k}^+$ are represented by the $m$ red points, and on these points $\ket{\downarrow_x}$ has a higher energy compared to $\ket{\uparrow_x}$. On the other points, i.e., points on $I^+_{m, k}\backslash A_{m, k}^+$, $\ket{\uparrow_x}$ has a higher energy compared to $\ket{\downarrow_x}$.}
    \label{fig: segments of block}
\end{figure}

To understand the excitation spectrum of $H_L$, we first understand the excitation spectrum of $H^+_{m, k}$ and then consider the excitations on $\mathcal{I}^c_L$. If a site $x\in I^+_{m, k}$ is to host a defect, as shown in Fig. \ref{fig: segments of block}, the term $4\hat P_x$ penalizes $\ket{\downarrow_x}$ and $\ket{\uparrow_x}$ in $A_{m, k}^+$ and $I_{m, k}^+\backslash A^+_{m, k}$, respectively. We call these penalized defects the ``wrong'' defects in a configuration of defects in this block. Then the energy of a configuration $F$ of defects in the block $I_{m, k}^+$ is
\begin{equation} \label{eq: block excitation energy}
E_{m, k}(F)=2c(F)+\frac{|F|}{km^4}+4w(F)>2,
\end{equation}
as long as there is any defect, where $c(F)$ is the total number of disjoint intervals that host defects on contiguous sites, $|F|$ is the total number of defects in the configuration $F$, and $w(F)$ is the total number of wrong defects in the configuration $F$. Note that the term $2c(F)$ includes the situation where one of these intervals contains an end point of $I^+_{m, k}$, and this simple expression results from the fact that the coefficients of the first three terms in Eq. \eqref{eq: 1d block Hamiltonian} are the same. 

On the other hand, any excitation on $\mathcal{I}^c_L$ costs energy at least 6, so Eq. \eqref{eq: block excitation energy} shows that the gap $\Delta$ of $\{H_L\}$ satisfies $\Delta\geqslant 2$.

Next, we turn to the spectrum of $H_L+V_{L, n}$. We will consider sufficiently large system sizes $L$ such that $N_L\geqslant n+2$, with $N_L$ given by Eq. \eqref{eq: number of blocks}.

On $\mathcal{I}^c_L$, any defect will now cost energy at least $6-\frac{4}{n+1}\geqslant 14/3$, so in the ground state the sites of $\mathcal{I}^c_L$ should all be in the vacuum state. On the blocks, for a configuration $F$ in the block $I^+_{m, k}$ discussed above, the perturbation $V_{L, n}$ changes its energy from Eq. \eqref{eq: block excitation energy} to
\begin{equation} \label{eq: perturbed block excitation energy}
    E'_{m, k}=2c(F)+\frac{|F|}{km^4}+4w(F)+\frac{4}{n+1}\left(-N^\uparrow+\frac{N^\downarrow}{2n}\right)
    =\frac{|F|}{km^4}+4w(F)+\sum_i\left[2+\frac{4}{n+1}\left(-N^\uparrow_i+\frac{N_i^\downarrow}{2n}\right)\right],
\end{equation}
where $N^\uparrow$ and $N^\downarrow$ are the total numbers of spin-up and spin-down defects in this configuration, respectively, the sum in the second equation is over all disjoint intervals with defects in the configuration $F$, and $N^{\uparrow,\downarrow}_i$ is the number of spin-up and spin-down defects in the $i$-th interval, respectively.

Eq. \eqref{eq: perturbed block excitation energy} shows that the unique ground state of $H_L+V_{L, n}$ is still $\otimes_x\ket{0_x}$, since $E'_{m, k}\geqslant 0$ and $E'_{m, k}=0$ only if no defect is present. To see it, we consider two cases separately. First, we consider the situation where $w(F)=0$, i.e., the defects on $A_{m, k}^+$ and $I^+_{m, k}\backslash A^+_{m, k}$ are spin-up and spin-down, respectively. In this case, we just need to show that $2+\frac{4}{n+1}\left(-N^\uparrow_i+\frac{N_i^\downarrow}{2n}\right)\geqslant 0$ for each disjoint interval $i$ with contiguous sites hosting defects. If $N_i^\uparrow=0$, this result clearly holds. If $0< N_i^\uparrow\leqslant m$, note that  $N_i^\downarrow\geqslant m(N_i^\uparrow-1)$, so $2+\frac{4}{n+1}\left(-N^\uparrow_i+\frac{N_i^\downarrow}{2n}\right)\geqslant\frac{2}{n(n+1)}\left[n(n+1)+(m-2n)N_i^\uparrow-m\right]\geqslant\frac{2}{n(n+1)}\min\{n(n-1), (m-n)(m-n-1)\}\geqslant 0$, with equal sign only if $m=n$ or $m=n+1$, and that all sites in the entire block $I^+_{m, k}$ host defects. In this case,
\begin{equation}
    \min_{F\neq\emptyset}\{E'_{m, k}(F)\}=\frac{m^2}{km^4}=\frac{1}{km^2}, \quad m\in\{n,n+1\}.
\end{equation}
The other case occurs when $w(F)>0$, and such a configuration has an energy that is at least $4-\frac{4}{n+1}(1+\frac{1}{2n})>0$ more than its counterpart configuration, which is obtained from this configuration by flipping all spins of the defects so that $w(F)=0$. 

Therefore, we conclude that for both the blocks and the remaining segment $\mathcal{I}^c_L$, the energy is minimized only if no defect is present. For $m=n$ or $m=n+1$ and all $1\leqslant k\leqslant N_L$, in the block $I_{m, k}^+$ the energy difference between the first excited state and the ground state is $1/(km^2)$. For other blocks, this energy difference is even larger. Moreover, for sites on $\mathcal{I}^c_L$, this energy difference is at least $6-\frac{4}{n+1}>0$. So the unique ground state is still $\otimes_x|0_x\rangle$.

Moreover, now the spectrum is quasi-continuous, because we can consider a state where most sites are in the vacuum state, but blocks $I^+_{m=n+1, k}$ are in their first excited state if $k\in K$, where $K$ is some subset of $\{1, 2, \cdots, N_L\}$. According to the above analysis, the energy of this state is
\begin{equation}
E_K=\frac{1}{(n+1)^2}\sum_{k\in K}\frac{1}{k}.
\end{equation}
This is a quasi-continuous spectrum because 1) $k$ can be made arbitrarily large for large enough $L$, so $E_K$ can be arbitrarily close to 0, and 2) the series $\sum_{k=1}^{N_L} 1/k$ grows without upper bound as $N_L$ increases, so any positive energy can be approximately reached by $E_K$ for sufficiently large $L$.

In summary, we have shown that although the unperturbed Hamiltonians are gapped and have a unique ground state, the perturbation, which is translation invariant, on-site and can be made arbitrarily small, will make the system gapless while retaining the unperturbed ground state as the unique ground state.

What are the key ingredients leading to this example? First, this example is made of multiple decoupled segments, where these segments can be arbitrarily long. Then we choose some of these segments, i.e., segments with $m=n$ or $m=n+1$, and use a perturbation to lower its excitation energy, which is possible again because there are excited states with many defects but only a constant energy. To further make the spectrum quasi-continuous, each block has an extra $k$ index that can also be arbitrarily large. The blocks with different values of $k$ then have a quasi-continuous spectrum under the perturbation.

\subsection{Infinite-size extensions of the two 1D examples}
\label{subsec:1d-infinite-ground-states}

In this final subsection of this section, we extend the discussion of the 1D example in the main text and the 1D example introduced in this section to their infinite-size counterparts, mostly focusing on the unperturbed Hamiltonians. The main purpose for performing this analysis is to show that the example introduced in this section has a single ground state even in the infinite-size setting. On the other hand, both the 1D example and 2D example in the main text have more than one ground state in their infinite-size versions, with the 1D example discussed in this subsection and the 2D example discussed in Sec. \ref{subsec:2d-infinite-ground-states}.

Because the example introduced in this section only has $\otimes_x|0_x\rangle$ as a unique ground state in both the finite-size and infinite-size settings, in both settings there is an adiabatic path connecting this example with Hamiltonian $H$ with another Hamiltonian, $H^{(0)}=-\sum_x|0_x\rangle\langle 0_x|$, where the path is given by $(1-s)H+sH^{(0)}$ for $s\in[0, 1]$ (here $H$ and $H^{(0)}$ may stand for either their finite-size version or infinite-size version). So under the usual definition, these two Hamiltonians are regarded as in the same gapped quantum phase. However, as we have seen, the gap of $H$ is unstable, while the gap of $H^{(0)}$ is stable (this stability of the gap of $H^{(0)}$ can be seen by applying the result of Ref. \cite{BravyiHastingsMichalakis2010}). So this example shows that the stability of gap is not a universal property of a gapped quantum phase of local Hamiltonians.

The rest of this subsection is organized as follows. In Sec. \ref{subsubsec: review of infinite lattice}, we briefly review the setup of infinite-size lattice spin systems. In Sec. \ref{subsubsec: infinite version of 1D main text}, we show that the infinite-size version of the 1D example in the main text always contains a gapped ground state and infinitely many non-gapped ground states. In Sec. \ref{subsubsec: infinite version 1D this section}, we show that the infinite-size version of Eq. \eqref{eq: 1d unperturbed} has a unique ground state and it is gapped.

\subsubsection{Review of the setup of infinite-size lattice spin systems} \label{subsubsec: review of infinite lattice}

We first briefly review the setup of infinite-size lattice spin systems. More details can be found in, for example, Ref. \cite[Appendix A]{Liu2024} and the references therein.

We suppose that the spins live on the sites of a $d$-dimensional lattice $\Gamma=\mathbb Z^d$ ($d\in\mathbb Z^+$), such that the spin at site $x$ is described by a Hilbert space $\mathcal{H}_x$. For any finite subset $\Lambda\subset\Gamma$, let
$\mathcal A_\Lambda=\bigotimes_{x\in\Lambda}\mathcal B(\mathcal H_x)$ be the algebra of bounded operators supported on $\Lambda$,
with inclusions given by tensoring with identities. The quasi-local
algebra is the norm closure
\[
 \mathcal A_\Gamma=
 \overline{\bigcup_{\Lambda\subset\Gamma\text{ finite}}\mathcal A_\Lambda}^{\|\cdot\|},
\]
where $||\cdot||$ denotes the operator norm.

A state $\omega$ is a positive normalized linear functional
on this algebra, which specifies the expectation values of all operators in this algebra and thus the density matrices
on all finite regions.

A Hamiltonian of such an infinite-size system is specified by its
interaction $\Phi(X)=\Phi(X)^{\dagger}$, rather than by a sum of operators, which often does not converge. For the bounded finite-range interactions below, define
\begin{equation} \label{eq: derivation}
\delta_\Phi(A)=\sum_{X:\,X\cap\supp A\ne\varnothing}[\Phi(X),A]
 \qquad(A\text{ local}).
\end{equation}
This sum is finite, and $\delta_\Phi(A)$ specifies the Heisenberg dynamics of the operator $A$.

We call $\omega$ a ground state of $\Phi$ if
\begin{equation}
 \omega\bigl(A^{\dagger}\delta_\Phi(A)\bigr)\ge0
 \quad\text{for every local }A.
 \label{eq:infinite-ground-state-condition}
\end{equation}
For finite-size systems, this condition means that no local operation can lower the
energy of the state $\omega$, so it is a natural generalization of the notion of ground states to infinite-size systems \cite[Sec.~7.2]{NSYstability2022}. A Hamiltonian has a unique ground state if there is a single state $\omega$ that satisfies Eq. \eqref{eq:infinite-ground-state-condition}.

A ground state $\omega$ is \emph{gapped in the algebraic sense}
if there is a constant $\gamma>0$ such that
\begin{equation}
 \omega\bigl(A^{\dagger}\delta_\Phi(A)\bigr)
 \ge\gamma\bigl[\omega(A^{\dagger}A)-|\omega(A)|^2\bigr]
 \quad\text{for every local }A.
 \label{eq:infinite-algebraic-gap}
\end{equation}
The constant $\gamma$ should be independent of $A$. Equivalently, one may test a local operator $A$ with $\omega(A)=0$, so that the right-hand side is
$\gamma\omega(A^{\dagger}A)$~\cite[Sec.~7.4]{NSYstability2022}.
For finite-size systems, this condition means that every state orthogonal to the reference state $\omega$ must cost a
uniformly positive energy that is at least $\gamma$, and it is a natural generalization of the notion of gap to infinite-size systems. Note that, for infinite-size systems, this condition excludes any local operation that creates a component ``orthogonal'' to the reference state at zero energy, but it does not exclude the existence of other ground state that may or may not be gapped in the sense of Eq. \eqref{eq:infinite-algebraic-gap}, which cannot be obtained from $\omega$ by local operations.

\subsubsection{1D example in the main text} \label{subsubsec: infinite version of 1D main text}

Now we extend our discussion of the 1D example in the main text to the infinite-size setting, and we will see that the infinite-size version of the unperturbed Hamiltonian has one gapped ground state and infinitely many non-gapped ground states.

The infinite-size version of the 1D example in the main text are qutrits on an infinite chain $\mathbb Z$. The infinite-size Hamiltonian can still be formally written as
\begin{equation}
 H=J\sum_{x\in\mathbb Z}(Q_x-Q_{x+1})^2+h_0(1-Q_0),
 \label{eq:infinite-pinned-interaction}
\end{equation}
where $Q_x=|0_x\rangle\langle 0_x|$, and this time we let $J, h_0>0$ without restricting $h_0<2J$ for generality here. Also, the above Hamiltonian contains an additional constant compared to the one in the main text, which is unimportant. Note that Eq. \eqref{eq:infinite-pinned-interaction} should not be understood as a convergent sum. Instead, only the commutator of this Hamiltonian with local operators is meaningful, as in Eq. \eqref{eq: derivation}.

Depending on the values of $J$ and $h_0$, there are different cases for the ground state of this Hamiltonian, as described by the following proposition.

\begin{proposition} \label{prop:infinite-pinned-states}
The vacuum functional $\omega_{\rm vac}$, given by $\omega_{\rm vac}(|0_x\rangle\langle 0_x|)=1$ for all $x\in\mathbb{Z}$, is a ground state of Eq.
\eqref{eq:infinite-pinned-interaction}. For any unit vector $\ket{\psi}\in\mathbb C^2$,
let $\omega_\psi^{\rm occ}$ be the product state with defect that has spin degree of freedom described by $\ket{\psi}$ at
every site, and let $\omega_\psi^{\rm hole}$ have vacuum at site $0$ and a
defect with spin $\ket{\psi}$ at every other site. Then
\[
 \begin{array}{ll}
 \omega_\psi^{\rm occ}\text{ is a ground state}
   &\Leftrightarrow\ h_0\le2J,\\
 \omega_\psi^{\rm hole}\text{ is a ground state}
   &\Leftrightarrow\ h_0\ge2J.
 \end{array}
\]
At $h_0=2J$, both families are ground states. In particular, the ground
state functional is non-unique for every $h_0>0$.
\end{proposition}
\begin{proof} For notational convenience, we write Eq. \eqref{eq:infinite-pinned-interaction} as $H=\sum_xq_x$, where $q_x=J(Q_x-Q_{x+1})^2$ if $x\neq 0$, and $q_x=J(Q_x-Q_{x+1})^2+h_0(1-Q_0)$ if $x=0$.

We first show that $\omega_{\rm vac}$ is always a ground state of Eq. \eqref{eq:infinite-pinned-interaction}. Our strategy is to use the locality to convert the problem of an infinite lattice into a problem of finite lattices.

For any local operator $A$, denote the sites connected by the bonds that intersect the support of $A$ by $\operatorname{bs}(A)$. Note that $\operatorname{bs}(A)$ is always larger than the support of $A$ by at least two sites. Denote $\overline{\operatorname{bs}}(A) = \bigcup_{x\in\operatorname{bs}(A)}\{x,x+1\}$. Then
\begin{equation}
    \omega_{\rm vac}(A^\dag\delta_\Phi (A))=\omega_{\rm vac}(A^\dag\sum_{x\in\mathbb{Z}}[q_x, A])=\omega_{\rm vac}(A^\dag\sum_{x\in \operatorname{bs}(A)}[q_x, A])=\sum_{x\in \operatorname{bs}(A)}
    \left(
    \omega_{\rm vac}(A^\dag q_xA)
    -
    \omega_{\rm vac}(A^\dag Aq_x)
    \right).
\end{equation}
The first term above is non-negative, because the operator $A^\dag q_xA$ is non-negative. The second term vanishes because $\omega_{\rm vac}(A^\dag Aq_x)=(\otimes_{x_1\in \overline{\operatorname{bs}}(A)}\langle 0_{x_1}|)A^\dag Aq_x(\otimes_{x_2\in \overline{\operatorname{bs}}(A)}|0_{x_2}\rangle)=0$, where we have used that $q_x(|0_{x}\rangle\otimes\ket{0_{x+1}})=0$ for all $x\in\mathbb{Z}$ in the last step.

Therefore, according to Eq. \eqref{eq:infinite-ground-state-condition}, $\omega_{\rm vac}$ is always a ground state of Eq. \eqref{eq:infinite-pinned-interaction}.

Next, we examine whether $\omega_\psi^{\rm occ}$ is a ground state. Our strategy is still to use the locality to convert the problem of an infinite lattice into a problem of finite lattices.

For the above local operator $A$,
\begin{equation}
\begin{split}
    \omega_\psi^{\rm occ}(A^\dag\delta_\Phi(A))
    =&\sum_{x\in \operatorname{bs}(A)}
    \left(
    \omega^{\rm occ}_\psi(A^\dag q_xA)
    -
    \omega^{\rm occ}_\psi(A^\dag Aq_x)
    \right)\\
    =&
    \sum_{x\in \operatorname{bs}(A)}
    \left[
    \otimes_{x_1\in {\overline{\operatorname{bs}}}(A)}\langle\psi_{x_1}|A^\dag q_xA\otimes_{x_2\in {\overline{\operatorname{bs}}}(A)}|\psi_{x_2}\rangle
    -
    \otimes_{x_1\in {\overline{\operatorname{bs}}}(A)}\langle\psi_{x_1}|A^\dag Aq_x\otimes_{x_2\in {\overline{\operatorname{bs}}}(A)}|\psi_{x_2}\rangle
    \right].
\end{split}
\end{equation}
Note that the state $\otimes_{x\in {\overline{\operatorname{bs}}}(A)}|\psi_x\rangle$ is an eigenstate of $\sum_{x\in \operatorname{bs}(A)}q_x$, and the above expression is identical to the change of energy when the operator $A$ is applied to the state $\otimes_{x\in {\overline{\operatorname{bs}}}(A)}|\psi_x\rangle$, if the Hamiltonian is $\sum_{x\in \operatorname{bs}(A)}q_x$. If $0\notin {\operatorname{bs}}(A)$, then $\otimes_{x\in {\overline{\operatorname{bs}}}(A)}|\psi_x\rangle$ is a ground state of $\sum_{x\in \operatorname{bs}(A)}q_x$, so $\omega_\psi^{\rm occ}(A^\dag\delta_\Phi(A))\geqslant 0$. If $0\in \operatorname{bs}(A)$, a ground state of $\sum_{x\in \operatorname{bs}(A)}q_x$ is $\otimes_{x\in {\overline{\operatorname{bs}}}(A)}|0_x\rangle$, and $\otimes_{x\in {\overline{\operatorname{bs}}}(A)}|\psi_x\rangle$ is a degenerate excited state with excitation energy $h_0$. Note that ${\operatorname{bs}}(A)$ always contains at least 2 more sites compared to the support of $A$, so applying $A$ to $\otimes_{x\in {\overline{\operatorname{bs}}}(A)}|\psi_x\rangle$ in an occupation sector will either lead to another degenerate excited state if $A$ does not remove any defect, or result in a state with energy at least $2J$ if $A$ removes some defects. So $\omega_\psi^{\rm occ}(A^\dag\delta_\Phi(A))\geqslant 0$ for every local operator $A$ is equivalent to that $2J\geqslant h_0$.

Therefore, we conclude that $\omega^{\rm occ}_\psi$ is a ground state if and only if $h_0\leqslant 2J$.

Finally, we turn to the state $\omega^{\rm hole}_\psi$. Similar reasoning as above implies that it is a ground state if and only if $h_0\geqslant 2J$.
\end{proof}

It should be obvious that neither $\omega^{\rm occ}_\psi$ nor $\omega^{\rm hole}_\psi$ is gapped in the sense of Eq. \eqref{eq:infinite-algebraic-gap}, because $|\psi\rangle$ is an arbitrary state of the spin degree of freedom, so choosing $A$ to be any unitary within the spin-1/2 space of any defect will result in $\omega^{\rm occ}_\psi(A^\dag\delta_\Phi(A))=\omega^{\rm hole}_\psi(A^\dag\delta_\Phi(A))=0$. On the other hand, $\omega_{\rm vac}$ is gapped in the sense of Eq. \eqref{eq:infinite-algebraic-gap} with $\gamma=2J$, which can be seen using a similar argument as in the proof of Proposition \ref{prop:infinite-pinned-states}.

\subsubsection{1D example introduced in this section} \label{subsubsec: infinite version 1D this section}

Finally, we turn to the infinite-size version of the example introduced in this section. 

First, we define this infinite-size generalization precisely. We use the blocks $I^{\pm}_{m, k}$ with unbounded $m$ and $k$ to partition the infinite lattice $\Lambda=\mathbb{Z}$, in a way similar to that discussed in Eq. \eqref{eq: partition using blocks}. In other words, there are no remaining sites or analog of $\mathcal{I}^c_L$ in this infinite lattice. The unperturbed Hamiltonian for this infinite lattice, denoted by $\Phi$, can be formally thought of as the sum of $H^\pm_{m, k}$ for all blocks. The perturbed Hamiltonian, denoted by $\Phi'$, can be formally thought of as the sum of $H^\pm_{m, k}$ for all blocks and the on-site perturbation in Eq. \eqref{eq: 1d perturbation} for all sites. Indeed, these infinite-size Hamiltonians are the thermodynamic limit of their finite-size versions in the sense of Eq. \eqref{eq:local-dynamics-compatible}.

With similar reasoning as in Sec. \ref{subsubsec: infinite version of 1D main text}, we can see that the vacuum state $\omega_{\rm vac}$ with $\omega_{\rm vac}(|0_x\rangle\langle 0_x|)=1$ for all $x\in\mathbb{Z}$ is a ground state of $\Phi$ with a gap $\gamma=2$.{\footnote{In passing, we note that $\omega_{\rm vac}$ is also the only ground state of $\Phi'$, but without a nonzero gap.}} Below we prove that $\omega_{\rm vac}$ is the only ground state of $\Phi$, which physically stems from the fact that $\otimes_{x\in I^\pm_{m, k}}|0_x\rangle$ is the only ground state of $H^\pm_{m, k}$, for all $m$ and $k$.

\begin{proposition}\label{prop:infinite-block-unique}
The vacuum state $\omega_{\rm vac}$ with $\omega_{\rm vac}(|0_x\rangle\langle 0_x|)=1$ for all $x\in\mathbb{Z}$ is the only ground state of $\Phi$.
\end{proposition}
\begin{proof}

Let $\omega$ be a ground state, i.e., $\omega$ satisfies Eq. \eqref{eq:infinite-ground-state-condition}.
For any excited normalized eigenvector $\ket{e}$ of a block Hamiltonian
$H_{m,k}^{\pm}$ with eigenvalue $E>0$, let the local operator
$A=\otimes_{x\in I^\pm_{m, k}}|0_x\rangle\bra{e}$ on that entire block. Since there are no interactions
joining this block to its complement, $\delta_\Phi(A)=[H^{\pm}_{m, k},A]=-EA$, so
\[
 0\le\omega\bigl(A^{\dagger}\delta_{\Phi}(A)\bigr)
   =-E\omega(\ket{e}\bra{e}).
\]
Because $|e\rangle\langle e|$ is non-negative, $\omega(|e\rangle\langle e|)\geqslant 0$. Because $E>0$, the above result means that $\omega(|e\rangle\langle e|)=0$, which implies that $\omega=\omega_{\rm vac}$, i.e., $\omega_{\rm vac}$ is the only ground state of $\Phi$.
\end{proof}

\section{Proof of Theorem \ref{thm:main} and Corollary \ref{corollary: continuity of gap}: The spin part} \label{sec: proof of main theorem}

In this section, we prove Theorem \ref{thm:main} and Corollary \ref{corollary: continuity of gap}, assuming that our system is a spin system. The extension of the proof to fermionic systems is in Sec. \ref{sec:fermions}, where many ideas in this section are also used, but appropriate modifications are made to incorporate the fermionic nature of the systems.

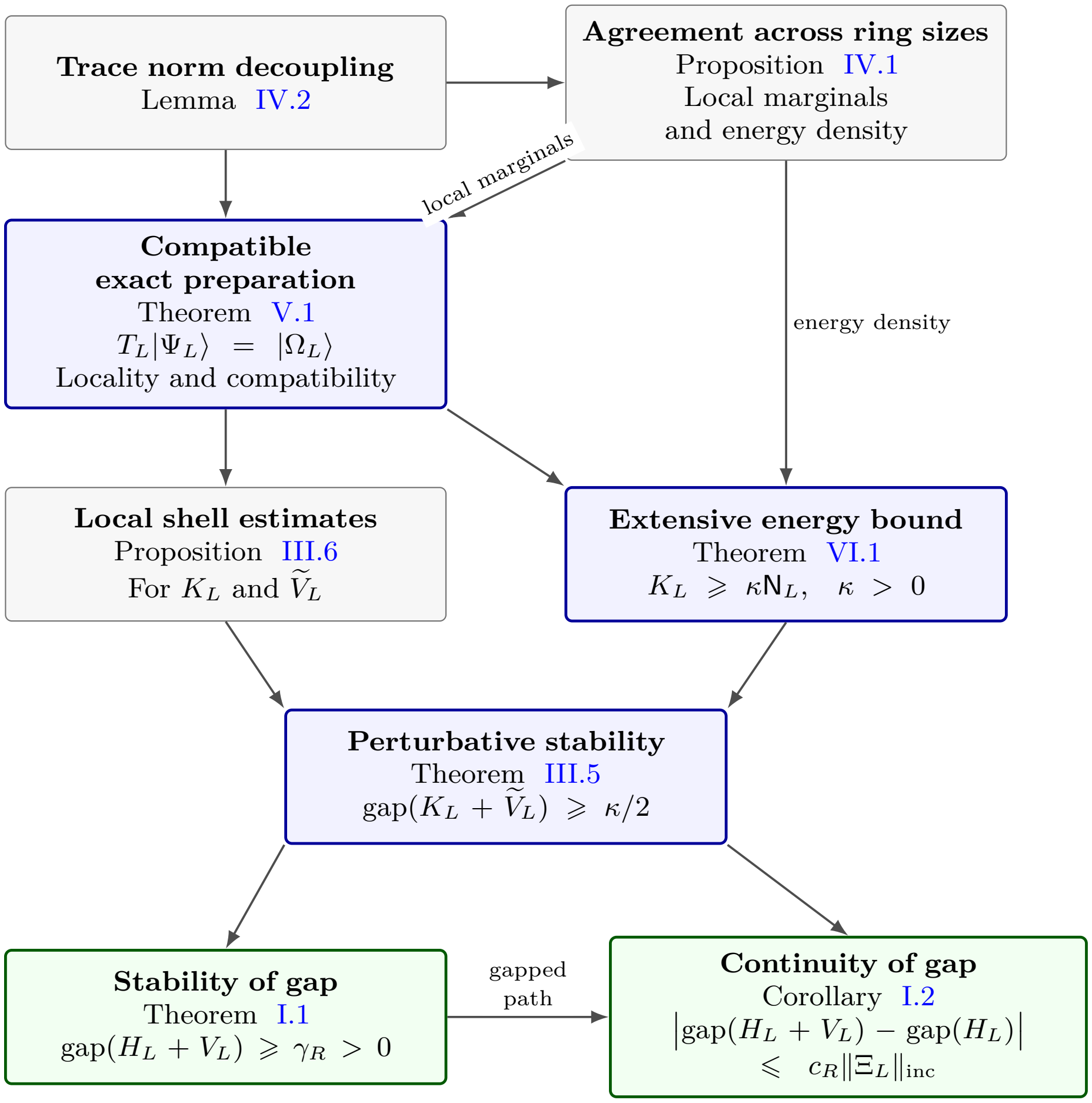
\begin{figure}[tb]
\centering
\begingroup
\providecommand{\dependencyref}[2]{
  \ifcsname r@#1\endcsname\ref{#1}\else#2\fi}

\begin{tikzpicture}[
  x=1cm,y=1cm,
  every node/.style={font=\small,align=center},
  box/.style={
    draw=black!55,
    rounded corners=2pt,
    fill=black!3,
    text width=46mm,
    minimum height=15mm,
    inner sep=5pt
  },
  key/.style={
    box,
    draw=blue!60!black,
    fill=blue!5,
    line width=.9pt
  },
  conclusion/.style={
    key,
    draw=green!35!black,
    fill=green!5
  },
  arr/.style={
    -{Latex[length=2mm]},
    line width=.7pt,
    draw=black!70
  },
  edge label/.style={
    font=\scriptsize,
    fill=white,
    inner sep=2pt
  }
]

\node[box] (dec) at (0,0)
  {\textbf{Trace norm decoupling}\\
   Lemma~\dependencyref{lem:decoupling}{IV.2}};

\node[box] (compare) at (6.3,0)
  {\textbf{Agreement across ring sizes}\\
   Proposition~\dependencyref{prop:circumference}{IV.1}\\
   Local marginals and energy density};

\node[key] (prep) at (0,-2.6)
  {\textbf{Compatible}\\
   \textbf{exact preparation}\\
   Theorem~\dependencyref{thm:prep-exact}{V.1}\\
   $T_L\ket{\Psi_L}=\ket{\Omega_L}$\\
   Locality and compatibility};

\node[box] (shell) at (0,-5.3)
  {\textbf{Local shell estimates}\\
   Proposition~\dependencyref{prop:shelltransfer}{III.6}\\
   For $K_L$ and $\widetilde V_L$};

\node[key] (energy) at (6.3,-5.3)
  {\textbf{Extensive energy bound}\\
   Theorem~\dependencyref{thm:energy}{VI.1}\\
   $K_L\ge\kappa \mathsf{N}_L$,\quad $\kappa>0$};

\node[key] (stability) at (3.15,-7.8)
  {\textbf{Perturbative stability}\\
   Theorem~\dependencyref{thm:productstability}{III.5}\\
   $\operatorname{gap}(K_L+\widetilde V_L)\ge\kappa/2$};

\node[conclusion] (main) at (0,-10.5)
  {\textbf{Stability of gap}\\
   Theorem~\dependencyref{thm:main}{I.1}\\
   $\operatorname{gap}(H_L+V_L)\ge\gamma_R>0$};

\node[conclusion,text width=50mm] (continuity) at (7.0,-10.5)
  {\textbf{Continuity of gap}\\
   Corollary~\dependencyref{corollary: continuity of gap}{I.2}\\
   $\bigl|\operatorname{gap}(H_L+V_L)
          -\operatorname{gap}(H_L)\bigr|$\\
   $\le c_R\|\Xi_L\|_{\mathrm{inc}}$};

\draw[arr] (dec) -- (compare);

\draw[arr] (dec) -- (prep);

\draw[arr] (compare.south west) --
  node[edge label,sloped,above]{local marginals}
  (prep.north east);

\draw[arr] (compare.south) --
  node[edge label,right]{energy density}
  (energy.north);

\draw[arr] (prep) -- (shell);

\draw[arr] (prep.south east) -- (energy.north west);

\draw[arr] (shell.south) -- (stability.north west);

\draw[arr] (energy.south) -- (stability.north east);

\draw[arr] (stability.south west) --
  (main.north);

\draw[arr] (stability.south east) --
  (continuity.north);

\draw[arr] (main.east) --
  node[edge label,above]{\shortstack{gapped\\path}}
  (continuity.west);

\end{tikzpicture}
\endgroup

\caption{Main logical steps in the proof of Theorem~\ref{thm:main}
and Corollary~\ref{corollary: continuity of gap}.
Technical lemmas and the fermionic adaptations are omitted.}
\label{fig:dependencies}
\end{figure}

The proof in this section consists of some intermediate steps, which are summarized in Fig.~\ref{fig:dependencies}. The core idea is to reduce the problem of the stability of gap of general one dimensional local translation invariant Hamiltonians in Theorem~\ref{thm:main} into a series of problems related to a special class of Hamiltonians, which we refer to as {\it normal-form Hamiltonians} that will be defined in Definition \ref{def: normal-form Hamiltonians}. In Sec. \ref{subsec:intermediate_step_summary}, we present three important intermediate steps, with their detailed proofs given in Sec. \ref{sec:across_sizes}-\ref{sec:coercivity}. In Sec. \ref{subsec:normal_form_gap_stability}, we present the proof of the stability of gap of normal-form Hamiltonians. Finally, in Sec. \ref{subsec:main_proof}, we give the proof of the spin part of Theorem \ref{thm:main} and Corollary \ref{corollary: continuity of gap}, which apply to general local translation invariant gapped Hamiltonians with a unique ground state.

\subsection{Key intermediate results}\label{subsec:intermediate_step_summary}

With the setup in Sec. \ref{sec: setup and summary}, in this subsection, we summarize the three intermediate results used in the proof of Theorem~\ref{thm:main} and Corollary \ref{corollary: continuity of gap}, under the assumption of the uniform gap condition with a unique ground state discussed around Eq. \eqref{eq:assumption}.
The more detailed statements and their proofs are given in
Sec. \ref{sec:across_sizes}-\ref{sec:coercivity}. We only present the spin formulation here, while the fermionic adaptations are discussed in Sec. \ref{sec:fermions}.

\subsubsection{Agreement between systems with different sizes}

One of the most nontrivial aspects of Theorem \ref{thm:main} and Corollary \ref{corollary: continuity of gap} lies in the requirement that the gap should be stable for arbitrarily large systems. Therefore, it is important to understand the relation between systems with different sizes. Our first intermediate result is about the energies and reduced density matrices at different system sizes.

\medskip\noindent\textbf{Agreement across ring sizes
(Proposition~\ref{prop:circumference}).}
{\it The ground state energy densities and local reduced density matrices agree
exponentially well between sufficiently large rings.
Specifically, there is a sufficiently large $N$, such that for $L\ge N$ with $1\le\ell\le\lfloor N/12\rfloor$,
\begin{equation}\label{eq:agreement_across_rings}
 \left|\frac{E_L}{L}-\frac{E_N}{N}\right|\le C_E e^{-c_EN},
 \qquad
 \|\rho_{L,I}-\rho_{N,I}\|_1\le C_{\rm marg} e^{-c_E N/4 },
\end{equation}
where $E_L$ and $E_N$ are the total ground state energies of the rings with length $L$ and $N$, respectively, $\rho_{L,I}$ and $\rho_{N, I}$ are the reduced density matrices of
$\rho_L=\ket{\psi_L}\bra{\psi_L}$ and $\rho_N=\ket{\psi_N}\bra{\psi_N}$ to $I=\{0, 1, \cdots, \ell-1\}$, and the corresponding
on-site Hilbert spaces are identified in the two rings.
The constants $C_E, C_{\rm marg}, c_E>0$ are independent of $L,N$, and $I$.
}

The agreement of energy densities is
used in the proof of the energy-defect inequality in Theorem~\ref{thm:energy}. The {agreement of} local density matrices (a.k.a. marginals) allows the preparation circuits below to be chosen consistently across different sizes. The proof of Proposition~\ref{prop:circumference} uses the translation invariance of the unperturbed Hamiltonian explicitly.

\subsubsection{Disentangling the ground state}

Another nontrivial aspect of Theorem \ref{thm:main} is that this theorem covers a large class of Hamiltonians, i.e., all local translation invariant gapped Hamiltonians with a unique ground state. Next, we present a theorem that reduces our problem of the stability of gap of such Hamiltonian to the corresponding problem in a special class of Hamiltonians, which we call {\it normal-form Hamiltonians}. To introduce this notion, we first introduce some convenient notations.

First, to prove Theorem~\ref{thm:prep-exact}, we will enlarge our system. It is possible to improve the argument below so that this enlargement is not needed for proving the spin part of Theorem \ref{thm:main}, but it is still necessary for the fermionic part. To have a unified treatment of both the spin and fermionic parts, we enlarge both types of systems. 

Concretely, at each site with a qudit, we introduce another copy of this qudit, so that the dimension of Hilbert space at this site becomes $\widehat D= D^2$. We use a bar to stand for entry-wise conjugation in the qudit basis, and set $H^c_L = \overline{H}_L$ and $\ket{\psi_L^c} = \ket{\overline\psi_L}$. Then in this doubled Hilbert space, we define
\begin{equation}\label{eq:enlargement}
    \widehat{H}_L = H_L\otimes \one + \one\otimes H_L^c,\qquad
    \ket{\Psi_L}=\ket{\psi_L}\otimes\ket{\psi_L^c},\qquad
 \hgamma=\gamma_0.
\end{equation}
The state $\ket{\Psi_L}$ is the unique ground state of $\widehat{H}_L$. $\widehat{H}_L$ can be written as sum of local terms $\widehat h_{x, L}$, with $\widehat{h}_{x, L}$ being the enlarged local term $h_{x, L}\otimes\one + \one\otimes h_{x, L}^c$. The ground state energy of $\widehat{H}_L$ is $2E_L$, and $\gap(\widehat H_L)\geq \widehat\gamma$. The stability condition $\gap(H_L+V_L)\geq \gamma_R$ is proved if we can show that $\hH_L + V_L\otimes\one$ has an energy gap at least $\gamma_R$, with $V_L\otimes\one$ the perturbation acting only on the original physical chain, because the physical chain and its conjugate copy are decoupled and $\gap(\widehat{H}_L + V_L\otimes\one)=\min\{\gap(H_L+V_L),\ \gap(H_L^c)\}\leqslant\gap(H_L+V_L)$.

The important result of Theorem~\ref{thm:prep-exact} is that $|\Psi_L\rangle$ can be disentangled in a specific way. For simplicity, denote by $\ket{e_x} = \ket{0}^{\otimes 2}$ the vacuum state of the two on-site qudits, where $|0\rangle$ is a basis state of the qudit that is interpreted as the on-site vacuum. We use $\ket{\Omega_L} = \bigotimes_{x\in \T_L} \ket{e_x}$ to denote the product vacuum state. 
Define
\begin{equation}\label{eq:defects}
 P_X^\Omega=\bigotimes_{x\in X}\ket{e_x}\bra{e_x},\qquad
 Q_x=\one-\ket{e_x}\bra{e_x},\qquad \mathsf{N}_X=\sum_{x\in X}Q_x,
 \qquad \mathsf{N}_L=\mathsf{N}_{\T_L}.
\end{equation}
The \emph{defect number} $\mathsf{N}_L$ counts sites that are not in the on-site vacuum, or sites occupied by defects. Then we have the following theorem.

\medskip\noindent\textbf{Exact state preparation
(Theorem~\ref{thm:prep-exact}).}
{\it
There are unitaries $T_L$ on the enlarged rings such that
\[
 T_L\ket{\Psi_L}=\ket{\Omega_L}.
\]
Both conjugations by $T_L$ and $T_L^{\dagger}$ satisfy the exponential locality
estimate given by Eq. \eqref{eq:uniform_locality}, with constants independent of $L$.
The unitaries can be chosen compatibly across ring sizes: their actions
on local observables agree up to an error that decays exponentially
with the distance from the closing seam of periodic chain.
Moreover, they admit local circuit approximations with exponentially
small errors on local observables. Note that $T_L$ is in general NOT translation invariant.
}

The precise statements on the compatibility and circuit approximation are
given in Theorem~\ref{thm:prep-exact}, and they are used in the argument in Sec.~\ref{sec:coercivity}. Here \emph{exponential locality} means that
conjugated local observables admit exponentially accurate local approximations:
\begin{equation}
    \label{eq:uniform_locality}
\inf_{A_r\in\A_{X(r)}} \norm{T_L A T_L^{\dagger} - A_r }\leq C|X| \norm{A} e^{-cr},
\qquad A\in\A_X,\quad r\ge0,
\end{equation}
where $X(r)=\{y: d_L(y, X)\leqslant r\}$. The constants $C,c>0$ are independent of $L,X,A$ and $r$.
Throughout this supplemental material, the meaning of exponential locality includes this independence
of the circumference $L$. As mentioned above, this exponential locality estimate also holds with $T_L$ replaced by $T_L^{\dagger}$ in Eq. \eqref{eq:uniform_locality}.

We remark that the existence of the above disentangling unitary $T_L$ does not {rely on} the translation symmetry of $H_L$. {Therefore}, Theorem~\ref{thm:prep-exact} implies that the non-degenerate ground states of gapped one dimensional local Hamiltonians are invertible. This result has been anticipated for a long time in the physics literature \cite{Chen2010, Fidkowski2010, Schuch2010}, and recently it has been rigorously proved for infinite quantum spin chains in Ref. \cite{deroeck2025}, which did not even enlarge the system and actually proved a stronger result that these states are short-range entangled. Our result here is the finite-size version of Ref. \cite{deroeck2025}, which is more relevant to real physical systems, and we also have a fermionic counterpart of this result in Sec. \ref{sec:fermions}.

By conjugation of $T_L$, the Hamiltonian $\widehat{H}_L$ can be transformed into a ``frame'' where the ground state is the vacuum. For convenience, we will refer to the original frame as the physical frame, and call the frame with the vacuum as the ground state the product-vacuum frame. Let $K_L = T_L (\widehat H_L - 2E_L) T_L^\dagger$, then $K_L$ is made up of exponentially local terms and its unique ground state is the vacuum $\ket{\Omega_L}$. $K_L$ is in what we call a {\it normal-form Hamiltonian}, whose complete definition is in Definition \ref{def: normal-form Hamiltonians}. Since $K_L$ has the same excitation spectrum as $\widehat H_L$, the problem of {the stability of gap of general Hamiltonians} {can be} reduced {to the corresponding problem of normal-form Hamiltonians}.

\subsubsection{The energy-defect inequality}

After relating the original Hamiltonian to the normal-form Hamiltonian, our next key step is to prove the following relation between the excitation energy and the number of defects in the product-vacuum frame.

\medskip\noindent\textbf{Energy-defect inequality
(Theorem~\ref{thm:energy}).}
{\it
For the unitaries constructed in Theorem~\ref{thm:prep-exact}, there is
a constant $\kappa>0$, independent of $L$, such that
\begin{equation}\label{eq:mechanism}
 K_L=T_L(\widehat H_L-2E_L\one)T_L^{\dagger}\ge\kappa \mathsf{N}_L
\end{equation}
for all sufficiently large $L$.}

\medskip  
Theorem~\ref{thm:energy} implies that the excitation energy
{is controlled by} the number of defects.  Note that the spectral gap alone gives only
$K_L\ge(\hgamma/L)\mathsf{N}_L$, and the uniform coefficient $\kappa$ in Eq.
\eqref{eq:mechanism} is nontrivial. Intuitively, this theorem rules out those states with an arbitrarily small excitation energy per defect. If those states existed, a weak but extensive perturbation could pull down these excited states in energy and the energy gap would be closed. Indeed, this is a common mechanism behind all our examples with an unstable gap, and proving the non-existence of such states is a key ingredient in our theorem on the stability of gap.

\subsection{{Stability of gap} of normal-form Hamiltonians}
\label{subsec:normal_form_gap_stability}

In this subsection, we will prove the {theorem on the stability of gap} {of} the special class {of Hamiltonians}, normal-form Hamiltonians, under the assumptions of stretched-exponential locality of the terms in these Hamiltonians and the energy-defect bound in Eq. \eqref{eq:mechanism}. To prove this stability, we introduce the notions of shell representation, filtering of operators and spectral flow in Section~\ref{subsubsec:locality_and_filtering}. Then in Section~\ref{subsubsec:normal-form-bootstrap}, we present the proof of {the stability of gap} of normal Hamiltonians with a bootstrap argument. In Section~\ref{subsubsec:product-vacuum-locality}, we verify that $K_L$ defined in Eq. \eqref{eq:mechanism} is indeed a normal-form Hamiltonian with the relevant locality requirement.

\subsubsection{Preparation: Shell representation, filtering, locality control and quasi-adiabatic continuation}
\label{subsubsec:locality_and_filtering}

To prepare for the proofs in the next two subsections, in this subsection we introduce some necessary notions, including a shell representation of Hamiltonians, filtering of operators, a Lieb-Robinson-type bound about locality and quasi-adiabatic continuation.

\medskip
\paragraph{Stretched-exponential function and its constants.}
The following notion of a {\it stretched-exponential function} and its constants is applied throughout the later part of this supplemental material. A non-negative real function $f(r)$ is called a stretched-exponential function if $f(r)\leqslant Je^{-ar^\alpha}$ for some constants $J, a>0$ and $0<\alpha\leqslant 1$. Later we may use ``constants in $f(r)$'' to refer to the parameters $J, a$ and $\alpha$ for the function $f(r)$, and these parameters determine how fast this function decays to zero as $r$ increases.

\medskip
\paragraph{Shell representation.} Suppose the Hamiltonian of a ring with circumference $L$ is $G_L$, which may or may not be gapped. We say that $G_L$ has an exact stretched-exponential shell representation if {there exists} a decomposition of $G_L$ into terms grouped by a center site $x$ and a radius $r$ {, with the following locality condition}:
\begin{equation}\label{eq:shell}
 G_L=\sum_{x\in\T_L}\sum_{r=0}^{M_L}g_L(x,r),\qquad
 g_L(x,r)=g_L(x,r)^{\dagger}\in\A_{B_L(x,r)},\qquad
 \|g_L(x,r)\|\le J e^{-ar^\alpha},
\end{equation}
where $J,a>0$ and $0<\alpha\le1$ are independent of $L$. The sum includes $r=M_L:=\lfloor L/2\rfloor$, for which $B_L(x,M_L)=\T_L$. 
Proposition~\ref{prop:shelltransfer} will establish that the
transformed Hamiltonian $K_L$ defined in Eq. \eqref{eq:mechanism} and the transformed perturbation {have such a shell representation}.

\medskip
\paragraph{Filtering.}
Next, we introduce an exact filter of operators defined by a Hamiltonian $G_L$ with a shell representation and a function $w_g(t)$ to be introduced below, based on
Ref.~\cite[Sec.~2]{BachmannMichalakisNachtergaeleSims2012}. 

Let $\tau_t^{G_L}(A)=e^{itG_L}Ae^{-itG_L}$ be the time evolution of an operator $A$.
For a {\it threshold} $g>0$, there exists a real even integrable
function $w_g$ such that
\[
 \widehat w_g(0)=1,\qquad\supp\widehat w_g\subset(-g,g),\qquad
 |w_g(t)|\le C_g e^{-c_g\sqrt{|t|}},
\]
where $\widehat w_g$ is the Fourier transformation of $w_g(t)$, which is generally defined as $\widehat f(E)=\int f(t)e^{itE}\,dt$ for a function $f(t)$. An example of such $w_g$ can be found in Ref.~\cite[Lemma~2.3]{BachmannMichalakisNachtergaeleSims2012}. The {\it filtered operator} $F_{G_L, w_g}(A)$ with the Hamiltonian $G_L$ and the filtering function $w_g(t)$ is
\begin{equation}
F_{G_L, w_g}(A)=\int w_g(t)\tau_t^{G_L}(A)dt.
\end{equation}
Since $\widehat w_g(E)=0$ for $|E|\ge g$, the filtered
operator $F_{G_L, w_g}(A)$ has vanishing matrix elements between
eigenstates of $G_L$ whose energies differ by at least $g$. In particular, if we {\it further assume} that the ground
state of $G_L$ is separated from the rest of the spectrum by a gap greater than
$g$, $F_{G_L, w_g}(A)$ commutes with the projector onto the ground state subspace.

There is another way to express the filtered operator $F_{G_L, w_g}(A)$. Let
\begin{equation}\label{eq:filter}
 W_g(t)=\int_{-\infty}^t w_g(u)\,du-\one_{[0,\infty)}(t),\quad
 \quad
 \mathcal D_{G_L, w_g}(A)=\int W_g(t)\tau_t^{G_L}(A)\,dt.
\end{equation}
Here $\one_{[0,\infty)}$ is the indicator function, i.e., it can be thought of as the step function $\theta(x)$ with $\theta(0)=1$. The real kernel
$W_g$ is odd almost everywhere (except $t=0$), and it obeys
$|W_g(t)|\le C'_g e^{-c'_g\sqrt{|t|}}$
for some constants $C'_g,c'_g>0$.
All integrals above converge in  the operator norm, and integration by parts yields the other way to express $F_{G_L, w_g}(A)$:
\begin{equation}\label{eq:exactfilter}
 F_{G_L, w_g}(A):=A+i[\mathcal D_{G_L, w_g}(A),G_L].
\end{equation}

\medskip
\paragraph{Locality of transported filtered operators.} The following locality condition is used in the proofs of statements such as Lemma \ref{lem:transported-gap}. Suppose for each $0\leqslant s\leqslant 1$ the operator $\mathcal{D}_L(s)$ is a Hamiltonian that has a shell representation as in Eq.~\eqref{eq:shell}, and the unitary $U_L(s)$ is an inverse time evolution operator generated by $\mathcal D_L(s)$ i.e.,
\begin{equation} \label{eq: transport unitary}
    \dot{U}_L(s) = i U_L(s) \mathcal{D}_L(s),
    \quad U(0)=\one.
\end{equation}
Then its conjugation on local operators also preserves locality accordingly. For example, for a set of Hermitian operators $v_L(x,r)\in\mathcal{A}_{B_L(x, r)}$, where $B_L(x, r):=\{y|d_L(x, y)\leqslant r\}$ and $\norm{v_{L}(x,r)}\leq J_v e^{-ar^\alpha}$, the locality of the {\it transported filtered operator}
\begin{equation}\label{eq:anchor}
 A_{x,L}(s):=\sum_r U_L(s)F_{G_L(s), w_g}(v_L(x,r))U_L(s)^{\dagger}
\end{equation}
can be shown to satisfy the following locality condition:
\begin{equation}
    \norm{A_{x,L}(s)} \leq C J_v, \quad \norm{A_{x,L}(s) - \E^{\tr}_{B_L(x,m)} A_{x,L}(s) } \leq C' J_v e^{-c' m^\eta},
\end{equation}
with $C, C', c',\eta$ positive constants, uniformly in $L,s$ (see Sec. \ref{subsec:anchored-locality} for the details). Here $G_L(s)$ is also assumed to obey the locality condition in Eq.~\eqref{eq:shell} with all constants independent of $L$ or $s$, and $\E^{\tr}_{X}O=\frac{\tr_{X^c}(O)}{{\rm dim}(\mathcal{H}_{X^c})}\otimes\one_{X^c}$ for any subregion $X\subseteq\mathbb{T}_L$ and operator $O\in\mathcal{A}_{\mathbb{T}_L}$, which can be thought of as the part of $O$ that is only supported within $X$. For example, for a chain of qubits any operator $O$ can be decomposed into a sum of Pauli strings, and $\E^{\tr}_{X}O$ singles out the sum of those Pauli strings that are only supported within $X$. We stress that these locality estimates do not require $\mathcal{D}_L(s)$ or $G_L(s)$ to be gapped.

\medskip
\paragraph{Transportation of the ground state.}
Finally, we review the notion of ground state transportation, which is also known as spectral flow, quasi-adiabatic continuation, automorphic equivalence, etc \cite{Hastings2005, BachmannMichalakisNachtergaeleSims2012}. This result will be frequently used below when we relate the ground states of various different Hamiltonians.

For a path of Hamiltonians $G_L(s)=G_L+sV_L$, $0\leqslant s\leqslant 1$, which can be thought of as slowly turning on the perturbation $V_L$ to the Hamiltonian $G_L$, we take the $\mathcal{D}_L(s)$ in Eq. \eqref{eq: transport unitary} as $\mathcal{D}_L(s)=\mathcal{D}_{G_L(s), w_g}(V_L)$ and use it to define a unitary $U_L(s)$ as in Eq. \eqref{eq: transport unitary}. Suppose that the ground state is unique and the gap is greater than $g$ along the entire path with $0\leqslant s\leqslant 1$, then $U_L(s)$ transports the ground projection $P_L(s)$ back to $P_L(0)$, i.e.,
$U_L(s) P_L(s) U_L(s)^\dagger = P_L(0)$ (see Section~\ref{subsec:anchored-locality} for a detailed proof).

\subsubsection{Normal-form Hamiltonians and the stability of their gap}
\label{subsubsec:normal-form-bootstrap}

With the above preparation, now we give a formal definition of normal-form Hamiltonians, and discuss how filtering can help us construct their shell representations where each shell term annihilates the vacuum. Then a bootstrap argument allows us to show the stability of gap of normal-form Hamiltonians, as stated in Theorem~\ref{thm:productstability}.

\begin{definition}[normal-form Hamiltonians] \label{def: normal-form Hamiltonians}
    A family of Hamiltonians $\{G_L\}$ are normal-form Hamiltonians, if each of them admits a uniformly stretched-exponential shell representation, $\ket{\Omega_L}$ is the unique ground state of $G_L$, the ground state energy of $G_L$ is zero, and $\inf_L\gap(G_L)>0$.
\end{definition}

Before proceeding, we present a useful lemma for later use. It is a product-vacuum version of the
bounds used in Refs.~\cite[Proposition~2]{MichalakisZwolak2013}
and~\cite[Theorem~3.8]{NSYstability2022}. In the following, for a region $X\subseteq\mathbb{T}_L$, we define $|\Omega_X\rangle :=\bigotimes_{x\in X}|e_x\rangle$, the conditional expectation map for an operator $A$ as $\E^\Omega_X (A) := \ev{\Omega_{X^c}|A|\Omega_{X^c}}\otimes\one_{X^c}$ with $A$ an operator supported on $X\cup X^c$. Note that another conditional expectation $\E^{\tr}_X(A) := \frac{\tr_{X^c}(A)}{{\rm dim}(\mathcal{H}_{X^c})}\otimes\one_{X^c}$ has already been introduced above, and both conditional expectation maps are norm-contracting, i.e., $\|\E^\Omega_X (A) \|\leqslant\|A\|$ and $\|\E^{\tr}_X (A) \|\leqslant\|A\|$.

\begin{lemma}\label{lem:vacuum}
Let $P^\Omega_X = \ket{\Omega_X}\bra{\Omega_X}$. If $a=a^{\dagger}\in\A_X$ and $aP_X^\Omega=0$, then
\begin{equation}\label{eq:vacuumform}
 -\|a\|\mathsf{N}_X\le a\le\|a\|\mathsf{N}_X,
\end{equation}
where $\mathsf{N}_X$ is introduced in Eq. \eqref{eq:defects}. If $a=a^{\dagger}$ annihilates $\ket{\Omega_L}$, then $\E_X^\Omega(a)P_X^\Omega=0$.
\end{lemma}

\begin{proof}
For the first claim, note that $P^\Omega_Xa=P^\Omega_Xa^\dag=(aP_X^\Omega)^\dag=0$, so $a=(\one-P_X^\Omega)a(\one-P_X^\Omega)+P_X^\Omega a(\one-P_X^\Omega)+(\one-P_X^\Omega)aP_X^\Omega+P_X^\Omega aP_X^\Omega=(\one-P_X^\Omega)a(\one-P_X^\Omega)$. Moreover, it is easy to check that $\one-P_X^\Omega\le \mathsf{N}_X$. Therefore, Eq. \eqref{eq:vacuumform} holds.

To prove the second claim, note that
$\bigl(\bra{u}\otimes\langle v|\bigr)\E_X^\Omega(a)\bigl(\ket{\Omega_X}\otimes|v\rangle\bigr)=\bigl(\bra{u}\otimes\langle \Omega_{X^c}|\bigr)\E_X^\Omega(a)\bigl(\ket{\Omega_X}\otimes|\Omega_{X^c}\rangle\bigr)=\bigl(\bra{u}\otimes\bra{\Omega_{X^c}}\bigr)a\ket{\Omega_L}=0$ for every normalized $\ket{u}\in\mathcal{H}_X$ and $\ket{v}\in\mathcal{H}_{X^c}$, so $\E_X^\Omega(a)P_X^\Omega=0$.
\end{proof}

We call a bound \emph{relative to $\mathsf N_L$}
for a self-adjoint operator $R_L$ if it is of the form
$-C\varepsilon\mathsf N_L\le R_L\le C\varepsilon\mathsf N_L$
with $C$ a constant independent of $L$. The above lemma shows an example of such relative bounds.

With Lemma \ref{lem:vacuum} in hand, we prove two properties of normal-form Hamiltonians that will lead to the stability of gap of these Hamiltonians.

The first property is the following lemma, which gives a shell representation of normal-form Hamiltonians and its upper bound relative to $\mathsf{N}_L$. 

\begin{lemma}[Shell representation and relative upper bound of normal-form Hamiltonians]\label{lem:normal-form-transport}
Let $G_L$ be a normal-form Hamiltonian, which, in particular, satisfies Eq. \eqref{eq:shell}. Then it admits a shell representation
\begin{equation}\label{eq:normal-form-representation}
 G_L=\sum_{x,r}\theta_L(x,r),\qquad
 \theta_L(x,r)=\theta_L(x,r)^{\dagger}\in\A_{B_L(x,r)},\qquad
 \theta_L(x,r)\ket{\Omega_{B_L(x,r)}}=0,
\end{equation}
with uniform stretched-exponential shell bounds, possibly with different
constants and decay exponent. Moreover, 
\begin{equation}
0\le G_L\le C_0\mathsf N_L
\end{equation}
for a constant $C_0$ independent of $L$.
\end{lemma}

\begin{proof}
We use the filtering construction of
Ref.~\cite[Proposition~D.1]{Kitaev2006}. Choose a fixed
$0<g<\inf_L\gap(G_L)$ in the filter function $w_g(t)$. We group the original shell terms at each
center $x$ into $g_{x,L}:=\sum_r g_L(x,r)$, and set
\begin{equation}\label{eq:b_xL_locality}
 b_{x,L}:=F_{G_L, w_g}(g_{x,L})
 -\bra{\Omega_L}g_{x,L}\ket{\Omega_L}\one.
\end{equation}
It is straightforward to check that the $\langle m|b_{x, L}|\Omega_L\rangle=0$ for each $b_{x, L}$ and each energy eigenstate $|m\rangle$, so each $b_{x,L}$ annihilates $\ket{\Omega_L}$. Since
$F_{G_L, w_g}(G_L)=G_L$ and the ground-state energy is zero, $\sum_xb_{x,L}=F_{G_L, w_g}(G_L)-\langle\Omega_L|G_L|\Omega_L\rangle=G_L$. Moreover, the Lieb-Robinson and filter estimates in
Section~\ref{subsec:anchored-locality}, summed over the input shells,
give
\[
\|b_{x,L}\|\le C,\qquad
\|b_{x,L}-\E_{B_L(x,r)}^{\tr}b_{x,L}\|\le f_0(r)
\]
for a uniform (i.e., $L$-independent) stretched-exponential function $f_0$.

With the above preparation, now we construct the targeted shell representation of $G_L$. For $L\geqslant 2$, set
\begin{equation}\label{eq:normal-form-telescoping}
\begin{aligned}
 \theta_L(x,0)&=\E_{B_L(x,0)}^\Omega b_{x,L},\\
 \theta_L(x,r)&=(\E_{B_L(x,r)}^\Omega-\E_{B_L(x,r-1)}^\Omega)b_{x,L},
 &&1\le r<M_L,\\
 \theta_L(x,M_L)&=b_{x,L}-\E_{B_L(x,M_L-1)}^\Omega b_{x,L}.
\end{aligned}
\end{equation}
Then $b_{x,L}=\sum_{r}\theta_L(x, r)$, and each $\theta_L(x, r)\in\mathcal{A}_{B_L(x, r)}$. Moreover, the second part of Lemma \ref{lem:vacuum} ensures that
every shell annihilates the vacuum on its support, i.e., $\theta_L(x, r)|\Omega_{B_L(x, r)}\rangle=0$.

To see the locality of $\theta_L(x,r)$, note that for $1\leqslant r\leqslant M_L$, $\theta_L(x, r)=(\E^\Omega_{B_L(x, r)}-\E^\Omega_{B_L(x, r-1)})(b_{x, L}-\E^{\tr}_{B_L(x, r-1)}(b_{x, L}))$, so $\|\theta_L(x, r)\|\leqslant 2\|b_{x, L}-\E^{\tr}_{B_L(x, r-1)}(b_{x, L})\|\leqslant 2f_0(r-1)$. Moreover, $\|\theta_L(x, 0)\|\leqslant\|b_{x, L}\|\leqslant C$. So the decay of the strength of $\theta_L(x, r)$ is also bounded by a stretched-exponential function.

Therefore, the above $\theta_L(x, r)$'s satisfy Eq. \eqref{eq:normal-form-representation}.

Finally, we show that $0\leqslant G_L\leqslant C_0\mathsf N_L$. Because $G_L$ annihilates the ground state and the ground state energy is zero, $G_L\geqslant 0$. To see that $G_L\leqslant C_0\mathsf N_L$, we apply Lemma \ref{lem:vacuum}, which shows that $\theta_L(x, r)\leqslant\|\theta_L(x, r)\|\mathsf{N}_{B_L(x, r)}$, shell by shell, and obtain $G_L=\sum_{x, r}\theta_L(x, r)\leqslant\sum_{x,r}\|\theta_{L}(x, r)\|\mathsf{N}_{B_L(x, r)}$. Suppose a stretched-exponential function that bounds the operator norm of $\theta_{L}(x, r)$ is $\tilde f(r)$, i.e., $\|\theta_L(x, r)\|\leqslant \tilde f(r)$. Then $G_L\leqslant\sum_r\tilde f(r)\sum_x\mathsf{N}_{B_L(x, r)}\leqslant \sum_r\tilde f(r)(2r+1)\mathsf{N}_L=C_0\mathsf{N}_L$, with $C_0=\sum_r\tilde f(r)(2r+1)$, which is independent of $L$.

\end{proof}

The second property is that, if we assume that a normal-form Hamiltonian $G_L$ satisfies an energy-defect bound $G_L\geq \kappa \mathsf N_L$, then under a quasi-adiabatic continuation, the transformed Hamiltonian also obeys an energy-defect bound. As we have emphasized, this is a key property that implies that a sufficiently weak local perturbation cannot close the gap of a normal-form Hamiltonian.

\begin{lemma}[Transported energy-defect bound]\label{lem:transported-gap}
Let $\{G_L\}$ be a family of normal-form Hamiltonians satisfying
$G_L\ge\kappa\mathsf N_L$ for an $L$-independent $\kappa>0$. Suppose for an $L$-independent $0<\varepsilon\leq 1$,
\[
 V_L=\sum_{x,r}v_L(x,r),\qquad
 v_L(x,r)=v_L(x,r)^{\dagger}\in\A_{B_L(x,r)},\qquad
 \|v_L(x,r)\|\le\varepsilon f_0(r)
\]
where $f_0$ is an $L$-independent stretched-exponential function.
Fix $0<g<\kappa$, a filter function $w_g(t)$ and $0<s_*\le1$. Assume that
$G_L(s)=G_L+sV_L$ has a unique ground state and a gap greater than $g$
for $0\le s\le s_*$. Let $e_L(s)$ be the ground state energy of $G_L(s)$, and let
$U_L(s)$ be the spectral flow with this filter threshold, defined by
\[
 \dot U_L(s)=iU_L(s)\mathcal D_{G_L(s), w_g}(V_L),\qquad U_L(0)=\one,
\]
where $\mathcal{D}_{G_L(s), w_g}(V_L)$ is defined through Eq. \eqref{eq:filter}. Then there is a constant $C_*$, depending only on the uniform shell bounds (i.e., $J$, $a$ and $\alpha$ in Eq. \eqref{eq:shell}),
$f_0$, $g$ and the filter function $w_g$, such that
\begin{equation}\label{eq:transported-energy-defect-bound}
 U_L(s)(G_L(s)-e_L(s)\one)U_L(s)^{\dagger}
 \ge(\kappa-sC_*\varepsilon)\mathsf N_L,
 \qquad 0\le s\le s_*.
\end{equation}
Moreover, whenever $\kappa-sC_*\varepsilon>0$,
$\gap(G_L(s))\ge\kappa-sC_*\varepsilon$.
The constant $C_*$ is independent of $L$, $s_*$, and $\varepsilon$.
\end{lemma}
\begin{proof}
Let $P_L(s)$ be the ground state projector along the path. Lemma~\ref{lem:spectral-flow-transport}
gives $U_L(s)P_L(s)U_L(s)^{\dagger}=P_L(0)=P_{\T_L}^{\Omega}$.
Define
\begin{equation} \label{eq: define tilde A}
 \mathcal G_L(s):=U_L(s)(G_L(s)-e_L(s)\one)U_L(s)^{\dagger},
 \qquad
 \widetilde A_{x,L}(s):=A_{x,L}(s)
 -\bra{\Omega_L}A_{x,L}(s)\ket{\Omega_L}\one,
\end{equation}
where $A_{x,L}(s)$ is transported filtered shell term given by Eq. \eqref{eq:anchor} for the shells of $V_L$. The important step in this proof is to control the derivative $\frac{d\mathcal{G}_L(s)}{ds}:=\dot{\mathcal{G}}_L(s)$, which turns out to be related to $\tilde A_{x, L}(s)$.

Each filtered shell term used above, i.e., $F_{G_L(s), w_g}(v_L(x, r))$, commutes with $P_L(s)$ because the entire path of Hamiltonians is assumed to have a unique ground state and a gap larger than $g$, so each $A_{x,L}(s)$ commutes
with $P_{\T_L}^{\Omega}$ and
$\widetilde A_{x,L}(s)\ket{\Omega_L}=0$.
For the ground state $\ket{\psi_L(s)} = U_L(s)^\dagger\ket{\Omega}$, the
Hellmann-Feynman theorem and the preservation of diagonal matrix elements
by the filter give
\begin{equation} \label{eq: calculating e'}
 \frac{de_L(s)}{ds}=\bra{\psi_L(s)}V_L\ket{\psi_L(s)} 
 =\sum_{x,r}\ev{\psi_L(s)|F_{G_L(s)}(v_L(x,r))|\psi_L(s)}
 =\sum_x\bra{\Omega_L}A_{x,L}(s)\ket{\Omega_L}.
\end{equation}
Differentiating $\mathcal G_L$ and using Eqs. \eqref{eq:exactfilter}, \eqref{eq: define tilde A} and \eqref{eq: calculating e'} therefore yield
\begin{equation}\label{eq:transported-derivative}
 \dot{\mathcal G}_L(s)
 =U_L(s)F_{G_L(s)}(V_L)U_L(s)^{\dagger}-e_L'(s)\one
 =\sum_x\widetilde A_{x,L}(s).
\end{equation}
So controlling $\dot{\mathcal{G}}_L(s)$ amounts to controlling $\sum_x\tilde A_{x, L}(s)$.

The locality estimate of Eq. \eqref{eq:anchor} now reads
\begin{equation}\label{eq:transported_perturbation_bound}
 \|A_{x,L}(s)\|\le C\varepsilon,\qquad
 \|A_{x,L}(s)-\E_{B_L(x,r)}^{\tr}A_{x,L}(s)\|
 \le\varepsilon f(r),
\end{equation}
where $f(r)$ is a stretched-exponential function and all its constants are uniform in $L$ and
$0\le s\le s_*$. These constants can be chosen independently of
$\varepsilon\le1$, since $g_L(x, r)+sv_L(x, r)$ has stretched-exponential locality for any $0<\varepsilon\leqslant 1$ and $0\leqslant s\leqslant 1$, where $g_L(x, r)$ is the shell term of $G_L$ in Eq. \eqref{eq:shell}.

Both $\E_X^\Omega$ and $\E_X^{\tr}$ are contractions, so
\begin{equation}
\begin{split}
 \| A_{x,L}(s) -\E_X^{\Omega} A_{x,L}(s) \| & = \|A_{x,L}(s) - \E^{\tr}_X(\E_X^{\Omega} A_{x,L}(s))\|\\
 &\leq \norm{A_{x,L}(s) - \E^\tr_X A_{x,L}(s) } + \norm{\E^\tr_X \big(A_{x,L}(s) - \E^\Omega_X A_{x,L}(s) \big) } \\
 &\le 2\| A_{x,L}(s) -\E_X^{\tr}A_{x,L}(s)\|,
\end{split}
\end{equation}
so
\begin{equation}\label{eq:Atilde}
\|\widetilde A_{x,L}(s)\|\le2C\varepsilon,\qquad
 \|\widetilde A_{x,L}(s)
   -\E_{B_L(x,r)}^{\Omega}\widetilde A_{x,L}(s)\|
 \le2\varepsilon f(r).
\end{equation}

We can now construct the shell representation for $\tilde A_{x,L}(s)$ while preserving the vacuum annihilation property.
Set $\E_{x,r}=\E_{B_L(x,r)}^{\Omega}$ for $0\le r\le M_L$,
with $\E_{x,-1}=0$ and $\E_{x,M_L}$ the identity map, and define
\[
 a_{x,r}(s):=(\E_{x,r}-\E_{x,r-1})\widetilde A_{x,L}(s).
\]
This sum telescopes to $\widetilde A_{x,L}(s)$. Each shell is self-adjoint,
is supported on $B_L(x,r)$, and annihilates its local vacuum by
Lemma~\ref{lem:vacuum}. The preceding estimates give
$\|a_{x,r}(s)\|\le\varepsilon f_1(r)$, where
\[
 f_1(0)=2C,\qquad f_1(r)=2\bigl(f(r)+f(r-1)\bigr)\quad(r\ge1).
\]
Applying Eq. \eqref{eq:vacuumform} to these shells, using
$\sum_x\mathsf N_{B_L(x,r)}\le(2r+1)\mathsf N_L$ and substituting Eq. \eqref{eq:transported-derivative} give
\begin{equation}\label{eq:transported-relative-bound}
 -C_*\varepsilon\mathsf N_L\le\dot{\mathcal G}_L(s)
 \le C_*\varepsilon\mathsf N_L,\qquad
 C_*:=\sum_{r\ge0}(2r+1)f_1(r)<\infty.
\end{equation}
Since $\mathcal G_L(0)=G_L$, integrating the lower bound gives
\begin{equation}\label{eq:gl_kappaN}
 \mathcal G_L(s)\ge G_L-sC_*\varepsilon\mathsf N_L
 \ge(\kappa-sC_*\varepsilon)\mathsf N_L.
\end{equation}

Finally, $\mathcal G_L(s)$ has ground state $\ket{\Omega_L}$ and ground-state
energy zero. In the Hilbert subspace orthogonal to $\ket{\Omega_L}$, we have $\mathsf N_L\ge\one$, so the Eq. \eqref{eq:gl_kappaN} gives $\gap(\mathcal G_L(s))\geq \kappa-sC_*\varepsilon$
whenever $\kappa-sC_*\varepsilon>0$. Therefore, $\gap(G_L(s))=\gap(\mathcal{G}_L(s))\geqslant\kappa-sC_*\varepsilon$.
\end{proof}

The above results are very close to proving the stability of gap of normal-form Hamiltonians, but as stated they do not prove the desired stability yet, because the constant $s_*$ above may decrease as the system size increases. Below we use a bootstrap-type argument to finish proving the stability of gap of normal-form Hamiltonians.

\begin{theorem}[Stability of gap of normal-form Hamiltonians]\label{thm:productstability}
Let $\{G_L\}$ be a family of normal-form Hamiltonians, namely, each $G_L$ admits a shell representation as in Eq.~\eqref{eq:shell} and
$G_L\ket{\Omega_L}=0$. Suppose that
$G_L\ge\kappa \mathsf{N}_L$ for a fixed $\kappa>0$.  Let $\{V_L\}$ be a family of perturbations with a shell representation and satisfy
\begin{equation}
    V_L = \sum_{x,r} v_{L}(x,r),\qquad \norm{v_L(x,r)}\leq \varepsilon f_0(r)
\end{equation}
with $\varepsilon$ an $L$-independent constant and $f_0(r)$ an $L$-independent stretched-exponential function in $r$. 
Then there is an $L$-independent $\varepsilon_*>0$, such that
$\gap(G_L+V_L)\ge\kappa/2$ whenever $\varepsilon<\varepsilon_*$.  The ground state of $G_L+V_L$ remains unique.
\end{theorem}

\begin{proof}
For a fixed length $L$, since $\|V_L\|$ is finite, there exists an $s_*>0$ such that $G_L(s) = G_L + sV_L$ always has a gap greater than $\kappa/2$ for $s\leq s_*$ according to the Weyl inequality. Now we apply Lemma~\ref{lem:transported-gap} with the filter threshold $g=\kappa/2$, and we conclude that there is a constant $C_*$ independent of $L,s_*$ and $\varepsilon$, such that the condition
\[
U_L(s) (G_L(s) - e_L(s) \one ) U_L(s)^\dagger \geq (\kappa-sC_*\varepsilon) \mathsf{N}_L
\]
holds. From the proof of Lemma~\ref{lem:transported-gap}, we see that this condition ensures that $G_L(s)$ is still gapped. But in general $s_*$ depends on $L$, and we need to prove that there exists an $\varepsilon_*$, such that for any $0<\varepsilon<\varepsilon_*$, the above condition holds, and $\kappa-sC_*\varepsilon>\kappa/2$ for all $L$ and all $0\leq s\leq1$. 

To get this, we use the gap-continuity argument of
Ref.~\cite[Sec.~7]{BravyiHastings2011}.
Choose
\begin{equation}\label{eq:choose_epsilon}
\varepsilon_*\le\min\{1,\kappa/(4\max(C_*,1))\}.
\end{equation}  If
$\varepsilon<\varepsilon_*$ and $0\le s\le1$, then
\[
 \kappa-sC_*\varepsilon\ge\frac{3\kappa}{4}.
\]
Moreover,
\[
 \mathsf{N}_L\ge\one-P_{\T_L}^\Omega,
\]
because $\mathsf{N}_L$ has eigenvalue zero only on $\ket{\Omega_L}$ and has eigenvalue at
least one on every other product-basis vector.  Therefore, wherever the
preceding spectral-flow calculation is valid,
\begin{equation}
 U_L(s)(G_L(s)-e_L(s)\one)U_L(s)^{\dagger}
 \ge\frac{3\kappa}{4}(\one-P_{\T_L}^\Omega).
 \label{eq:continuation-gap}
\end{equation}
This inequality says that the ground state is unique and that the gap is at
least $3\kappa/4$, which is strictly larger than the filter threshold
$g=\kappa/2$.

We now use the above analysis to justify that the calculation actually applies along the entire path with $0\leqslant s\leqslant 1$. In other words, now we take the above $s_*$ to be $s_*=\sup\{t\in[0, 1]: \gap(G_L(s))>\kappa/2\ {\rm for\ every\ }s\in[0, t]\}$, and we will prove that $s_*=1$.

Suppose that $s_*<1$. Then Eq. \eqref{eq:continuation-gap} holds for every
$s<s_*$. Note that the matrices $G_L(s)$ and $U_L(s)$, their eigenvalues, and the
ground energy $e_L(s)$ are continuous in $s$.  Taking $s\uparrow s_*$ in Eq.
\eqref{eq:continuation-gap} therefore shows that $G_L(s_*)$ still has a
unique ground state and a gap at least $3\kappa/4$. The continuity of the eigenvalues then
keeps the gap above $g=\kappa/2$ on a neighborhood to the right of $s_*$, contradicting
the definition of $s_*$.  Hence $s_*=1$, and Eq.
\eqref{eq:continuation-gap} at $s=1$ proves the asserted bound of gap, i.e., $\gap(G_L+V_L)\ge\kappa/2$.

Finally, $C_*$ depends only on the uniform shell and locality constants, so
$\varepsilon_*$, which is defined by Eq. \eqref{eq:choose_epsilon}, can be taken independently of $L$. Namely, although the initial interval used
in the continuation argument generally depends on $L$, the same threshold
$\varepsilon_*$ applies for every $L$.

\end{proof}

\subsubsection{Normal-form and locality in product-vacuum frame}
\label{subsubsec:product-vacuum-locality}

Theorem \ref{thm:productstability} establishes the stability of gap of normal-form Hamiltonians. The Hamiltonians considered in Theorem \ref{thm:main} are general gapped local translation invariant Hamiltonians with a unique ground state, which may not be normal-form Hamiltonians. In the following, we will see that the transformed Hamiltonian $K_L$ in Eq. \eqref{eq:mechanism} is a normal-form Hamiltonian and has the same excitation spectrum as $\widehat{H}_L$. Moreover, $K_L$ satisfies the locality hypothesis needed in Theorem \ref{thm:productstability}, and so does the perturbation after being transformed into the product-vacuum frame as in Eq. \eqref{eq:mechanism}. These results ultimately lead to Theorem \ref{thm:main}.

\begin{proposition}[Shell representation of $K_L$ and perturbations]\label{prop:shelltransfer}
Suppose $T_L$ is a unitary such that $T_L\ket{\Psi_L}=\ket{\Omega_L}$, and $T_L$ is exponentially local
in the sense of Eq. \eqref{eq:uniform_locality}.
Then $K_L$ in Eq. \eqref{eq:mechanism} is a normal-form Hamiltonian and satisfies $0\le K_L\le C \mathsf{N}_L$.

If
$V_L=\sum_X\Xi_L(X)$ has range $R$, then its lift
$\widehat V_L:=V_L\otimes\one$ satisfies
$T_L\widehat V_LT_L^{\dagger}=\sum_{x,m}\widetilde v_L(x,m)$, where
\begin{equation}
    \tilde v_L(x, m)=\tilde v_L(x, m)^\dag\in\mathcal{A}_{B_L(x, m)},
    \quad
    \norm{\tilde v_L(x,m)} \leq C_R \norm{\Xi_L}_{\rm inc} e^{-bm},
\end{equation}
where $C_R>0$ and $b>0$ are $L$-independent constants.
\end{proposition}
\begin{proof}

We will prove this proposition by constructing the shell representations of $K_L$ and $V_L$ that satisfy the required locality condition. The strategy to construct the shell representations of both $K_L$ and $V_L$ is the same. In both cases, we will find a set of operators $A=A^{\dagger}\in\A_{B_L(x,r_0)}$ with fixed $r_0$, and set
$B=T_LAT_L^{\dagger}$. Then such operators $B$ will be used to construct the shell terms satisfying the relevant locality condition, as discussed below.

We first establish a general result to be used below. For any $A=A^{\dagger}\in\A_{B_L(x,r_0)}$ with a fixed $r_0$ and $B=T_LAT_L^{\dagger}$, the exponential locality in Eq. \eqref{eq:uniform_locality}
and contractivity of the normalized partial trace give
\[
 \|B-\E_{B_L(x,m)}^{\tr}B\|
 \le \inf_{A_m\in \A_{B_L(x,m)}}(\norm{B-A_m} + \norm{\E^{\tr}_{B_L(x,m)}[A_m-B] }) \leq C_{r_0}\|A\|e^{-bm},\qquad 0\le m\le M_L,
\]
with $C_{r_0},\, b>0$ independent of $L$ . For $m\ge r_0$, the last inequality comes directly from Eq.
\eqref{eq:uniform_locality} with the enlargement radius $m-r_0$. For $m<r_0$, as $\|B-\E_{B_L(x,m)}^{\tr}B\|$ is bounded, we can always choose $C_{r_0}$ so that the above inequality still holds.

Now we define
\begin{equation} \label{eq: shells of transformed Hamiltonian}
 \Delta_{x,0}(B) = \E_{B_L(x,0)}^{\tr}(B),\qquad
 \Delta_{x,m}(B):=
 (\mathcal \E_{B_L(x,m)}^{\tr}-\E_{B_L(x,m-1)}^{\tr})B,\quad 1\leq m\leq M_L.
\end{equation}
Then $\sum_{m=0}^{M_L}\Delta_{x,m}(B)=B$. Every shell, i.e., $\Delta_{x, m}(B)$ for $0\leqslant m\leqslant M_L$, is self-adjoint and supported in $B_L(x,m)$. Note that for $m\geqslant 1$, $\|\Delta_{x, m}(B)\|=\|(\mathcal \E_{B_L(x,m)}^{\tr}-\E_{B_L(x,m-1)}^{\tr})(B-\E^{\tr}_{B_L(x, m-1)}B)\|\leqslant 2C_{r_0}\|A\|e^{-b(m-1)}$. Moreover, $\|\Delta_{x, 0}(B)\|\leqslant\|A\|$. So
\[
 \|\Delta_{x, m}(B)\|=\|\Delta_{x,m}(T_LAT_L^{\dagger})\|
 \le C'_{r_0}\|A\|e^{-bm}.
\]
with $C'_{r_0}=\max\{1, 2C_{r_0}e^b\}$. Note that both $C'_{r_0}$ and $b$ are $L$-independent.

Now we apply the above strategy to construct a shell representation of $K_L$. Here we take the operators $A$ above as $A_x=\widehat h_{x,L}-(2E_L/L)\one$ and sum over $x$.
Each $A_x$ has a fixed range and a uniformly bounded norm, since
$\|\widehat h_{x,L}\|\le2\|\Phi_0\|$ and $|2E_L/L|\le2\|\Phi_0\|$ with $\Phi_0$ below Eq.~\eqref{eq: unperturbed Hamiltonian family}. Moreover, $\sum_{x\in\mathbb{T}_L}T_L A_xT_L^\dag=K_L$. According to the above discussion, $K_L$ has a uniform exponential shell
representation, where each shell term is $\Delta_{x, m}(T_LA_xT_L^\dag)$ defined via Eq. \eqref{eq: shells of transformed Hamiltonian}. By construction, $K_L$ has a unique ground
state $\ket{\Omega_L}$, zero ground state energy, and a gap at least $\widehat\gamma$. Hence $K_L$ is a normal-form Hamiltonian, and
Lemma~\ref{lem:normal-form-transport} gives $0\le K_L\le C\mathsf N_L$.

For the perturbation, choose a site $x(X)\in X$ for each nonzero
interaction term and group the lifted terms by this site:
\[
 v_{x,L}:=\sum_{X:\,x(X)=x}\Xi_L(X)\otimes\one.
\]
The range condition implies $v_{x,L}\in\A_{B_L(x,R)}$, and the incidence
norm gives $\|v_{x,L}\|\le\|\Xi_L\|_{\inc}$.
Now we take the operator $A$ to be $A_x=v_{x,L}$. Then the above discussion shows that
$\widetilde v_L(x,m)=\Delta_{x, m}(T_LA_xT_L^\dag)=\Delta_{x,m}(T_Lv_{x,L}T_L^{\dagger})$ satisfies
\[
\tilde v_L(x, m)=\tilde v_L(x, m)^\dag\in\mathcal{A}_{B_L(x, m)},
\qquad
 T_L\widehat V_LT_L^{\dagger}=\sum_{x,m}\widetilde v_L(x,m),\qquad
 \|\widetilde v_L(x,m)\|\le C_R\|\Xi_L\|_{\inc}e^{-bm},
\]
where $C_R$ and $b$ are
independent of $L$.
\end{proof}

\subsection{Proof of the spin part of Theorem \ref{thm:main} and Corollary \ref{corollary: continuity of gap}}
\label{subsec:main_proof}

With all the preparations above, we can now prove the spin part of Theorem \ref{thm:main} and Corollary \ref{corollary: continuity of gap}. We start with Theorem \ref{thm:main}, which does not assume that the perturbation is translation invariant.

\begin{proof}[Proof of Theorem~\ref{thm:main} for spins]
Fix the perturbation range $R$, and let
$\widehat V_L=V_L\otimes\one$ act on the physical copy of the enlarged
spin system. Theorem~\ref{thm:prep-exact}, which will be proved later, supplies an exponentially local
unitary $T_L$ such that $T_L\ket{\Psi_L}=\ket{\Omega_L}$. Define
\begin{equation}\label{eq:KL}
 K_L=T_L(\widehat H_L-2E_L\one)T_L^{\dagger},
 \qquad
 \widetilde V_L=T_L\widehat V_LT_L^{\dagger}.
\end{equation}

Now we will apply the stability theorem in Theorem~\ref{thm:productstability} to these $K_L$ and $\widetilde V_L$. Its four hypotheses in the present
setting are:
\begin{enumerate}[label=(\roman*),leftmargin=*,nosep]
 \item $K_L$ has an exact stretched-exponential shell representation with
 constants uniform in $L$;
 \item $K_L\ket{\Omega_L}=0$;
 \item $K_L\ge\kappa \mathsf{N}_L$ with $\kappa>0$ independent of $L$; and
 \item $\widetilde V_L$ has a stretched-exponential shell representation
 whose strength is smaller than the threshold $\varepsilon_*$.
\end{enumerate}
Proposition~\ref{prop:shelltransfer} proves (i), (ii),
and the required shell representation in (iv), with shell strength at most
$C_R\|\Xi_L\|_{\inc}$. Theorem~\ref{thm:energy} proves (iii). By Theorem~\ref{thm:productstability}, there is a small threshold $\varepsilon_*$, such that when $C_R\norm{\Xi_L}_{\rm inc}<\varepsilon_*$ the gap of $K_L$ is stable in the sense of this theorem. Therefore, we can choose
\[
 \varepsilon_R=\varepsilon_*/C_R,
 \qquad \gamma_R=\kappa/2.
\]
Let $L_0$ be the threshold in the original unique-ground-state and uniform-gap assumption, and let $L_{\mathrm{ED}}$ be the size threshold for $K_L\geq \kappa \mathsf N_L$ in Theorem~\ref{thm:energy}, we choose
$L_*=\max\{L_0,L_{\mathrm{ED}}\}$. Then the assumption that
$\|\Xi_L\|_{\rm inc}<\varepsilon_R$ implies that
\[
 \gap(K_L+\widetilde V_L)\ge\kappa/2
\]
for any $L>L_*$.

Therefore, the uniform gap of $H_L+V_L$ can be proved, as
\[
\gamma_R=\kappa/2\leqslant\gap(K_L+\widetilde V_L)
 =\gap(\widehat H_L+\widehat V_L)
 =\min\{\gap(H_L+V_L),\gap(H_L)\}\leqslant\gap(H_L+V_L)
\]
for any $L>L_*$. Note that the eigenvalues in the definition of $\gap$ are
counted with multiplicity, so
$\gap(H_L+V_L)=\lambda_2(H_L+V_L)-\lambda_1(H_L+V_L)>0$ also implies that
the perturbed ground eigenspace is one dimensional. This proves all claims
of Theorem~\ref{thm:main} for spin systems.
\end{proof}

Next, we turn to Corollary \ref{corollary: continuity of gap}.

\begin{corollary}[Continuity of gap, spin part]
\label{cor:gap-continuity_corollary}
Under the assumptions of Theorem~\ref{thm:main}, fix a perturbation range
$R$. There are constants $L_*,\eta_R,c_R>0$ such that, for every $L>L_*$
and every interaction $\Xi_L$ with range at most $R$ and strength
$\|\Xi_L\|_{\inc}<\eta_R$,
\begin{equation}
 \bigl|\gap(H_L+V_L)-\gap(H_L)\bigr|
 \le c_R\|\Xi_L\|_{\inc},
 \label{eq:gap-continuity-2}
\end{equation}
where $V_L=\sum_X\Xi_L(X)$. The constants are independent of $L$ and whether the
perturbation is translation invariant. 
\end{corollary}

\begin{proof}
This proof follows from the path of the spectral flow of the Hamiltonians from the unperturbed Hamiltonian to the perturbed Hamiltonian, which is differentiable. For notational convenience, let $\eta=\norm{\Xi_L}_{\rm inc}$ and $\Delta_L = \gap(H_L)$. We want to bound the $\gap(H_L+V_L)$ by $\Delta_L$ up to some correction that is controlled by perturbation strength $\eta$. 

To get this bound, consider the perturbation on both copies,
\[
 \widehat V_L=V_L\otimes\one+\one\otimes V_L^c,\qquad
 \widehat H_L+s\widehat V_L=(H_L+sV_L)+(H_L+sV_L)^c.
\]
The two copies have the identical spectra and act independently, so the gap of the total Hamiltonian is $\gap(H_L+sV_L)$. By Theorem~\ref{thm:main}, the $\gap(H_L +sV_L)\geq \kappa/2$ for $\eta<\varepsilon_R, 0\leq s\leq 1$.

We turn to the product-vacuum frame to monitor the energy gap along the path. Let $K_L = T_L(\widehat H_L-2E_L)T_L^\dagger$, and $W_L = T_L \widehat V_L T_L^\dagger$. Theorem~\ref{thm:energy} and Proposition~\ref{prop:shelltransfer} give
$\kappa\mathsf N_L\le K_L\le C_0\mathsf N_L$, with uniform
$\kappa,C_0>0$. Let $W_L=\sum_{x,r} w_L(x,r)$ be the shell representation constructed analogously to Proposition~\ref{prop:shelltransfer}, then 
\begin{equation}
    \norm{w_L(x,r)} \leq 2C_R \norm{\Xi_L}_{\rm inc} e^{-br} = 2C_R\eta e^{-br}.
\end{equation}
with constants $C_R, b$ in Proposition~\ref{prop:shelltransfer}.

Choose $\eta_R$ small enough so that the continuation argument in Theorem~\ref{thm:productstability} applies for all $0\leq s\leq 1$. A sufficient choice, as will be clear, is $\eta_R\leq \min\{\varepsilon_R, \frac{1}{2C_R}, \frac{\kappa}{4C_R C_*}\}$ (in fact, $\eta_R\leqslant\min\{\varepsilon_R, \frac{1}{2C_R}\}$ is sufficient for invoking Theorem~\ref{thm:productstability}, and the condition $\eta_R\leqslant\frac{\kappa}{4C_R C_*}$ will be used later to go from Eq. \eqref{eq:KL_inequality} to Eq. \eqref{eq: many-body Weyl}). Now we define the operator of excitation energy along the flow,
\begin{equation}
    \mathcal{K}_L(s) := U_L(s) (K_L + sW_L - e_L(s)) U_L(s)^\dagger,
\end{equation}
where $U_L(s)$ is the spectral flow unitary satisfying 
\[
\dot U_L(s) = i U_L(s)\mathcal D_{K_L + sW_L, w_g}(W_L),\quad U_L(0)=\one,
\] 
and $e_L(s)$ is the ground state energy of $K_L + sW_L$. Here the filter threshold $g$ is taken to be $g=\kappa/4$. Notice that the ground state of $\mathcal{K}_L(s)$ is $\ket{\Omega_L}$ as discussed in the transportation of the ground state, and $\mathcal K_L(0)= K_L$. Using the argument similar to Eqs. (\ref{eq:transported-derivative}-\ref{eq:transported-relative-bound}), we can get 
\begin{equation}
    - 2C_* C_R\eta \mathsf{N}_L \leq \frac{d}{ds} \mathcal{K}_L(s) \leq 2 C_* C_R \eta \mathsf N_L.
\end{equation}
Hence, integrating from $s=0$ to $s=1$, and using $\mathsf N_L\leq K_L/\kappa$, we get
\begin{equation}\label{eq:KL_inequality}
    (1- a_R \eta ) K_L \leq \mathcal K_L(1) \leq (1+ a_R \eta ) K_L,\qquad a_R = 2C_R C_*/\kappa.
\end{equation}

To obtain a bound of gap, we use an argument similar to those in Refs. \cite{Kuwahara2019, Liu2024}. We take a normalized product vector $\ket{v}$
with exactly one site orthogonal to its vacuum. Since $\ket{v}\perp\ket{\Omega_L}$
and $\bra{v}\mathsf N_L\ket{v}=1$, then we get $\Delta_L\le\bra{v}K_L\ket{v}\le C_0$ since $K_L\leq C_0 \mathsf N_L$. Minimizing the quadratic form in Eq. \eqref{eq:KL_inequality} with respect to states orthogonal to $\ket{\Omega_L}$ and using $\gap(\mathcal K_L(1)) = \gap(H_L +V_L)$, we get
\begin{equation} \label{eq: many-body Weyl}
    (1-a_R\eta)\Delta_L \leq \gap(H_L + V_L) \leq (1+a_R \eta)\Delta_L.
\end{equation}
Thus Eq. \eqref{eq:gap-continuity} follows with $c_R=a_R C_0$. Note that the condition $\eta<\frac{\kappa}{4C_RC_*}$ implies that $1-a_R\eta>0$, which is needed to go from Eq. \eqref{eq:KL_inequality} to Eq.~\eqref{eq: many-body Weyl}.

\end{proof}

\section{Agreements across ring sizes}
\label{sec:across_sizes}

In this section, we present the complete statement of Proposition \ref{prop:circumference} and its proof. Intuitively, every $H_L$ is built from the same local interaction $\Phi_0$, so we expect that the reduced density matrices of local regions in the ground state of systems with different sizes are similar to each other, and systems with different sizes have similar ground state energy densities. Proposition~\ref{prop:circumference} below proves that their energy densities and reduced density matrices on a common short interval differ by exponentially small amounts when both circumferences are large.

Throughout this section, $H_L$ denotes the family in Eq.~\eqref{eq: unperturbed Hamiltonian family}, with the scalar $c_L$ in Eq.~\eqref{eq: decomposing translation invariant Hamiltonian} removed.

\begin{proposition}[Agreement across sizes]\label{prop:circumference}
Let $H_L$ be a local translation invariant Hamiltonian satisfying the uniform gap condition with a unique ground state on a ring of length $L$, as discussed around Eq. \eqref{eq:assumption}. For such a length-$L$ ring, denote by $E_L$ the ground state energy and by $\rho_L=|\psi_L\rangle\langle\psi_L|$ the density matrix of the ground state. There are constants $\mathcal E_\infty\in\mathbb R$, $C_E,c_E,C_{\rm marg}>0$, and $L_1$ such that
for $L,N\ge L_1$,
\begin{equation}\label{eq:energyconvergence}
 \left|\frac{E_L}{L}-\frac{E_N}{N}\right|\le C_E e^{-c_E \min(L,N)},
 \qquad |E_L-L\mathcal E_\infty|\le C_ELe^{-c_E L}.
\end{equation}
For $L\geq N\geq L_1$ and $1\le \ell\leq\lfloor N/12\rfloor$,
let $I=\{0,\ldots,\ell-1\}$ in both rings. Identifying the corresponding on-site Hilbert spaces, then the following inequality holds:
\begin{equation}\label{eq:marginalagreement}
 \|\rho_{L,I}-\rho_{N,I}\|_1\le C_{\rm marg}e^{-c_E N/4},
\end{equation}
where $\rho_{L, I}$ and $\rho_{N, I}$ are the reduced density matrices of $\rho_L$ and $\rho_N$ on the region $I$.
\end{proposition}

The idea to prove Proposition~\ref{prop:circumference} is to change the ring size while keeping the reduced density matrices on short intervals almost unchanged. This is possile because Lemma~\ref{lem:decoupling} shows that
distant intervals are nearly unentangled. Then we can take the reduced density matrix of one interval from a longer ring, and put a few copies of this interval on disjoint intervals on a shorter ring and use Lemma~\ref{lem:erasure}
to recover the intermediate regions between these intervals. The so-constructed state serves as a state that is close to both marginals of the longer ring and the shorter ring when restricted to an interval. Hence the inequalities in the proposition can be proved.

\subsection{Properties of gapped ground states}
\label{subsec:property_gapp_ground_state}

To prove Proposition~\ref{prop:circumference}, in this subsection we first establish two properties of the gapped ground states $\ket{\psi_L}$. The first property, Lemma \ref{lem:decoupling}, says that regions far away are nearly unentangled in $|\psi_L\rangle$.

\begin{lemma}[Trace-norm decoupling]\label{lem:decoupling}
Denote by $\rho_{L, X}$ the reduced density matrix of the ground state $\rho_L$ on a region $X\subseteq\mathbb{T}_L$, and by $\lambda_j(\rho_{L, X})$ the $j$-th eigenvalue of $\rho_{L, X}$, ordered as $\lambda_1\geqslant\lambda_2\geqslant\cdots$. There are $L$-independent constants $C,c,a>0$, such that for every interval $X$, the Schmidt tail satisfies
\begin{equation}\label{eq:schmidttail}
\sum_{j>K}\lambda_j(\rho_{L,X})\le CK^{-a},\qquad \forall\ K\ge1.
\end{equation}
Moreover, for disjoint intervals $X$ and $Y$ with a distance at least $r$,
\begin{equation}\label{eq:decoupling}
 \|\rho_{L,XY}-\rho_{L,X}\otimes\rho_{L,Y}\|_1\le Ce^{-cr}.
\end{equation}
\end{lemma}

\begin{proof}

This proof consists of three steps. First, we discuss the Schmidt tail in Eq. \eqref{eq:schmidttail}, which follows from Hastings' estimation. Then we discuss the clustering property of correlation functions of operators. In third step, we combine the results of the two previous steps to get Eq. \eqref{eq:decoupling}.

{\it Step 1: Schmidt tail.} To get Eq. \eqref{eq:schmidttail}, we use Hastings' open-chain result \cite[Eqs.~(2), (27), and (28)]{Hastings} for our periodic chain, by a folding trick similar to the ones in Refs. \cite[Section 3.1.4]{Verstraete_2008} and \cite[Appendix B. 7]{Liu2024}.
 
For an open chain with a uniform upper bound $d_*$ of on-site dimensions, local interactions, and a unique ground state with a uniform spectral gap, Hastings' estimation gives constants $k_*, C_*, \xi_*>0$, such that the eigenvalues at each spatial cut satisfy
\begin{equation} \label{eq: Hastings's Schmidt tail}
    \sum_{j>k_* d_*^{2m} } \lambda_j \leq C_* e^{-2m/\xi_*},\qquad m\geq 0.
\end{equation}
Here the constants depend only on the local bounds of onsite dimension $d_*$, the interaction strength and the spectral gap, independent of the length of the chain or the position of the cut.

\begin{figure}
    \centering
    \includegraphics[width=0.6\linewidth]{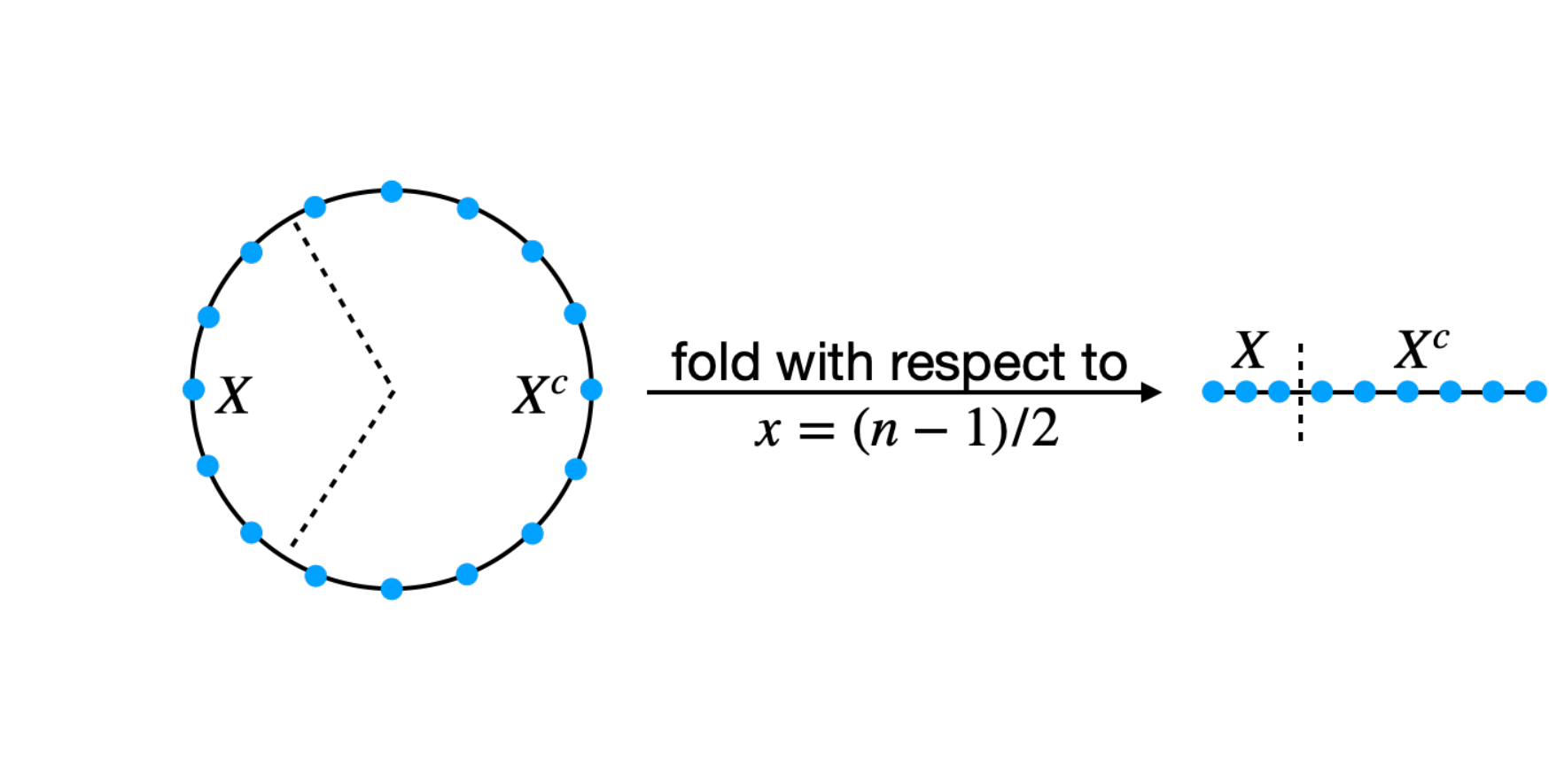}
    \caption{The ring is folded into a chain with open ends under the reflection $j\mapsto n-1-j\pmod L$. Each new unit cell contains two original sites, except at fixed points satisfying $2j\equiv n-1\pmod L$, which give single-site cells.
    }
    \label{fig: folding}
\end{figure}

To apply this estimate to an interval $X=\{0,1,\cdots, n-1\}$ on a ring, as in Fig. \ref{fig: folding}, we fold the ring with respect to $x=(n-1)/2$. In other words, we reflect the ring by $j\mapsto n-1-j\ ({\rm mod}\ L)$. Each reflection orbit contains one or two ring sites. We regard each orbit as a new cell, and order these cells by their distance from the midpoint of $X$. This order of cells gives an open chain in which $X$ contains the first sites.

Since each interaction term acts on at most $R_0+1$ consecutive ring sites, we can group consecutive cells into blocks, such that 1) the original Hamiltonian becomes nearest-neighbor under the blocking, 2) the cut between $X$ and $X^c$ is between two blocks. One way is to take $b=\max\{1,R_0\}$, and group every $b$ consecutive cells to the left of the cut $X | X^c$, and merge any remaining cells near the cut into the adjacent block. Similarly for the $X^c$ side. Under such grouping, the largest block contains at most $b+(b-1)=2b-1$ cells.  Hence we can use Hastings' estimate above to get constants $k_*, C_*, \xi_*$ for an on-site dimension at most $d_* = D^{4b-2}$. In particular, if $K\geqslant K_0:=k_*d_*^2$ so that $\log(K/k_*)/(2\log d_*)\geq 1$, then we take $m=\lfloor\log(K/k_*)/(2\log d_*)\rfloor$, Eq. \eqref{eq: Hastings's Schmidt tail} implies Eq. \eqref{eq:schmidttail} with $a=1/((4b-2)\xi_*\log D)$ and $C= C_* e^{2/\xi_*} k_*^a$. If $K<K_0$, then $\sum_{j>K}\lambda_j=\sum_{K<j\leqslant K_0}\lambda_j+\sum_{j> K_0}\lambda_j\leqslant 1+CK_0^{-a}$. So in all cases, Eq. \eqref{eq:schmidttail} holds with $a=1/((4b-2)\xi_*\log D)$ and $C=K_0^a+C_*e^{2/\xi_*}k_*^a$ for all $K\geqslant 1$.

{\it Step 2: operator correlations.} To obtain Eq. \eqref{eq:decoupling}, we first prove a weaker version of it, which concerns the correlation estimate of operator product, as shown in Eq.~\eqref{eq:covariance}. This result follows from the standard exponential-clustering argument of Refs.~\cite{Hastings-Koma_2006,NachtergaeleSims2006}, and we briefly discuss its procedures for completeness. The finite-range Lieb-Robinson bound gives, for $A\in\A_X$ and $B\in\A_Y$ with distance $r$,
\[
 \|[\tau_t(A),B]\|\le C\|A\|\|B\|(e^{v|t|}-1)
       \sum_{x\in X,y\in Y}e^{-\mu d_L(x,y)}
 \le C'\|A\|\|B\|(e^{v|t|}-1)e^{-\mu r},
\]
where $\tau_t(A)=e^{iH_Lt}Ae^{-iH_Lt}$, $v>0$ is a constant, and the other constant $C'>0$ does not depend on $|X|$ or $|Y|$.\footnote{To see that $C'$ does not depend on $|X|$ or $|Y|$, we just need to show that $S_{X,Y} = \sum_{x\in X,y\in Y} e^{-\mu d_L(x,y)}\leqslant C'e^{-\mu r}$ with $C'$ independent of $|X|$ or $|Y|$. Note that there are two directions of counting distances on a ring, so $S_{X,Y}$ is upper bounded by the sum of two items, where one of them is summing along clockwise direction and the other is counter-clockwise direction. For instance, denote the distance between $X$ and $Y$ along clockwise direction by $r_1$, then the clockwise item is $e^{-\mu r_1} \left(\sum_{i=0}^{|X|-1} e^{-\mu i}\right)\left(\sum_{j=0}^{|Y|-1} e^{-\mu j}\right)$, which is upper bounded by $e^{-\mu r_1}/(1-e^{-\mu})^2$. The counter-clockwise item is similarly upper bounded by $e^{-\mu r_2}$, with $r_2$ the  counterclockwise distance. So $S_{X, Y}\leqslant C'e^{-\mu r}$, with $C'=2/(1-e^{-\mu})^2$ and $r=\min\{r_1, r_2\}$. We see that $C'$ indeed does not depend on $|X|$ or $|Y|$.} Moreover, $C'$, $v$ and $\mu$ do not depend on the operators $A$ and $B$. To get Eq. \eqref{eq:covariance}, we first introduce \(A'=A-\Tr(\rho_LA)\one\) and \(B'=B-\Tr(\rho_LB)\one\), so the left hand side of Eq. \eqref{eq:covariance} is $\Tr(\rho_LA'B')$. The way to bound $\Tr(\rho_LA'B')$ is to note that, for $\eta>0$, we can define
\[
 A^-_\eta=\tfrac12 A'+\frac{i}{2\pi}\operatorname{pv}\!\int_{\mathbb R}
             e^{-\eta t^2}\frac{\tau_t(A')}{t}\,dt,
\]
where $\operatorname{pv}$ is the symmetric principal value. Then we can use the spectral gap to show that $|\Tr(\rho_L A'B')-\Tr(\rho_L[A^-_\eta,B'])|\leqslant C\|A'\|\|B'\|e^{-\gamma_0^2/(4\eta)}$, where $\gamma_0$ is the lower bound of the spectral gap \cite[Section 3, Eq.(3.23)]{Hastings-Koma_2006}.  
Since $X\cap Y=\varnothing$, we have $[A',B']=0$, and therefore
$$
 \Tr(\rho_L [A^-_\eta,B'])
 =\frac{i}{2\pi}\int_{\mathbb R}
 e^{-\eta t^2}\frac{\Tr(\rho_L[\tau_t(A'),B'])}{t}\,dt.
$$
In the above, the principal value is removed because the Lieb-Robinson factor $e^{v|t|}-1$ makes the integrand
bounded near $t=0$. The standard way to control the above quantity is to use the Lieb-Robinson bound for $|t|\leqslant T$, and use
$\|[\tau_t(A'),B']\|\le2\|A'\|\|B'\|$ for $|t|>T$.
Altogether, we get a new constant $C>0$ and the following bound
$$
 |\Tr(\rho_L A'B')|
 \le C\|A'\|\|B'\|
 \left(
 e^{-\mu r+vT}
 +\frac{e^{-\eta T^2}}{\eta T^2}
 +e^{-\gamma_0^2/(4\eta)}
 \right).
$$
The three terms bound the short-time integral, the long-time
Gaussian tail, and the spectral error, respectively.
Taking $T=\mu r/(2v)$ and $\eta=\gamma_0/(2T)$, it is straightforward to show the existence of constants $C, c_0>0$ such that

\begin{equation}\label{eq:covariance}
 |\Tr(\rho_L AB)-\Tr(\rho_L A)\Tr(\rho_L B)|=|\Tr(\rho_LA'B')|\le Ce^{-c_0r}\|A\|\|B\|,
\end{equation}
with $C$ and $c_0$ independent of $L,|X|,|Y|$.

{\it Step 3: trace-norm.} Note that Eq. \eqref{eq:covariance} is weaker than Eq. \eqref{eq:decoupling}, because Eq. \eqref{eq:decoupling} means that not only the variance of operators of the form $AB$ is small, which is dictated by Eq. \eqref{eq:covariance}, but the variance of operators of the form $\sum_jA_jB_j$ is also small, where $A_j\in\mathcal{A}_X$ and $B_j\in\mathcal{A}_Y$. This is exactly how Eq. \eqref{eq:schmidttail} can help, because Eq. \eqref{eq:schmidttail} effectively limits the number of terms that can appear in $\sum_jA_jB_j$. Therefore, we can combine the results of the two previous steps to get Eq. \eqref{eq:decoupling}. The precise strategy, which follows the proof of Dalzell and Brand\~ao
\cite[Lemma~7]{DalzellBrandao2019}, is to use the Schmidt tail Eq. \eqref{eq:schmidttail} to construct a state that is close to both $\rho_{L,XY}$ and $\rho_{L,X}\otimes \rho_{L,Y}$.

Let $P$ be the projector to the eigenvectors for the $K$ largest eigenvalues of $\rho_{L,X}$, and set 
\begin{equation}
    q = \Tr(P\rho_{L,X}) >0,\qquad \sigma=\frac{1}{q} (P\otimes\one_Y) \rho_{L,XY} (P\otimes\one_Y).
\end{equation}
We will see that $\sigma$ is close to $\rho_{L, XY}$.

We also define the projection of the original ground state:
\[
 \ket{\psi_{L,K}}
 :=\frac{(P\otimes\one_{X^c})\ket{\psi_L}}{\sqrt q},
 \qquad
 \braket{\psi_L}{\psi_{L,K}}=\sqrt q.
\]
Note that $\sigma=\Tr_{(XY)^c}(\ket{\psi_{L, K}}\bra{\psi_{L, K}})$. The monotonicity of trace norm under partial trace and the Fuchs-van de Graaf inequality in Eq. \eqref{eq:matrixfacts} give
\[
 \begin{aligned}
 \|\sigma-\rho_{L,XY}\|_1
 &\le\bigl\|\ket{\psi_{L,K}}\bra{\psi_{L,K}}
              -\ket{\psi_L}\bra{\psi_L}\bigr\|_1\\
 &\le 2\sqrt{1-|\braket{\psi_L}{\psi_{L,K}}|^2}
 =2\sqrt{1-q}\le 2\sqrt{C}K^{-a/2}.
 \end{aligned}
\]
The last step uses the Schmidt-tail bound Eq. \eqref{eq:schmidttail}, since
$1-q=\sum_{j>K}\lambda_j(\rho_{L,X})\le CK^{-a}$. Below we will abuse the notation and write the constant $2\sqrt{C}$ as $C$, and in the remainder of this proof the symbol $C$ means a large enough constant so that all previously established bounds are applicable.

Next, we show that $\sigma$ satisfies an analog of Eq. \eqref{eq:decoupling}. We start by noting that for $q\geq 1/2$, $A=PAP\in \A_X$ and $B\in \A_Y$,
\begin{equation}\label{eq:sigma_AB}
\begin{split}     
|\Tr(\sigma AB)-\Tr(\sigma A)\Tr(\sigma B)|
=&\left|\frac{1}{q}\Tr(\rho_L AB)-\frac{\Tr(\rho_LA)}{q}\frac{\Tr(\rho_L PB)}{q}\right|\\
=&\left|\frac{1}{q}[\Tr(\rho_LAB)-\Tr(\rho_LA)\Tr(\rho_LB)]-\frac{\Tr(\rho_LA)}{q^2}[\Tr(\rho_LPB)-\Tr(\rho_LP)\Tr(\rho_LB)]\right|\\
\le& C(q^{-1}+q^{-2})e^{-c_0r}\|A\|\|B\|
 \le6Ce^{-c_0r}\|A\|\|B\|,
\end{split}
\end{equation}
where the triangle inequality and Eq. \eqref{eq:covariance} are used in the second last step.

We next use the trace-norm estimate of Brand\~ao and
Horodecki~\cite[Lemma~20]{BrandaoHorodecki2015} to establish the analog of Eq. \eqref{eq:decoupling} for $\sigma$. Note that for any operator $O\in\mathcal{B}(P\mathcal{H}_X\otimes\mathcal{H}_Y)$ with $\|O\|=1$, it can be written as $O=\sum_{i}A_iB_i$, where $A_i\in\mathcal{B}(P\mathcal{H}_X)$ (i.e., $PA_iP=A_i$) is a complete set of orthonormal basis of operators (e.g. matrix units $\ket{v_a}\bra{v_b}$ with $\ket{v_a}$ an orthonormal basis of $P\mathcal H_X$.) in $\mathcal{B}(P\mathcal{H}_X)$, and $B_i\in\mathcal{B}(\mathcal{H}_Y)$. Importantly, in this decomposition of $O$ there are at most $K^2$ terms, and each term has a norm at most 1. Therefore,
\[
 \|\sigma-\sigma_X\otimes\sigma_Y\|_1 = \sup_{\norm{O}\leq 1} |\Tr[(\sigma-\sigma_X\otimes\sigma_Y)O]| \le 6CK^2e^{-c_0r}.
\]
Namely, $\sigma$ is close to $\sigma_X\otimes\sigma_Y$.

On the other hand, the contractive property of partial trace leads to 
\begin{equation}
\begin{split}
    \norm{\sigma_X - \rho_{L,X}}_1 &\leq \norm{\sigma -\rho_{L,XY}}_1\leq C K^{-a/2},\\ \norm{\sigma_Y - \rho_{L,Y}}_1 &\leq \norm{\sigma -\rho_{L,XY}}_1\leq C K^{-a/2},
\end{split}
\end{equation}
which means that $\sigma_X$ and $\sigma_Y$ are close to $\rho_{L, X}$ and $\rho_{L, Y}$, respectively. The above results together also imply that $\sigma$ is close to $\rho_{L, X}\otimes\rho_{L, Y}$.

Finally, we can use triangle inequalities to bound the trace-norm:
\begin{equation}
 \begin{aligned}
 \|\rho_{L,XY}-\rho_{L,X}\otimes\rho_{L,Y}\|_1
 &\leq \norm{\rho_{L,XY} - \sigma}_1 + \norm{\sigma - \sigma_{X}\otimes\sigma_Y}_1\\
 & + \norm{\sigma_X\otimes (\sigma_Y -\rho_{L,Y})}_1 + \norm{(\sigma_X-\rho_{L,X})\otimes\rho_{L,Y}}_1
 \\
 &\le3\|\sigma-\rho_{L,XY}\|_1
       +\|\sigma-\sigma_X\otimes\sigma_Y\|_1\\
 &\le C\bigl(K^{-a/2}+K^2e^{-c_0r}\bigr).
 \end{aligned}
\end{equation}
Choosing $K=\lceil e^{2c_0r/(a+4)}\rceil$ makes both terms decay
exponentially in $r$ and ensures that $q\ge1/2$ for all sufficiently large
$r$.  Increasing the constants to cover the remaining bounded values of
$r$ proves Eq. \eqref{eq:decoupling}.
\end{proof}

After proving that regions far away are nearly unentangled in $|\psi_L\rangle$ just now, another property we will establish is that as long as a chain is long enough, and it hosts a pure translation invariant state where far away subregions are nearly unentangled, then it can be used to build up another translation invariant state on a chain of any large enough size, such that the local reduced density matrices of the new state are nearly identical to those of the original state. In particular, our ground state $|\psi_L\rangle$ has this property.

\begin{lemma}[Local comparison between rings]
\label{lem:local-ring-comparison}
Let $r\ge1$ be an integer, and let $\eta$ be a pure
translation invariant state on a ring of length $L\ge12r$.
Suppose
\[
 \|\eta_{XY}-\eta_X\otimes\eta_Y\|_1\le\delta
\]
for all disjoint intervals $X,Y$ with
$d_L(X,Y)\ge r$. Then, for every $N\ge8r$, there is a translation invariant state
$\sigma_N$ on the ring of length $N$ such that
\begin{equation}\label{eq:loca_ring_construction}
 \|(\sigma_N)_I-\eta_I\|_1
 \le\delta+4\sqrt{\delta}
\end{equation}
for every interval $I$ with $|I|\le r$.
Here $\eta_I$ denotes the marginal on a matching interval
of the seed ring with length $L$, with sites identified in the same
forward order.
\end{lemma}

We will call the length-$L$ ring a seed ring, and length-$N$ ring a target ring. The above lemma is proved by explicitly constructing the state $\sigma_N$ on the target ring. Note that $\sigma_N$ is in general a mixed state.

\begin{proof}

\emph{Step 1: start with independent blocks.} We first prepare disjoint marginals on the target ring. 
Let $M=\lfloor N/(4r)\rfloor\ge2$.  We arrange blocks $A_i$ and holes $B_i$ in order as $A_1,B_1,\ldots,A_M,B_M$, with
$3r\le|A_i|\le5r$ and $|B_i|=r$ (see Fig. \ref{fig:periodization-partition}).
One way to get such a partition is as follows: start with $|A_i|=3r$ and distribute the
$N-4rM<4r$ remaining sites among the blocks $A_i$.
On each $A_i$, we prepare a copy of the seed marginal $\eta_{A_i}$. Prepare these copies independently and leave the holes empty for now. Thus the initial state is
\[
 A_{\rm ret}:=\bigcup_{i=1}^M A_i,\qquad
 \omega_{\rm ret}^{(0)}:=\bigotimes_{i=1}^M\eta_{A_i}
 \quad\text{on }\mathcal H_{A_{\rm ret}}.
\]
Within each block the seed marginal is reproduced exactly, but the correlations
between different blocks are absent at this stage.

\emph{Step 2: recover the holes.}
Next, we use Lemma~\ref{lem:erasure} to get approximate recovery channels to fill the holes and prepare a correlated state on the target ring. To fill the holes $B_i$, we denote the $r$ sites immediately to its left by $C_i^-$, and the $r$ sites
immediately to its right by $C_{i}^+$. $C_i^-$ is at the end of $A_i$ and $C_i^+$ at the beginning of $A_{i+1}$ (with $i+1$ modulo $M$). $C_i^\pm$ are called collar regions.
We denote $C_i= C_i^- C_i^+$ and $W_i = C_i^- B_i C_i^+$.
Each block $A_i$ has the form $A_i=C_{i-1}^+D_iC_i^-$, where the remaining core is denoted by $D_i$ and has at least $r$ sites, i.e., $D_i=A_i\backslash (C_{i-1}^+\cup C_i^{-})$.  
The cores separate neighboring recovery regions $W_i$, so these regions are disjoint (see Figure~\ref{fig:periodization-partition}(a)).

We choose the recovery channel using the seed state.  Take a matching
$3r$-site region $W_i$ on the seed ring and let
$F_i=\T_L\setminus W_i$.  The two collars separate the hole $B_i$ from
$F_i$ by at least $r$ sites.  Hence the decoupling hypothesis of this lemma gives
$\|\eta_{B_iF_i}-\eta_{B_i}\otimes\eta_{F_i}\|_1\le\delta$.
Lemma~\ref{lem:erasure}, applied to the pure seed state on $B_iC_iF_i$,
then supplies a channel $\mathcal R_i:C_i\to B_iC_i$ that restores the
erased hole with trace-norm error at most $4\sqrt\delta$.

We copy this channel to the target ring and apply it to the two collars
around $B_i$.  Since the recovery regions are disjoint, all $M$ channels
can be applied together, leaving the cores unchanged.  Writing
$D_{\rm core}:=\bigcup_{i=1}^M D_i$, the input and output systems are
\[
 A_{\rm ret}=D_{\rm core}\sqcup\bigsqcup_{i=1}^M C_i,\qquad
 \T_N=D_{\rm core}\sqcup\bigsqcup_{i=1}^M W_i.
\]
The resulting state is defined on the whole target ring:
\[
 \omega_N=
 \left(\id_{D_{\rm core}}\otimes
       \bigotimes_{i=1}^M\mathcal R_i\right)
       \left(\omega_{\rm ret}^{(0)}\right).
\]

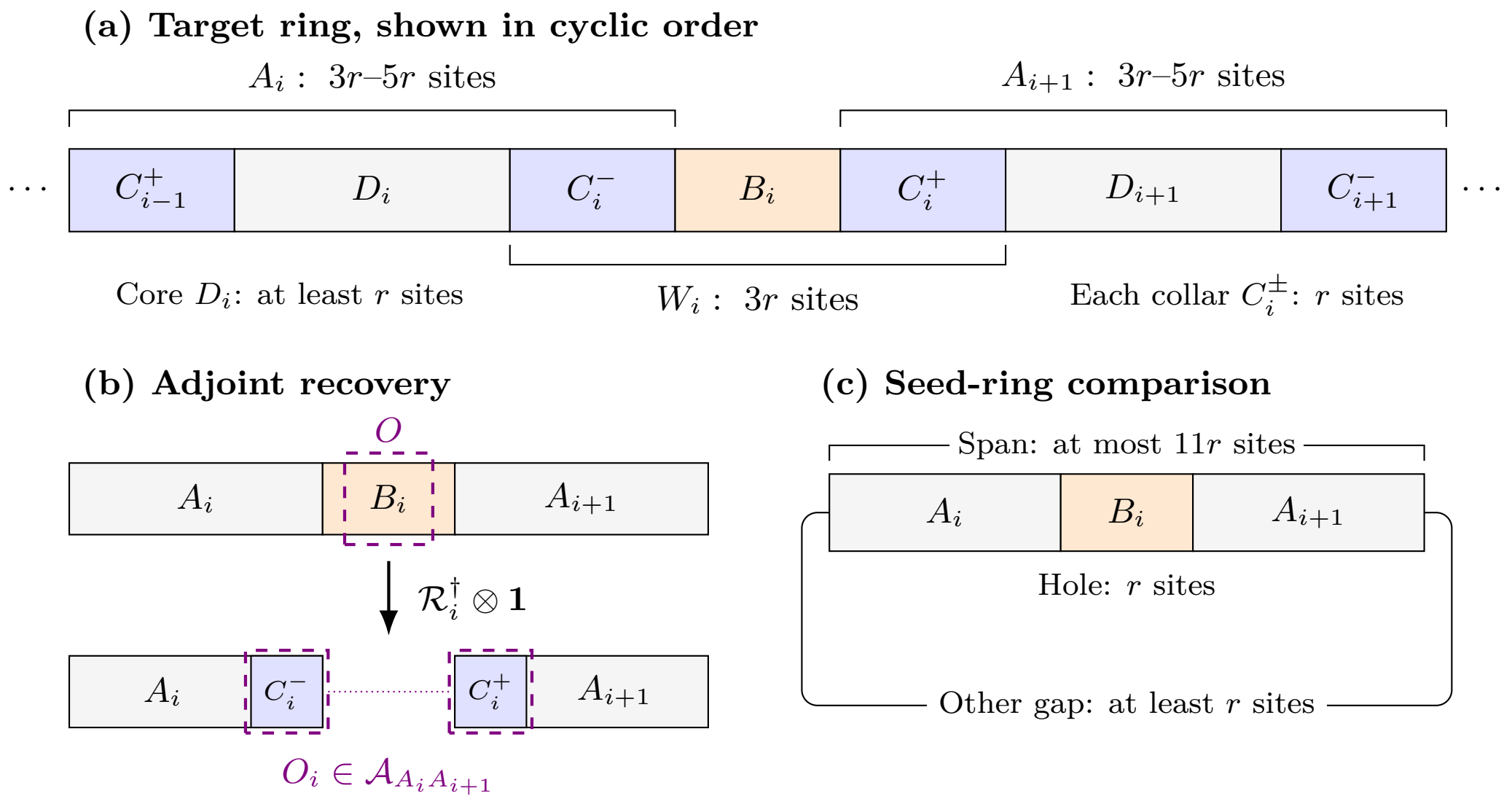
\begin{figure}[tb]
\centering
\begin{tikzpicture}[x=1cm,y=1cm,font=\small,
 collar/.style={draw,fill=blue!12},
 core/.style={draw,fill=black!4},
 hole/.style={draw,fill=orange!18}]
\node[anchor=west,font=\small\bfseries] at (0,1.85) {(a) Target ring, shown in cyclic order};
\draw[collar] (0,0) rectangle (1.5,.75);
\draw[core] (1.5,0) rectangle (4,.75);
\draw[collar] (4,0) rectangle (5.5,.75);
\draw[hole] (5.5,0) rectangle (7,.75);
\draw[collar] (7,0) rectangle (8.5,.75);
\draw[core] (8.5,0) rectangle (11,.75);
\draw[collar] (11,0) rectangle (12.5,.75);
\node at (.75,.38) {$C_{i-1}^+$};
\node at (2.75,.38) {$D_i$};
\node at (4.75,.38) {$C_i^-$};
\node at (6.25,.38) {$B_i$};
\node at (7.75,.38) {$C_i^+$};
\node at (9.75,.38) {$D_{i+1}$};
\node at (11.75,.38) {$C_{i+1}^-$};
\node at (-.35,.38) {$\cdots$};
\node at (12.85,.38) {$\cdots$};
\draw (0,.95)--(0,1.1)--(5.5,1.1)--(5.5,.95);
\draw (7,.95)--(7,1.1)--(12.5,1.1)--(12.5,.95);
\node at (2.75,1.4) {$A_i:\ 3r\text{--}5r$ sites};
\node at (9.75,1.4) {$A_{i+1}:\ 3r\text{--}5r$ sites};
\draw (4,-.12)--(4,-.3)--(8.5,-.3)--(8.5,-.12);
\node at (6.25,-.62) {$W_i:\ 3r$ sites};
\node[font=\footnotesize,align=center] at (2,-.58) {Core $D_i$: at least $r$ sites};
\node[font=\footnotesize,align=center] at (10.6,-.58) {Each collar $C^\pm_i$: $r$ sites};
\node[anchor=west,font=\small\bfseries] at (0,-1.4) {(b) Adjoint recovery};
\draw[core] (0,-2.75) rectangle (2.3,-2.1);
\draw[hole] (2.3,-2.75) rectangle (3.5,-2.1);
\draw[core] (3.5,-2.75) rectangle (5.8,-2.1);
\node at (1.15,-2.425) {$A_i$};
\node at (2.9,-2.425) {$B_i$};
\node at (4.65,-2.425) {$A_{i+1}$};
\draw[violet,thick,dashed] (2.5,-2.84) rectangle (3.3,-2.01);
\node[text=violet] at (2.9,-1.8) {$O$};
\draw[-{Latex},thick] (2.9,-2.99)--(2.9,-3.67);
\node[anchor=west] at (3.05,-3.33) {$\mathcal R_i^{\dagger}\otimes\id$};
\draw[core] (0,-4.5) rectangle (2.3,-3.85);
\draw[core] (3.5,-4.5) rectangle (5.8,-3.85);
\draw[collar] (1.65,-4.5) rectangle (2.3,-3.85);
\draw[collar] (3.5,-4.5) rectangle (4.15,-3.85);
\node at (.85,-4.175) {$A_i$};
\node at (4.95,-4.175) {$A_{i+1}$};
\node[font=\scriptsize] at (1.975,-4.175) {$C_i^-$};
\node[font=\scriptsize] at (3.825,-4.175) {$C_i^+$};
\draw[violet,thick,dashed] (1.6,-4.55) rectangle (2.35,-3.8);
\draw[violet,thick,dashed] (3.45,-4.55) rectangle (4.2,-3.8);
\draw[violet,densely dotted] (2.35,-4.175)--(3.45,-4.175);
\node[text=violet] at (2.9,-4.95) {$O_i\in\A_{A_iA_{i+1}}$};
\node[anchor=west,font=\small\bfseries] at (6.7,-1.4) {(c) Seed-ring comparison};
\draw[core] (6.9,-2.9) rectangle (9,-2.2);
\draw[hole] (9,-2.9) rectangle (10.2,-2.2);
\draw[core] (10.2,-2.9) rectangle (12.3,-2.2);
\node at (7.95,-2.55) {$A_i$};
\node at (9.6,-2.55) {$B_i$};
\node at (11.25,-2.55) {$A_{i+1}$};
\draw (6.9,-2.08)--(6.9,-1.94)--(12.3,-1.94)--(12.3,-2.08);
\node[fill=white,inner sep=2pt,font=\footnotesize] at (9.6,-1.94)
 {Span: at most $11r$ sites};
\node[font=\footnotesize] at (9.6,-3.2) {Hole: $r$ sites};
\draw[rounded corners=4pt] (6.9,-2.55)--(6.65,-2.55)--(6.65,-4.3)
 --(12.55,-4.3)--(12.55,-2.55)--(12.3,-2.55);
\node[fill=white,inner sep=3pt,font=\footnotesize] at (9.6,-4.3)
 {Other gap: at least $r$ sites};
\end{tikzpicture}
\caption{Local recovery and comparison between rings.
(a) The channel $\mathcal R_i$ fills $B_i$ using its two blue collars.
The cores separate neighboring recovery regions.
(b) Adjoint recovery pulls $O$ back to $O_i$ on $A_i\cup A_{i+1}$.
The illustrated $O$ is supported inside the hole, so $O_i$ is supported
on its two collars.
(c) The blocks $A_i$ have gaps of at least $r$ sites in both
cyclic directions on the seed ring.}
\label{fig:periodization-partition}
\end{figure}

\emph{Step 3: compare with the seed state.} We will see that $\omega_N$ and $\eta$ are close on local regions. Concretely,
let $I$ be a cyclic interval with $|I|\le r$, and let $O\in\A_I$.
The core between neighboring recovery regions contains at least $r$ sites,
so $I$ meets at most one $W_i$.  If it meets none, it lies in a 
core, and its expectation is exactly $\eta(O)$ and $\|(\omega_N)_I-\eta_I\|_1=0$.  If it meets $W_i$, only
the channel $\mathcal R_i$ matters. So we can check the error of recovery by considering the segment $A_iB_iA_{i+1}$.

Define
\[
O_i:=(\mathcal R_i^\dagger\otimes\mathrm{id})(O)
\in\mathcal A_{A_iA_{i+1}},
\qquad \|O_i\|\le\|O\|,
\]
where \(\mathrm{id}\) is an identity channel acting on \((A_i\cup A_{i+1})\setminus C_i\). Since the initial blocks are independent, we get
\[
\Tr(\omega_NO)
=\Tr\!\left[(\eta_{A_i}\otimes\eta_{A_{i+1}})O_i\right].
\]

There are two sources of difference between $\Tr(\omega_NO)$ and
$\eta(O)$. The first is the missing correlation between $A_i$ and
$A_{i+1}$.  Place matching copies of $A_i,B_i,A_{i+1}$ consecutively on
the seed ring, as in Figure~\ref{fig:periodization-partition}(c).
The segment contains at most $5r+r+5r=11r$ sites. Since $L\ge12r$, the two
blocks $A_i$ and $A_{i+1}$ are separated by at least $r$ sites in both cyclic directions:
one gap is $B_i$, and the other is the remainder of the seed ring.
The decoupling hypothesis of this lemma therefore ensures
$\|\eta_{A_i}\otimes\eta_{A_{i+1}}-\eta_{A_iA_{i+1}}\|_1\leqslant\delta$.
Replacing the independent input by the seed marginal changes the
expectation of $O_i$ by at most $\delta\|O_i\|$.

The second difference is the imperfection in the recovery of $B_i$.  Applied to the
seed marginal $\eta_{A_iA_{i+1}}$, the same channel restores the seed
expectation of $O$ up to $4\sqrt\delta\|O\|$, by
Lemma~\ref{lem:erasure}.  Combining the two errors gives
\[
 \begin{aligned}
 |\Tr(\omega_NO)-\eta(O)|
 &\le
 \bigl|\Tr[(\eta_{A_i}\otimes\eta_{A_{i+1}}
              -\eta_{A_iA_{i+1}})O_i]\bigr|\\
 &\quad+\bigl|\Tr(\eta_{A_iA_{i+1}}O_i)-\eta(O)\bigr|\\
 &\le\delta\|O_i\|+4\sqrt\delta\|O\|
 \le(\delta+4\sqrt\delta)\|O\|,
 \end{aligned}
\]
so $\|(\omega_N)_I-\eta_I\|_1\leqslant\delta+4\sqrt{\delta}$.

\emph{Step 4: restore translation invariance.}
The placement of the blocks singles out positions on the target ring.
For the use in next subsection, we average $\omega_N$ over all cyclic translations to obtain a translation-invariant state $\sigma_N:=\sum_{x\in\mathbb{T}_N}\tau_x(\omega_N)/N$.  Each translated state $\tau_x(\omega_N)$ obeys
the same local bound, because $\eta$ is translation invariant and the
bound holds for every interval of length at most $r$.  Hence
\begin{equation}\label{eq:periodization}
 |\Tr(\sigma_NO)-\eta(O)|\le(\delta+4\sqrt\delta)\|O\|,
 \qquad |I|\le r,
\end{equation}
so $\|(\sigma_N)_I-\eta_I\|_1\leqslant \delta+4\sqrt{\delta}$.
    
\end{proof}

For the ground states in Lemma~\ref{lem:decoupling}, we may take
$\delta=Ce^{-cr}$, so the error above is exponentially small in $r$.
It does not grow as the target length $N$ increases, since each short observable
encounters at most one recovery channel.  This is the local comparison
we need between rings of different lengths.

\subsection{Proof of Proposition~\ref{prop:circumference}}
\label{subsec:proof_of_agreement}

With the above preparation, we can prove Proposition~\ref{prop:circumference}. Let $h = \Phi_0$ be the term in \eqref{eq: unperturbed Hamiltonian family}.

\begin{proof}
We first identify the limiting energy density $\mathcal{E}_\infty$. Let $H_n^{\rm op}$ be the sum of all translates of $h_{0,L}$ (c.f. Eq.\ref{eq: unperturbed Hamiltonian family}) whose supports are contained in the open interval of $n$ sites (say $\{0,1,\cdots, n-1\}$), and denote $b_n = \lambda_1(H^{\rm op}_n)$ (let $b_n=0$ for $n\leq R_0$). We would like to show that $\{b_n/n\}$ form a Cauchy sequence. To do this, we join two open intervals of lengths $n, m$ and get 
\begin{equation}
 H^{\rm op}_{n+m} = H^{\rm op}_{n} + H_m^{\rm op} + B_{n,m}, \qquad \norm{B_{n,m}}\leq R_0\norm{h},
\end{equation}
with $H_m^{\rm op}$ supported on $\{n,n+1,\cdots,n+m-1\}$. Let us denote $A = H_n^{\rm op} + H_{m}^{\rm op}$ and $B = B_{n,m}$ for brevity. It is clear that $\lambda_1(A) = b_n + b_m$. On the other hand, with the operator inequality $A-\norm{B}\leq A + B\leq A + \norm{B}$, we get $|\lambda_1(A+B)-\lambda_1(A)|\leq \norm{B}$. \footnote{For a self-adjoint matrix $\mathfrak{h}$, the smallest eigenvalue is $\lambda_1(\mathfrak{h}) = \min_{\norm{u}=1}\ev{u,\mathfrak hu} $. Using the relation $|\ev{u,B u}|\leq \norm{B}$, we take the minimum over all normalized states, which gives $\min_{v'}\ev{v',Av'}- \norm{B}\leq \min_{u}\ev{u,(A+B)u}\leq \min_{v}\ev{v, Av}+\norm{B}$. Generalizing this argument proves the Weyl inequality.} Together with the fact that $\norm{B}\leq R_0 \norm{h}$, this proves
\begin{equation}\label{eq:b_nm_difference}
    |b_{n+m} - b_{n} - b_m| \leq R_0 \norm{h}.
\end{equation}
Iterating Eq. \eqref{eq:b_nm_difference} for a sequence of open intervals, we get
\begin{equation}
    |b_{kn} - k b_n| \leq (k-1) R_0 \norm{h},
\end{equation}
so
\begin{equation}
\begin{split}
 \left|\frac{b_n}{n}-\frac{b_m}{m}\right|
 &=\frac{|mb_n-nb_m|}{mn} \\
 &\le\frac{|mb_n-b_{mn}|+|b_{mn}-nb_m|}{mn}
 \le\frac{R_0\norm{h}}{n}+\frac{R_0\norm{h}}{m}.
 \label{eq:b_cauchy_sequence}
 \end{split}
\end{equation}
We thus have shown that $\{b_n/n\}$ is a Cauchy sequence. Denote the limit of this sequence by $\mathcal{E}_\infty$, then $|b_n-n\mathcal{E}_\infty| \leq R_0 \norm{h}$. To form a closed chain, we add at most $R_0$ terms, so on the ring of length $L$, $\norm{H_L - H_L^{\rm op}}\leq R_0 \norm{h}$. Hence, we get
\begin{equation}\label{eq:ring-energy-rough-limit}
    |E_L - L\mathcal E_\infty|\leq |E_L - b_L|+|b_L - L\mathcal{E}_\infty|\leq 2R_0\norm{h}.
\end{equation}
This establishes the thermodynamic limit of the energy density on rings, and the following construction will improve the bound to be exponentially small in the ring circumference.

Before proving Eq. \eqref{eq:energyconvergence}, we set some constants. Fix large $L,N$ and set
\[
 r=\left\lfloor\frac{\min(L,N)}{12}\right\rfloor,
 \qquad
 \delta_r=C_0e^{-c_0r},
 \qquad
 \epsilon_r=\delta_r+4\sqrt{\delta_r}.
\]
Here $C_0,c_0$ are chosen uniformly in the circumference as given by Lemma~\ref{lem:decoupling}, and 
$\delta_r$ denotes the decoupling error in Lemma \ref{lem:local-ring-comparison}.
Then $\epsilon_r\le C_1e^{-c_1r}$ with
$C_1=C_0+4\sqrt{C_0}$ and $c_1=c_0/2$, also independent of $L,N$.
Choose $L_1$ large enough so that
$r\ge R_0+1$ whenever $L,N\geq L_1$. Since $L,N\geq 12r$, either $L$-ring or $N$-ring may serve as the seed ring.

Use the ground state $\rho_L$ as the seed state and the $N$-ring as the target in Eq.
\eqref{eq:periodization}. According to Lemma \ref{lem:local-ring-comparison}, this produces a translation invariant state
$\sigma_N$ such that, for every operator $O$ supported on at most $r$
consecutive sites,
\begin{equation}
 |\Tr(\sigma_NO)-\rho_L(O)|\le\epsilon_r\|O\|.
 \label{eq:periodized-local-comparison}
\end{equation}
Note that every $h_{x,N}$ has a support with length at most $R_0+1\le r$, 
 
and translation invariance of
$\rho_L$ gives
\[
 \rho_L(h_{x,L})=\frac1L\rho_L(H_L)=\frac{E_L}{L}.
\]
So summing Eq. \eqref{eq:periodized-local-comparison} over the $N$ target terms
and using the variational principle therefore yield
\begin{equation}
 \frac{E_N}{N}
 \le\frac1N\Tr(\sigma_NH_N)
 \le\frac{E_L}{L}+\epsilon_r\|h\|.
 \label{eq:periodized-energy-upper}
\end{equation}
Interchanging the seed and target rings gives the reverse inequality, i.e., the inequality with $L$ and $N$ swapped.
Since $r\ge\min(L,N)/12-1$, we obtain
\[
 \left|\frac{E_L}{L}-\frac{E_N}{N}\right|
 \le\epsilon_r\|h\|
 \le C_1\|h\|e^{-c_1r}
 \le C_1\|h\|e^{c_1}e^{-(c_1/12)\min(L,N)}.
\]
Thus the first estimate in Eq. \eqref{eq:energyconvergence} holds with
$C_E=C_1\|h\|e^{c_1}$ and $c_E=c_1/12$.
These constants depend only on the fixed interaction and the uniform
decoupling constants, not on $L$ or $N$.
We keep these values of $C_E,c_E$ fixed throughout the rest of this proof. For any fixed $L\ge L_1$, let $N\to\infty$.
Equation~\eqref{eq:ring-energy-rough-limit} implies
$E_N/N\to\mathcal E_\infty$.  Since $\min(L,N)=L$ for $N\ge L$,
the same constants give
\[
 \left|\frac{E_L}{L}-\mathcal E_\infty\right|\le C_E e^{-c_E L}.
\]
Multiplication by $L$ proves the second estimate in Eq.
\eqref{eq:energyconvergence}, with $C_E,c_E$ still independent of $L$.

The remaining task is to prove Eq. \eqref{eq:marginalagreement}, the comparison of the ground-state marginals of an arbitrary interval.

To proceed, assume that $L\ge N$,
take $r=\lfloor N/12\rfloor$, and again construct $\sigma_N$ from the
seed $\rho_L$.  Subtracting $E_N/N$ in Eq. 
\eqref{eq:periodized-energy-upper} and using the first estimate in Eq.
\eqref{eq:energyconvergence} give
\begin{equation}
 \begin{aligned}
 0\le\Tr[\sigma_N(H_N-E_N\one)]
 &\le N\left(
   \left|\frac{E_L}{L}-\frac{E_N}{N}\right|
   +\epsilon_r\|h\|\right)\\
 &\le 2C_E N e^{-c_E N}.
 \end{aligned}
 \label{eq:periodized-excess-energy}
\end{equation}
The assumed gap inequality
$H_N-E_N\one\ge\gamma_0(\one-\rho_N)$ then implies
\[
 1-\Tr(\sigma_N\rho_N)
 \le\frac{2C_E}{\gamma_0}Ne^{-c_E N}.
\]
Because $\rho_N$ is pure, the squared root fidelity between $\rho_N$ and
$\sigma_N$ is
$F(\sigma_N,\rho_N)^2=\Tr(\sigma_N\rho_N)$.  Eq.
\eqref{eq:matrixfacts} therefore gives
\begin{equation}
 \begin{aligned}
 \|\sigma_N-\rho_N\|_1
 &\le2\sqrt{1-\Tr(\sigma_N\rho_N)}
 \le2\sqrt{\frac{2C_E}{\gamma_0}}\sqrt N e^{-c_E N/2}\\
 &\le4\sqrt{\frac{C_E}{c_E\gamma_0}}e^{-c_E N/4}.
 \end{aligned}
 \label{eq:periodized-ground-state-distance}
\end{equation}
The last step uses
$\sqrt N e^{-c_E N/4}\le\sqrt{2/c_E}$, which follows from
$(c_E N/2)e^{-c_E N/2}\le1$.
The square root and the factor $\sqrt N$ weaken the decay rate to $c_E/4$.

Finally let $I$ be an interval with $|I|\le r$ and let
$O\in\A_I$ be a Hermitian contraction.  Eqs.
\eqref{eq:periodized-local-comparison} and
\eqref{eq:periodized-ground-state-distance} give
\[
 \begin{aligned}
 |\Tr[(\rho_{L,I}-\rho_{N,I})O]|
 &\le |\Tr(\rho_LO)-\Tr(\sigma_NO)|
      +|\Tr[(\sigma_N-\rho_N)O]| \\
 &\le\epsilon_r+\|\sigma_N-\rho_N\|_1
 \le C_1e^{c_1}e^{-c_E N}
       +4\sqrt{\frac{C_E}{c_E\gamma_0}}e^{-c_E N/4}\\
 &\le C_{\rm marg}e^{-c_E N/4},
 \qquad
 C_{\rm marg}:=C_1e^{c_1}+4\sqrt{\frac{C_E}{c_E\gamma_0}}.
 \end{aligned}
\]
Taking the supremum over Hermitian contractions and using the trace-norm
duality then prove \eqref{eq:marginalagreement}.
The separate prefactor $C_{\rm marg}$ depends only on the already fixed
constants and $\gamma_0$, so it is independent of $L,N$.

\end{proof}

\section{Exact ground state preparation}
\label{sec:state_preparation}

In this section, we construct a unitary $T_L$ that sends the doubled ground state $\ket{\Psi_L}$
to the product vacuum state $\ket{\Omega_L}$ and whose conjugation preserves the locality of an operator up to exponentially smaller errors. 

Consider a general translation invariant, pure state $\ket{\phi_L}$ on the single copy of the physical spin ring, and denote $\varrho_L = \ket{\phi_L}\bra{\phi_L}$. The state on the enlarged system is denoted by $\ket{\Phi_L} = \ket{\phi_L}\otimes\ket{\overline{\phi_L}}$. We suppose that there are fixed constants $C,c>0$, such that for any finite disjoint intervals $X$ and $Y$ with $r=d_L(X,Y)$ and an interval $I$, the state $\varrho_L$ satisfies
\begin{align}
\|\varrho_{L,XY}-\varrho_{L,X}\otimes\varrho_{L,Y}\|_1
 &\le Ce^{-cr},\label{eq:prep-input-decouple}\\
 \|\varrho_{L,I}-\varrho_{N,I}\|_1
 &\le Ce^{-cN}\qquad
 (L\ge N,\ |I|\le\lfloor N/12\rfloor).
 \label{eq:prep-input-compare}
\end{align}
Here $\varrho_{L,X}$ is the reduced density matrix of $\varrho_L$ on $X$, and the intervals on different rings are identified by the order of their sites.

By our assumption of $H_L$, Proposition~\ref{prop:circumference}
and Lemma~\ref{lem:decoupling}, the ground state $\ket{\psi_L}$ clearly satisfies the above conditions. Yet, the following theorem of state preparation applies to more general states $\ket{\phi_L}$ as described.

\begin{theorem}[Exact preparation of doubled states]\label{thm:prep-exact}
Let $\{\varrho_L=\ket{\phi_L}\bra{\phi_L}\}$ be a family of
translation invariant pure spin states satisfying Eqs.
\eqref{eq:prep-input-decouple} and \eqref{eq:prep-input-compare}.
There are unitaries $\{T_L\}$ that send their doubled states to the product vacuum:
\[
 T_L\ket{\Phi_L}=\ket{\Omega_L},\qquad
 \ket{\Phi_L}=\ket{\phi_L}\otimes\ket{\overline{\phi_L}},
\]
and $L$-independent constants $C,\mu>0$ with the following properties.
\begin{enumerate}
\item Both conjugations by $T_L$ and $T_L^{\dagger}$ preserve locality up to
exponentially small errors. Namely, for every nonempty set $X$ and every $A\in\mathcal A_X$ and $n\ge0$,
\begin{equation}\label{eq:prep-locality}
 \|T_L A T_L^{\dagger}-\E_{X(n)}^{\tr}(T_L A T_L^{\dagger})\|
 \le C|X|\|A\|e^{-\mu n}.
\end{equation}

\item The unitaries can be chosen so that their action on local observables is approximately independent of the ring size, away from the closing seam. More concretely,
for $L\ge N$, let $\iota_{N,L}$ embed operators on a ring with size $N$ into a ring with size $L$ by identifying sites $0,\ldots,N-1$ in the fixed on-site bases.
For a nonempty interval $I\subset[0,N-1]$, define its distance from the
ends of this ordered segment by
\[
 d(I,N)=\min\{\min I,N-1-\max I\}.
\]
Then, for every $A\in\mathcal A_I$,
\begin{equation}\label{eq:prep-compatible}
 \|T_L\iota_{N,L}(A)T_L^{\dagger}
       -\iota_{N,L}(T_NAT_N^{\dagger})\|
 \le C|I|\|A\|e^{-\mu d(I,N)}.
\end{equation}

\item There are constants $C_{\rm stop}, b>0$ and $L$-independent initial length scale $r_0$, such that for every $s=2^kr_0, k\geq 0$ with
$r_0\le s\le L/C_{\rm stop}$, the inverse $\mathcal U_L=T_L^{\dagger}$ has a truncated circuit approximation $\mathcal U_{L,<s}$, such that, for $A\in\mathcal A_I$,
\[
 \mathcal U_{L,<s}A\mathcal U_{L,<s}^{\dagger}\in\mathcal A_{I(C_1s)},\quad
 \mathcal U_{L,<s}^{\dagger}A\mathcal U_{L,<s}
 \in\mathcal A_{I(C_1s)}, C_1 =1832.
\]
and
\begin{equation}\label{eq:prep-truncation}
 \max\left\{
 \left\|\mathcal U_LA\mathcal U_L^{\dagger}
       -\mathcal U_{L,<s}A\mathcal U_{L,<s}^{\dagger}\right\|,
 \left\|\mathcal U_L^{\dagger}A\mathcal U_L
       -\mathcal U_{L,<s}^{\dagger}A\mathcal U_{L,<s}\right\|
 \right\}
 \le C|I|\|A\|e^{-bs}.
\end{equation}
Away from a length-$(Cs)$ neighborhood of the closing seam, the gates of
$\mathcal U_{L,<s}$ have period $448s$ and are the same matrices for every $L$.
\end{enumerate}

Both estimates in 1 and 2 hold for the inverse conjugations, with $T_L,T_N$
replaced by $T_L^{\dagger},T_N^{\dagger}$.
\end{theorem}

In fact, we can show that the constructed $\{T_L\}$ in Theorem \ref{thm:prep-exact} are time evolutions generated by exponentially local interactions, as discussed in Theorem \ref{thm:prep-lga}. So all such states $|\phi_L\rangle$, including our ground state $|\psi_L\rangle$, are invertible. In fact, even if the Hamiltonian is not translation invariant, as long as it is local, gapped and has a unique ground state, in Sec. \ref{subsec:prep-lga} we show that the ground state is invertible.

The proof of Theorem~\ref{thm:prep-exact} proceeds in four steps:
\begin{enumerate}
    \item Lemma~\ref{lem:prep-insert} uses trace-norm decoupling to approximately restore an erased region in the canonical purification, using a unitary on that region and its neighboring collars in the doubled system.
    \item Lemma~\ref{lem:prep-approx} combines these recovery gates into a two-layer circuit that approximately prepares $\ket{\Phi_L}$ from $\ket{\Omega_L}$, using gates of size $O(r)$ with exponentially small local errors.
    \item Comparing the preparations at scales $r$ and $2r$, Lemma~\ref{lem:prep-correct} yields a bounded-depth correction circuit that connects them up to phase. Each correction gate has support of size $O(r)$ and is exponentially close to the identity.
    \item We compose these corrections at dyadic scales up to a scale proportional to $L$, then remove the residual error by a whole-ring rotation exponentially close to the identity. The truncation proves exponential locality, while the common bulk gates give compatibility across circumferences.
\end{enumerate}

\subsection{Lemmas}
\label{subsec:preparation-lemmas} 

In this subsection we introduce some necessary lemmas. For a density matrix $\sigma$, we define the \emph{canonical purification} of $\sigma$ by
\begin{equation}
 \ket{\operatorname{can}(\sigma)}=
 \sum_{i,j}(\sqrt\sigma)_{ij}\ket{i}\otimes\ket{\bar j}
\end{equation}
with $\ket{i},\ket{\bar j}$ the basis states in the physical and auxiliary space, respectively.
For an isometry $W$ and density matrices $\sigma,\tau$, we have the following relations
\begin{equation}\label{eq:prep-canonical}
 \ket{\operatorname{can}(W\sigma W^{\dagger})}
  =(W\otimes\bar W)\ket{\operatorname{can}(\sigma)},\qquad
 \|\ket{\operatorname{can}(\sigma)}-\ket{\operatorname{can}(\tau)}\|
  \le\sqrt{\|\sigma-\tau\|_1}.
\end{equation}
The first identity follows by taking square roots {within the image of $W$}. The second inequality follows from the Powers--St{\o}rmer inequality (see Eq.\eqref{eq:powers-stormer}).

The following lemma shows that, if a subregion is nearly uncorrelated with distant sites, a local gate acting on that subregion and its neighbors can restore the whole state approximately in the doubled space.

\pagebreak[3]
\begin{lemma}[Recovery unitary]\label{lem:prep-insert}
On a spin ring, let $B$ be an interval, let
$C=C_-\cup C_+$ be the two adjacent collars, and let
$F=(B\cup C)^c$ be the remaining sites.  Let
$\sigma_{BCF}=\ket{\psi_\sigma}\bra{\psi_\sigma}$ be a pure reference state satisfying
\begin{equation}\label{eq:mutual_correlation_relation}
 \|\sigma_{BF}-\sigma_B\otimes\sigma_F\|_1\le\delta.
\end{equation}
Let $G$ be any finite exterior system
and let $\tau_{BCG}$ be any density matrix on $BCG$.  Regard
$\tau_{BCG}$ as the state to be reconstructed from
$\tau_{CG}:=\Tr_B\tau_{BCG}$, and assume
\begin{equation}\label{eq:tau_BC_sigma_BC}
 \|\tau_{BC}-\sigma_{BC}\|_1\le\epsilon.
\end{equation}
Then there is a unitary
$U_{\rm loc}$ on $BC\bar B\bar C$ that depends only on the reference
state $\sigma_{BCF}$ but does not depend on the exterior system $G$ or the target state $\tau_{BCG}$, such that
$U=U_{\rm loc}\otimes\one_{G\bar G}$ satisfies
\begin{equation}\label{eq:prep-insert}
 \left\|U\bigl(\ket{0}_{B\bar B}\otimes
       \ket{\operatorname{can}(\tau_{CG})}\bigr)
       -\ket{\operatorname{can}(\tau_{BCG})}\right\|
 \le 4(\delta+\epsilon/2)^{1/4}.
\end{equation}
\end{lemma}

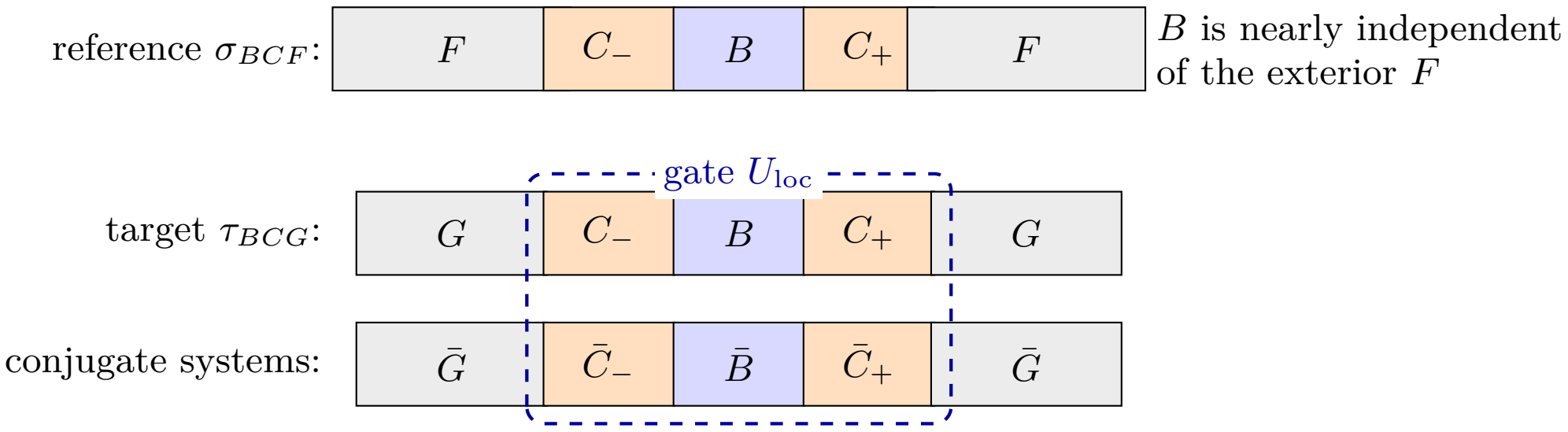
\begin{figure}[hb]
\centering
\begin{tikzpicture}[x=1.15cm,y=1cm,font=\small]
\tikzset{
 region/.style={draw,minimum height=7mm,inner sep=1pt},
 exterior/.style={region,fill=gray!15},
 collar/.style={region,fill=orange!25},
 hole/.style={region,fill=blue!15}
}
\node[anchor=east] at (-.15,2.1) {reference $\sigma_{BCF}$:};
\node[exterior,minimum width=20mm] at (.7,2.1) {$F$};
\node[collar,minimum width=11mm] at (1.85,2.1) {$C_-$};
\node[hole,minimum width=11mm] at (2.8,2.1) {$B$};
\node[collar,minimum width=11mm] at (3.75,2.1) {$C_+$};
\node[exterior,minimum width=20mm] at (4.9,2.1) {$F$};
\node[anchor=west,align=left] at (5.75,2.1)
 {$B$ is nearly independent\\of the exterior $F$};

\node[anchor=east] at (-.15,.55) {target $\tau_{BCG}$:};
\node[exterior,minimum width=16mm] at (.7,.55) {$G$};
\node[collar,minimum width=11mm] at (1.85,.55) {$C_-$};
\node[hole,minimum width=11mm] at (2.8,.55) {$B$};
\node[collar,minimum width=11mm] at (3.75,.55) {$C_+$};
\node[exterior,minimum width=16mm] at (4.9,.55) {$G$};

\node[anchor=east] at (-.15,-.55) {conjugate systems:};
\node[exterior,minimum width=16mm] at (.7,-.55) {$\bar G$};
\node[collar,minimum width=11mm] at (1.85,-.55) {$\bar C_-$};
\node[hole,minimum width=11mm] at (2.8,-.55) {$\bar B$};
\node[collar,minimum width=11mm] at (3.75,-.55) {$\bar C_+$};
\node[exterior,minimum width=16mm] at (4.9,-.55) {$\bar G$};

\draw[blue!60!black,dashed,rounded corners,thick]
 (1.25,-1.05) rectangle (4.35,1.05);
\node[blue!60!black,fill=white,inner sep=2pt] at (2.8,1.05)
 {gate $U_{\rm loc}$};
\end{tikzpicture}
\caption{Reference and target systems in Lemma~\ref{lem:prep-insert}.
The target $\tau_{BCG}$ is
reconstructed at the level of canonical purifications after $B$ is erased.
Here $C=C_-\cup C_+$; the dashed region is the support
$BC\bar B\bar C$ of $U_{\rm loc}$. The exterior $G\bar G$ is untouched,
and the circuit uses only the physical and conjugate copies.}
\label{fig:insertion-gate}
\end{figure}

In this lemma, the exterior $G$ need not equal the reference exterior $F$: it may be a
subregion of $F$ or contain additional degrees of freedom. Fig.~\ref{fig:insertion-gate} shows the geometric structure of these regions schematically.

We need to find a local unitary $U$ that only acts on $BC\bar B\bar C$ nontrivially but can recover the region $B\bar B$ approximately. To do it, we need to construct some operator to decouple $\ket{\can(\tau_{BCG})}$ into subregions $B\bar B$ and $CG\bar C\bar G$. Then this operator can be turned into a unitary up to some controllable error.
\begin{proof}
{\it Step 1: constructing decoupling operator.} We first construct an operator that disentangles the regions $B\bar B$ and $CG\bar C\bar G$ approximately. To do this, we consider the two reduced density matrices of $\sigma_{BF}$ on $B,F$, denoted by $\sigma_B,\sigma_F$. Let $d_B=\dim\mH_B$ and $d_C=\dim\mH_C$, and introduce spaces
$C_1,C_2$ of dimensions $d_B,d_Bd_C$, respectively. We choose purifications
$\ket{\beta}_{BC_1}$ of $\sigma_B$ and $\ket{\xi}_{C_2F}$ of $\sigma_F$, which exist because $\operatorname{rank}\sigma_B\leqslant d_B$ and 
$\operatorname{rank}\sigma_F\le d_Bd_C$ (using that $\sigma_{BCF}$ is pure).
By Corollary~\ref{cor:appendix-decoupled-purification}, we have an isometry
$W:\mH_C\to\mH_{C_1}\otimes\mH_{C_2}$ such that for
\[
 \ket{\widetilde\psi_\sigma}:=(\one_B\otimes W\otimes\one_F)\ket{\psi_\sigma},
\] 
we have
\begin{equation}\label{eq:enlarge_space_product_approximation}
 \|\ket{\widetilde\psi_\sigma}-\ket{\beta}\otimes\ket{\xi}\|\le\sqrt\delta.
\end{equation} 
We use this $W$ to define the transformed target state
\[
\widetilde\tau := (\one_B\otimes W\otimes \one_G) \tau_{BCG} (\one_B\otimes W^\dagger\otimes \one_G).
\]
and will show that $\widetilde\tau$ is close to a product state between $BC_1$ and $C_2G$. The physical picture of $W$ is as follows. In the state $\tau_{BCG}$, there is correlation between $B$ and $C$, and correlation between $C$ and $G$. The operator $W$ effectively distributes these two types of correlations into the correlation within $BC_1$ and $C_2G$. In other words, $\tau_{BCG}$ is effectively disentangled in some sense.

Concretely, let $P_{\beta} = \ket{\beta}\bra{\beta}_{BC_1}\otimes\one_{C_2}$, and let $\ket{\widetilde\phi} = (\one_B\otimes W \otimes\one_{GR})\ket{\phi}$ with $\ket{\phi}$ a purification of $\tau_{BCG}$ with an auxiliary space $R$. By definition, we have $\widetilde\tau = \Tr_R(\ket{\widetilde \phi}\bra{\widetilde \phi})$.
Next, we define
\begin{equation}
    \ket{\chi}_{C_2GR} = \frac{1}{\sqrt{ \ev{\widetilde\phi | P_\beta\otimes\one_{GR}|\widetilde\phi} }} \bra{\beta}_{BC_1} \big(P_\beta\otimes\one_{GR}\ket{\widetilde\phi}\big),\qquad \omega_{C_2G} = \Tr_R (\ket{\chi}\bra{\chi}_{C_2GR}).
\end{equation}
We claim that, for $\eta := \delta+\epsilon/2$ \footnote{If \(\eta\geq1\), choose any density matrix \(\omega_{C_2G}\); the two trace-norm bounds below hold trivially. Hence we assume \(\eta<1\).}, the following two inequalities hold:
\begin{align}
 \|\widetilde\tau
    -\ket{\beta}\bra{\beta}_{BC_1}\otimes\omega_{C_2G}\|_1
 &\le2\sqrt\eta,\label{eq:prep-project-full}\\
 \|(W\otimes1_G)\tau_{CG}(W^{\dagger}\otimes1_G)
    -\beta_{C_1}\otimes\omega_{C_2G}\|_1
 &\le2\sqrt\eta,
 \label{eq:prep-project-reduced}
\end{align}
where $\beta_{C_1}=\Tr_B(\ket{\beta}\bra{\beta})$. These two inequalities imply that $\tilde\tau$ is close to a product state between $BC_1$ and $C_2G$. If $\eta=1$, the above inequalities are obvious, so in the following we assume $\eta<1$.

Indeed, the following normalized state 
\[
\ket{\widehat\phi} := \frac{1}{\sqrt{q}} P_\beta\otimes\one_{GR}\ket{\widetilde\phi},\quad q =  \ev{\widetilde\phi | P_\beta\otimes\one_{GR}|\widetilde\phi},
\]
is factorized as $\ket{\widehat\phi} = \ket{\beta}_{BC_1}\otimes\ket{\chi}_{C_2GR}$, and the overlap $\ev{\widehat\phi|\widetilde\phi} = \sqrt{q}$.
If we can show that $q\geq 1-\eta$, then we get
\begin{equation}
\bigl\|\ket{\widetilde\phi}\bra{\widetilde\phi}
       -\ket{\widehat\phi}\bra{\widehat\phi}\bigr\|_1
 =2\sqrt{1-q}\le2\sqrt\eta,
\end{equation}
and tracing out $R$ and further tracing out $B$ lead to Eqs. \eqref{eq:prep-project-full} and \eqref{eq:prep-project-reduced}. In the following, we will assume that $\eta\leqslant 1$, otherwise $q\geqslant 1-\eta$ and the to-be-proved lemma trivially hold.

To see that $q\geq 1-\eta$, notice that if we denote $M = (\one_B\otimes W^\dagger) P_\beta (\one_B\otimes W)$, then $q = \Tr(M\tau_{BC})$. Using the fact that
\begin{equation}
    \Tr[(1-M)\sigma_{BC}] = \norm{(1-P_\beta)\ket{\widetilde\psi_\sigma}}^2
    =\|(1-P_\beta)\big(|\widetilde\psi_\sigma\rangle-|\beta\rangle\otimes|\xi\rangle\big)\|^2
    \leq \norm{\ket{\widetilde\psi_\sigma} - \ket{\beta}\otimes\ket{\xi}}^2\leq\delta,
\end{equation}
we obtain
\begin{equation}
\begin{split}
 \Tr[(\id-M)\tau_{BC}]
 =&\Tr[(\id-M)(\sigma_{BC}+\tau_{BC}-\sigma_{BC})]
 =\Tr[(\id-M)\sigma_{BC}]+\Tr[\big(\frac{\id}{2}-M\big)(\tau_{BC}-\sigma_{BC})]\\
 \le& \Tr[(\id-M)\sigma_{BC}]
      +\tfrac12\|\tau_{BC}-\sigma_{BC}\|_1
 \le\delta+\epsilon/2= \eta.
\end{split}
\end{equation}
For the second equation, we have replaced the \(\id-M\) in $\Tr[(\id-M)(\tau_{BC}-\sigma_{BC})]$ by \(\frac12\id-M\) as $\Tr(\tau_{BC}-\sigma_{BC})=0$. The next inequality follows from the fact that $|\Tr(AB)|\leq \norm{A} \norm{B}_1$ and $\|\frac{\id}{2}-M\|\leqslant\frac{1}{2}$. Therefore, $q\geqslant 1-\eta$, and we have proved Eqs. \eqref{eq:prep-project-full} and \eqref{eq:prep-project-reduced}.

{\it Step 2: constructing unitary.} It remains to construct a unitary to get Eq. \eqref{eq:prep-insert}. Denote
\[
 \ket{v_{\rm in}}=\ket{0}_{B\bar B}\otimes\ket{\operatorname{can}(\tau_{CG})},
 \qquad \ket{v_{\rm out}}=\ket{\operatorname{can}(\tau_{BCG})},
\]
and let $\mathcal W=W\otimes\bar W$, with identities on
$B\bar B G\bar G$ understood. Applying Eq. \eqref{eq:prep-canonical} to Eqs.
\eqref{eq:prep-project-reduced} and \eqref{eq:prep-project-full} gives
\begin{align}
 \|\mathcal W\ket{v_{\rm in}}-\ket{a_{\rm in}}\|&\le\sqrt2\eta^{1/4},
 &\ket{a_{\rm in}}&=\ket{0}_{B\bar B}\otimes
   \ket{\operatorname{can}(\beta_{C_1})}\otimes\ket{\operatorname{can}(\omega_{C_2G})},\label{eq:w_vin}\\
 \|\mathcal W\ket{v_{\rm out}}-\ket{a_{\rm out}}\|&\le\sqrt2\eta^{1/4},
 &\ket{a_{\rm out}}&=(\ket{\beta}_{BC_1}\otimes\ket{\overline{\beta}}_{\bar B\bar C_1})\otimes
   \ket{\operatorname{can}(\omega_{C_2G})}. \label{eq:w_vout}
\end{align}
Here canonical purification respects tensor products, and
$\ket{\operatorname{can}(\ket{\beta}\bra{\beta}_{BC_1})}=\ket{\beta}_{BC_1}\otimes\ket{\overline{\beta}}_{\bar B\bar C_1}$.
We can choose a unitary $G_0$ on
$B\bar B C_1\bar C_1$ such that
\[
 G_0\bigl(\ket{0}_{B\bar B}\otimes
          \ket{\operatorname{can}(\beta_{C_1})}\bigr)
 =\ket{\beta}_{BC_1}\otimes\ket{\overline{\beta}}_{\bar B\bar C_1}.
\]
This choice
uses only $\ket{\beta}$, so $G_0$ is the same for every target $\tau_{BCG}$ and exterior $G$.
Since $G_0\ket{a_{\rm in}}=\ket{a_{\rm out}}$, the triangle inequality gives
\begin{equation}\label{eq:prep-isometric-recovery}
 \|G_0\mathcal W\ket{v_{\rm in}}-\mathcal W\ket{v_{\rm out}}\|
 \le\|\mathcal W\ket{v_{\rm in}}-\ket{a_{\rm in}}\|
     +\|\ket{a_{\rm out}}-\mathcal W\ket{v_{\rm out}}\|
 \le2\sqrt2\eta^{1/4}.
\end{equation}
Hence, we can compress $G_0$ to the original doubled space:
\[
 K=\mathcal W^{\dagger}G_0\mathcal W:\quad \mH_{BCG\bar B\bar C\bar G}\to\mH_{BCG\bar B\bar C\bar G}.
\]
The operator $K$ nearly achieves Eq. \eqref{eq:prep-insert}. However, instead of being a unitary, it is a contraction supported on $BC\bar B\bar C$ with trivial action on $G\bar G$. Since $K$ is a square finite-dimensional matrix, we can do a polar decomposition such that $K=U_{\rm loc} |K|$, where $U_{\loc}$ is unitary and $|K|=(K^\dagger K)^{1/2}$. This $U_{\loc}$ will be shown to serve as the unitary in Lemma \ref{lem:prep-insert}. Notice the construction of $U_{\rm loc}$ only relies on the reference state $\sigma_{BC}$, but not the exterior region $G$ or the target state $\tau_{BCG}$. 

Replacing $K$ by $U_{\rm loc}$ induces some errors. To estimate these errors, we use the projector $\mathcal{W}\mathcal{W}^\dagger$ to split $G_0\mathcal{W}\ket{v_{\rm in}}-\mathcal{W}\ket{v_{\rm out}}$ into two orthogonal pieces, i.e., $G_0\mathcal{W}\ket{v_{\rm in}}-\mathcal{W}\ket{v_{\rm out}}=\mathcal{W}\mathcal{W}^\dagger(G_0\mathcal{W}\ket{v_{\rm in}}-\mathcal{W}\ket{v_{\rm out}})+(1-\mathcal{W}\mathcal{W}^\dagger)(G_0\mathcal{W}\ket{v_{\rm in}}-\mathcal{W}\ket{v_{\rm out}})=\mathcal{W}(K|v_{\rm in}\rangle-|v_{\rm out}\rangle)+(1-\mathcal W\mathcal W^
 \dagger) G_0\mathcal W\ket{v_{\rm in}}$, and denote their norm as
\[
 d_{\parallel}=\|K\ket{v_{\rm in}}-\ket{v_{\rm out}}\|,
 \qquad
 d_{\perp}=\|(1-\mathcal W\mathcal W^
 \dagger) G_0\mathcal W\ket{v_{\rm in}}\|.
\]
Then Eq. \eqref{eq:prep-isometric-recovery} gives
\[
 d_{\parallel}^2+d_{\perp}^2\le8\sqrt\eta,
 \qquad
 d_{\perp}^2=\bra{v_{\rm in}}(1-K^{\dagger}K)\ket{v_{\rm in}}.
\]
Since $0\le|K|\le1$, we have $(1-|K|)^2\le1-K^{\dagger}K$, and hence
\[
 \|(U_{\rm loc}-K)\ket{v_{\rm in}}\|^2
 =\bra{v_{\rm in}}(1-|K|)^2\ket{v_{\rm in}}
 \le d_{\perp}^2.
\]
Finally, let $U = U_{\rm loc}\otimes\one_{G\bar G}$, we get
\[
 \|U\ket{v_{\rm in}}-\ket{v_{\rm out}}\|
 \le d_{\parallel}+d_{\perp}
 \le\sqrt{2(d_{\parallel}^2+d_{\perp}^2)}
 \le4\eta^{1/4}.
\]
This proves Eq. \eqref{eq:prep-insert} on the doubled system, uniformly in
$\tau$ and $G$. 

\end{proof}

The proof of Lemma~\ref{lem:prep-insert} also provides the following bound. Under the decoupling assumption Eq. \eqref{eq:mutual_correlation_relation}, for every $F_0\subset F$ the canonical purifications of $\sigma_{BCF_0}$
and $\sigma_{CF_0}$ have $F_0\bar F_0$ marginals differing by at most $4\sqrt2\,\delta^{1/4}$,
\begin{equation}\label{eq:prep-erasure}
 \left\|
 \bigl(\ket{\operatorname{can}(\sigma_{BCF_0})}
       \bra{\operatorname{can}(\sigma_{BCF_0})}\bigr)_{F_0\bar F_0}
 -
 \bigl(\ket{\operatorname{can}(\sigma_{CF_0})}
       \bra{\operatorname{can}(\sigma_{CF_0})}\bigr)_{F_0\bar F_0}
 \right\|_1
 \le 4\sqrt2\,\delta^{1/4}.
\end{equation}
To see it, take $\tau_{BCG}=\sigma_{BCF_0}$ in the above proof so that $\epsilon=0$, we can get 
\begin{equation}
\begin{split}
     \bigl(\ket{\operatorname{can}(\sigma_{BCF_0})}
       \bra{\operatorname{can}(\sigma_{BCF_0})}\bigr)_{F_0\bar F_0} &= \Tr_{B\bar B C_1\bar C_1 C_2\bar C_2}[\mathcal{W}\ket{v_{\rm out}}\bra{v_{\rm out}}\mathcal W^\dagger], \\
       \bigl(\ket{\operatorname{can}(\sigma_{CF_0})}
       \bra{\operatorname{can}(\sigma_{CF_0})}\bigr)_{F_0\bar F_0} &= \Tr_{B\bar B C_1\bar C_1 C_2\bar C_2}[\mathcal{W}\ket{v_{\rm in}} \bra{v_{\rm in}}\mathcal{W}^\dagger],
\end{split}
\end{equation}
and 
\begin{equation}
    \Tr_{B\bar B C_1 \bar C_1 C_2 \bar C_2 }[\ket{a_{\rm in}}\bra{a_{\rm in}}]=\Tr_{B\bar BC_1 \bar C_1 C_2 \bar C_2}[\ket{a_{\rm out}}\bra{a_{\rm out}}].
\end{equation}
Using the monotonicity of the trace distance under partial trace, the triangle inequality, Eqs. \eqref{eq:w_vin} and \eqref{eq:w_vout}, and the Fuchs-van de Graaf inequality, we get Eq. \eqref{eq:prep-erasure}.
The same estimate holds when \(B\) is a union of several intervals, since the proof does not require \(B\) to be a single interval.

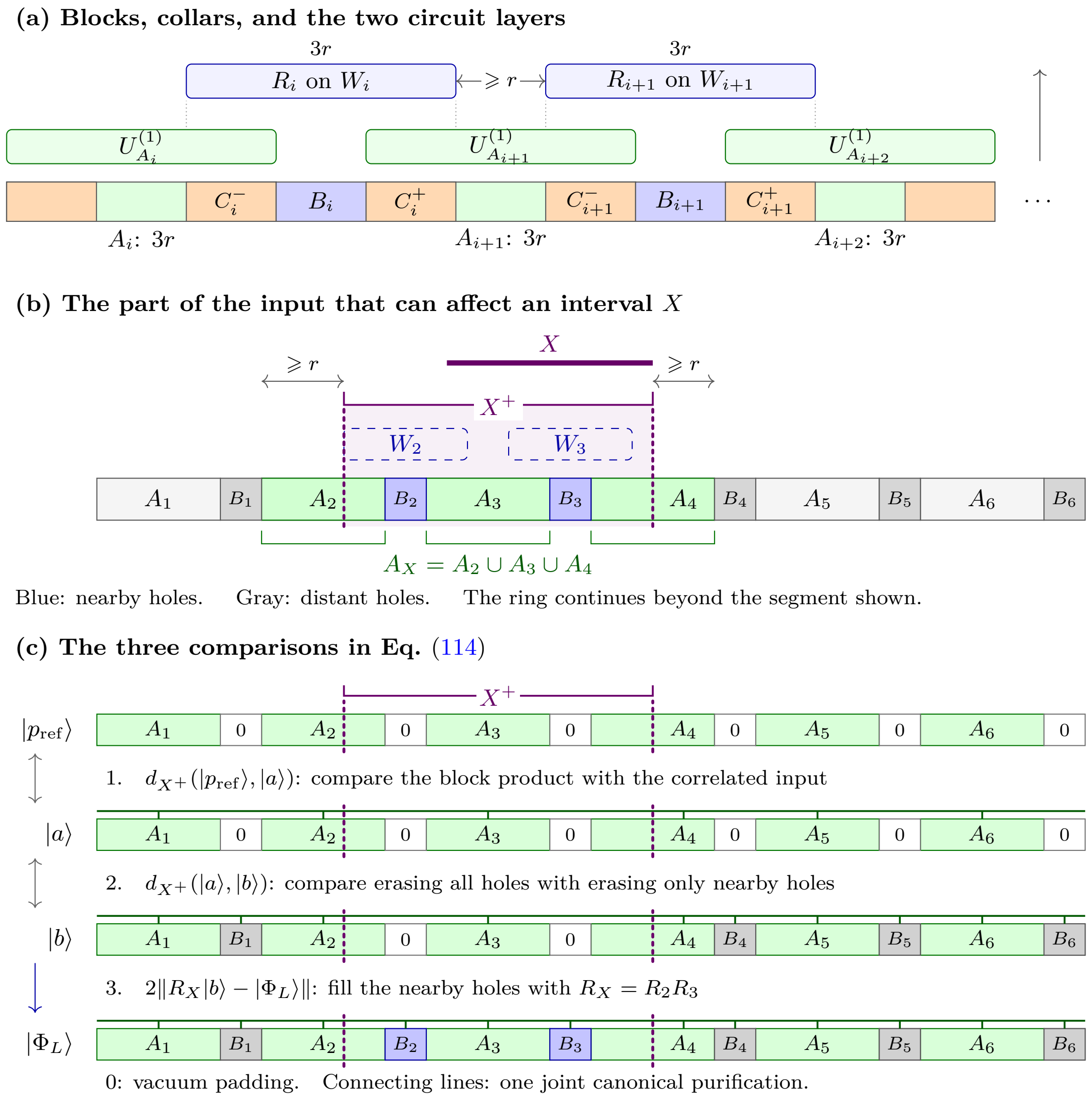
\begin{figure}[!htbp]
\centering
\begin{tikzpicture}[x=.60cm,y=.70cm,font=\small,
  block/.style={draw=black!65,fill=green!12},
  collar/.style={draw=black!65,fill=orange!30},
  hole/.style={draw=black!65,fill=blue!18},
  gateone/.style={draw=green!45!black,fill=green!8,rounded corners=2pt},
  gatetwo/.style={draw=blue!65!black,fill=blue!5,rounded corners=2pt}]
\path[use as bounding box] (0,-1.7) rectangle (24,3.4);
\node[anchor=west,font=\small\bfseries] at (0,3.1)
  {(a) Blocks, collars, and the two circuit layers};
\foreach \x/\lab in {0/i,8/{i+1},16/{i+2}} {
  \draw[gateone] (\x,.35) rectangle ++(6,.65);
  \node at (\x+3,.675) {$U^{(1)}_{A_{\lab}}$};
  \node at (\x+3,-1.1) {$A_{\lab}$: $3r$};
  \draw[block] (\x,-.75) rectangle ++(6,.75);
  \draw[collar] (\x,-.75) rectangle ++(2,.75);
  \draw[collar] (\x+4,-.75) rectangle ++(2,.75);
}
\draw[hole] (6,-.75) rectangle (8,0);
\draw[hole] (14,-.75) rectangle (16,0);
\node at (7,-.375) {$B_i$};
\node at (15,-.375) {$B_{i+1}$};
\node[font=\footnotesize] at (5,-.375) {$C_i^-$};
\node[font=\footnotesize] at (9,-.375) {$C_i^+$};
\node[font=\footnotesize] at (13,-.375) {$C_{i+1}^-$};
\node[font=\footnotesize] at (17,-.375) {$C_{i+1}^+$};
\node at (23,-.375) {$\cdots$};
\foreach \x/\lab in {4/i,12/{i+1}} {
  \draw[black!40,densely dotted] (\x,1.05)--(\x,1.55);
  \draw[black!40,densely dotted] (\x+6,1.05)--(\x+6,1.55);
  \draw[gatetwo] (\x,1.6) rectangle ++(6,.65);
  \node at (\x+3,1.925) {$R_{\lab}$ on $W_{\lab}$};
  \node[font=\footnotesize] at (\x+3,2.55) {$3r$};
}
\draw[<->,black!65] (10,1.925)--(12,1.925);
\node[fill=white,inner sep=1pt,font=\footnotesize] at (11,1.925) {$\ge r$};
\draw[->,black!70] (23,.4)--(23,2.15);
\end{tikzpicture}

\par\medskip
\begin{tikzpicture}[x=.55cm,y=.70cm,font=\small,
  boundary/.style={draw=violet!85!black,dotted,line width=1.1pt,line cap=round}]
\path[use as bounding box] (-2.2,-10.9) rectangle (24,4.4);
\begin{scope}[overlay]
\node[anchor=west,font=\small\bfseries] at (-2.2,4.1)
  {(b) The part of the input that can affect an interval $X$};
\fill[violet!6] (6,-.12) rectangle (13.5,2.2);
\draw[violet!80!black,line width=2pt] (8.5,3.0)--(13.5,3.0);
\node[violet!80!black,anchor=south] at (11,3.05) {$X$};
\draw[violet!85!black,line width=.7pt] (6,2.45)--(6,2.2)--(13.5,2.2)--(13.5,2.45);
\node[fill=white,inner sep=2pt,text=violet!80!black] at (9.75,2.2) {$X^+$};
\draw[<->,black!65] (4,2.65)--(6,2.65);
\node[anchor=south,font=\footnotesize] at (5,2.65) {$\ge r$};
\draw[<->,black!65] (13.5,2.65)--(15,2.65);
\node[anchor=south,font=\footnotesize] at (14.25,2.65) {$\ge r$};
\foreach \x/\lab in {6/2,10/3} {
  \draw[blue!65!black,dashed,rounded corners=2pt]
    (\x,1.15) rectangle ++(3,.6);
  \node[text=blue!65!black] at (\x+1.5,1.45) {$W_{\lab}$};
}
\foreach \x/\lab in {0/1,4/2,8/3,12/4,16/5,20/6} {
  \draw[black!65,fill=black!4] (\x,0) rectangle ++(3,.8);
  \ifnum\lab=4
    \node at (\x+2.25,.4) {$A_{\lab}$};
  \else
    \node at (\x+1.5,.4) {$A_{\lab}$};
  \fi
}
\foreach \x/\lab in {4/2,8/3,12/4} {
  \draw[green!45!black,fill=green!18] (\x,0) rectangle ++(3,.8);
  \ifnum\lab=4
    \node at (\x+2.25,.4) {$A_{\lab}$};
  \else
    \node at (\x+1.5,.4) {$A_{\lab}$};
  \fi
}
\foreach \x/\lab in {3/1,7/2,11/3,15/4,19/5,23/6} {
  \draw[black!65,fill=black!16] (\x,0) rectangle ++(1,.8);
  \node[font=\scriptsize] at (\x+.5,.4) {$B_{\lab}$};
}
\foreach \x/\lab in {7/2,11/3} {
  \draw[blue!65!black,fill=blue!23] (\x,0) rectangle ++(1,.8);
  \node[font=\scriptsize] at (\x+.5,.4) {$B_{\lab}$};
}
\foreach \x in {4,8,12}
  \draw[green!45!black] (\x,-.2)--(\x,-.45)--(\x+3,-.45)--(\x+3,-.2);
\node[text=green!35!black] at (9.5,-.85)
  {$A_X=A_2\cup A_3\cup A_4$};
\foreach \x in {6,13.5}
  \draw[boundary] (\x,-.12)--(\x,2.2);
\node[anchor=west,font=\footnotesize] at (-2.2,-1.5)
  {Blue: nearby holes. \quad Gray: distant holes. \quad The ring continues beyond the segment shown.};
\end{scope}
\begin{scope}[shift={(0,-4.3)},overlay,
  block/.style={draw=green!45!black,fill=green!15},
  emptyhole/.style={draw=black!55,fill=white},
  farhole/.style={draw=black!65,fill=black!18},
  nearhole/.style={draw=blue!65!black,fill=blue!23},
  boundary/.style={draw=violet!85!black,dotted,line width=1.1pt,line cap=round},
  link/.style={draw=green!35!black,line width=.7pt}]
\node[anchor=west,font=\small\bfseries] at (-2.2,1.85)
  {(c) The three comparisons in Eq.~\eqref{eq:prep-local-three-errors}};
\foreach \y/\state in {0/{p_{\rm ref}},-2/a,-4/b,-6/{\Phi_L}} {
  \node[anchor=east] at (-.35,\y+.30) {$\ket{\state}$};
  \foreach \x/\lab in {0/1,4/2,8/3,12/4,16/5,20/6} {
    \draw[block] (\x,\y) rectangle ++(3,.6);
    \ifnum\lab=4
      \node[font=\footnotesize] at (\x+2.25,\y+.30) {$A_{\lab}$};
    \else
      \node[font=\footnotesize] at (\x+1.5,\y+.30) {$A_{\lab}$};
    \fi
  }
  \foreach \x in {3,7,11,15,19,23} {
    \draw[emptyhole] (\x,\y) rectangle ++(1,.6);
    \node[font=\scriptsize] at (\x+.5,\y+.30) {$0$};
  }
}
\foreach \y in {-2,-4,-6} {
  \draw[link] (0,\y+.75)--(24,\y+.75);
  \foreach \x in {1.5,5.5,9.5,14.25,17.5,21.5}
    \draw[link] (\x,\y+.6)--(\x,\y+.75);
}
\foreach \y in {-4,-6} {
  \foreach \x/\lab in {3/1,15/4,19/5,23/6} {
    \draw[farhole] (\x,\y) rectangle ++(1,.6);
    \node[font=\scriptsize] at (\x+.5,\y+.30) {$B_{\lab}$};
    \draw[link] (\x+.5,\y+.6)--(\x+.5,\y+.75);
  }
}
\foreach \x/\lab in {7/2,11/3} {
  \draw[nearhole] (\x,-6) rectangle ++(1,.6);
  \node[font=\scriptsize] at (\x+.5,-5.70) {$B_{\lab}$};
  \draw[link] (\x+.5,-5.4)--(\x+.5,-5.25);
}
\draw[violet!85!black,line width=.7pt] (6,1.15)--(6,.95)--(13.5,.95)--(13.5,1.15);
\node[text=violet!80!black,fill=white,inner sep=1pt] at (9.75,.95) {$X^+$};
\foreach \y in {0,-2,-4,-6} {
  \draw[boundary] (6,\y-.12)--(6,\y+.85);
  \draw[boundary] (13.5,\y-.12)--(13.5,\y+.85);
}
\draw[<->,black!60] (-1.5,-.15)--(-1.5,-1.10);
\node[anchor=west,font=\footnotesize] at (0,-.65)
  {$1.\quad d_{X^+}(\ket{p_{\rm ref}},\ket{a})$: compare the block product with the correlated input};
\draw[<->,black!60] (-1.5,-2.15)--(-1.5,-3.10);
\node[anchor=west,font=\footnotesize] at (0,-2.65)
  {$2.\quad d_{X^+}(\ket{a},\ket{b})$: compare erasing all holes with erasing only nearby holes};
\draw[->,blue!65!black] (-1.5,-4.15)--(-1.5,-5.10);
\node[anchor=west,font=\footnotesize] at (0,-4.65)
  {$3.\quad 2\|R_X\ket{b}-\ket{\Phi_L}\|$: fill the nearby holes with $R_X=R_2R_3$};
\node[anchor=west,font=\footnotesize] at (0,-6.45)
  {$0$: vacuum padding.\quad Connecting lines: one joint canonical purification.};
\end{scope}
\end{tikzpicture}
\caption{Geometry of the two-layer circuit in
Lemma~\ref{lem:prep-approx}.  
(a) Gate order runs upward: the first layer acts on the blocks $A_i$;
the second acts on
$W_i=C_i^-\cup B_i\cup C_i^+$, using the orange collars to fill the blue
holes.  The unused middle part of each block separates neighboring
recovery gates.
(b) A representative observation interval $X$ meets two recovery
supports.  Their union with $X$ is $X^+$; the three green blocks form
$A_X$.  The remaining holes are at distance at least $r$ from $X^+$.
(c) The three terms of \eqref{eq:prep-local-three-errors}, using the same
blocks and holes as in (b).
The first comparison replaces the product of reference block
purifications $\ket{p_{\rm ref}}$ by the joint target purification $\ket{a}$, with
all holes erased.  The second compares $\ket{a}$ with $\ket{b}$, in which only
the nearby holes $B_2,B_3$ are erased.
Both comparisons concern the $X^+$ marginal before applying $R_X$.
The third applies $R_X=R_2R_3$ to $\ket{b}$ and compares the result with
$\ket{\Phi_L}$.
Panels (b) and (c) share the same horizontal scale; the violet dotted
guides mark the same endpoints of $X^+$ in (b) and all four rows of (c).
Zeros mark erased holes padded with vacuum; colored holes are included
in the target marginal.  Connecting lines indicate one joint canonical
purification, without specifying pairwise entanglement.}
\label{fig:prep-scale-geometry}
\end{figure}

The above results allow us to construct a depth-2 quantum circuit that approximately disentangles $|\Phi_L\rangle$, where each gate in this circuit is supported on an interval with a uniformly bounded length.

\begin{lemma}[Preparation at scale $r$]\label{lem:prep-approx}
Let $\{\ket{\Phi_L}\}$ and $\{\varrho_L\}$ be the sequences of states satisfying assumptions of Theorem \ref{thm:prep-exact}. For any integers $r\geq 1$ and $L\ge96r$, there are
depth-two unitary circuits $U_{L,r}$, with gate intervals of length at most $7r$, and constants $C,c>0$ such that
\begin{align}
 \|U_{L,r}\ket{\Omega_L}-\ket{\Phi_L}\|&\le C(L/r)e^{-cr},
 \label{eq:prep-global}\\
 \|(U_{L,r}\ket{\Omega_L}\bra{\Omega_L}U_{L,r}^{\dagger})_X
-( \ket{\Phi_L}\bra{\Phi_L})_X\|_1
 &\le C(1+|X|/r)e^{-cr}
 \label{eq:prep-local}
\end{align}
for every interval $X$, including the whole ring.
The bulk gates and their locations
are independent of $L$, and only a single gate near the closing seam
has size dependent on $(L\mod 4r)$.
\end{lemma}

\begin{proof}
The two-layer construction adapts the use of decoupling and Uhlmann's
theorem in Dalzell and Brand\~ao~\cite[Theorem~2]{DalzellBrandao2019}.
We now construct the two layers, prove the estimation in Eq. \eqref{eq:prep-global} by three
comparisons, and then restrict those comparisons to an interval $X$ to prove Eq. \eqref{eq:prep-local}.
Throughout the proof, spatial supports are in the enlarged system.

{\it Step 1: construct two-layer circuit.} Let $M=\lfloor L/(4r)\rfloor$ and $s=L-4Mr$.  Starting at site $0$,
partition the ring in the cyclic order
$A_1,B_1,\ldots,A_M,B_M$.  The blocks $A_i$, $i<M$, have length $3r$,
the last block has length $3r+s<7r$, and every hole $B_i$ has length $r$.
Let $C_i^-$ be the last $r$ sites of $A_i$ and $C_i^+$ the first $r$
sites of $A_{i+1}$, with indices modulo $M$.  Set
$C_i=C_i^-\cup C_i^+$, $W_i=C_i^-\cup B_i\cup C_i^+$,
$A=\bigcup_i A_i$ and $B=\bigcup_i B_i$.
The intervals $W_i$ have length $3r$ and are disjoint, and the two collars
inside each $A_i$ leave at least $r$ sites between them (see Fig. ~\ref{fig:prep-scale-geometry}(a)).

Choose the gates using the reference ring of length $N=96r$.
For each block of length $\ell$ with $3r\leq\ell\leq 7r-1$, let $\nu_\ell = \varrho_{N, [0,\ell-1]}$
be the reference marginal of $\varrho_N$ on $[0,\ell-1]$.
We can find a unitary on this interval which sends the vacuum to
$\ket{\operatorname{can}(\nu_\ell)}$ on the physical and conjugate copies.
Placing a copy of this unitary on each $A_i$, we get the first-layer gate $U_{A_i}^{(1)}$.

To construct the second layer, we will invoke  Lemma~\ref{lem:prep-insert}. Specifically, here the pure state $\varrho_N$ on the reference ring will serve as the state $\sigma_{BCF}$ in Lemma~\ref{lem:prep-insert}, and the region $BC$ in Lemma~\ref{lem:prep-insert} here is a length-$3r$ interval. Then Lemma~\ref{lem:prep-insert} gives a recovery unitary on $W_i$, which is independent of the states to be recovered. We call these gates $R_i$. Thus both layers consist of gates on disjoint intervals, and
\[
 R:=\prod_{i=1}^M R_i,\qquad
 U_{L,r}:=R\prod_{i=1}^M U_{A_i}^{(1)}
\]
has depth two and gate lengths less than $7r$. By construction, the bulk gates and their locations
are independent of $L$, and only a single gate near the closing seam
has size dependent on $(L\mod 4r)$.

In the following, we will prove that this two-layer circuit satisfies Eqs. \eqref{eq:prep-global} and \eqref{eq:prep-local}.

For the next steps, we define two error bounds independent of $L$:
\[
 \delta_r=C e^{-c r},\qquad
 \epsilon_r=C e^{-c N} = C e^{-96 cr}
\]
with $C,c$ the constants from Eqs. \eqref{eq:prep-input-decouple} and \eqref{eq:prep-input-compare}.
$\delta_r$ bounds the decoupling error in Eq.
\eqref{eq:prep-input-decouple}, on both the target and reference rings. The second bounds the difference between reference and target marginals
on every $A_i$ and $W_i$ by Eq. \eqref{eq:prep-input-compare}, since their lengths are below $N/12=8r$. Now suppose that we start with a block-wise product state $\otimes_i \varrho_{L,A_i}$ from $\varrho_L=\ket{\phi_L}\bra{\phi_L}$, and, as in Lemma~\ref{lem:prep-insert}, use the gate $R_i$ to recover the hole $B_i$ for the target state $\tau_{B_iC_i}^{(i)} = \varrho_{L,W_i}$ and $\tau_{C_i G_i}^{(i)} = \varrho_{L, AB_1B_2\cdots B_{i-1}}$, then the recovery error will be bounded by
\begin{equation}\label{eq:prep-one-hole-error}
 \norm{R_i\big( \ket{0}_{B_i \bar B_i} \otimes \ket{\can(\varrho_{L,AB_1B_2\cdots B_{i-1}})} \big) - \ket{\can(\varrho_{L,AB_1B_2\cdots B_i})}}\leq \eta_r:=4(\delta_r+\epsilon_r/2)^{1/4}.
\end{equation}

{\it Step 2: prove state approximation.} We first prove Eq. \eqref{eq:prep-global}.

Introduce three pure states,
with the region $B\bar B$ in the vacuum implicitly:
\[
 \ket{p_{\rm ref}}:=\bigotimes_i\ket{\operatorname{can}(\nu_{|A_i|})},\qquad
 \ket{p_L}:=\bigotimes_i\ket{\operatorname{can}(\varrho_{L,A_i})},\qquad
 \ket{a}:=\ket{\operatorname{can}(\varrho_{L,A})}.
\]
The first-layer gates acting on $\ket{\Omega_L}$ produce $\ket{p_{\rm ref}}$.
Triangle inequalities then give
\begin{equation}\label{eq:prep-vector-three-errors}
 \begin{aligned}
 \|U_{L,r}\ket{\Omega_L}-\ket{\Phi_L}\|
 &=\|R\ket{p_{\rm ref}}-\ket{\Phi_L}\|\\
 &\le
 \underbrace{\|\ket{p_{\rm ref}}-\ket{p_L}\|}_{\text{reference block marginals}}
 +\underbrace{\|\ket{p_L}-\ket{a}\|}_{\text{correlations between blocks}}
 +\underbrace{\|R\ket{a}-\ket{\Phi_L}\|}_{\text{filling the holes}}.
 \end{aligned}
\end{equation}
The meanings of the above three terms are indicated. In the following we bound these three terms separately.

For the first term, replacing one tensor factor at a time and using Eq.
\eqref{eq:prep-canonical} give
\[
 \|\ket{p_{\rm ref}}-\ket{p_L}\|
 \le\sum_{i=1}^M
 \|\ket{\operatorname{can}(\nu_{|A_i|})}
       -\ket{\operatorname{can}(\varrho_{L,A_i})}\|
 \le M\sqrt{\epsilon_r}.
\]

For the second term, successive blocks can be separated at cost
$\delta_r$ each.  Indeed, for each $k>1$, the interval containing $A_1,\ldots,A_{k-1}$
is separated from the interval $A_k$ by $r$ sites.  With Eq. \eqref{eq:prep-input-decouple}, the monotonicity of the trace distance under partial trace, and the triangle inequality, we get
\[
 \begin{aligned}
 \left\|\varrho_{L,\bigcup_{i=1}^k A_i}
             -\bigotimes_{i=1}^k\varrho_{L,A_i}\right\|_1
 &\le
 \left\|\varrho_{L,\bigcup_{i=1}^k A_i}
       -\varrho_{L,\bigcup_{i=1}^{k-1} A_i}\otimes\varrho_{L,A_k}\right\|_1\\
 &\quad+
 \left\|\varrho_{L,\bigcup_{i=1}^{k-1} A_i}
             -\bigotimes_{i=1}^{k-1}\varrho_{L,A_i}\right\|_1\\
 &\le(k-1)\delta_r.
 \end{aligned}
\]
Consequently, $\|\ket{p_L}-\ket{a}\|\le\sqrt{(M-1)\delta_r}$ by Eq. \eqref{eq:prep-canonical}.

For the third term, let $\ket{a_k}$ be the canonical purification of
the reduced density matrix of $\varrho_L$ on $A\cup B_1\cup\cdots\cup B_k$.
Thus $\ket{a_0}=\ket{a}$ and $\ket{a_M}=\ket{\Phi_L}$.
At step $k$, the reduced density matrix of $\varrho_L$ on $A\cup B_1\cup\cdots\cup B_k$
contains all of $W_k$, and tracing out $B_k$ gives the previous one.
Each consecutive pair therefore satisfies $\|R_k \ket{a_{k-1}}-\ket{a_k}\|\le \eta_r$ by Eq.
\eqref{eq:prep-one-hole-error}.  Inserting these intermediate vectors
gives the explicit telescoping bound
\[
 \begin{aligned}
 \|R\ket{a}-\ket{\Phi_L}\|
 &=\left\|\sum_{k=1}^M
       R_M\cdots R_{k+1}(R_k \ket{a_{k-1}}-\ket{a_k})\right\|\\
 &\le\sum_{k=1}^M\|R_k \ket{a_{k-1}}-\ket{a_k}\|
 \le M \eta_r .
 \end{aligned}
\]
Combining the three estimates in Eq. \eqref{eq:prep-vector-three-errors} yields
\begin{equation}\label{eq:prep-vector-explicit}
 \|U_{L,r}\ket{\Omega_L}-\ket{\Phi_L}\|
 \le M\sqrt{\epsilon_r}+\sqrt{(M-1)\delta_r}+M \eta_r
 \le C_{\rm glob}(L/r)e^{-c_{\rm glob} r}.
\end{equation}
The constants are $C_{\rm glob} = \frac{\sqrt{C}}{2} + (3C/2)^{1/4}$ and $c_{\rm glob} = c/4$, where we have used $M\le L/(4r)$.
All vector phases are fixed by the canonical purifications and the
defining actions of the gates.

{\it Step 3: prove interval approximation.}
For inequality \eqref{eq:prep-local}, fix a nonempty interval $X$.
Call a hole $B_i$ \emph{nearby} if its recovery interval $W_i$ meets $X$.
All other holes are called \emph{distant}.  Let $m$ be the number of nearby
holes, and let $R_X$ be the product of their recovery gates.
Only these gates can affect the output state on $X$.
Enlarge $X$ just enough to include their supports:
\[
 X^+:=X\cup\bigcup_{W_i\cap X\ne\varnothing}W_i,
 \qquad |X^+|\le |X|+6r,\qquad
 m\le 2+|X|/(4r).
\]
$X^+$ is an interval, possibly the whole ring.
Let $A_X$ be the union of the blocks $A_i$ meeting $X^+$, and let $t$ be
their number.  These blocks are consecutive, and
$t\le m+2\le4+|X|/(4r)$ since only the two endpoint blocks can contribute
without an intervening nearby hole.
Fig. \ref{fig:prep-scale-geometry}(b) shows the geometry of $X$ and $X^+$.

Besides the input $\ket{a}$, in which all holes are erased, use the comparison
vector $\ket{b}$ in which only nearby holes are erased:
\[
 \ket{b}:=\ket{\operatorname{can}\!\left(
 \varrho_{L,\,\T_L\setminus
             \bigcup_{W_i\cap X\ne\varnothing}B_i}\right)}.
\]
Again, every erased hole is padded with vacuum in both copies.
For normalized states $\ket{u},\ket{v}$ in the enlarged space, write
\[
 d_Z(\ket{u},\ket{v}):=\|(\ket{u}\bra{u})_Z
                  -(\ket{v}\bra{v})_Z\|_1
 \le2\|\ket{u}-\ket{v}\|
\]
for any region $Z$.
Because $R_X$ is supported on $X^+$, tracing to $X$ after applying $R_X$
is a channel on the $X^+$ marginal.  The local triangle inequality,
illustrated in Fig.~\ref{fig:prep-scale-geometry}(c), is therefore
\begin{equation}\label{eq:prep-local-three-errors}
 \begin{aligned}
 d_X(U_{L,r}\ket{\Omega_L},\ket{\Phi_L})
 &=d_X(R_X\ket{p_{\rm ref}},\ket{\Phi_L})\\
 &\le d_X(R_X\ket{p_{\rm ref}},R_X\ket{a})
       +d_X(R_X\ket{a},R_X\ket{b})+d_X(R_X\ket{b},\ket{\Phi_L})\\
 &\le
 \underbrace{d_{X^+}(\ket{p_{\rm ref}},\ket{a})}_{\text{product-input}}
 +\underbrace{d_{X^+}(\ket{a},\ket{b})}_{\text{erasing distant holes}}
 +\underbrace{2\|R_X\ket{b}-\ket{\Phi_L}\|}_{\text{filling nearby holes}}.
 \end{aligned}
\end{equation}
We now estimate these three terms.

For the term $d_{X^+}(\ket{p_{\rm ref}},\ket{a})$, first separate the blocks in $A_X$ from
the remaining blocks, and then separate those $t$ blocks from each other.
If $A_X\ne A$, the intervals containing the two groups have holes of
length $r$ between them in both directions.  Thus
\[
 \|\varrho_{L,A}
       -\varrho_{L,A_X}\otimes\varrho_{L,A\setminus A_X}\|_1\le\delta_r,
 \qquad
 \left\|\varrho_{L,A_X}
       -\bigotimes_{A_i\subset A_X}\varrho_{L,A_i}\right\|_1
       \le(t-1)\delta_r .
\]
To get estimation of $d_{X^+}(\ket{p_{\rm ref}}, \ket{a})$,
we use the states
\[
 \ket{u}:=\ket{\operatorname{can}(\varrho_{L,A_X})}\otimes
             \ket{\operatorname{can}(\varrho_{L,A\setminus A_X})},\qquad
 \ket{v}:=\bigotimes_{A_i\subset A_X}\ket{\operatorname{can}(\varrho_{L,A_i})}
       \otimes\ket{\operatorname{can}(\varrho_{L,A\setminus A_X})}.
\]
They satisfy \[
\bigl\||u\rangle-|a\rangle\bigr\|\le\sqrt{\delta_r},
\qquad
\bigl\||v\rangle-|u\rangle\bigr\|
\le\sqrt{(t-1)\delta_r}.
\]
On $X^+$, the vectors $\ket{p_{\rm ref}}$ and $\ket{v}$ differ only in the $t$
block factors, each of which costs at most $\sqrt{\epsilon_r}$.
Hence
\begin{equation}\label{eq:prep-local-product-error}
 \begin{aligned}
 d_{X^+}(\ket{p_{\rm ref}},\ket{a})
 &\le d_{X^+}(\ket{p_{\rm ref}},\ket{v})+d_{X^+}(\ket{v},\ket{u})+d_{X^+}(\ket{u},\ket{a})\\
 &\le2t\sqrt{\epsilon_r}
       +2\sqrt{(t-1)\delta_r}+2\sqrt{\delta_r}.
 \end{aligned}
\end{equation}
If $A_X=A$, take $\ket{u}=\ket{a}$ and omit the last term.

Next we consider the distant-hole term $d_{X^+}(\ket{a}, \ket{b})$. In $ \ket{b} $, only the nearby holes have been erased, while in $\ket{a}$ the distant holes have been erased as well. So we need to show that erasing the distant holes has only a small effect on the reduced state inside $X^+$. These distant holes are separated from $X^+$ by at least $r$ sites, so the trace-norm decoupling relation controls their correlation with $X^+$ by $\delta_r$ \footnote{We explain why trace-norm decoupling of intervals implies the case of $B$ being the union of holes. Let $F=X^+$. Let $Y=\{y\in \T_L: d_L(y,F)\geq r\}$ be an interval. Thus $B\subset Y$, and the decoupling relation is $\norm{\varrho_{L,FY} - \varrho_{L,F}\otimes \varrho_{Y}}_1\leq \delta_r$. Taking the partial trace over $Y\setminus B$ and using contractivity of trace norm, we get $\norm{\varrho_{L,FB} - \varrho_{L,F}\otimes \varrho_{B}}_1\leq \delta_r$.}. We can now apply Eq. \eqref{eq:prep-erasure} with $B$ equal to the union of distant holes, $F=X^+$, $C=\mathbb{T}_L\setminus(B\cup F)$ and $F_0=X^+\cap A$. Then the two canonical purifications in Eq. \eqref{eq:prep-erasure} are precisely those defining $\ket{b}$ and $\ket{a}$. Hence we get
\begin{equation}
    d_{X^+} (\ket{a}, \ket{b})=d_{F_0} (\ket{a},\ket{b}) \leq 4\sqrt{2} \delta_r^{1/4},
\end{equation}
the first equation follows because both in $\ket{a}$ and $\ket{b}$ the {nearby} holes are in the vacuum.

Finally, we consider $2\norm{R_X\ket{b} - \ket{\Phi_L}}$ by starting from $\ket{b}$ and inserting the $m$ nearby holes one by one. For simplicity, we denote $\ket{b_0} = \ket{b}$, $\ket{b_k} = \ket{\can(\varrho_{L, AB_{\rm d} B_{(1)} B_{(2)} \cdots B_{(k)}})}$, where $B_{\rm d}$ denotes all distant holes and $B_{(i)}$ are nearby holes. Clearly, $\ket{b_m} = \ket{\Phi_L}$.
Then we get
\begin{equation}\label{eq:prep-local-insertion-error}
    2\norm{R_X \ket{b} - \ket{\Phi_L}} \leq 2\sum_{k} \norm{ R_{k}\ket{b_{k-1}} - \ket{b_k} }\leq 2m\eta_r.
\end{equation}

Substituting Eq. \eqref{eq:prep-local-product-error}--%
\eqref{eq:prep-local-insertion-error} into Eq.
\eqref{eq:prep-local-three-errors} gives
\begin{equation}
 \begin{aligned}
 d_X(U_{L,r}\ket{\Omega_L},\ket{\Phi_L})
 &\le
 \underbrace{2t\sqrt{\epsilon_r}
       +2\sqrt{(t-1)\delta_r}+2\sqrt{\delta_r}}_{\text{product input}}
 +\underbrace{4\sqrt2\,\delta_r^{1/4}}_{\text{distant holes}}
 +\underbrace{2m \eta_r}_{\text{nearby insertions}}\\
 &\le C(1+t+m)e^{-c_{\rm loc}r}
 \le C_{\rm loc}(1+|X|/r)e^{-c_{\rm loc}r}.
 \end{aligned}
\end{equation}
where $c_{\rm loc} = c/4$ and $C_{\rm loc} = 16\sqrt{C} + 4\sqrt{2}C^{1/4} + 16(3C/2)^{1/4}$.

For the case $X=\T_L$, all holes are nearby and $A_X=A$. Then the distant-hole
comparison and separation from other blocks are absent.
Thus the argument also covers the whole ring.

To get a common group of constants for Lemma \ref{lem:prep-approx}, notice $C_{\rm loc}>C_{\rm glob}$ and $c_{\rm loc} = c_{\rm glob}$, we can take $C_{\rm loc}$ and $c_{\rm loc}$ as the common constants for the inequalities.
\end{proof}

The preceding lemmas give a depth-two approximate preparation circuit where each gate has a support of size at the scale $r$.
We next construct a correcting unitary circuit that removes the error. This construction is based on the following result of a general finite-depth quantum circuit $Q$.

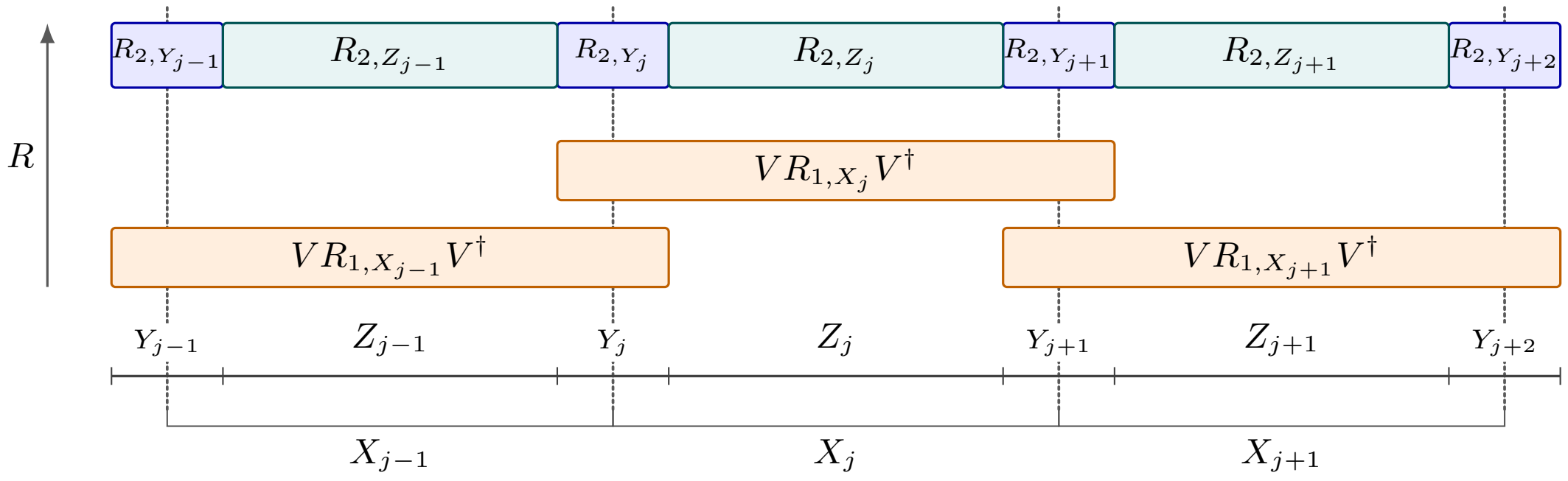
\begin{figure}[htbp]
\centering
\begin{tikzpicture}[x=.50cm,y=1cm,font=\small,
  blockgate/.style={draw=orange!75!black,fill=orange!13,
    rounded corners=1pt,line width=.6pt},
  coregate/.style={draw=teal!65!black,fill=teal!9,
    rounded corners=1pt,line width=.6pt},
  collargate/.style={draw=blue!65!black,fill=blue!9,
    rounded corners=1pt,line width=.6pt},
  flow/.style={-{Latex[length=1.8mm]},draw=black!65,line width=.6pt}]
\foreach \xleft/\xright/\regionlabel in
  {0/8/{X_{j-1}},8/16/{X_j},16/24/{X_{j+1}}} {
  \draw[black!65] (\xleft,-.85)--(\xleft,-1)--(\xright,-1)--(\xright,-.85);
  \node[below] at ({(\xleft+\xright)/2},-1) {$\regionlabel$};
}
\foreach \x in {0,8,16,24}
  {\draw[black!65,densely dotted,line width=.7pt,line cap=round]
    (\x,-.85)--(\x,2.78);}
\draw[black!75,line width=.65pt] (-1,-.55)--(25,-.55);
\foreach \x in {-1,1,7,9,15,17,23,25}
  {\draw[black!75] (\x,-.63)--(\x,-.47);}
\foreach \x/\regionlabel in
  {0/{Y_{j-1}},8/{Y_j},16/{Y_{j+1}},24/{Y_{j+2}}} {
  \node[above,font=\scriptsize,fill=white,inner sep=1.5pt]
    at (\x,-.43) {$\regionlabel$};
}
\foreach \xleft/\xright/\regionlabel in
  {1/7/{Z_{j-1}},9/15/{Z_j},17/23/{Z_{j+1}}} {
  \node[above,inner sep=1.5pt] at ({(\xleft+\xright)/2},-.43) {$\regionlabel$};
}
\draw[blockgate] (-1,.25) rectangle (9,.78);
\node at (4,.515) {$VR_{1,X_{j-1}}V^\dagger$};
\draw[blockgate] (15,.25) rectangle (25,.78);
\node at (20,.515) {$VR_{1,X_{j+1}}V^\dagger$};
\draw[blockgate] (7,1.03) rectangle (17,1.56);
\node at (12,1.295) {$VR_{1,X_j}V^\dagger$};
\foreach \x/\regionlabel in
  {0/{Y_{j-1}},8/{Y_j},16/{Y_{j+1}},24/{Y_{j+2}}} {
  \draw[collargate] ({\x-1},2.04) rectangle ({\x+1},2.62);
  \node[font=\scriptsize] at (\x,2.33) {$R_{2,\regionlabel}$};
}
\foreach \xleft/\xright/\regionlabel in
  {1/7/{Z_{j-1}},9/15/{Z_j},17/23/{Z_{j+1}}} {
  \draw[coregate] (\xleft,2.04) rectangle (\xright,2.62);
  \node at ({(\xleft+\xright)/2},2.33) {$R_{2,\regionlabel}$};
}
\draw[flow] (-2.15,.25)--(-2.15,2.62)
  node[midway,left] {$R$};
\end{tikzpicture}
\caption{The correction circuit $R$ in Lemma~\ref{lem:prep-correct}.
The figure shows a portion of the ring away from the closing seam, with blocks of
length $8da$.  The lower two gate rows implement
$\prod_j VR_{1,X_j}V^\dagger$; the top gate row implements $R_2$ on
disjoint cores and collars.  Rectangles show containing intervals of
gate supports.  For an odd number of blocks, one additional layer of
the enlarged block gates may be needed at the seam.}
\label{fig:prep-correction-circuit}
\end{figure}

Consider a general circuit $Q$ of depth $d$ whose gates have length at most $a$, acting on a ring with circumference $L\geq 24da$ (see Fig. \ref{fig:prep-correction-circuit}). Consider the following partition: Identify the ring with \(\mathbb Z/L\mathbb Z\), and set
\begin{equation}
\ell_0=8da,\qquad
M=\left\lfloor{L}/{\ell_0}\right\rfloor,\qquad
s=L-M\ell_0.
\end{equation}
Partition the ring into consecutive blocks \(X_1,\ldots,X_M\) from site \(0\), where
\begin{equation}\label{eq:da_partion}
X_j=\{(j-1)\ell_0,\ldots,j\ell_0-1\}
\quad(1\le j<M),\quad
X_M=\{(M-1)\ell_0,\ldots,L-1\}.
\end{equation}
Thus the first \(M-1\) blocks have length \(8da\), and the last has length \(8da+s<16da\). Let \(c_j=(j-1)\ell_0\) be the first site of \(X_j\). Define the collar at this boundary by
\begin{equation}\label{eq:da_partiaion_Y}
Y_j=\{c_j-da,\ldots,c_j+da-1\}\pmod L, \quad |Y_j| =2da, 
\end{equation}and the core of \(X_j\) by
\begin{equation}\label{eq:da_partion_Z}
Z_j=X_j\setminus(Y_j\cup Y_{j+1}),
\qquad Y_{M+1}:=Y_1. 
\end{equation}
The intervals \(Y_1,Z_1,\ldots,Y_M,Z_M\) form a partition of the ring.

\begin{lemma}\label{lem:prep-correct}
Let $\ket{\Omega}$ be the product-vacuum state on a ring with length $L$, and let $Q$ be a depth-$d$ circuit whose gates have length at most $a$, on a ring of circumference $L\geq 24da$. Consider the partition of ring as in Eqs. \eqref{eq:da_partiaion_Y} and \eqref{eq:da_partion_Z}. Suppose that, for every collar or core denoted as $I$,
\begin{equation}
\bra{\Omega_I}
\bigl(Q\ket{\Omega}\bra{\Omega}Q^{\dagger}\bigr)_I
\ket{\Omega_I}
\ge 1-\epsilon,
\qquad 0\le\epsilon<\frac16.
\end{equation}
Then there is a unitary circuit \(R\) such that
\begin{equation}
RQ\ket{\Omega}=e^{i\alpha}\ket{\Omega},\quad \alpha\in[0,2\pi).
\end{equation}
Its depth is at most four, and each gate \(G\) is supported on an interval of length at most $18da$ and satisfies
\begin{equation}
\norm{G-\mathbf1}\le \sqrt{6\epsilon}.
\end{equation}
The constants are independent of the onsite dimension and circumference. The correction gates can be chosen using only the gates of \(Q\) in neighborhoods of size \(O(da)\).
\end{lemma}

\begin{proof}
To get a correction circuit $R$, we take three steps. First, we reorganize the gates in the given circuit $Q$ according to the partition of cores and collars, then in second step we construct an intermediate state that is the product of collars and cores.
In Step 3 we do corrections in each core and collar separately and aggregate the gates.

{\it Step 1: reorganize $Q$. } We write $Q = Q_d\cdots Q_1$ with each $Q_t$ standing for a layer. Then we split each layer as $Q_t = K_t D_t$. $K_t$ contains the gates lying inside individual blocks $X_j$ and $D_t$ contains those crossing the boundaries between the blocks, so $K_t$ and $D_t$ commute. Define $W = K_d\cdots K_1$, then we can express $QW^\dagger$ as
\begin{equation}\label{eq:QW_boundary}
    QW^\dagger = \prod_{t=1}^d (Q_d\cdots Q_{t+1}) D_t (Q_d\cdots Q_{t+1})^\dagger =: V,
\end{equation}
where the factors are ordered by increasing $t$ from left to right, and we have used the fact that $D_t$ and $K_t$ commute in the above. By definition, each gate in $D_t$ crosses only one boundary. Conjugation by the remaining $d-t$ layers expands its support by at most $(d-t)a$ sites on either side, so the resulting support lies in the collar at that boundary (since each collar has radius $da$ from a boundary of $X_j$). Since these collars are disjoint, we group the factors in Eq.~\eqref{eq:QW_boundary} by boundary and
obtain
\begin{equation}
 Q=VW,\qquad V=\prod_j V_{Y_j},\qquad W=\prod_j W_{X_j},
\end{equation}
where $W_{X_j}$ are gates in blocks $X_j$ and $V_{Y_j}$ are gates only acting on collars. In the following, we also use the symbol $V_I$ with $I\in\{Y_j,Z_j\}$, where
for $I=Y_j$, we write $V_I=V_{Y_j}$, and for $I=Z_j$, we write $V_I=\one$.

{\it Step 2: approximation by product of blocks.} Now that $Q$ is reorganized as above, we need to understand its effect on $|\Omega\rangle$. First consider the state $W\ket{\Omega}=\bigotimes_j\ket{\phi_j}$ with
$\ket{\phi_j}=W_{X_j}\ket{\Omega_{X_j}}$, which is a product over blocks.  Denote
\[
\sigma = W\ket{\Omega}\bra{\Omega}W^\dagger,
\]
In the following, we will find a unitary to approximately convert $\sigma$ into a product state between blocks.

To do it, we split each block $X_j$ into the three pieces
$X_j\cap Y_j$, $Z_j$, and $X_j\cap Y_{j+1}$, so that the entire ring can be partitioned into such pieces. Denote by $E$ a piece of this
partition.  A collar consists of pieces from two adjacent blocks, whereas
a core is a single piece.  Hence, for every collar or core $I$,
\begin{equation}
 \sigma_I=\bigotimes_{E\subset I}\sigma_E,\qquad
 \bigl(Q\ket{\Omega}\bra{\Omega}Q^{\dagger}\bigr)_I=V_I\sigma_IV_I^{\dagger}.
\end{equation}
The assumption in Lemma~\ref{lem:prep-correct} therefore implies
\[
 1-\epsilon
 \le \bra{\Omega_I}V_I\sigma_I V_I^{\dagger}\ket{\Omega_I}
 \le \lambda_{\max}(\sigma_I)
 =\prod_{E\subset I}\lambda_{\max}(\sigma_E).
\]
Every factor $\lambda_{\max}(\sigma_E)$ is at most one, so each $\sigma_E$ has a largest eigenvalue at
least $1-\epsilon>1/2$.  So the largest eigenvalue of $\sigma_E$ is non-degenerate. We take the non-degenerate leading eigenstate  $\ket{\zeta_E}$ of $\sigma_E$ for each $E$, and define $\ket{\zeta}=\bigotimes_E\ket{\zeta_E}$ and $\ket{\zeta_{X_j}}=\bigotimes_{E\subseteq X_j}\ket{\zeta_E}$.

Next, we construct a circuit to rotate $W\ket{\Omega}$ to this $\ket{\zeta}$, which is a product state between the blocks. Note that\footnote{This can be seen as follows: let $P_1$, $P_2$ and $P_3$ denote the projection onto the three states $|\zeta_{E}\rangle$ in $X_j$. Then the left hand side is equal to $\ev{\phi_j|(1-P_1P_2P_3)|\phi_j}$. Using the inequality $1-P_1P_2P_3\leq (1-P_1)+(1-P_2)+(1-P_3)$, we get the first inequality in Eq. \eqref{eq:overlap_estimate}.}
\begin{equation}\label{eq:overlap_estimate}
 1-|\braket{\phi_j}{\zeta_{X_j}}|^2
 \le \sum_{E\subset X_j}
       \bigl(1-\bra{\zeta_E}\sigma_E\ket{\zeta_E}\bigr)
 \le 3\epsilon.
\end{equation}
Thus by Lemma~\ref{lem:rotation}, there is a unitary rotation $R_{1,X_j}$ that sends $\ket{\phi_j}$ to $\ket{\zeta_{X_j}}$ up to phase, with
\begin{equation}\label{eq:R1_gates}
 \|R_{1,X_j}-\mathbf1\|\le\sqrt{6\epsilon}.
\end{equation}
The product $R_1=\prod_jR_{1,X_j}$ consequently sends $W\ket{\Omega}$ to
$\ket{\zeta}$ up to an overall phase.

{\it Step 3: corrections in collars and cores.} Now that $R_1W\Omega\propto|\zeta\rangle$, we next find another shallow circuit $R_2$ such that $R_2V|\zeta\rangle\propto|\Omega\rangle$. The vector $\ket{\zeta_I}$ is the leading
eigenvector of $\sigma_I$, with eigenvalue
$p_I=\lambda_{\max}(\sigma_I)\ge1-\epsilon$, so the remaining positive part
of $\sigma_I$ has trace $1-p_I$ and
\[
 1-\epsilon
 \le \bra{\Omega_I}V_I\sigma_I V_I^{\dagger}\ket{\Omega_I}
 \le p_I|\bra{\Omega_I}V_I\ket{\zeta_I}|^2+(1-p_I).
\]
It follows that
\begin{equation}
 1-|\bra{\Omega_I}V_I\ket{\zeta_I}|^2
 \le \frac{\epsilon}{p_I}\le\frac{\epsilon}{1-\epsilon}.
\end{equation}
By Lemma~\ref{lem:rotation}, there is a second unitary rotation $R_{2,I}$ that sends $V_I\ket{\zeta_I}$ to $\ket{\Omega_I}$ up
to phase and satisfies
\begin{equation}\label{eq:R2_gates}
 \|R_{2,I}-\mathbf1\|
 \le\sqrt{\frac{2\epsilon}{1-\epsilon}}\le2\sqrt\epsilon.
\end{equation}
Since the collars and cores partition the ring, $R_2=\prod_I R_{2,I}$
sends $V\ket{\zeta}$ to $\ket{\Omega}$ up to phase.  Consequently, we will take the correcting unitary as
\[
 R:=R_2VR_1V^{\dagger},\qquad RQ\ket{\Omega}=e^{i\alpha}\ket{\Omega}.
\]

To see that $R$ is a circuit of gates with small sizes, we write $R$ as
\[
 R=\left(\prod_I R_{2,I}\right)
   \left(\prod_j VR_{1,X_j}V^{\dagger}\right).
\]
Fig. \ref{fig:prep-correction-circuit} shows the supports of the gates in $R$ schematically. 
Only the two collars adjacent to $X_j$ can enlarge the support of $R_{1,X_j}$, so
\[
 \operatorname{supp}(VR_{1,X_j}V^{\dagger})
 \subset X_j\cup Y_j\cup Y_{j+1},
 \qquad |X_j\cup Y_j\cup Y_{j+1}|<18da.
\]
These enlarged block supports overlap only for neighboring blocks.
The conjugated gates commute, and can therefore be arranged in two
alternating layers if $M$ is even, or in three layers if $M$ is odd
(by placing the last block in its own layer).  The gates $R_{2,I}$ form
one further layer. Thus $R$ has depth at most 4, and every gate has length
less than $18da$. 

On the other hand, every gate is within $\sqrt{6\epsilon}$ of the
identity, as shown in Eqs. \eqref{eq:R1_gates} and \eqref{eq:R2_gates}.

Finally, the correction gates are based only on local data as claimed. With the partition and vacuum fixed, the correction gates \(R_{2,Z_j}\), \(R_{2,Y_{j+1}}\), and \(VR_{1,X_j}V^\dagger\) can be chosen using only the gates of \(Q\) contained in \(X_{j-1}\cup X_j\cup X_{j+1}\). This region has length less than \(32da\), which proves the claim of the lemma. 

\end{proof}

\subsection{Proof of Theorem~\ref{thm:prep-exact}}
\label{subsec:proof-preparation} %

\begin{proof}
{\it Step 1: successive circuits and errors.} We first construct a circuit with local gates in successive scales $r, 2r, 2^2r,...$, and control the errors of this circuit preparation of $\ket{\Phi_L}$.

Under assumptions Eqs. \eqref{eq:prep-input-decouple} and
\eqref{eq:prep-input-compare}, Lemma~\ref{lem:prep-approx} constructs
depth-two circuits $U_r:=U_{L,r}$ that approximately prepare $\ket{\Phi_L}$ at scale $r$ and the following inequality holds for any interval $J$,
\begin{equation} \label{eq: preparation at scale r}
 \|(U_r\ket{\Omega_L}\bra{\Omega_L}U_r^{\dagger})_J-(\ket{\Phi_L}\bra{\Phi_L})_J\|_1
 \le C_{\rm loc}(1+|J|/r)e^{-a_pr},\qquad a_p>0,
\end{equation}
where $C_{\rm loc}, a_p$ are constants given in Lemma~\ref{lem:prep-approx}.
The $U_r$ has depth two, causal radius \footnote{The causal radius of a circuit $U$ is the smallest $R\ge0$ such that
both $UAU^\dagger$ and $U^\dagger AU$ are supported on $X(R)=\{x\in \T_L: d_L(x,X)\leq R \}$
whenever $A$ is supported on a set $X$.} at most $10r$, and
its bulk gates have period $4r$, starting at site 0. These bulk gates are independent of $L$, and only gates sitting in region $[L-8r-1,L-1]$ depend on $L$. We would like to find a circuit that prepares $|\Phi_L\rangle$ at the scale $2r$.

To do so, define a circuit $Q_r:=U_{2r}^{\dagger}U_r$, which has depth $d=4$ and its gate length is at most $a=14r$.  For convenience, we define the fixed integers
\[
 h_*:=4\cdot14=56,\qquad \ell_*:=8h_*=448.
\]
For the partition (the blocks $X_j$, collars $Y_j$ and cores $Z_j$) used in Lemma~\ref{lem:prep-correct},
the collar width on each side of a cut is at most  $da=h_*r$,
and each block except the last has length at most  $8da=\ell_*r$.  Assume $L\ge24h_*r=3\ell_*r$, as required
by Lemma~\ref{lem:prep-correct}.  
By computing the estimation of local observables in $J$, we claim that the following inequality holds:
\begin{equation}\label{eq:Qr_trace_norm}
 \|(Q_r\ket{\Omega_L}\bra{\Omega_L}Q_r^{\dagger})_J-(\ket{\Omega_J}\bra{\Omega_J})\|_1
 \le C_Q (1+|J|/r)e^{-a_pr},
\end{equation}
with a constant $C_Q$ proportional to $C_{\rm loc}$.
Indeed, for $O_J\in\mathcal A_J$ with $\|O_J\|\le1$, set
$O'=U_{2r}O_JU_{2r}^{\dagger}$.  Then $\|O'\|\le1$ and $O'$ is supported on
$J(20r)=\{x:d_L(x, J)\leqslant 20r\}$, whose size is at most $|J|+40r$.  Inserting $\ket{\Phi_L}$ gives
\[
\begin{aligned}
&|\bra{\Omega_L}Q_r^{\dagger}O_JQ_r\ket{\Omega_L}
       -\bra{\Omega_L}O_J\ket{\Omega_L}|\\
\quad\le&
|\bra{\Omega_L}U_r^{\dagger}O'U_r\ket{\Omega_L}
       -\bra{\Phi_L}O'\ket{\Phi_L}| +
|\bra{\Phi_L}O'\ket{\Phi_L}
       -\bra{\Omega_L}U_{2r}^{\dagger}O'U_{2r}\ket{\Omega_L}| \\
\quad\leq& C_{\mathrm{loc}}
 \left(1+\frac{|J|+40r}{r}\right)e^{-a_pr}
+
C_{\mathrm{loc}}
 \left(1+\frac{|J|+40r}{2r}\right)e^{-2a_pr}\\
\quad\le&
C_{\mathrm{loc}}
 \left(62+\frac{3|J|}{2r}\right)e^{-a_pr}
 \le
C_Q\left(1+\frac{|J|}{r}\right)e^{-a_pr},
\end{aligned}
\]
and taking the supremum over $O_J$ proves Eq.
\eqref{eq:Qr_trace_norm}.

To construct the correction circuit that takes $Q_r|\Omega_L\rangle$ back to $|\Omega_L\rangle$, we apply Lemma~\ref{lem:prep-correct} to $Q_r$, with depth bound
$d=4$ and gate-length bound $a=14r$, using the blocks $X_j$,
collars $Y_j$, and cores $Z_j$ defined there. Blocks have length less
than $16da=2\ell_*r$, collars have length $|Y_j|=2da=2h_*r$, and cores have length less
than $(2\ell_*-2h_*)r=14h_*r$.  For each collar or core denoted by $I$, their length is smaller than $(2\ell_*-2h_*)r=14h_*r$. Note that
\[
\begin{split}
1 - \bra{\Omega_I}\bigl(Q_r\ket{\Omega_L}\bra{\Omega_L}Q_r^{\dagger}\bigr)_I\ket{\Omega_I} &= \Tr[\ket{\Omega_I}\bra{\Omega_I} \big(\ket{\Omega_I}\bra{\Omega_I} - \bigl(Q_r\ket{\Omega_L}\bra{\Omega_L}Q_r^{\dagger}\bigr)_I  \big) ] \\
&\leq \norm{ \ket{\Omega_I}\bra{\Omega_I} - \bigl(Q_r\ket{\Omega_L}\bra{\Omega_L}Q_r^{\dagger}\bigr)_I  }_1,
\end{split}
\]
so Eq. \eqref{eq:Qr_trace_norm} implies
\begin{equation}
\bra{\Omega_I}\bigl(Q_r\ket{\Omega_L}\bra{\Omega_L}Q_r^{\dagger}\bigr)_I\ket{\Omega_I}
 \ge1-\varepsilon_r,\qquad \varepsilon_r:=(1+2\ell_*-2h_*)C_Q e^{-a_pr}.
\end{equation}
We choose an integer $r_0$
large enough so that $\varepsilon_{r_0}<1/6$ and consider $r\geq r_0$.
Then Lemma~\ref{lem:prep-correct} constructs a unitary
circuit $R_r$, of depth at most four, whose gates act on intervals of length $O(r)$ and $\norm{G-\one}\leq \sqrt{6\varepsilon_r}\sim e^{-a_pr/2}$.  Applying this correction returns the state
$Q_r\ket{\Omega_L}$ exactly to the product vacuum, up to an overall phase:
\[
 R_rQ_r\ket{\Omega_L}=e^{i\alpha_r}\ket{\Omega_L},\qquad \alpha_r\in\mathbb R.
\]
Conjugating $R_r$ by $U_{2r}$ therefore gives a correction circuit that converts a
scale-$r$ preparation into a scale-$2r$ preparation:
\begin{equation}\label{eq:prep-scale-correction}
 \mathcal C_r:=U_{2r}R_rU_{2r}^{\dagger},\qquad
 \mathcal C_rU_r\ket{\Omega_L}=e^{i\alpha_r}U_{2r}\ket{\Omega_L}.
\end{equation}
For any local region $J$, this state $U_{2r}|\Omega_L\rangle$ is potentially closer to $|\Phi_L\rangle$ compared to $U_{r}|\Omega_L\rangle$, with $r$ in Eq. \eqref{eq: preparation at scale r} replaced by $2r$.

To better characterize the circuit \(\mathcal C_r\), next we bound the difference of the gates in \(\mathcal C_r\) from the identity operator, the support length of each gate in \(\mathcal C_r\), and the number of gates meeting a given interval. These bounds will control the error from omitting correction circuits at large scales.
Using $R_r=R_2VR_1V^\dagger$ as in Lemma~\ref{lem:prep-correct}, these gates are
\[
 U_{2r}R_{2,I}U_{2r}^\dagger,
 \qquad
 U_{2r}VR_{1,X_j}V^\dagger U_{2r}^\dagger,
\]
where $I$ runs over collars and cores.
Each gate $G$ satisfies $\|G-\one\|\le C_{\rm gate} e^{-br}$ with $C_{\rm gate} = \sqrt{6(1+2\ell_*-2h_*)C_Q}$, with $b=a_p/2$.
Associate each block \(X_j\) with the following three gates of \(\mathcal C_r\):
\[
U_{2r}VR_{1,X_j}V^\dagger U_{2r}^\dagger,\qquad
U_{2r}R_{2,Z_j}U_{2r}^\dagger,\qquad
U_{2r}R_{2,Y_{j+1}}U_{2r}^\dagger.
\]
Their supports lie in
$X_j((h_*+20)r)$. 
Since $|X_j|<2\ell_*r$, each gate is supported on an interval
of length less than $(2\ell_*+2h_*+40)r$.
Lastly, we consider the number of gates meeting an interval. Denote the number of
these gates meeting an interval $J$ as $N_r(J)$. Since every block has length at least $\ell_*r$, we get $N_r(J)\leq 3\cdot \#\{j: X_j\cap J((h_*+20)r)\neq\emptyset\}$, which gives
\begin{equation}\label{eq:prep-gate-count}
 N_r(J)\le
 3\left(2+\frac{|J|+2(h_*+20)r}{\ell_*r}\right)
 \le 8(1+|J|/r).
\end{equation} 

We also discuss support growth of a local operator under conjugation by the whole
circuit $\mathcal C_r$. Use the factorization
\[
 \mathcal C_r=U_{2r}(R_2V)R_1V^\dagger U_{2r}^\dagger.
\]
The five factors enlarge supports by at most
$20r$, $(2\ell_*-2h_*)r$, $2\ell_*r$, $2h_*r$, and $20r$,
respectively. Thus conjugation by $\mathcal C_r$ or
$\mathcal C_r^\dagger$ enlarges supports by at most $C_1r$,
where $C_1=4\ell_*+40=1832$.

\medskip
{\it Step 2: assemble the exact preparation and its truncations.}
We compose correction circuits at dyadic scales, then remove the remaining
error by one final rotation.
The strategy of composing small local gates at increasing scales is also used in Refs.~\cite[Appendix~D]{Sopenko2023}
and~\cite[Appendix~B]{deroeck2025}.

For Lemmas~\ref{lem:prep-approx} and \ref{lem:prep-correct} to apply, we need to choose a scale $r$ satisfying $L\geq 3\ell_* r$. We set $C_{\rm stop}:=3\ell_*/2=672$ and
$L_{\rm prep}:=C_{\rm stop}r_0$, and first suppose $L\ge L_{\rm prep}$ (the case $L<L_{\rm prep}$ is discussed at the end of Step 3).
Let $r_j=2^jr_0$, with $r_K$ the largest scale not exceeding
$L/C_{\rm stop}$, i.e.,
\[
 \frac{L}{2C_{\rm stop}}<r_K\le\frac{L}{C_{\rm stop}}.
\]
Now we have $3\ell_*r_j=2C_{\rm stop}r_j\le C_{\rm stop}r_K\le L$ for any $j<K$, so every correction circuit $\mathcal C_{r_j}$ with $j<K$ is available.
For $s=r_k$ with $1\le k\le K$, define
\begin{equation}
 \mathcal U_{L,<r_k}:=\mathcal C_{r_{k-1}}\cdots \mathcal C_{r_0}U_{r_0},
 \qquad \ket{v_L}:=\mathcal U_{L,<r_K}\ket{\Omega_L},
\end{equation}
where $\mathcal U_{L,<r_0}:=U_{r_0}$.
By Eq. \eqref{eq:prep-scale-correction}, $\mathcal U_{L,<s}\ket{\Omega_L}$ equals
$U_s\ket{\Omega_L}$ up to a phase.  Applying the local preparation estimate Eq.
\eqref{eq:prep-local} to the whole ring at scale $r_K$, and using
$L/r_K<2C_{\rm stop}$, we get
\begin{equation}
\begin{split}
 \min_{|\xi|=1}\|\ket{v_L}-\xi\ket{\Phi_L}\|
 &\le 2^{-1/2}\|\ket{v_L}\bra{v_L}-\ket{\Phi_L}\bra{\Phi_L}\|_1 \\
 &\le  \frac{C_{\rm loc}}{\sqrt{2}}(1+\frac{L}{r_K})e^{-a_pr_K}\le C_F e^{-a_pL/(2C_{\rm stop})},\quad C_F = \frac{1+2C_{\rm stop}}{\sqrt2}C_{\rm loc}.
 \end{split}
\end{equation}
Choose a minimizing phase $\xi_L$.  By Lemma \ref{lem:rotation}, the plane rotation $F_L$ sending $\ket{v_L}$ to $\xi_L\ket{\Phi_L}$ satisfies
\begin{equation}
 \|F_L-\one\|=\|\ket{v_L}-\xi_L\ket{\Phi_L}\|
 \le C_F e^{-a_p r_K}\leq C_F e^{-a_p L/(2C_{\rm stop})}.
\end{equation}
The rotation $F_L$ may act on the whole ring, and its exponentially small
distance from the identity enters the truncation error in Step~3.  Define
\begin{equation}\label{eq:prep-dyadic}
 \mathcal U_L:=\overline{\xi_L}F_L\mathcal U_{L,<r_K},\qquad
 \mathcal U_L\ket{\Omega_L}=\ket{\Phi_L},\qquad
 T_L:=\mathcal U_L^{\dagger}.
\end{equation}
This finishes the construction of the circuit $T_L$ (and its inverse $\mathcal{U}_L$). Fig. \ref{fig:prep-exact-circuit} shows the full circuit, its truncation,
and two successive correction stages.

\begin{figure}[tbp]
\centering
\begin{tikzpicture}[x=1cm,y=.80cm,font=\small,
  stage/.style={draw=orange!70!black,fill=orange!12,
    rounded corners=1.5pt,minimum height=7mm,inner sep=3pt},
  flow/.style={-{Latex[length=1.8mm]},draw=black!65,line width=.6pt}]
\path[use as bounding box] (-.45,-1.5) rectangle (15.55,1.85);
\node[anchor=west,font=\small\bfseries] at (-.17,1.55)
  {(a) Exact preparation and the truncation at scale $s$};
\node (input) at (.15,0) {$\ket{\Omega_L}$};
\node[stage,draw=blue!65!black,fill=blue!8,minimum width=12mm]
  (initial) at (1.75,0) {$U_{r_0}$};
\node[stage,minimum width=11mm] (first) at (3.50,0) {$\mathcal C_{r_0}$};
\node (dotsone) at (4.75,0) {$\cdots$};
\node[stage,minimum width=12mm] (lastkept) at (6.00,0) {$\mathcal C_{s/2}$};
\node[stage,minimum width=11mm] (firstomitted) at (8.00,0) {$\mathcal C_s$};
\node (dotstwo) at (9.25,0) {$\cdots$};
\node[stage,minimum width=15mm] (last) at (10.75,0) {$\mathcal C_{r_{K-1}}$};
\node[stage,draw=violet!70!black,fill=violet!8,minimum width=16mm]
  (final) at (13.10,0) {$\overline{\xi_L}F_L$};
\node (output) at (15.25,0) {$\ket{\Phi_L}$};
\foreach \from/\to in {input/initial,initial/first,first/dotsone,
  dotsone/lastkept,lastkept/firstomitted,firstomitted/dotstwo,
  dotstwo/last,last/final,final/output}
  {\draw[flow] (\from.east)--(\to.west);}
\draw[violet!75!black,densely dashed,line width=.8pt]
  (7,-.70)--(7,.85);
\node[anchor=south,text=violet!75!black,font=\footnotesize]
  at (7,.85) {stop here};
\draw[blue!65!black,line width=.7pt]
  (1.05,-.55)--(1.05,-.77)--(6.70,-.77)--(6.70,-.55);
\node[anchor=north,text=blue!65!black] at (3.875,-.86)
  {$\mathcal U_{L,<s}$};
\draw[black!55,line width=.7pt]
  (7.35,-.55)--(7.35,-.77)--(14.05,-.77)--(14.05,-.55);
\node[anchor=north,font=\footnotesize] at (10.70,-.86)
  {omitted from $\mathcal U_{L,<s}$};
\end{tikzpicture}

\par\medskip
\begin{tikzpicture}[x=1.65cm,y=.82cm,font=\small,
  rotation/.style={draw=orange!70!black,fill=orange!14,
    rounded corners=1pt,line width=.6pt},
  guide/.style={draw=black!65,line width=.7pt,densely dotted,line cap=round},
  flow/.style={-{Latex[length=1.8mm]},draw=black!65,line width=.6pt},
  rowlabel/.style={anchor=east,font=\footnotesize},
  dim/.style={<->,>=Latex,draw=black!60,thin}]
\path[use as bounding box] (-1.35,-.82) rectangle (8.347,7.4);
\node[anchor=west,font=\small\bfseries] at (-1.18,7.0)
  {(b) Two successive corrections on the same spatial scale};
\draw[flow] (-1.03,.12)--(-1.03,5.58)
  node[midway,rotate=90,above,font=\footnotesize] {application order};
\begin{scope}
\clip (0,-.1) rectangle (6,5.67);
\foreach \j in {0,2,4,6}
  {\draw[rotation] ({\j-76/448},.12) rectangle ({\j+1+76/448},.54);}
\foreach \j in {-1,1,3,5}
  {\draw[rotation] ({\j-76/448},.75) rectangle ({\j+1+76/448},1.17);}
\foreach \j in {0,...,5}
  {\draw[rotation] ({\j+36/448},1.38) rectangle ({\j+412/448},1.80);}
\foreach \j in {0,...,6}
  {\draw[rotation] ({\j-76/448},2.01) rectangle ({\j+76/448},2.43);}
\foreach \j in {0,2}
  {\draw[rotation] ({2*\j-152/448},3.20) rectangle ({2*\j+2+152/448},3.62);}
\foreach \j in {-1,1,3}
  {\draw[rotation] ({2*\j-152/448},3.83) rectangle ({2*\j+2+152/448},4.25);}
\foreach \j in {0,1,2}
  {\draw[rotation] ({2*\j+72/448},4.46) rectangle ({2*\j+824/448},4.88);}
\foreach \j in {0,1,2,3}
  {\draw[rotation] ({2*\j-152/448},5.09) rectangle ({2*\j+152/448},5.51);}
\end{scope}
\foreach \x in {0,1,...,6}
  {\draw[guide] (\x,-.08)--(\x,2.57);}
\foreach \x in {0,2,4,6}
  {\draw[guide] (\x,3.05)--(\x,5.66);}
\draw[black!65,line width=.65pt]
  (0,-.13)--(0,-.28)--(1,-.28)--(1,-.13);
\node[anchor=north,font=\footnotesize,inner sep=1pt] at (.5,-.34) {$X_j^{(r)}$};
\draw[black!65,line width=.65pt]
  (0,5.62)--(0,5.72)--(2,5.72)--(2,5.62);
\node[anchor=south,font=\footnotesize,inner sep=1pt] at (1,5.74) {$X_j^{(2r)}$};
\foreach \y in {.645,3.725}
  {\node[rowlabel,align=right] at (-.10,\y) {block\\rotations};}
\foreach \y in {1.59,4.67}
  {\node[rowlabel] at (-.10,\y) {core};}
\foreach \y in {2.22,5.30}
  {\node[rowlabel] at (-.10,\y) {collar};}
\draw[orange!70!black,line width=.7pt]
  (6.10,.12)--(6.22,.12)--(6.22,2.43)--(6.10,2.43);
\node[anchor=west,align=left] at (6.37,1.28)
  {$\mathcal C_r$\\[2pt]$\|G-\one\|\le C_{\rm gate}e^{-br}$};
\draw[orange!70!black,line width=.7pt]
  (6.10,3.20)--(6.22,3.20)--(6.22,5.51)--(6.10,5.51);
\node[anchor=west,align=left] at (6.37,4.36)
  {$\mathcal C_{2r}$\\[2pt]$\|G-\one\|\le C_{\rm gate}e^{-2br}$};
\node at (3,-.46) {$U_r\ket{\Omega_L}$};
\draw[black!40,densely dashed] (0,2.82)--(6,2.82);
\node[fill=white,inner sep=2pt] at (3,2.82) {$U_{2r}\ket{\Omega_L}$};
\node at (3,5.93) {$U_{4r}\ket{\Omega_L}$};
\draw[dim] (0,6.43)--(1,6.43) node[midway,above,font=\footnotesize] {$\ell_*r$};
\draw[dim] (2,6.43)--(4,6.43) node[midway,above,font=\footnotesize] {$2\ell_*r$};
\node[anchor=west,font=\footnotesize] at (4.30,6.43) {coarse-block lengths};
\end{tikzpicture}
\caption{The preparation circuit in the proof of Theorem~\ref{thm:prep-exact}.
(a) At an intermediate scale $s=r_k$, the truncation stops before $\mathcal C_s$:
it prepares $U_s\ket{\Omega_L}$ up to
phase.  The remaining corrections and the final rotation $F_L$ complete
the exact preparation.
(b) A bulk window of $\mathcal C_{2r}$ and $\mathcal{C}_r$; states are shown
up to phase, and rectangles show containing intervals of gate supports.
From bottom to top, the first two rows of $\mathcal C_r$ contain the conjugated
block rotations $U_{2r}VR_{1,X_j}V^{\dagger}U_{2r}^{\dagger}$ on alternating coarse blocks
$X_j$.  The third and fourth rows contain $U_{2r}R_{2,I}U_{2r}^{\dagger}$ for
cores and collars $I$, respectively.  
The construction of $\mathcal C_{2r}$ is the same at
scale $2r$.
Dotted lines mark the block boundaries; $X_j^{(r)}$ and $X_j^{(2r)}$
label representative blocks at the two scales.
}
\label{fig:prep-exact-circuit}
\end{figure}
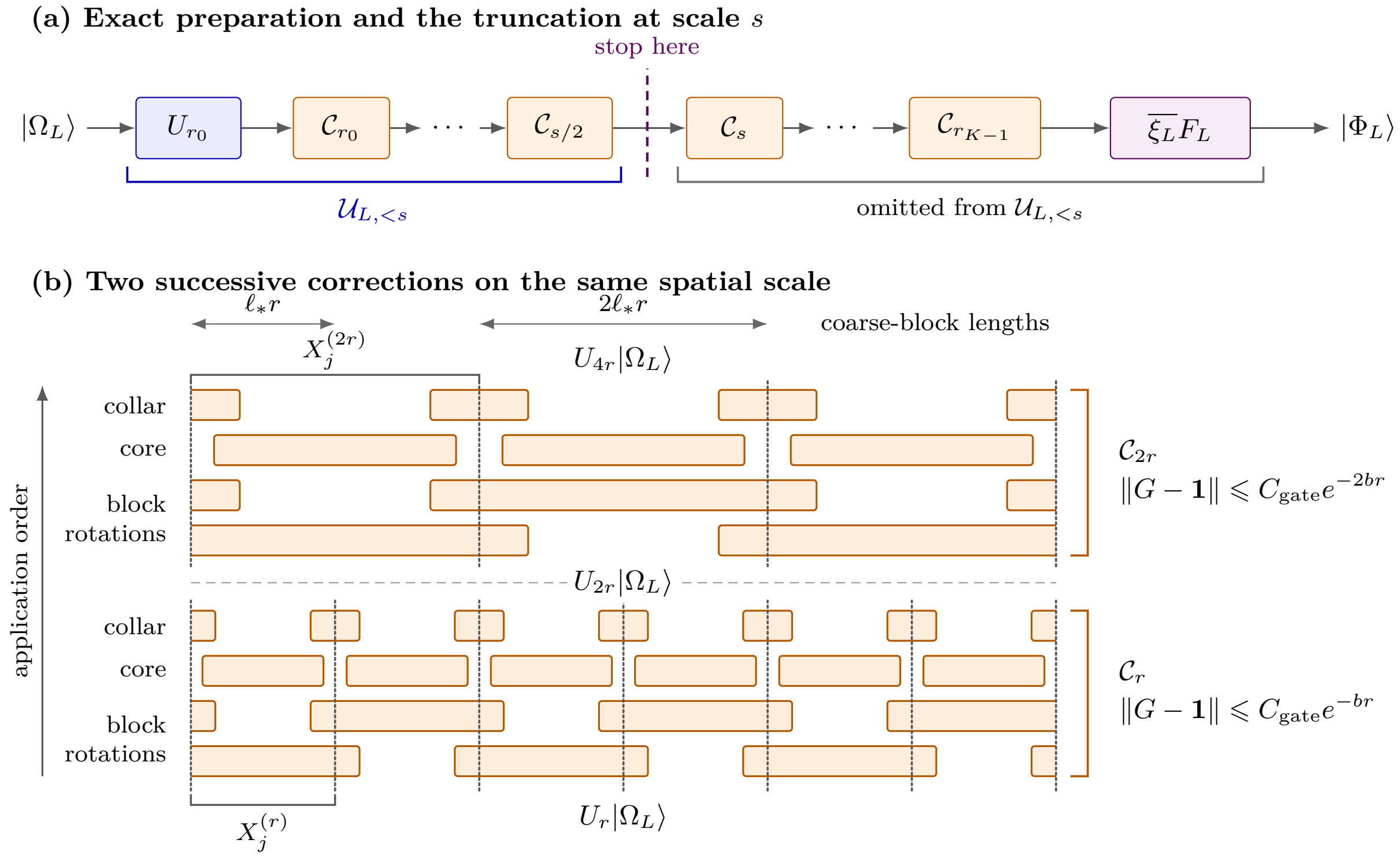

For $s=r_k$, the truncated circuit $\mathcal U_{L,<s}$ contains
$U_{r_0}$ followed by the corrections $\mathcal C_{r_0},\ldots,\mathcal C_{r_{k-1}}$.
Their causal radii are at most $10r_0$ and $C_1r_j$, respectively.
Hence, conjugating by $\mathcal{U}_{L,<s}$ or $\mathcal{U}_{L,<s}^\dagger$ enlarges supports by at most
\[
 10r_0+C_1\sum_{j<k}r_j
 =10r_0+C_1(s-r_0)\le C_1s,
\]
since $C_1>10$.  Each layer has common bulk period $\ell_*s$, since
$4r_0$ and $\ell_*r_j$ divide $\ell_*s$ for $j<k$.
The initial circuit has exceptional gates (i.e. gates depending on system size) only within $8r_0$ of the seam,
and the corrections contribute at most
$(2\ell_*+h_*+20)r_j\le(2\ell_*+h_*+20)s/2$.
Thus all exceptional gates lie within $l_{\rm seam}s$ of the seam, where
\[
 l_{\rm seam}:=\max\left\{8,\frac{2\ell_*+h_*+20}{2}\right\}=486.
\]

\medskip
{\it Step 3: truncation error, locality, and compatibility.}
We bound the effect of the omitted gates from $\mathcal{U}_{L,<s}$ and prove Eqs. \eqref{eq:prep-locality}-\eqref{eq:prep-truncation} as claimed in Theorem~\ref{thm:prep-exact}. We first assume \(L\ge L_{\rm prep}\). 
Smaller rings are discussed at the end of this proof.

The following fact is useful: for a product of unitary gates $Z=G_m\cdots G_1$, there is an equation
\[
 [Z,A]=\sum_{j=1}^mG_m\cdots G_{j+1}[G_j,A]G_{j-1}\cdots G_1,
\]
and $\|[G_j,A]\|=\|[G_j-\one, A]\|\le2\|A\|\|G_j-\one\|$ gives
\begin{equation}\label{eq:prep-discrete-tail}
 \|Z^{\dagger}AZ-A\|=\|ZAZ^{\dagger}-A\|
 \le2\|A\|\sum_{j:\,\supp G_j\cap X\ne\varnothing}
                         \|G_j-\one\|,
 \qquad A\in\mathcal A_X.
\end{equation}

For $s=r_k$, define the omitted part by
\[
 \mathcal U_{L,\ge s}:=F_L\mathcal C_{r_{K-1}}\cdots\mathcal C_{r_k},
\]
so that $\mathcal U_L=\overline{\xi_L}\mathcal U_{L,\ge s}\mathcal U_{L,<s}$.
We now estimate how the omitted circuit \(\mathcal U_{L,\ge s}\) changes an operator \(A\in\A_J\) supported on an interval \(J\). Applying Eq. \eqref{eq:prep-discrete-tail}, and then using the gate count $N_r(J)$ in \eqref{eq:prep-gate-count} and the bounds of $\norm{G-\one}$ and \(\norm{F_L-\one}\), we get
\begin{equation}\label{eq:estimation_conjugate}
\begin{aligned}
&
\|\mathcal U_{L,\ge s}^{\dagger}A\mathcal U_{L,\ge s}-A\|=
\|\mathcal U_{L,\ge s}A\mathcal U_{L,\ge s}^{\dagger}-A\|
\\
&\le
2\|A\|\left[
\sum_{j=k}^{K-1}
N_{r_j}(J)C_{\mathrm{gate}}e^{-br_j}
+C_Fe^{-a_pr_K}
\right]\\
&\le
2\|A\|\left[
8C_{\mathrm{gate}}
\sum_{j=k}^{K-1}
\left(1+\frac{|J|}{r_j}\right)e^{-br_j}
+C_Fe^{-a_pr_K}
\right]\\
&\le
2C_{\mathrm{tail}}
\left(1+\frac{|J|}{s}\right)\|A\|e^{-bs},
\qquad A\in\mathcal A_J,
\end{aligned}
\end{equation}
where $C_{\mathrm{tail}}
:=\frac{8C_{\mathrm{gate}}}{1-e^{-b}}+C_F.$
The last inequality uses \(r_j\ge s+(j-k)\), \(r_K\ge s\), and \(a_p=2b\).

We now prove Eq. \eqref{eq:prep-truncation}. Fixing a nonempty interval  $I$ and considering any operator $A\in\mathcal A_I$, we define $\mathcal B_s[A]=\mathcal U_{L,<s}A\mathcal U_{L,<s}^{\dagger}$.
By the unitary invariance of the norm, the two errors in Eq.
\eqref{eq:prep-truncation} are, respectively,
\begin{equation}\label{eq:two_truncation_errors}
\begin{split}
\norm{ \mathcal U_L A \mathcal U_L^\dagger - \mathcal U_{L,<s} A \mathcal U_{L,<s}^\dagger} &= \norm{ \mathcal{U}_{L,\geq s} \mathcal{B}_{s}[A] \mathcal{U}_{L,\geq s}^\dagger - \mathcal{B}_s[A] } \\
\norm{ \mathcal U_L^\dagger A \mathcal U_L - \mathcal U_{L,<s}^\dagger A \mathcal U_{L,<s}} &= \norm{ \mathcal{U}_{L,<s}^\dagger \big( \mathcal{U}_{L,\geq s}^\dagger A \mathcal{U}_{L,\geq s} - A \big) \mathcal{U}_{L,<s} } \\
&=\norm{\mathcal{U}_{L,\geq s}^\dagger A \mathcal{U}_{L,\geq s} - A}
\end{split}
\end{equation}
Both $A$ and $\mathcal B_s[A]$ are supported on $I(C_1s)$,
and $\|\mathcal B_s[A]\|=\|A\|$.
Applying Eq. \eqref{eq:estimation_conjugate} with
$J=I(C_1s)$, whose length is at most $|I|+2C_1s$,
bounds each error in Eq. \eqref{eq:two_truncation_errors} by
\[
 2C_{\rm tail}
 \left(1+2C_1+\frac{|I|}{s}\right)\|A\|e^{-bs}.
\]
Since $s\ge1$ and $|I|\ge1$,
\[
 1+2C_1+\frac{|I|}{s}\le(2+2C_1)|I|.
\]
Thus both errors are at most
$C_{\rm tr}|I|\|A\|e^{-bs}$, where
\begin{equation}
 C_{\rm tr}:=2(2+2C_1)C_{\rm tail}.
\end{equation}
This proves Eq. \eqref{eq:prep-truncation}.

Then we consider an arbitrary support $X$. We cover $X(C_1s)$ by the $|X|$ intervals
$B_{L}(x, C_1s)$, $x\in X$.
Each has length at most $2C_1s+1$ and hence meets at most $8(2+2C_1)$ gates at every omitted scale $r\ge s$, by Eq. \eqref{eq:prep-gate-count}, namely
\[
N_r\bigl(X(C_1s)\bigr)
\le\sum_{x\in X}N_r\bigl(B_{L}(x,C_1s)\bigr)
\le8(2+2C_1)|X|.
\]
Then we can estimate that for $A\in\A_X$, the two errors in Eq. \eqref{eq:two_truncation_errors} are bounded by
$C_{\rm tr}|X|\|A\|e^{-bs}$, again by Eq. \eqref{eq:prep-discrete-tail}.

To prove Eq. \eqref{eq:prep-locality}, we consider any operator $A\in \A_X$, and we want to choose a truncation $\mathcal B_s[A]$ whose conjugated observable is supported inside $X(n)$ for a given $n$.  If $X(n)=\T_L$, the locality
error is zero.  So we only need to consider $n<L/2$. We first suppose $n\ge C_1r_0$. Take the largest dyadic scale $s\le n/C_1$, which implies $s>n/(2C_1)$.  This scale is sufficient for Lemma \ref{lem:prep-correct} because $2C_1>C_{\rm stop}$ and thus $C_{\rm stop} s< 2n<L$. Denote $A'$ for $\mathcal{U}_L A \mathcal{U}_L^\dagger$ (the estimation for $\mathcal{U}_L^\dagger A \mathcal U_L$ is similar, just redefine the conjugation in definition of $\mathcal B_{s}[A]$), then $A'$ has a truncated approximation $\mathcal B_s[A]\in\mathcal A_{X(C_1 s)}\subset\A_{X(n)}$ as we desired. Inserting $\mathcal B_s[A]=\E^{\tr}_{X(n)}(\mathcal B_s[A])$ and using contractive property of $\E^{\tr}_{X(n)}$, we get
\begin{equation}
 \|A'-\E_{X(n)}^{\tr}(A')\|
 \le2\|A'-\mathcal B_s[A]\|
 \le 2C_{\rm tr}|X|\|A\|e^{-bn/(2C_1)}.
\end{equation}
For the case $n< C_1 r_0$, the trivial bound $\norm{A'-\E^{\rm tr}_{X(n)}(A')}\leq 2\norm{A} \leq 2e^{br_0/2} \norm{A} e^{-bn/(2C_1)}$. By taking a larger prefactor from the case $n<C_1 r_0$ and $n\geq C_1 r_0$, we prove Eq. \eqref{eq:prep-locality}.

For \eqref{eq:prep-compatible}, we consider an interval $I$ and let $A\in\A_I$, and need to estimate 
\[
\norm{ \mathcal{U}_L^\dagger\iota_{N,L}(A) \mathcal{U}_L - \iota_{N,L}(\mathcal{U}_N^\dagger A \mathcal{U}_N )}. 
\]
Suppose we choose a truncation scale $s$ such that the gates near the seam in $\mathcal{U}_{L,<s}$ do not act on $I$, then we can use this $\mathcal{B}'_{N,s}[A]:= \mathcal U_{N,<s}^\dagger A\mathcal U_{N,<s}$ to get the triangular inequality for Eq. \eqref{eq:prep-compatible}. Let $L\geq N\geq L_{\rm prep}$, a sufficient condition for getting such a scale $s$ is to suppose $d=d(I,N) \geq (C_1+l_{\rm seam}+1)r_0 =: C_2 r_0$. Under this condition, we choose $s$ to be the largest dyadic scale $2^kr_0$ satisfying $s\leq d/C_2$. Thus $d/(2C_2)\leq s\leq d/C_2$. We can show $r_0\leq s\leq \frac{d}{C_2}<\frac{N}{2C_2}<\frac{N}{C_{\rm stop}}<\frac{L}{C_{\rm stop}}$, which implies the existence of $\mathcal{U}_{L,<s}$ (if $s=r_0$, take $\mathcal{U}_{L,r_0}$ instead). Due to $s\leq d/C_2$, the gates in $\mathcal{U}_{L,<s}$ acting on $I$ have their distances from the seam of the $N$-ring at least $d-C_1s> l_{\rm seam}s$, hence these gates are size-independent. Consequently, we have
\[
\mathcal{U}_{L,<s}^\dagger \iota_{N,L}(A) \mathcal{U}_{L,<s} = \iota_{N,L}(\mathcal{B}'_{N,s}[A]).
\]
Then by the proof of \eqref{eq:prep-truncation} we get
\begin{equation}
\begin{split}
\norm{ \mathcal{U}_L^\dagger\iota_{N,L}(A) \mathcal{U}_L - \iota_{N,L}(\mathcal{U}_N^\dagger A \mathcal{U}_N )} &\leq \norm{\mathcal{U}_L^\dagger\iota(A) \mathcal{U}_L - \mathcal{U}_{L,<s}^\dagger \iota_{N,L}(A) \mathcal{U}_{L,<s} } \\
&+ \norm{ \iota_{N,L}( \mathcal{B}_{N,s}'[A] ) - \iota_{N,L}( \mathcal{U}_N^\dagger A\mathcal{U}_N ) } \\
&\leq 2C_{\rm tr} |I| \norm{A} e^{-bs} \leq 2C_{\rm tr} |I| \norm{A} e^{-bd/(2C_2)}.
\end{split}
\end{equation}
Thus by denoting $\mu = b/(2C_2)$ we get the proof for the $d\geq C_2 r_0$ case. 

We discuss the remaining cases of Eq. \eqref{eq:prep-compatible} by tuning the prefactor but keeping the exponent $\mu$. For the case $d<C_2 r_0$, we use the trivial bound $\norm{ \mathcal{U}_L^\dagger\iota_{N,L}(A) \mathcal{U}_L - \iota_{N,L}(\mathcal{U}_N^\dagger A \mathcal{U}_N )}\leq 2\norm{A}$, and use $2\norm{A}\leq 2 e^{\mu C_2 r_0 } |I| \norm{A} e^{-\mu d}$. For the case $N\leq L_{\rm prep}$, use $2\norm{A}\leq 2 e^{\mu L_{\rm prep}/2}|I|\norm{A} e^{-\mu d}$. Picking the largest constant proves these cases.

For the cases \(L<L_{\rm prep}\), the condition $L\geq 3\ell_*r_0$ of Lemma~\ref{lem:prep-correct} is not satisfied, but we can choose any unitary \(T_L\) sending \(\ket{\Phi_L}\) to \(\ket{\Omega_L}\). The error is bounded since $L$ is bounded, so it can be absorbed to the prefactor of inequalities \eqref{eq:prep-locality}--\eqref{eq:prep-truncation}.
\end{proof}

\subsection{Further properties of the preparation}
\label{subsec:prep-lga}

In this last subsection, we discuss some further properties of the preparation circuits.

First, we can relax the requirement of translation invariance. The following corollary shows that if translation invariance is not assumed, then states satisfying Eq. \eqref{eq:prep-input-decouple} can still be prepared by circuits satisfying the good locality properties as above, and the expense is to lose the compatibility between the circuits on rings with different sizes.

\begin{corollary}[Preparation from decoupling]
\label{cor:prep-decoupling}
Let $\{\varrho_L=\ket{\phi_L}\bra{\phi_L}\}$ be pure spin states with
fixed on-site dimension $D$, satisfying trace-norm decoupling inequality \eqref{eq:prep-input-decouple}
for disjoint intervals, with constants
independent of $L$. Translation invariance and Eq.
\eqref{eq:prep-input-compare} are not required. Then
there are unitaries $\{T_L\}$ such that
\[
 T_L\ket{\Phi_L}=\ket{\Omega_L},\qquad
 \ket{\Phi_L}=\ket{\phi_L}\otimes\ket{\overline{\phi_L}},
\]
satisfying Eq. \eqref{eq:prep-locality} and admitting
truncated circuits $\mathcal U_{L,<s}$ with the support bound and
error estimate Eq. \eqref{eq:prep-truncation} as in
Theorem~\ref{thm:prep-exact}.  All constants and the initial scale $r_0$
can be chosen independently of $L$, although the gates may depend on
position and $L$.
\end{corollary}

\begin{proof}
Since Lemma~\ref{lem:prep-insert} and Lemma~\ref{lem:prep-correct} do not use the translation symmetry of states, we only need to modify Lemma~\ref{lem:prep-approx}. In the proof of Lemma~\ref{lem:prep-approx}, use $\varrho_L$ itself as the reference state:
prepare $\ket{\operatorname{can}(\varrho_{L,A_i})}$ on each block $A_i$ instead of $\ket{\varrho_{N,[0,\ell-1]}}$ in Step 1,
and choose each recovery gate from Lemma~\ref{lem:prep-insert} with
$\sigma=\varrho_L$ instead of the choice $\sigma=\varrho_N$ in the translation invariant case. Then $\epsilon_r=0$ and
$\eta_r=4\delta_r^{1/4}$, so the same proof gives Eqs.
\eqref{eq:prep-global} and \eqref{eq:prep-local} without Eq.
\eqref{eq:prep-input-compare}.
Lemma~\ref{lem:prep-correct} and Steps~1--3 of
Theorem~\ref{thm:prep-exact} then give Eqs. \eqref{eq:prep-locality} and
\eqref{eq:prep-truncation} with constants independent of $L$.
The gates are chosen separately on each ring, and this construction does
not assert a common bulk period or Eq. \eqref{eq:prep-compatible}.
\end{proof}

The next corollary shows that if a local Hamiltonian is gapped and has a unique ground state, then its ground state can be prepared by circuits discussed in Corollary \ref{cor:prep-decoupling}, even if translation invariance is not assumed.

\begin{corollary}[Preparation without translation invariance]
\label{cor:prep-nonti}
Let $H_L=\sum_{x\in\T_L}h_{x,L}$ be spin Hamiltonians with a fixed
on-site dimension $D$, where $h_{x,L}$ is supported on the cyclic
interval $\{x,\ldots,x+R_0\}$ and $\|h_{x,L}\|\le J$, for fixed
$R_0,J$.  Suppose each $H_L$ has a unique ground state
$\ket{\phi_L}$ and a spectral gap at least $\gamma_0>0$, uniformly in
$L$. No translation invariance is assumed.
Then the doubled ground states satisfy the conclusions of
Corollary~\ref{cor:prep-decoupling}, with constants depending only on
$D,R_0,J,\gamma_0$.
\end{corollary}

\begin{proof}
In the proof of Lemma~\ref{lem:decoupling}, neither the folding
argument for Eq. \eqref{eq:schmidttail} nor the correlation estimate Eq.
\eqref{eq:covariance} uses translation invariance. Their constants
depend only on $D,R_0,J,\gamma_0$, so the trace-norm decoupling property Eq. \eqref{eq:decoupling} holds
uniformly in $L$. Hence, we can apply Corollary~\ref{cor:prep-decoupling} and get the disentangler $T_L$.
\end{proof}

We next show that $T_L$ is indeed generated by a quasi-local Hamiltonian dynamics with the time independent of $L$. Therefore, states satisfying the trace-norm decoupling condition Eq. \eqref{eq:prep-input-decouple}, including the ground states of local gapped Hamiltonians without ground state degeneracy, are invertible states.

Recall that for a self-adjoint
interaction $\Phi(Z)\in\mathcal A_Z$ on the doubled ring, the following commonly used norm is defined as
\begin{equation}\label{eq:prep-lga-norm}
 \|\Phi\|_{\lambda,L}
 :=\sup_{x\in\T_L}\sum_{\varnothing\ne Z\subseteq\T_L:\,x\in Z}
 |Z|e^{\lambda\operatorname{diam}_L(Z)}\|\Phi(Z)\|,
 \qquad
 \operatorname{diam}_L(Z):=\max_{x,y\in Z}d_L(x,y).
\end{equation}
The induced automorphism of the quasi-local operator algebra is locally generated
if it arises in a fixed time from interactions whose norm defined in Eq.
\eqref{eq:prep-lga-norm} is bounded uniformly in $L$ and time.
This is the finite-ring version of a locally generated automorphism
(LGA). For one-dimensional infinite chain, disentangler LGA of states satisfying the trace-norm decoupling condition has been discussed in Ref.~\cite[Theorem~2.1]{deroeck2025}, even without doubling the system. This successive circuit construction is also discussed in Ref.~\cite[Theorem~5.9]{Ranard_2022}.

\begin{theorem}[Local generation]\label{thm:prep-lga}
The unitaries $T_L$ constructed in Theorem~\ref{thm:prep-exact} and
Corollary~\ref{cor:prep-nonti} are uniformly locally generated.
More precisely, there are $\lambda>0$, constants $J_m<\infty$ for
every integer $m\ge0$, and smooth self-adjoint interactions
$\Phi_{L,t}$, $0\le t\le1$, such that
\begin{equation}\label{eq:prep-lga-uniform}
 \sup_{L,t}\|\partial_t^m\Phi_{L,t}\|_{\lambda,L}\le J_m.
\end{equation}
Their Hamiltonians $H_L^{\rm gen}(t)=\sum_Z\Phi_{L,t}(Z)$ generate
unitaries satisfying
\begin{equation}\label{eq:prep-lga-evolution}
 \frac{d}{dt}\mathcal V_L(t)=-iH_L^{\rm gen}(t)\mathcal V_L(t),
 \qquad \mathcal V_L(0)=\one,\qquad \mathcal V_L(1)=T_L.
\end{equation}
The same statement holds for $T_L^{\dagger}$.
\end{theorem}

\begin{proof}
Here we construct a generator with bounds uniform in the ring size for
the doubled-state preparation $\mathcal U_L = T_L^\dagger = \overline{\xi_L} F_L \mathcal{C}_{r_{K-1}}\cdots\mathcal{C}_{r_{0}} U_{r_0}$ from Eq. \eqref{eq:prep-dyadic}, where each gate in $\mathcal C$ has support of length $O(r)$ and distance $O(e^{-br})$ from the identity.

For a unitary gate $G\in\mathcal{A}_Z$, we choose its spectral logarithm $h_G\in\A_Z$ with
\begin{equation}
e^{-ih_G} = G, \qquad \norm{h_G} \leq \pi,\qquad \norm{h_G}\leq \frac{\pi}{2}\norm{G-\one}.
\end{equation}
The last bound follows from $|\theta|\leq (\pi/2)|e^{i\theta}-1|$ for $-\pi<\theta\leq \pi$. Next, at each scale, we write the two commuting families from Step 1 in the proof of Theorem~\ref{thm:prep-exact} as
\[
 \begin{aligned}
 \mathcal G_r^{(1)}
 &=\{U_{2r}VR_{1,X_j}V^{\dagger}U_{2r}^{\dagger}: j=1,2,\cdots, M\},\\
 \mathcal G_r^{(2)}
 &=\{U_{2r}R_{2,Y_j}U_{2r}^{\dagger}, U_{2r}R_{2,Z_j}U_{2r}^{\dagger}: j=1,2,\cdots M\}.
 \end{aligned}
\]
Then $\mathcal C_r=(\prod_{\mathcal G_r^{(2)}}G)
(\prod_{\mathcal G_r^{(1)}}G)$.
For each block $X_j$, the three gates
\[
 U_{2r}VR_{1,X_j}V^\dagger U_{2r}^\dagger,\qquad
 U_{2r}R_{2,Z_j}U_{2r}^\dagger,\qquad
 U_{2r}R_{2,Y_{j+1}}U_{2r}^\dagger
\]
are supported in $X_j((h_*+20)r)$.
For each of these gates $G$, set $Z_G:=X_j((h_*+20)r)$.
The support and counting bounds give
\begin{equation}\label{eq:prep-lga-gates}
 \begin{gathered}
 |Z_G|\le A_*r,\qquad \|h_G\|\le Ce^{-br},\qquad
 A_*:=2\ell_*+2h_*+40=1048,\quad b=a_p/2,\\
 \sup_x\#\{G\in\mathcal G_r^{(1)}\cup\mathcal G_r^{(2)}:
 x\in Z_G\}\le16.
 \end{gathered}
\end{equation}
Recall that $b$ in the above is exponent in $O(e^{-br})$, the distance of this unitary gate from identity. As for the last unitary $F_L$, its logarithm satisfies
\[
 F_L = e^{-ih_{F_L}},\qquad \|h_{F_L}\|\le Ce^{-\kappa_F L},\qquad
 \kappa_F:=a_p/(2C_{\rm stop}).
\]

Fix a function $f\in C_c^\infty((0,1))$, $f\ge0$, with $\int f=1$.
For each interval $I=(a,a+\tau)$, which will be interpreted as the evolution time later, define
\begin{equation}
 f_I(t)=
 \begin{cases}
  \tau^{-1}f((t-a)/\tau),&t\in I,\\
  0,&t\notin I,
 \end{cases}
 \qquad \int_I f_I(t)\,dt=1.
\end{equation}
For $r_j=2^jr_0$, set
\[
 t_j=\frac34-2^{-j-1},\qquad
 \tau_j=2^{-j-3},\qquad
 t_{j+1}-t_j=2\tau_j,
\]
and write $\overline{\xi_L}=e^{-i\theta_L}$ with
$|\theta_L|\le\pi$.
Assign the following gate families to the indicated intervals,
in application order:
\[
\begin{array}{c|c}
 I & \mathcal G_I\\ \hline
 (0,1/8) & \text{first layer of }U_{r_0}\\
 (1/8,1/4) & \text{second layer of }U_{r_0}\\
 (t_j,t_j+\tau_j) & \mathcal G_{r_j}^{(1)}
                  \quad(0\le j<K)\\
 (t_j+\tau_j,t_{j+1}) & \mathcal G_{r_j}^{(2)}
                  \quad(0\le j<K)\\
 (3/4,7/8) & \{F_L\}\\
 (7/8,1) & \{\overline{\xi_L}\one\}
\end{array}
\]
For the initial gates, take $Z_G$ to be their gate supports.
Set $Z_{F_L}=\T_L$, and assign the scalar logarithm
$\theta_L\one$ to $\{0\}$.
Define
\[
 \widetilde\Phi_{L,t}(Z)
 =\sum_I f_I(t)
   \sum_{\substack{G\in\mathcal G_I\\Z_G=Z}}h_G,
\]
where the outer sum runs over the intervals listed above.
Since the logarithms within each family commute, the evolution
over each interval is
\[
 \exp\!\left(-i\int_I\sum_Z
             \widetilde\Phi_{L,t}(Z)\,dt\right)
 =\exp\!\left(-i\sum_{G\in\mathcal G_I}h_G\right)
 =\prod_{G\in\mathcal G_I}G.
\]
Let $\widetilde{\mathcal V}_L(t)$ be the evolution generated by
$\sum_Z\widetilde\Phi_{L,t}(Z)$, with
$\widetilde{\mathcal V}_L(0)=\one$.  Then
\[
 \widetilde{\mathcal V}_L(1)
 =e^{-i\theta_L}F_L
   \mathcal C_{r_{K-1}}\cdots\mathcal C_{r_0}U_{r_0}
 =\mathcal U_L.
\]

It remains to prove the smoothness and Eq. \eqref{eq:prep-lga-uniform}. Choose
$0<\lambda<\min\{b/A_*,\kappa_F\}$.
During a scale-$r_j$ pulse, its $m$th time derivative is bounded by
\begin{equation}\label{eq:prep-lga-scale-bound}
 \|\partial_t^m\widetilde\Phi_{L,t}\|_{\lambda,L}
 \le C_m\tau_j^{-m-1}r_j e^{-(b-\lambda A_*)r_j}
 \le C'_m r_j^{m+2}e^{-(b-\lambda A_*)r_j}.
\end{equation}
Here we have used Eq. \eqref{eq:prep-lga-gates},
$\operatorname{diam}_L(Z_G)\le|Z_G|$, and
$\tau_j^{-1}=8r_j/r_0$.
This bound is uniform in $j,L,t$ and tends to zero as $j\to\infty$;
at any time only one family is active.
The initial layers have a uniform bound because $r_0$ is fixed.
The final rotation contributes at most
$C_mL e^{-(\kappa_F-\lambda)L}$, and the scalar pulse has a uniform
one-site bound.  All pulses vanish with every derivative at their
endpoints, and only finitely many occur for each $L$.
These observations prove smoothness and Eq.
\eqref{eq:prep-lga-uniform} for $\widetilde\Phi$.
For $L<L_{\rm prep}$, a single smooth pulse of a spectral logarithm
of the chosen $\mathcal U_L$ has the same uniform bounds after
increasing $J_m$, since those circumferences are bounded.

If $\widetilde{\mathcal V}_L$ is the resulting path to
$\mathcal U_L$, then
$\mathcal V_L(t)=\widetilde{\mathcal V}_L(1-t)\mathcal U_L^{\dagger}$
runs from $\one$ to $T_L$ and is generated by
$\Phi_{L,t}=-\widetilde\Phi_{L,1-t}$.
Then we get
\begin{equation}
\begin{split}
\frac{d}{dt} \mathcal{V}_L(t) &= i \left(\sum_{Z} \widetilde\Phi_{L,1-t}(Z)\right)\mathcal{V}_{L}(t) \\
&= -i H_L^{\rm gen}(t) \mathcal{V}_L(t),\\
\mathcal{V}_L(0) &= \one,\qquad \mathcal{V}_L(1) = \mathcal{U}_L^\dagger = T_L
\end{split}
\end{equation}
This preserves all the bounds and proves the theorem.
\end{proof}

In the LGA we have just constructed, the generator need not be translation invariant, even if the Hamiltonian $H_L$ and ground state $\ket{\psi_L}$ are translation-invariant.

\section{Energy-defect bound}
\label{sec:coercivity}

In this section, we present the following essential theorem of an energy-defect bound and its proof. Intuitively, it rules out those states with a low energy but an extensive number of excitations or defects, which is what protects a gap from being closed by an arbitrarily weak local perturbation.

\begin{theorem}[energy-defect inequality]\label{thm:energy}
In the spin setting, assume that the translation invariant Hamiltonian $H_L$
has a unique ground state and a uniform gap as in Eq. \eqref{eq:assumption}.
Let $\widehat H_L$ be the enlarged Hamiltonian from Eq.
\eqref{eq:enlargement}, with ground state energy $\widehat E_L=2E_L$ and
ground state vector $\ket{\Psi_L}$.  Fix the compatible unitaries $T_L$ constructed
in Theorem~\ref{thm:prep-exact}, so that
$T_L\ket{\Psi_L}=\ket{\Omega_L}=\bigotimes_x \ket{e_x}$, and define
\[
 K_L=T_L(\widehat H_L-\widehat E_L\one)T_L^{\dagger},\qquad
 \mathsf{N}_L=\sum_{x\in\T_L}Q_x,\qquad Q_x=\one-\ket{e_x}\bra{e_x}.
\]
Here $\ket{e_x}=\ket{0}^{\otimes2}$ is the enlarged on-site vacuum, and $Q_x$
is embedded by the identity on all other sites.

Then there exist $\kappa>0$ and $L_1$, depending only on the fixed Hamiltonian
family $H_L$ and preparation circuit $T_L$, such that
\begin{equation}\label{eq:energy_defect_bound}
 K_L\ge\kappa \mathsf{N}_L\qquad\text{for every }L\ge L_1.
\end{equation}
Equivalently, every density matrix $\sigma$ on enlarged Hilbert space satisfies
\[
 \Tr(\sigma K_L)\ge\kappa\sum_{x\in\T_L}\Tr(\sigma Q_x).
\]
\end{theorem}

\begin{figure}[htbp]
\centering
\resizebox{\textwidth}{!}{%
\begingroup
\definecolor{edblue}{RGB}{45,82,126}
\definecolor{edteal}{RGB}{35,112,109}
\definecolor{edgold}{RGB}{151,99,31}
\definecolor{edink}{RGB}{40,46,54}
\begin{tikzpicture}[
  x=1cm,y=1cm,
  every node/.style={font=\fontsize{8}{9.5}\selectfont,align=center,text=edink},
  box/.style={draw=edblue!65,fill=edblue!4,rounded corners=1.5pt,
    line width=.55pt,text width=22mm,minimum height=12mm,inner sep=4pt},
  repair/.style={box,draw=edteal!75,fill=edteal!4},
  bound/.style={box,draw=edgold!80,fill=edgold!6,text width=25mm,
    minimum height=17mm},
  arr/.style={-{Latex[length=1.6mm,width=1.1mm]},draw=edink!65,line width=.55pt},
  flow/.style={arr,dashed,draw=edblue!85},
  lab/.style={font=\fontsize{7.5}{9}\selectfont,fill=white,inner sep=2pt},
  heading/.style={font=\fontsize{8}{10}\selectfont\bfseries,anchor=west}
]
\node[heading,text=edblue] at (-.35,5.95) {EXTRACT A SOURCE STATE};
\node[heading,text=edteal] at (-.35,2.95) {SEAM REPAIR};
\node[heading,text=edgold] at (12.1,5.95) {ENERGY CONTRADICTION};

\node[box,text width=19mm] (vac) at (.85,4.1)
  {\textbf{Lemma \ref{lem:vacuum}}  $a\leq \norm{a}\cdot\mathsf{N}$ };
\node[box] (block) at (3.7,4.8)
  {\textbf{Lemma \ref{lem:block-compression-energy}}\\ $K_L - \frac{1}{n}\sum G_{L,I}\leq c\mathsf{N}_L$ };
\node[box] (trans) at (3.7,3.4)
  {\textbf{Lemma \ref{lem:translation-defects-energy}}\\ defects non-concentration};
\node[box,text width=24mm,minimum height=16mm] (failure) at (7,4.8)
  {\textbf{Lemma \ref{lem:failure-states-energy}}\\Bounded energy\\Extensive defects \\$\mathcal E_L\le A_*$\\$\bar n\ge\delta_*$};

\node[repair,text width=19mm] (adjust) at (.85,1.9)
  {\textbf{Lemma \ref{lem:marginal-adjustment-energy}}\\Marginal\\ error correction};
\node[repair] (seam) at (3.7,1.9)
  {\textbf{Lemma \ref{lem:seam-repair-energy}}\\One-seam repair};
\node[repair] (entropy) at (7,2.65)
  {\textbf{Lemma \ref{lem:entropy-selection-energy}}\\Entropy bands\\Small CMI};
\node[repair] (recur) at (7,.45)
  {\textbf{Lemma \ref{lem:oriented-recurrence-energy}}\\Block\\exchange};
\node[repair,text width=24mm,minimum height=20mm] (short) at (10.25,1.9)
  {\textbf{Lemma \ref{lem:shorter-ring-variational}}\\Shorter ring\\Small seam cost\\$(\chi_M)_I=\rho_I$\\$|I|\ge M/3$};
\node[box,text width=24mm] (retain) at (10.25,0)
  {\textbf{Lemma \ref{lem:retained-defects-energy}}\\Defects\\across the seam};

\node[bound] (upper) at (13.4,4.8)
  {\textbf{Upper bound}\\Closing cost\\+ energy density\\$\mathcal E_M<3g_*/4$};
\node[bound] (lower) at (13.4,0)
  {\textbf{Lower bound}\\Defects + gap\\$\mathcal E_M\ge g_*$};
\node[box,draw=edink,fill=edink!4,text width=22mm,
  minimum height=21mm,line width=.8pt] (theorem) at (16.6,2.4)
  {\textbf{Contradiction}\\[3pt]\textbf{Theorem \ref{thm:energy}}\\$K_L\ge\kappa \mathsf{N}_L$};

\draw[arr] (vac.east) -- (block.west);
\draw[arr] (vac.east) -- (trans.west);
\draw[arr] (block.east) -- (failure.west);
\draw[arr] (trans.east) -- (5.35,3.4) |- (failure.south west);
\draw[arr] (adjust.east) -- (seam.west);
\draw[arr] (seam.east) -- (short.west);
\draw[arr] (entropy.east) -- (short.north west);
\draw[arr] (recur.east) -- (8.5,.45) -- (short.south west);
\draw[arr] (failure.east) -- (upper.west);
\draw[arr] (short.north east) -- (11.95,2.95) -- (11.95,4) -- (upper.south west);
\draw[arr] (short.south) -- (retain.north);
\draw[arr] (retain.east) -- (lower.west);
\draw[arr,draw=edblue!75] (trans.south) -- (3.7,2.73) -- (-.62,2.73)
  -- (-.62,-1.13) -- (13.4,-1.13) -- (lower.south);
\node[lab,text=edblue] at (5.35,-1.13)
  {Lemma \ref{lem:translation-defects-energy}: interval defect bound $\sum_{x\in I}n_x\ge c_f\bar n|I|-C_b$};

\draw[flow] (failure.south) -- (7,3.75) -- (10.25,3.75) -- (short.north);
\node[lab,text=edblue] at (9.45,3.75) {source state $\rho_L$};
\draw[arr] (upper.east) -- (15.1,4.8) -- (theorem.north);
\draw[arr] (lower.east) -- (15.1,0) -- (theorem.south);
\end{tikzpicture}
\endgroup
}
\caption{Dependencies in the contradiction proof of
Theorem~\ref{thm:energy} for $R_0\ge0$.
Solid arrows indicate uses of estimates and constructions, and the dashed
arrow supplies the source state. Both energy bounds concern the same
shorter-ring state $\chi_M$. CMI denotes conditional mutual information.
The on-site case with $R_0=0$ follows directly from
Lemma~\ref{lem:translation-defects-energy}.}
\label{fig:energy-bound-dependencies}
\end{figure}
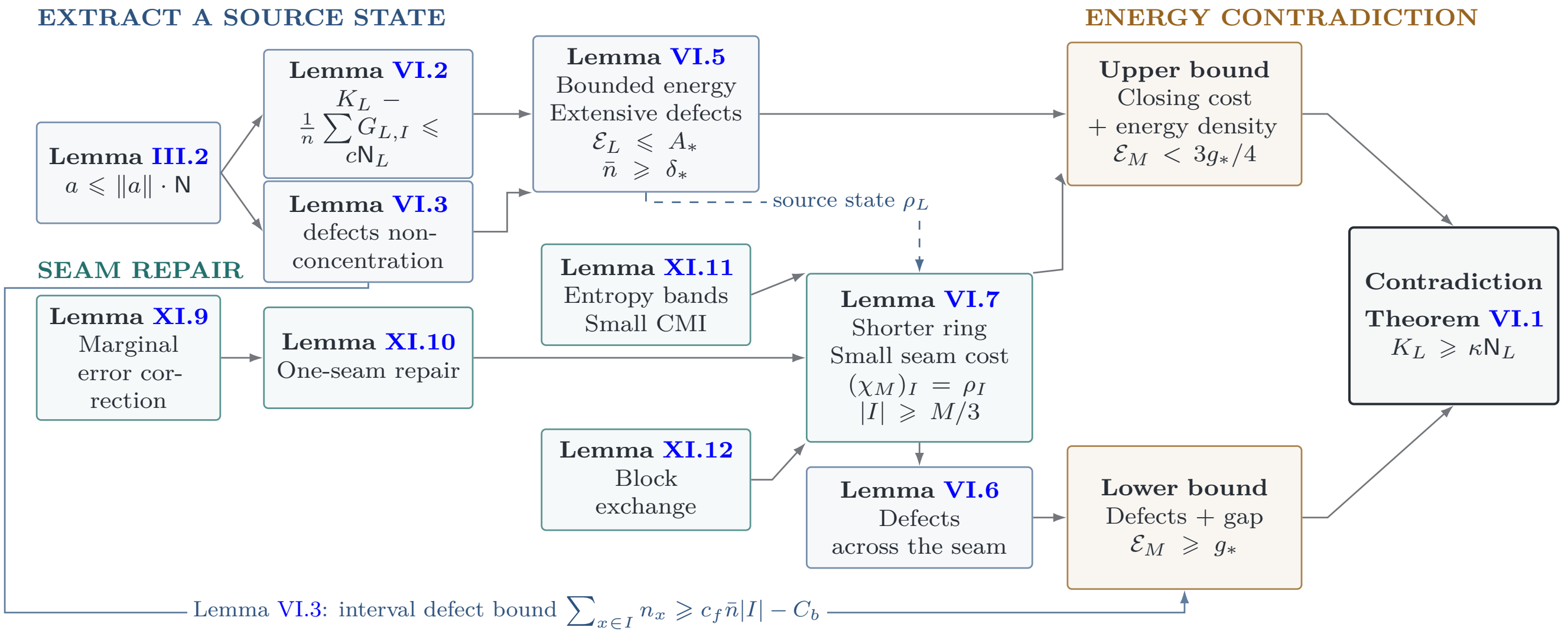

The proof of Theorem \ref{thm:energy} is via contradiction. Assume that the claim fails, then we are able to construct a state on a ring whose energy is larger than some threshold, but at same time we would also show that its energy is smaller than a smaller threshold, leading to a contradiction. In Secs. \ref{subsec:energy-local}-\ref{subsec:energy-contradiction}, we present a few necessary lemmas, and in Sec. \ref{subsec:proof-energy-defect-bound} we show the proof details of the energy-defect bound. Fig. \ref{fig:energy-bound-dependencies} shows the dependencies of lemmas and also the proof logic.

\subsection{Local compression and defect non-concentration}
\label{subsec:energy-local}

In this subsection, we present two lemmas about Hamiltonian truncation and defect concentration. {Specifically, we will define a notion of truncation of the Hamiltonian $K_L$ to a block $I$, which is denoted by $G_{L, I}$.} Lemma~\ref{lem:block-compression-energy} compares {$K_L$ and a translation averaged version of $G_{L, I}$, and shows that their difference} comes from interactions crossing the block boundaries and is proportional to number of defects.
Lemma~\ref{lem:translation-defects-energy} prevents {the defects from concentrating near one site, after the physical ground state is transformed into the product-vacuum frame.}

We first set up some conventions for convenience. Fix the exact shell representation of $K_L$ supplied by
Proposition~\ref{prop:shelltransfer} and Lemma~\ref{lem:normal-form-transport}, and denote its shells by
$\theta_L(x,r)$, with $x\in\T_L,\ 0\leq r\leq M_L=\lfloor L/2\rfloor$ and
\[
K_L = \sum_{x,r} \theta_L(x,r), \quad\theta_L(x,r)^\dagger = \theta_L(x,r),\quad \theta_L(x,r) P^{\Omega}_{B_L(x,r)} = 0.
\]
Recall that the operator $\theta_{L}(x,r)$ is the shell term supported on the interval $B_{L}(x,r)$ centered at $x$ with radius $r$.  With this notation, define
\begin{equation}
 \begin{gathered}
C_\partial:=3\sup_{L,y}\sum_{\substack{x,r:\\
  y\in B_{L}(x,r)}}
 (|B_{L}(x,r)|-1)\|\theta_{L}(x,r)\|<\infty.
 \end{gathered}
 \label{eq:energy-shells}
\end{equation}
 Distinct $(x,r)$ are counted separately even when supports of $\theta_{L}(x,r)$ coincide,
since each $(x,r)$ labels a separate summand.  The stretched-exponential shell bound
makes $C_\partial$ finite.

For $I_{x,n}=\{x,\ldots,x+n-1\}$, $1\le n<L$, define the isometry
$J_I:\mathcal H_I\to\mathcal H_{\T_L}$ by
\[
 J_I\ket{u}=\ket{u}\otimes\ket{\Omega_{I^c}},\qquad \ket{u}\in\mathcal H_I,
\]
using cyclic tensor-factor order. Define the \emph{block compressed} Hamiltonian \footnote{A compression refers to a projection of operator to a subspace.}
\[
G_{L,I}:=J_I^{\dagger}K_LJ_I,
\]obtained by fixing the sites outside \(I\) to the vacuum. We also use \(G_{L,I}\) for its extension to the whole ring \(G_{L,I}\otimes\mathbf1_{I^c}=\mathbb E_I^\Omega(K_L)\). Also recall that the energy gap of $H_L$ and thus of $K_L$ is at least $\gamma_0$. The following lemma compares $K_L$ and a translation averaged version of $G_{L, I}$, and shows that their difference is bounded by the number of defects in Eq. \eqref{eq:block-average-energy}.

\begin{lemma}[Block compression]
\label{lem:block-compression-energy}
For every proper cyclic interval $I\subset\T_L$, the compressed operator
$G_{L,I}$ satisfies
\begin{equation}
 G_{L,I}\ge\gamma_0(1-P_I^\Omega),\qquad G_{L,I}P_I^\Omega=0.
 \label{eq:block-compression-gap}
\end{equation}
Moreover, for every $1\le n<L$,
\begin{equation}
 -\frac{C_\partial}{n}\mathsf{N}_L
 \le K_L-\frac1n\sum_{x\in\T_L}G_{L,I_{x,n}}
 \le\frac{C_\partial}{n}\mathsf{N}_L.
 \label{eq:block-average-energy}
\end{equation}
\end{lemma}
 
\begin{proof}
The first inequality in Eq. \eqref{eq:block-compression-gap} follows from $K_L\geq \gamma_0(\one-P^{\Omega}_{\T_L})$, since for any $\ket{u}$ we have $\bra{u}J_I^\dagger K_L J_I\ket{u}\geq \gamma_0\ev{u|(1-P^{\Omega}_I)|u}$ by the definition of positive definiteness. $G_{L,I}P^{\Omega}_I = 0$ because $J_I$ preserves the vacuum by definition.

For Eq. \eqref{eq:block-average-energy}, which is estimating the energy difference of $K_L$ and its truncated version, we can use the shell property of $K_L=\sum_{x,r}\theta(x,r)$. Let $\ell = |B_L(x,r)|$, and define
\begin{equation}
    D_{\theta(x,r)} := \theta_L(x,r) - \frac{1}{n}\sum_{y\in\T_L}\E^\Omega_{I_{y,n}} \theta_L(x,r).
\end{equation}
We will prove Eq. \eqref{eq:block-average-energy} by bounding $D_{\theta(x,r)}$, as $ K_L-\frac1n\sum_{x\in\T_L}G_{L,I_{x,n}}=\sum_{x, r}D_{\theta(x, r)}$.

First consider the case $\ell\leq n$, then exactly $n-\ell+1$ length-$n$ intervals contain $B_L(x,r)$, and at most other $2(\ell-1)$ intervals meet $B_L(x,r)$ but do not contain it. Hence
\begin{equation}
    D_{\theta(x,r)} = \frac{\ell-1}{n}\theta_L(x,r) - \frac{1}{n}\sum_{\substack{I_{y,n}\cap B_{L}(x,r)\neq\emptyset, \\ B_L(x,r)\not\subset I_{y,n} }} \E^{\Omega}_{I_{y,n}}\theta_L(x,r),
\end{equation}
which further implies
\begin{equation}
    \norm{ D_{\theta(x,r)} }\leq \frac{3(\ell-1)}{n} \norm{\theta_L(x,r)}.
\end{equation}
If $\ell>n$, no length-$n$ interval contains $B_L(x,r)$, and at most
$\min\{L,n+\ell-1\}$ intervals meet $B_L(x,r)$.  Contractivity of $\E^{\Omega}_I$ and $\ell-1\ge n$ then give
\[
 \|D_{\theta(x,r)}\|
 \le\left(1+\frac{n+\ell-1}{n}\right)\|\theta_L(x,r)\|
 \le\frac{3(\ell-1)}n\|\theta_{L}(x,r)\|.
\]
 Every term in
$D_{\theta(x,r)}$ is supported in $B_L(x,r)$ and annihilates $\ket{\Omega_{B_L(x,r)}}$, so
Lemma~\ref{lem:vacuum} yields
\[
 -\frac{3(\ell-1)}n\|\theta_L(x,r)\|\mathsf{N}_{B_L(x, r)}
 \le D_{\theta(x,r)}\le
 \frac{3(\ell-1)}n\|\theta_L(x,r)\|\mathsf{N}_{B_L(x, r)}.
\]
Finally,
\[
 K_L-\frac1n\sum_{x\in\T_L}G_{L,I_{x,n}}
 =\sum_{x,r}D_{\theta(x,r)},
\]
so
\[
-\frac{C_\partial\mathsf{N}_L}{n}\leqslant-\sum_{x, r} \frac{3(\ell-1)}n\|\theta_L(x,r)\|\mathsf{N}_{B_L(x, r)}\leqslant K_L-\frac1n\sum_{x\in\T_L}G_{L,I_{x,n}}\leqslant\sum_{x, r} \frac{3(\ell-1)}n\|\theta_L(x,r)\|\mathsf{N}_{B_L(x, r)}\leqslant \frac{C_\partial\mathsf{N}_L}{n},
\]
where we have used $0\leqslant\mathsf{N}_{B_L(x, r)}\leqslant\mathsf{N}_L$. Thus inequality
\eqref{eq:block-average-energy} is proved.
\end{proof}

The next lemma controls the distribution of the number of defects at different locations, after the ground state is transformed into the product-vacuum frame. In particular, Eq. \eqref{eq:defect-density-energy} shows that the defect cannot concentrate at some specific sites in the product-vacuum frame.

Although the Hamiltonian and ground states are translation invariant, the unitary $T_L$ bringing $\ket{\Psi_L}$ to product state may not commute with one-site translation.  Write
$S_s$ for cyclic translation by $s$ sites and define
\[
 T_{L,s}=S_sT_LS_s^{\dagger},\qquad W_{L,s}=T_LT_{L,s}^{\dagger}=T_LS_sT_L^{\dagger}S_s^{\dagger}.
\]
Since the unique ground state $\ket{\Psi_L}$ is translation invariant up to phase,
both $T_L$ and $T_{L,s}$ send $\ket{\Psi_L}$ to $\ket{\Omega_L}$ up to phase.  Hence
$W_{L,s}$ preserves $\ket{\Omega_L}$ up to phase. On the other hand, since $T_L$ and $T_{L,s}^\dagger$ both have exponential locality in the sense of Eq. \eqref{eq:uniform_locality}, $W_{L,s}$ is also exponentially local. 

\begin{lemma}[Defect non-concentration]
\label{lem:translation-defects-energy}
Let $W$ be a unitary on the ring that preserves $\ket{\Omega_L}$ up to phase and
suppose that conjugation by both $W$ and $W^{\dagger}$ is exponentially local
in the sense of Eq. \eqref{eq:uniform_locality}.  There are uniform $C_2, C,\nu>0$ such that
\begin{equation}
 W^{\dagger}Q_xW\le C_2\sum_y e^{-\nu d_L(x,y)}Q_y,\qquad
 C^{-1}\mathsf{N}_L\le W^{\dagger}\mathsf{N}_LW\le C\mathsf{N}_L.
 \label{eq:vacuum-unitary-comparison}
\end{equation}

For the ground state disentangling unitaries \(T_L\), the preceding estimate has the following consequences.
There are constants
$C_f,c_f>0$ and $C_b<\infty$, independent of $L$ and $s$, with the following properties.
For an arbitrary state $\rho$, put $\sigma=T_L\rho T_L^{\dagger}$ and define
its physical translation average by
\[
 \overline\rho=\frac1L\sum_{s\in\T_L}S_s\rho S_s^{\dagger},\qquad
 \overline\sigma=T_L\overline\rho T_L^{\dagger}.
\]
Then
\begin{equation}
 \Tr(\overline\sigma K_L)=\Tr(\sigma K_L),\qquad
 \Tr(\overline\sigma \mathsf{N}_L)\ge C_f^{-1}\Tr(\sigma \mathsf{N}_L).
 \label{eq:translation-average-energy}
\end{equation}
If $\rho$ is translation invariant and 
\[
 n_x=\Tr(\rho T_L^{\dagger}Q_xT_L),\qquad
 \bar n=L^{-1}\sum_{x\in\T_L}n_x,
\]
then
\begin{equation}
 n_x\le C_f\bar n,\qquad
 \sum_{x\in I}n_x\ge c_f\bar n\,|I|-C_b
 \quad\hbox{for every interval }I.
 \label{eq:defect-density-energy}
\end{equation}
\end{lemma}

\begin{proof}
To prove Eq. \eqref{eq:vacuum-unitary-comparison}, telescope $W^{\dagger}Q_xW$ with
the vacuum expectations on the balls
$B_L(x,r)$.  For convenience, let $A=W^{\dagger}Q_xW$, $B_r=B_L(x,r)$, and
\[
 A_r=\E_{B_r}^{\Omega}(A),\qquad
 D_0=A_0,\qquad D_r=A_r-A_{r-1}\quad(1\le r\le M_L).
\]
Since $B_{M_L}=\T_L$,
\[
 A=A_{M_L}=A_0+\sum_{r=1}^{M_L}(A_r-A_{r-1})
   =\sum_{r=0}^{M_L}D_r.
\]
Each term $D_r$ is a self-adjoint operator supported on $B_r$. The assumed exponential locality of $W$ gives
$\|A-A_r\|\le C_0e^{-\nu r}$ for some uniform constants $C_0, \nu>0$, so, for $r\ge1$,
\[
 \|D_r\|\le\|A_r-A\|+\|A-A_{r-1}\|
 \le C_0(1+e^\nu)e^{-\nu r}.
\]
Together with $\|D_0\|\le1$, this yields $\|D_r\|\le C_1 e^{-\nu r}$
for all $r$, with $C_1 =\max\{1,C_0(1+e^\nu)\}$.

Moreover, $W\ket{\Omega_L}\propto|\Omega_L\rangle$,
so $A\ket{\Omega_L}=0$.  Vacuum compression $\E^{\Omega}_{B_r}$ then gives $A_r\ket{\Omega_{B_r}}=0$
and $D_r\ket{\Omega_{B_r}}=0$, with each operator viewed on its support.
Lemma~\ref{lem:vacuum} therefore proves the first inequality of Eq.
\eqref{eq:vacuum-unitary-comparison}:
\begin{equation}
\begin{split}
 W^{\dagger}Q_xW
 =A=\sum_{r=0}^{M_L}D_r &\le C_1\sum_{r=0}^{M_L}e^{-\nu r}\mathsf{N}_{B_L(x,r)} = C_1\sum_{y} \left(\sum_{r=d_L(x,y)}^{M_L} e^{-\nu r}\right) Q_y \\
 & \le C_2\sum_y e^{-\nu d_L(x,y)}Q_y,\qquad C_2 = \frac{C_1}{1-e^{-\nu}}.
\end{split}
\end{equation}
Summing over $x$ proves the upper total bound in the second inequality of Eq.
\eqref{eq:vacuum-unitary-comparison}:
\begin{equation}
    W^\dagger \mathsf N_L W \leq C_2\sum_{x,y} e^{-\nu d_L(x,y)} Q_y \leq C_2(1+2\sum_{r=1}^\infty e^{-\nu r}) \sum_{y}Q_y \leq \frac{1+e^{-\nu}}{1-e^{-\nu}} C_2 \mathsf N_L:=C\mathsf{N}_L.
\end{equation}
Applying the same bound to $W^{\dagger}$ and
conjugating prove the lower total bound. Also, the first inequality in Eq.
\eqref{eq:vacuum-unitary-comparison} is still satisfied with the above $C_2$ replaced by $C=C_2(1+e^{-\nu})/(1-e^{-\nu})$.

Next, we consider the case of the ground state disentangling unitary $T_L$. The exponential locality of $T_L$ and $T_L^{\dagger}$ in Eq.
\eqref{eq:prep-locality} shows that $W_{L,s}$ satisfies Eq. 
\eqref{eq:vacuum-unitary-comparison} uniformly in $s$ and $L$.  Moreover,
\[
 \overline\sigma=T_L\overline\rho T_L^{\dagger}
 =\frac1L\sum_{s\in\T_L}W_{L,s}S_s\sigma S_s^{\dagger}W_{L,s}^{\dagger}.
\]
Translation invariance of $\widehat H_L$ gives
$\Tr(\overline\sigma K_L)=\Tr(\sigma K_L)$, while Eq.
\eqref{eq:vacuum-unitary-comparison} and translation invariance of $\mathsf{N}_L$
give
\begin{equation}
\begin{split}
    \Tr(\overline\sigma \mathsf{N}_L) &= \frac{1}{L} \sum_{s}\Tr(\sigma S_s^{\dagger} W_{L,s}^{\dagger} \mathsf{N}_LW_{L,s} S_s) \\
    &\geq \frac{C^{-1}}{L} \sum_{s}\Tr(\sigma S_s^{\dagger} \mathsf{N}_L S_s) = C^{-1}\Tr(\sigma \mathsf{N}_L).
\end{split}
\end{equation}
This proves Eq. \eqref{eq:translation-average-energy} for any $C_f\ge C$.

It remains to prove the two bounds for a translation invariant $\rho$.
Note that $n_x=\Tr(\sigma Q_x)$.
Translation invariance of $\rho$ gives
$T_{L,s}\rho T_{L,s}^{\dagger}=S_s\sigma S_s^{\dagger}$, and therefore
\[
 \sigma=W_{L,s}(S_s\sigma S_s^{\dagger})W_{L,s}^{\dagger},\quad \forall s\in \mathbb{T}_L.
\]
Take the expectation of the first inequality in Eq.
\eqref{eq:vacuum-unitary-comparison}, with $W=W_{L,s}$, in the positive
density matrix $S_s\sigma S_s^{\dagger}$.  Cyclicity of the trace yields
\begin{equation}\label{eq:nx_leq_sum}
 \begin{aligned}
 n_x
 &=\Tr\!\left[(S_s\sigma S_s^{\dagger})W_{L,s}^{\dagger}Q_xW_{L,s}\right]\\
 &\le C\sum_y e^{-\nu d_L(x,y)}\Tr[(S_s\sigma S_s^{\dagger})Q_y]\\
 &=C\sum_y e^{-\nu d_L(x,y)}n_{y-s}.
 \end{aligned}
\end{equation}
The last equality uses $S_s^{\dagger}Q_yS_s=Q_{y-s}$, with indices taken modulo
$L$. For each fixed $y$, the site $y-s$ runs through the entire ring as
$s$ varies, so
\[
 \frac1L\sum_{s\in\T_L}n_{y-s}
 =\frac1L\sum_{z\in\T_L}n_z=\bar n.
\]
Averaging the bound for $n_x$ over $s$ therefore gives
\[
 n_x\le C\sum_y e^{-\nu d_L(x,y)}\bar n
 \le C\left(1+2\sum_{r\ge1}e^{-\nu r}\right)\bar n.
\]
There is one site at distance zero from $x$ and at most two sites at each
positive distance, which proves the last inequality. Taking
$C_f=C(1+2\sum_{r\ge1}e^{-\nu r})\ge C$ proves the first inequality in Eq.
\eqref{eq:defect-density-energy} and preserves Eq.
\eqref{eq:translation-average-energy}. Note that $C_f$ is independent of $L$.

For the second inequality in Eq.
\eqref{eq:defect-density-energy}, in Eq. \eqref{eq:nx_leq_sum} set $s=x-z$ with a fixed $z$, and average over $x$
instead to obtain
$\bar n\le C\sum_u e^{-\nu d_L(z,u)}n_u$.
Summing over $z\in I$ yields
\[
 |I|\bar n
 \le C\left(1+2\sum_{r\ge1}e^{-\nu r}\right)
       \sum_{u\in I}n_u
   +4C\sum_{r\ge1}re^{-\nu r}.
\]
For the second term, we used $0\le n_u\le1$ and the fact that at most
$4r$ ordered pairs at distance $r$ cross the two boundaries of $I$.
Rearranging gives the second inequality in Eq.
\eqref{eq:defect-density-energy} with fixed $c_f>0$ and $C_b<\infty$.
\end{proof}

\subsection{Consequences of failure of the energy-defect bound}
\label{subsec:failure-energy-defect-bound} 
In this subsection, we discuss what consequences can be drawn if the energy-defect bound fails. Suppose there is no constant $\kappa$ such that $K_L\geq \kappa \mathsf N_L$ for all sufficiently large $L$, in Lemma \ref{lem:failure-states-energy} we will construct translation 
invariant states on arbitrarily large rings whose excitation energy stays bounded, while their expected
density of defects is of constant order.

To prove Lemma~\ref{lem:failure-states-energy}, we will transfer
an $n$-site block state from a ring of length $L$ to a ring of length comparable to $n$.
We first prove a relation in Eq. \eqref{eq:compressed-comparison-energy}, which
controls the energy change under this transfer independently of both ring sizes.

Consider two intervals of the same length on two rings of different circumferences. To compare their energy, we choose both intervals away from the closing seam, so that they meet the same gates of the truncated unitary from $T_L$. Denote 
\[
I = \{a, a+1, \cdots, a+n-1\}\subset \T_L,\qquad I'=\{a',a'+1,\cdots a'+n-1\}\subset \T_{M},
\]
so neither interval crosses the seam. Let $d(I,L) = \min\{a, L-a-n\}$ which is the  distance to the closing seam, and similarly $d(I',M)$. At a dyadic scale $s$ (as described in Theorem \ref{thm:prep-exact}), the truncated preparation circuit $\mathcal{U}_{L,<s}$ in Theorem~\ref{thm:prep-exact} has period $p_0s,\ p_0:=\ell_* = 448$. Thus shifting a block by a
multiple of $p_0 s$ preserves its alignment with this pattern.

\begin{lemma}\label{lem:compressed_interaction_size_agreement}
Suppose $I,I'$ are two intervals as stated above. Then there are constants $C_g,C_{\rm cmp},c_{\rm cmp}>0$, independent of
$L,M,n,s$, such that, for every sufficiently large dyadic scale $s$ and
$L,M\ge C_gs$, the conditions 
\[
 d(I,L)\ge C_gs,\qquad d(I',M)\ge C_gs,\qquad
 a'-a\in p_0s\mathbb Z
\]
imply that
\begin{equation}
 \|{\iota_{I\to I'}}(G_{L,I})-G_{M,I'}\|\le C_{\rm cmp}(n+s)e^{-c_{\rm cmp}s},
 \label{eq:compressed-comparison-energy}
\end{equation}
where ${\iota_{I\to I'}}(O)=V_{I\to I'}OV_{I\to I'}^{\dagger}$ where $V_{I\to I'}$ identifies site $a+j$ with site $a'+j$ in the fixed
onsite bases for $0\le j<n$.
\end{lemma}

Throughout the following proof, $\mu$ and $b$ are the fixed exponents in Eqs.
\eqref{eq:prep-locality} and \eqref{eq:prep-truncation}, respectively. We define $
C_0:=C_{\rm prep}\sup_{L,x}
|\supp\widehat h_{x,L}|\,\|\widehat h_{x,L}\|$,
with $C_{\rm prep}$ denoting the constant $C$ in Theorem \ref{thm:prep-exact}. This lemma uses translation invariance of $\widehat H_L$.
\begin{proof}
The estimate is rooted in the locality of unitary $T_L$ and translation invariance of Hamiltonian $\widehat H_L$. 
Write $B_{x,L}=T_L\widehat h_{x,L}T_L^{\dagger}$ and introduce the centered
compression
\begin{equation}
 \mathcal G_{L,I}(B)
 :=J_I^{\dagger}BJ_I-\bra{\Omega_L}B\ket{\Omega_L}\one_I.
\end{equation}
Translation invariance of the ground-state density matrix gives
$\bra{\Omega_L}B_{x,L}\ket{\Omega_L}=\widehat E_L/L$ and hence
\begin{equation}
 G_{L,I}=\sum_x\mathcal G_{L,I}(B_{x,L}).
\end{equation}
Splitting all $x$ into those far from $I$ and those close to $I$, then the left hand side of Eq. \eqref{eq:compressed-comparison-energy} is bounded by
\begin{equation}\label{eq:compression_embedding_error}
\begin{split}
\norm{\iota_{I\to I'}(G_{L,I}) - G_{M,I'}} &\leq \sum_{x: d_L(x,I)\leq s+R_0} \norm{\iota_{I\to I'}(\mathcal G_{L,I}(B_{x,L})) - \mathcal G_{M,I'}(B_{x',M}) } \\
& + \sum_{x: d_L(x,I)>s+R_0} \norm{\mathcal G_{L,I}(B_{x,L})} + \sum_{y: d_M(y, I')>s+R_0} \norm{ \mathcal G_{M,I'}(B_{y,M})}
\end{split}
\end{equation}
Below we bound each of the three terms above.

{We start with the case where $x$ satisfies} $q_x := d_L(x,I)>s+R_0$ with the dyadic scale $s$ {and $R_0$ the range of the interaction $\widehat h_{x, L}$}.  Then consider a truncation of $B_{x,L}$, denoted by $\widetilde B_{x,L}$ whose support is in $\tilde X_x = \{ y: d_L(y, \supp \widehat h_{x,L})\leq q_x-R_0-1\}$ ($\tilde X_x$ therefore has no overlap with $I$). Concretely, $\widetilde B_{x,L} = \E^{\tr}_{\tilde X_x} B_{x,L}.$ The exponential locality of $T_L$ in Theorem \ref{thm:prep-exact} shows $\norm{B_{x,L} - \widetilde B_{x,L}} \leq C_0 e^{\mu(R_0+1)} e^{-\mu q_x}$. Since $\widetilde B_{x,L}$ is outside $I$,
we have $\mathcal G_{L,I} (\widetilde B_{x,L}) = 0$. Hence we get 
\begin{equation}
\|\mathcal G_{L,I}(B_{x,L})\|=\|\mathcal{G}_{L, I}(B_{x, L}-\widetilde B_{x, L})\|
\le2\|B_{x,L}-\widetilde B_{x,L}\|
\le2C_0e^{\mu(R_0+1)}e^{-\mu q_x},
\end{equation}
and there exists a constant $C_\mu$ such that
\begin{equation}
     \sum_{x:\,d_L(x,I)>s+R_0}\|\mathcal G_{L,I}(B_{x,L})\|
 \le 4C_0 e^{\mu(R_0+1)}\sum_{q>s+R_0}e^{-\mu q}\le C_\mu  e^{-\mu s}.
\end{equation}
with $C_\mu = {4C_0}/({1-e^{-\mu}})$.
The same bound holds on the $M$-ring. Hence we get bounds for the last two terms in Eq. \eqref{eq:compression_embedding_error}.

Next we consider $x$ close to $I$. For each pair of $x\in \T_L$ and $x' = x+a'-a\in \T_M$, we consider the approximation by truncated circuits $B^{(s)}_{x,L} = \mathcal{U}_{L,<s}^\dagger \widehat h_{x,L} \mathcal{U}_{L,<s}$. According to Eq. \eqref{eq:prep-truncation}, the error of this approximation is given by $\|B_{x,L}-B_{x,L}^{(s)}\|\le C_0 e^{-bs}.$ Under the compression we get 
\begin{equation}
    \norm{\mathcal{G}_{L,I}( B_{x,L} - B_{x,L}^{(s)} ) }\leq 2 \norm{ B_{x,L}-B_{x,L}^{(s)} } \leq 2C_0 e^{-bs}.
\end{equation}
Notice that $a'-a$ is a multiple of the period of the bulk gates in $\T_L$ and $\T_M$, and we can choose $C_g$ sufficiently large such that $I$ is away from the seam region: $C_g s - ((C_1+1)s+2R_0)>l_{\rm seam} s$, where $C_1$ and $l_{\rm seam}$ are constants appearing in proof of Theorem \ref{thm:prep-exact} (Step 1 and 2). Then the gates in $B^{(s)}_{x,L}$ and $B^{(s)}_{x',M}$ are identical. So if we let $Z\supset I$ contain the support of $B^{(s)}_{x,L}$, and $Z' = Z+a'-a$, then   
$\iota_{Z\to Z'} (B_{x,L}^{(s)} ) = B_{x',M}^{(s)}$. This equality is preserved by centered compression,
\begin{equation}
\iota_{I\to I'}\bigl(\mathcal G_{L,I}(B_{x,L}^{(s)})\bigr)
=
\mathcal G_{M,I'}(B_{x',M}^{(s)}).
\end{equation}
Therefore, for these $x$ close to $I$, the contribution to Eq. \eqref{eq:compression_embedding_error} comes only from truncation errors of $T_L, T_M$,
\begin{equation}
\begin{split}
&\norm{\iota_{I\to I'}\bigl(\mathcal G_{L,I}(B_{x,L})\bigr)
-\mathcal G_{M,I'}(B_{x',M})} \\
\le& \norm{\iota_{I\to I'}\bigl(\mathcal G_{L,I}(B_{x,L} -B_{x,L}^{(s)})\bigr)
-\mathcal G_{M,I'}(B_{x',M} - B_{x',M}^{(s)})}
\le2C_0e^{-bs}+2C_0e^{-bs}
= 4C_0e^{-bs}.
\end{split}
\end{equation}
There are at most $n+2s+2R_0$ such $x$'s.

Finally, we put these estimations together. The right hand side of Eq. \eqref{eq:compression_embedding_error} is bounded by $4(n+2s+2R_0)C_0 e^{-bs} + 2C_{\mu} e^{-\mu s} \leq C_{\rm cmp} (n+s)e^{-c_{\rm cmp}s},$ with $C_{\rm cmp}:=4C_0(R_0+2)+2C_\mu$ and $c_{\rm cmp}=\min\{b,\mu\}$. Hence the lemma is proved.
\end{proof}

Now we can prove the following Lemma \ref{lem:failure-states-energy}. Assuming that Theorem \ref{thm:energy} is false, later this lemma will be used to construct a state that has an energy larger than a threshold (according to the second inequality in Eq. \eqref{eq:failure-states-energy}), but also smaller than an even smaller threshold (according to the first inequality in Eq. \eqref{eq:failure-states-energy}).

\begin{lemma}[Bounded energy and positive defect density]
\label{lem:failure-states-energy}
Let $\{T_L\}$ be the family of unitaries constructed in
Theorem~\ref{thm:prep-exact} for different $L$, with $T_L\ket{\Psi_L}=\ket{\Omega_L}$.
If Theorem~\ref{thm:energy} fails, then there are $A_*<\infty$ and $\delta_*>0$
such that, for every sufficiently large $n$, there is a circumference
$2n\le L_n\le3n$ and a translation-invariant physical-frame state $\rho_n$ on the whole $L_n$-ring
satisfying
\begin{equation}
 \operatorname{Tr}[\rho_n(\widehat H_{L_n}-\widehat E_{L_n})]\le A_*,
 \qquad
 \operatorname{Tr}(T_{L_n}\rho_nT_{L_n}^{\dagger}\mathsf{N}_{L_n})\ge\delta_*L_n.
 \label{eq:failure-states-energy}
\end{equation}
\end{lemma}

\begin{proof}
We start from a ring with sufficiently large $L$, select an $n$-site block with small energy per defect, remove its
vacuum component, and transfer it to a ring of size $L_n$ comparable to $n$.

Fix a sufficiently large $n$ and let $C_f\ge1$ be the constant from
Lemma~\ref{lem:translation-defects-energy}. Choose a dyadic
$s=O(\log n)$ with $C_{\rm cmp}(n+s)e^{-c_{\rm cmp}s}\le1$. Put
$m=\lceil C_gs\rceil$ and $p=p_0s$, the seam margin and circuit period
required by Eq. \eqref{eq:compressed-comparison-energy}.

\smallskip\noindent\emph{Step 1: select a block away from the seam.}
Failure of Theorem~\ref{thm:energy} gives a ring with
$L>3n$, $L\ge2C_f(n+2m)$, and a state $\sigma^{(0)}$ such that
\begin{equation}
 \operatorname{Tr}(\sigma^{(0)}\mathsf{N}_L)>0,\qquad
 \frac{\operatorname{Tr}(\sigma^{(0)}K_L)}
      {\operatorname{Tr}(\sigma^{(0)}\mathsf{N}_L)}<\frac1{C_fn}.
\end{equation}
To control where the defects occur without changing the energy,
we take average in the physical frame:
\begin{equation}
 \rho^{(0)}=T_L^{\dagger}\sigma^{(0)}T_L,\qquad
 \rho=\frac1L\sum_{a\in\T_L}S_a\rho^{(0)}S_a^{\dagger},\qquad
 \sigma=T_L\rho T_L^{\dagger}.
\end{equation}
By Eq. \eqref{eq:translation-average-energy},
\begin{equation}
 \mathcal N:=\Tr(\sigma \mathsf{N}_L)>0,\qquad
 \Tr(\sigma K_L)\le\frac{\mathcal N}{n}.
\end{equation}
For length-$n$ blocks put energy
$F_x=\Tr(\sigma G_{L,I_{x,n}})$ and defect number
$D_x=\Tr(\sigma \mathsf{N}_{I_{x,n}})$.
Equations \eqref{eq:block-average-energy} and
\eqref{eq:defect-density-energy} give
\[
 \sum_xF_x\le(1+C_\partial)\mathcal N,\qquad
 \sum_xD_x=n\mathcal N,\qquad
 D_x\le\frac{nC_f\mathcal N}{L}.
\]
The last inequality requires the translation invariance of $T_L^{\dagger}\sigma T_L$, which is the reason why $\sigma$ cannot be replaced simply by $\sigma^{(0)}$.

We now select a block $I_{x,n}$ away from the closing seam. Counting those blocks crossing the seam or having an endpoint less than $m$
sites from it, there are at most $n+2m$ such blocks. Their total $D_x$ is at most $n\mathcal N/2$, as $L\geq 2C_f(n+2m)$.
The remaining blocks therefore carry at least half the defects.
Hence for the set  $\mathcal G:=\{x\in \T_L: d(I_{x,n},L)\geq m\}$, we have
\begin{equation}
    \sum_{x\in\mathcal G} F_x\leq (1+C_\partial) \mathcal{N} \leq \frac{2(1+C_\partial)}{n} \sum_{x\in\mathcal G} D_x,
\end{equation}
which implies that there exists at least one site $x$ such that 
\begin{equation}
D_x>0,\quad 
    F_x\leq \frac{A_{\rm blk}}{n} D_x,\quad A_{\rm blk} := 2(1+C_\partial).
\end{equation}
In the following, we denote by $I$ the block $I_{x, n}$ with the $x$ satisfying the above relations.

\smallskip\noindent\emph{Step 2: remove the vacuum component.}
The reduced state $\sigma_I$ may still have a large probability of being in the
vacuum. Define its normalized projection to non-vacuum components:
\begin{equation}\label{eq:orthogonal-projection}
 P=1-P_I^\Omega,\qquad q=\Tr(P\sigma_I)>0,\qquad
 \omega_I=\frac{P\sigma_IP}{q}.
\end{equation}
Here $q>0$ because $D_x>0$. Both $G_{L,I}$ and $\mathsf{N}_I$ annihilate the
vacuum, so 
\[
 \Tr(\omega_IG_{L,I})=\Tr(\frac{P\sigma_{I}P}{q}G_{L, I})=\frac{1}{q}\Tr(\sigma_IPG_{L, I}P)=\frac{F_x}{q}
 \le\frac{A_{\rm blk}}n\frac{D_x}{q}
 =\frac{A_{\rm blk}}n\Tr(\frac{\sigma_IP\mathsf{N}_IP}{q})
 =\frac{A_{\rm blk}}n\Tr(\omega_I\mathsf{N}_I).
\]
The gap bound Eq. \eqref{eq:block-compression-gap} gives
$\Tr(\omega_IG_{L,I})\ge\gamma_0$, while $\mathsf{N}_I\le n\one$ by definition. Hence
\[
 \Tr(\omega_IG_{L,I})\le A_{\rm blk},\qquad
 \Tr(\omega_I\mathsf{N}_I)\ge\frac{\gamma_0 n}{A_{\rm blk}}.
\]
Thus the projection in Eq.~\eqref{eq:orthogonal-projection} preserves the small energy per defect, and the gap
forces an expected defect number to be at least proportional to $n$.

\smallskip\noindent\emph{Step 3: transfer to a smaller ring.}
To turn this defect number into a positive density, set
$L_n=2n+2m+p$. Choose $y\in\{m,\ldots,m+p-1\}$ congruent to
$\min I$ modulo $p$, and put $I'=[y,y+n-1]$ on the $L_n$-ring.
Both intervals have seam margin at least $m$, and their starting
positions agree modulo $p$. Copy $\omega_I$ to $I'$ and fill the
remaining sites with vacuum:
\[
 \widetilde\sigma={\iota_{I\to I'}}(\omega_I)\otimes P_{(I')^c}^\Omega.
\]
By vacuum compression and Eq. \eqref{eq:compressed-comparison-energy},
this increases the energy by at most one and preserves the defect number:
\[
 \begin{aligned}
 \Tr(\widetilde\sigma K_{L_n})
 &=\Tr[{\iota_{I\to I'}}(\omega_I)G_{L_n,I'}]
 \le\Tr(\omega_IG_{L,I})+1\le A_{\rm blk}+1,\\
 \Tr(\widetilde\sigma \mathsf{N}_{L_n})
 &=\Tr(\omega_I\mathsf{N}_I)\ge\frac{\gamma_0 n}{A_{\rm blk}}.
 \end{aligned}
\]
Finally, define the translation-invariant state on the whole ring by
\begin{equation}
 \rho_n=\frac1{L_n}\sum_{a\in\T_{L_n}}
 S_aT_{L_n}^{\dagger}\widetilde\sigma T_{L_n}S_a^{\dagger}.
\end{equation}
Since $m,p=O(\log n)$, we have $2n\le L_n\le3n$ for large $n$.
Applying Eq. \eqref{eq:translation-average-energy} proves both inequalities
in \eqref{eq:failure-states-energy}:
\[
 \begin{aligned}
 \Tr[\rho_n(\widehat H_{L_n}-\widehat E_{L_n})]
 &=\Tr(\widetilde\sigma K_{L_n})\le A_{\rm blk}+1=:A_*,\\
 \Tr(T_{L_n}\rho_nT_{L_n}^{\dagger}\mathsf{N}_{L_n})
 &\ge C_f^{-1}\Tr(\widetilde\sigma \mathsf{N}_{L_n})
 \ge\frac{\gamma_0 n}{A_{\rm blk}C_f}
 \ge\frac{\gamma_0}{3A_{\rm blk}C_f}L_n=:\delta_*L_n.
 \end{aligned}
\]
The constants $A_*<\infty$ and $\delta_*>0$ are independent of $n$.
\end{proof}

\subsection{Choosing the boundaries and repairing the seam}
\label{subsec:energy-repair}

In order to prove the energy-defect bound in Theorem \ref{thm:energy}, we will first assume that Theorem \ref{thm:energy} is false and then prove that there is a state whose energy is larger than a threshold, but also smaller than an even smaller threshold, which leads to a contradiction. Specifically, we will need to construct a state on a shorter ring from a given reduced density matrix of an interval on a longer ring, and use Lemma \ref{lem:failure-states-energy} to conclude that its energy is small, but its number of defects is large so its energy is also large. Directly putting the reduced density matrix on the shorter ring would join the two ends of the reduced density matrix, then the local interaction on the shorter ring crossing the seam will give an energy penalty that cannot be ignored in general. Since we want the constructed state on the shorter ring to have a very low energy for the contradiction proof, we have to repair the state.

In the following, we discuss the strategy to repair the state near that seam while preserving its interior
marginal.
We first describe the repair region and bound its energy cost,
then explain how to choose block widths that ensure these two conditions.
The general lemmas used below are proved in
Sections~\ref{app:marginal-repair}--\ref{app:boundary-selection}. The precise construction of the state on the shorter ring is described in Lemma \ref{lem:shorter-ring-variational}.

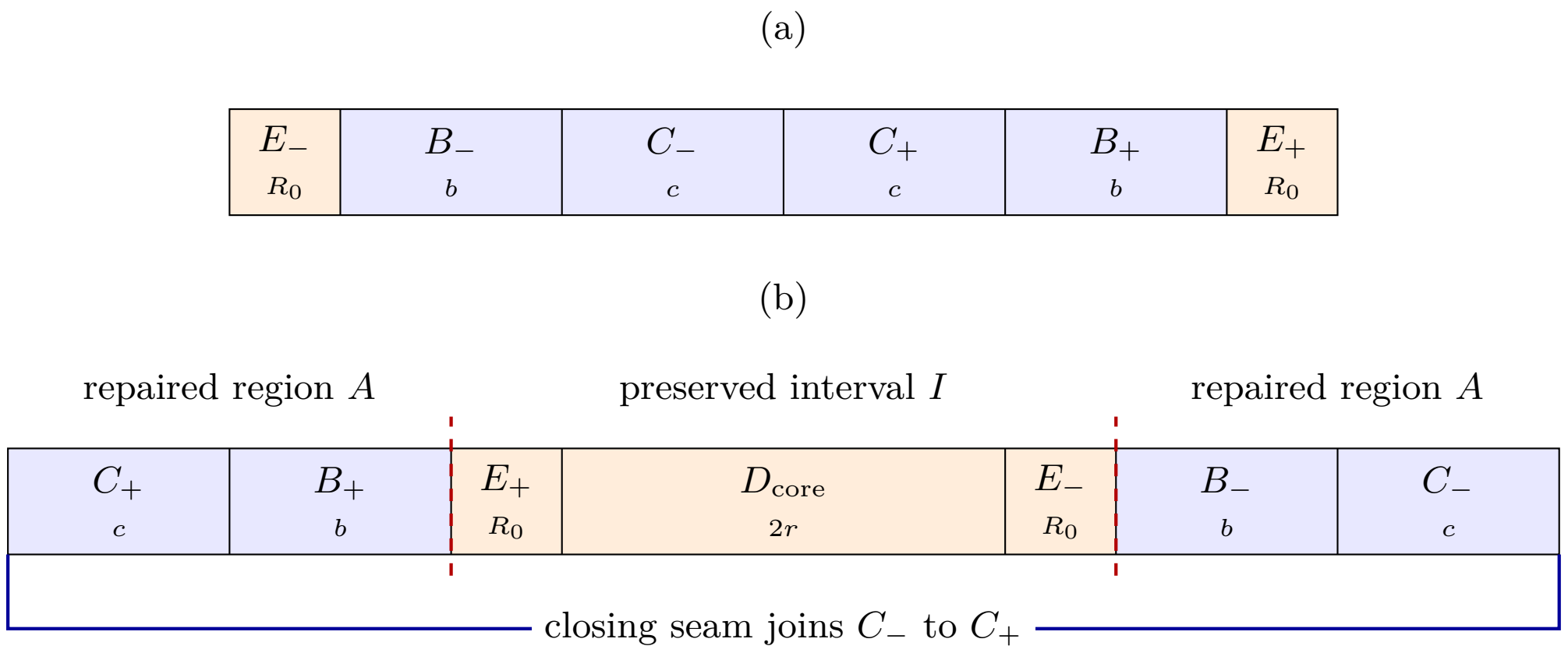
\begin{figure}[ht]
\centering
\begin{tikzpicture}[x=.94cm,y=.9cm,every node/.style={font=\small,align=center}]
\node at (7,4.95) {(a)};
\fill[orange!14] (2,3.2) rectangle (3,4.2);
\fill[blue!9] (3,3.2) rectangle (11,4.2);
\fill[orange!14] (11,3.2) rectangle (12,4.2);
\draw (2,3.2) rectangle (12,4.2);
\foreach \x in {3,5,7,9,11}{\draw (\x,3.2)--(\x,4.2);}
\node at (2.5,3.7) {$E_-$\\$\scriptstyle R_0$};
\node at (4,3.7) {$B_-$\\$\scriptstyle b$};
\node at (6,3.7) {$C_-$\\$\scriptstyle c$};
\node at (8,3.7) {$C_+$\\$\scriptstyle c$};
\node at (10,3.7) {$B_+$\\$\scriptstyle b$};
\node at (11.5,3.7) {$E_+$\\$\scriptstyle R_0$};
\node at (7,2.4) {(b)};
\fill[blue!9] (0,0) rectangle (4,1);
\fill[orange!14] (4,0) rectangle (10,1);
\fill[blue!9] (10,0) rectangle (14,1);
\draw (0,0) rectangle (14,1);
\foreach \x in {2,4,5,9,10,12}{\draw (\x,0)--(\x,1);}
\node at (1,.5) {$C_+$\\$\scriptstyle c$};
\node at (3,.5) {$B_+$\\$\scriptstyle b$};
\node at (4.5,.5) {$E_+$\\$\scriptstyle R_0$};
\node at (7,.5) {$D_{\rm core}$\\$\scriptstyle 2r$};
\node at (9.5,.5) {$E_-$\\$\scriptstyle R_0$};
\node at (11,.5) {$B_-$\\$\scriptstyle b$};
\node at (13,.5) {$C_-$\\$\scriptstyle c$};
\draw[red!70!black,thick,dashed] (4,-.2)--(4,1.3);
\draw[red!70!black,thick,dashed] (10,-.2)--(10,1.3);
\node at (2,1.55) {repaired region $A$};
\node at (7,1.55) {preserved interval $I$};
\node at (12,1.55) {repaired region $A$};
\draw[blue!60!black,thick] (0,0)--(0,-.7)--(14,-.7)--(14,0);
\node[fill=white,inner sep=3pt] at (7,-.7) {closing seam joins $C_-$ to $C_+$};
\end{tikzpicture}
\caption{Geometry of the seam repair. (a) The reference interval.
(b) The shorter ring, obtained by joining the displayed ends.
The channel acts on $A=BC$ (blue), preserves the marginal on $I=A^c$,
and uses the exterior collars $E_\pm$ to control interactions across
the two dashed boundaries.}
\label{fig:seam}
\end{figure}

We start with a longer ring of length $L$ and a shorter ring of length $M<L$. On the $L$-ring, we consider consecutive blocks $E_-B_-C_-C_+B_+E_+$ with
$|E_\pm|=R_0$, $|B_\pm|=b$, and $|C_\pm|=c$ (see Fig.~\ref{fig:seam}(a)). 
Set $E=E_-E_+$, $B=B_-B_+$, $C=C_-C_+$, and $A=BC$.
The blocks $E_\pm$ are collars wide enough to contain every
range-$R_0$ interaction window that crosses a boundary of region $A$. We use the same name of blocks on the $M$-ring, but arranged as $C_+ B_+ E_+ D_{\rm core} E_- B_- C_-$ with an extra region $D_{\rm core}$ of length $2r$, with $t=2R_0+2b+2c$ and $t+2r = M$.

Let $\rho_{EBC}$ be the reference marginal that comes from the longer ring, $\rho_{L, [0,t-1]}$. Let $\sigma = \rho_{L,[0,M-1]}$ be the input marginal, and its reduced density matrix  $\sigma_{EBC} = (\rho_{L, [0,M-1]})_{EBC}$. 
We place $\sigma$ on the $M$-ring, and want to repair its seam $E_-B_- C_-| C_+ B_+ E_+$ to get a state that is close to $\rho_{EBC}$ in trace distance. 
Indeed, by Lemma~\ref{lem:seam-repair-energy}, we have a channel acting only on the $A$ region in the state $\sigma_{EBC}$ whose
output state $\zeta$ satisfies
\[
 \zeta_E=\sigma_E, \qquad
 \zeta_A=\rho_A,\qquad
 \|\zeta_{EBC}-\rho_{EBC}\|_1
 \le\Delta_{\rm seam}(\epsilon,\eta)
 :=2\sqrt\eta+\sqrt{2\epsilon}+\tfrac32\epsilon,
\]
where $\epsilon=\|\rho_{EB}-\sigma_{EB}\|_1$ and
$\eta=I(E:C\mid B)_\rho$ is the conditional mutual information.
The channel preserves the input marginal outside $A$ and restores the
reference marginal inside $A$ exactly. Thus only interaction terms
crossing the boundary of $A$ can introduce an energy difference between $\zeta_{EBC}$ and $\rho_{EBC}$.

For later energy accounting, let $\partial_{R_0}A$ denote the starting points whose
interaction windows meet both $A$ and $A^c$,
\[
\partial_{R_0}A := \{ x: [x,x+R_0]\cap A\neq\emptyset,\, [x,x+R_0]\cap A^{c}\neq\emptyset \}.
\]
A window $[x,x+R_0]$
crossing the boundary between sites $j$ and $j+1$ must have
$x\in\{j-R_0+1,\ldots,j\}$, with site labels read cyclically.
Thus there are at most $R_0$ crossing windows at each of the two boundaries
of $A$, so $|\partial_{R_0}A|\le2R_0$.
Every window contained in $A$ has exactly the reference expectation
because $\zeta_A=\rho_A$.  Every crossing window lies in $EBC$, since
the exterior collars $E_\pm$ have length $R_0$.  Hence the bound $\|\widehat h_x\|\le J$ and Eq.
\eqref{eq:seam-repair-energy} give, for each $x\in\partial_{R_0}A$,
\[
 \left|\Tr\bigl[(\zeta_{EBC}-\rho_{EBC})\widehat h_x\bigr]\right|
 \le\|\zeta_{EBC}-\rho_{EBC}\|_1\|\widehat h_x\|
 \le J\Delta_{\rm seam}(\epsilon,\eta),
\]
where we have used the inequality
$|\Tr(XY)|\le\|X\|_1\|Y\|$ of any two operators $X,Y$.
Summing these bounds and using the triangle inequality yield
\begin{equation}
 \left|\sum_{x\in\partial_{R_0}A}
 \operatorname{Tr}\bigl[(\zeta_{EBC}-\rho_{EBC})\widehat h_x\bigr]\right|
 \le |\partial_{R_0}A|J\Delta_{\rm seam}(\epsilon,\eta)
 \le2R_0J\Delta_{\rm seam}(\epsilon,\eta).
 \label{eq:seam-cost-energy}
\end{equation}

The bound Eq. \eqref{eq:seam-cost-energy} reduces the energy estimate to the estimate of $\epsilon$ and $\eta$. 
Starting from a translation invariant state on the longer ring, we take the input $\sigma_{EBC}$ and reference marginals $\rho_{EBC}$ from this state. The two lemmas in Section~\ref{app:boundary-selection} let us choose the block widths so that the input and reference marginals on \(EB\) are close (hence $\epsilon$ is controlled) and the reference conditional mutual information \(I(E:C\mid B)_\rho\) is small (hence $\eta$ is controlled).
We now discuss how these widths $b,c,r$ are chosen.

For the conditional mutual information estimate (small $\eta$), we write $S(j)$ for the 
length-$j$ interval entropy of state $\rho_L$, and $s_j=S(j)-S(j-1)$, with $S(0)=0$.
Lemma~\ref{lem:entropy-selection-energy} bounds
$\eta=I(E:C\mid B)$ by $2R_0(s_{b+1}-s_{2b+2c+2R_0})$ and selects
widths $b,c$ such that this bound is small. Concretely, for arbitrary positive integer $m$, we take $m$ pairs of candidates of $(b_i, c_i),\, 1\leq i\leq m$ with $t_i = 2b_i +2 c_i +2R_0$ satisfying the condition $b_{i+1} +1 \geq t_i$ for $i<m$. Then Lemma~\ref{lem:entropy-selection-energy} shows that, there must be a pair $(b_i,c_i)$ such that $s_{b_i + 1} - s_{t_i} \leq 2\log d/m$. Hence if we take this pair $(b_i, c_i)$ as our choice of widths $b,c$, we get 
\begin{equation}\label{eq:eta_bound}
    \eta = I(E:C|B) \leq \frac{4R_0\log d}{m}, \quad\text{ with onsite Hilbert space dimension } d.
\end{equation}

For boundary agreement (small $\epsilon$), we need to estimate $\norm{\rho_{EB} - \sigma_{EB}}_1$. Both $\rho_{EBC}$ and $\sigma_{EBC}$ are marginals of $\rho_L$, but their sites are identified by name of blocks, which is up to a swap operator. Concretely, let $q_{u} = (\rho_{L})|_{[0,\ell-1]\cup [\ell+2u, 2\ell-1+2u]}$ be a reduced density matrix of two joint length-$\ell$ intervals. Since $\rho_L$ is translation invariant by assumption, the starting site is not important. Then take $\ell = b+R_0$ even, we have
\[
\rho_{E_-B_- B_+ E_+} = q_{c}, \quad \sigma_{E_- B_- B_+ E_+} = \mathsf S q_{r} \mathsf S^\dagger
\]
where $\mathsf S$ swaps the two length-$\ell$ intervals. Lemma \ref{lem:oriented-recurrence-energy} states that for each $\alpha>0$, there is an integer $U\geq b$ such that we can get two integers $c, r$ with $b\leq c<r\leq U$ and 
\begin{equation}
    \epsilon = \norm{q_c - \mathsf S q_r \mathsf S^\dagger}_1 \leq \alpha.
\end{equation}

We still need to make both estimates of $\epsilon$ and $\eta$ hold at the same widths. This can be done iteratively. Fix the desired accuracy $\alpha$ and number $m$, we first take $b_1$ such that $b_1 +R_0$ is even. Then we use Lemma~\ref{lem:oriented-recurrence-energy} to get an $U_1$. Regard this $U_1$ as constant $c$ for next round. Take $b_2$ satisfying $b_2+1> 2b_1 +2U_1+2R_0$ and $b_2+R_0$ even, and obtain $U_2$ by Lemma \ref{lem:oriented-recurrence-energy} again. Continue with this iteration, $\cdots\to b_i\to U_i \to b_{i+1}\to\cdots $,  we get $m$ pairs of $(b_i, U_i)$. Finally apply Lemma \ref{lem:entropy-selection-energy}, we get one of the $m$ pairs, $(b_i, U_i)$ satisfying 
\begin{equation}\label{eq:entropy_bound_ss}
    s_{b_i +1 } - s_{2b_i + 2U_i+ 2R_0}\leq \frac{2 \log d}{m}.
\end{equation}
This pair $(b_i, U_i)$ also gives integers $b_i\leq c_i< r_i\leq U_i$, and we use width $b_i, c_i, r_i$ for the $BCE$ blocks. We have the relation $\eta\leq 2R_0(s_{b_i +1 } - s_{2b_i + 2c_i+ 2R_0})\leq 2R_0(s_{b_i +1 } - s_{2b_i + 2U_i+ 2R_0})\leq \frac{4R_0 \log d}{m}$,
and $\epsilon\leq \alpha$. 

\subsection{Precise construction of the state on the shorter ring}
\label{subsec:energy-contradiction}
In this section, we first show that preserving a long interior marginal also preserves number of defects in the shorter ring up to a bounded boundary loss in Lemma \ref{lem:retained-defects-energy}. Then, with the strategy in Section \ref{subsec:energy-repair}, in Lemma \ref{lem:shorter-ring-variational} we construct the state on the shorter ring for the proof of Theorem~\ref{thm:energy}.

\begin{lemma}[Defects through the seam]
\label{lem:retained-defects-energy}
Let $L\ge M$ be sufficiently large for the estimates of
Theorem~\ref{thm:prep-exact} to apply.  Suppose a physical-frame state
$\zeta_M$ on the $M$-ring has
the same marginal on $I=[a,M-a-1]$, $1\le a<M/2$, as a
physical-frame state $\rho_L$ on the $L$-ring. 
The state $\rho_L$ may be either translation invariant or
translation non-invariant.
Then there is a constant $C_s\ge0$, independent of the states and lengths, such that
\begin{equation}
 \operatorname{Tr}(T_M\zeta_MT_M^{\dagger}\mathsf{N}_M)
 \ge\sum_{x\in I}\operatorname{Tr}(\rho_LT_L^{\dagger}Q_xT_L)-C_s.
 \label{eq:retained-defects-energy}
\end{equation}
\end{lemma}

\begin{proof}
For $x\in I$, let 
\[r_x = \min\{x-a, M-a-1-x\},\quad A_{M,x} = T_M^\dagger Q_x T_M,\quad A_{L,x} = T_L^\dagger Q_x T_L.
\]
$A_{L,x}$ and $A_{M,x}$ are quasi-local defect operators of defects in the physical frame.
We denote $B_x = [x-r_x, x+r_x]$, and approximate the quasi-local defect operators by tracing out the region outside $B_x$,
\[
\widetilde A_{M,x} = \E^{\tr,M}_{B_x}(A_{M,x}),\qquad \widetilde A_{L,x} = \E^{\tr,L}_{B_x}(A_{L,x}).
\]
Applying Eq.
\eqref{eq:prep-locality} to $Q_x$ gives
\[
 \|A_{M,x}-\widetilde A_{M,x}\|
 +\|A_{L,x}-\widetilde A_{L,x}\|
 \le C_{T}e^{-\mu r_x}
\]
with constant $C_T$.

We want to further compare $\widetilde A_{M,x}$ and $\widetilde A_{L,x}$, based on the natural identification of the $B_x$'s defined on the two rings by their ordered site coordinates. Notice that the distance between $x$ and the closing seam on the ring with length $M$ is $a+r_x$, so by Eq. \eqref{eq:prep-compatible} we have 
\[
\norm{\iota_{M,L}(A_{M,x}) - A_{L,x}} \leq C_{T} e^{-\mu (a+r_x)}. 
\]
The normalized partial trace of $\iota_{M,L}(A_{M,x}) $ onto $B_x$ is exactly $\tilde A_{M,x}$, so together with the contractivity of partial trace we get
\begin{equation}
 \begin{aligned}
 \|\widetilde A_{M,x}-\widetilde A_{L,x}\|
 &=\bigl\|\E_{B_x}^{\tr,L}
       \bigl(\iota_{M,L}(A_{M,x})-A_{L,x}\bigr)\bigr\|\\
 &\le\|\iota_{M,L}(A_{M,x})-A_{L,x}\|
 \le C_Te^{-\mu(a+r_x)}.
 \end{aligned}
 \label{eq:seam-local-comparison}
\end{equation}

Since $B_x\subset I$ and $(\zeta_M)_I=(\rho_L)_I$, we have
$\Tr(\zeta_M\widetilde A_{M,x})
 =\Tr(\rho_L\widetilde A_{M,x})$. Inserting these local approximations
and using $|\Tr(\sigma O)|\le\|O\|$ for a density matrix $\sigma$ yield
\[
 \begin{aligned}
 &\bigl|\Tr(\zeta_M A_{M,x})-\Tr(\rho_L A_{L,x})\bigr|\\
 \le&
   \bigl|\Tr[\zeta_M(A_{M,x}-\widetilde A_{M,x})]\bigr|
   +\bigl|\Tr[\rho_L(\widetilde A_{M,x}-\widetilde A_{L,x})]\bigr|
   +\bigl|\Tr[\rho_L(\widetilde A_{L,x}-A_{L,x})]\bigr|\\
\le& C_T e^{-\mu r_x}+C_T e^{-\mu(a+r_x)}.
 \end{aligned}
\]
For each nonnegative integer $r$, at most two sites of $I$ have $r_x=r$,
one measured from each endpoint. Consequently,
\[
 \sum_{x\in I}\bigl(e^{-\mu r_x}+e^{-\mu(a+r_x)}\bigr)
 \le2(1+e^{-\mu a})\sum_{r\ge0}e^{-\mu r}
 \le\frac4{1-e^{-\mu}}.
\]
Finally, every $A_{M,x}$ is positive and $\mathsf{N}_M=\sum_{x\in\T_M}Q_x$.
With $C_s=4C_T/(1-e^{-\mu})$, we obtain
\[
 \begin{aligned}
 \Tr(T_M\zeta_MT_M^{\dagger}\mathsf{N}_M)
 &=\sum_{x\in\T_M}\Tr(\zeta_M A_{M,x})
 \ge\sum_{x\in I}\Tr(\zeta_M A_{M,x})\\
 &\ge\sum_{x\in I}\Tr(\rho_L A_{L,x})-C_s,
 \end{aligned}
\]
which is Eq. \eqref{eq:retained-defects-energy}.

\end{proof}

Now we move to construct the state on the shorter ring. As we will see, the second part of Eq. \eqref{eq:shorter-ring-variational} essentially means that the constructed state on the shorter ring has the same energy density as the original state on the longer ring. This property will be used in the final proof of Theorem \ref{thm:energy} in Sec. \ref{subsec:proof-energy-defect-bound}.

\begin{lemma}[Shorter-ring state]
\label{lem:shorter-ring-variational}
Assume $R_0\ge1$ and let $H_L$ be the translation invariant Hamiltonian. For every energy $\varepsilon_{\rm cl}>0$ and integer
$M_{\min}>2R_0$, there is a finite integer
$M_{\max}\ge M_{\min}$, depending only on $\widehat D$ (on-site Hilbert space dimension of enlarged system), $R_0$ (interaction range of $H_L$), $J$ (interaction strength $\norm{\widehat h_{x,L}} \leq J$), and
$\varepsilon_{\rm cl},M_{\min}$, with the following property.
For every $L>M_{\max}$ and every translation invariant density matrix
$\rho$ on the enlarged $L$-ring, there exist integers $M,a$ and a density
matrix $\chi_M$ on the enlarged $M$-ring, such that
\[
 M_{\min}\le M\le M_{\max},\qquad
 1\le a\le\lfloor M/3\rfloor,\qquad I=[a,M-a-1],
\]
and
\begin{equation}
 (\chi_M)_I=\rho_I,\qquad
 \Tr(\chi_M\widehat H_M)
 \le\frac ML\Tr(\rho\widehat H_L)+\varepsilon_{\rm cl}.
 \label{eq:shorter-ring-variational}
\end{equation}
In particular, $|I|=M-2a\ge M/3$.  The bound $M_{\max}$ is independent
of the source circumference $L$ and the source state $\rho$.
\end{lemma}

{\bfseries Remark.} The translation invariance of the Hamiltonian $H_L$ is very important in this lemma. Without this condition, we cannot ensure the validity of inequality Eq. \eqref{eq:shorter-ring-variational}.

\begin{proof}
We first identify the sufficient conditions for the conclusions, then obtain them from the entropy lemma \ref{lem:entropy-selection-energy} and exchange lemma \ref{lem:oriented-recurrence-energy}.

{\it Step 1: reduce the energy estimate.} For the moment, we suppose that integers $b,c$ have been chosen to satisfy $R_0+1\leq b\leq c<r$, and define 
\[
t = 2R_0 + 2b +2c ,\quad M = t+2r< L, \quad a=b+c.
\]
On the $L$-ring, we use the reference marginal $\omega_{EBC} = \rho_{[0,t-1]}$ with the ordered partition of $[0,t-1]$ (schematically shown in Fig. \ref{fig:seam}(a))
\[
E_-\ B_-\ C_-\ C_+\ B_+\ E_+,
 \qquad |E_\pm|=R_0,\quad |B_\pm|=b,\quad |C_\pm|=c.
\]
On the $M$-ring, we partition $[0,M-1]$ in the order (schematically shown in Fig. \ref{fig:seam}(b))
\[
 C_+\ B_+\ E_+\ D_{\rm core}\ E_-\ B_-\ C_-,
 \qquad |D_{\rm core}|=2r,
\]
and place $\tau_{M} = \rho_{[0,M-1]}$ on it. To compare the marginal $(\tau_M)_{EBC}$ with the reference $\omega_{EBC}$, we identify the equally named blocks site by site preserving their internal site order. On the $M$-ring, the region $A=B_-C_-C_+ B_+$ crosses the new closing bond which is between $M-1$ and $0$, and its complement is $A^c = E_+ D_{\rm core} E_- = [a, M-a-1]$. {We will identify $A^c$ as the interval $I$ in the theorem to be proved.} Since $|A^c| = 2R_0 + 2r\geq b+c = a$ and $M=2a+|A^c|$, we get $1\leq a\leq \lfloor M/3\rfloor$.

In order to prove Eq. \eqref{eq:shorter-ring-variational} in the theorem, it suffices to find a channel $\Lambda_A$ for which 
\begin{equation}\label{eq:repair_criterions}
    \chi_M = (\Lambda_A\otimes\one_I)(\tau_M),\qquad (\chi_M)_A = \omega_A,\qquad \norm{(\chi_M)_{EBC} - \omega_{EBC}}_1\leq \Delta,\qquad 2R_0J\Delta<\varepsilon_{\rm cl}.
\end{equation}
Indeed, the channel leaves $A^c$ untouched, giving $(\chi_M)_{A^c} = \rho_{A^c}$, which proves the first part of Eq. \eqref{eq:shorter-ring-variational}.

For the energy comparison as in the second part of Eq. \eqref{eq:shorter-ring-variational}, let $e_{y,L} := \Tr(\rho \widehat h_{y,L})$ for $y\in\T_L$, and let $X_{x,M} = \{x,x+1,\cdots, x+R_0\}$ for $x\in\T_M$ (with coordinates modulo $M$). We match this $X_{x,M}$ with $R_0+1$ sites starting from a site $j(x)$ on $L$-ring as follows: if $X_{x,M}\subset A^c$, take $j(x)=x$; otherwise $X_{x,M}$ must have overlap with region $A$, hence a subset of $EBC$, so we identify $X_{x,M}$ with the corresponding window on the $L$-ring based on the $EBC$ region correspondence. We denote the corresponding operator map from the $M$-ring to the $L$-ring as $\beta_x$. By our assumption, the Hamiltonian terms $h_{x,L}$ are independent of ring circumference, hence
\begin{equation}
    \beta_x(\widehat h_{x,M}) = \widehat h_{j(x),L}.
\end{equation}
The exact marginals on $A$ and $A^c$ give $\Tr(\chi_M\widehat h_{x,M}) = e_{j(x),L}$ for windows $X_{x,M}$ contained in either region. At most $2R_0$ windows cross the bond between $A$ and $A^c$, and they all lie in $EBC$ as mentioned, so Eq. \eqref{eq:repair_criterions} bounds the difference between $\Tr(\chi_M\hat h_{x, M})$ and
$e_{j(x),L}$ by $J\Delta$. Summing over $x$ gives
\begin{equation}\label{eq:shorter-ring-selected-energies}
    \Tr(\chi_M \widehat H_{M}) =\left(\sum_{X_{x,M}\subset A} + \sum_{X_{x,M}\subset A^c} + \sum_{\substack{X_{x,M}\cap A\neq\emptyset,\\ X_{x,M}\cap A^c\neq\emptyset}} \right)\Tr(\chi_M \widehat h_{x,M}) \leq \sum_{x\in\T_M} e_{j(x),L} + 2R_0J\Delta.
\end{equation}

It remains to replace the selected sum {, $\sum_{x\in\T_M} e_{j(x),L}$,} by $M/L$ times the $L$-ring energy. By translation invariance of the Hamiltonian $\widehat H_L$, we have $\hat h_{y,L} = S_y \hat h_{0,L} S_y^\dagger$, and translation invariance of $\rho$ gives $S_{y}^\dagger\rho S_y = \rho$. Hence
\[
e_{y,L} = \Tr(S_y^\dagger \rho S_y \widehat h_{0,L}) = \Tr(\rho \widehat h_{0,L}) =: \bar h_\rho  \qquad\text{for every }y,
\]
and consequently
\begin{equation}
 \sum_{x\in\T_M}e_{L, j(x)}
 =M\bar h_\rho
 =\frac ML\sum_{y\in\T_L}e_{L, y}
 =\frac ML\Tr(\rho\widehat H_L).
 \label{eq:shorter-ring-energy-average}
\end{equation}
Substituting Eq. \eqref{eq:shorter-ring-energy-average} into Eq.
\eqref{eq:shorter-ring-selected-energies} and using
$2R_0J\Delta<\varepsilon_{\rm cl}$ proves the required energy bound as in the second part of Eq. \eqref{eq:shorter-ring-variational}.
It therefore remains to construct a channel \(\Lambda_A\) satisfying the repair estimates in Eq. \eqref{eq:repair_criterions} and to choose the widths so that \(M_{\min}\le M\le M_{\max}\), with \(M_{\max}\) independent of \(L\) and \(\rho\). 

{\it Step 2: choose the width and} $M_{\max}$. 
Lemma~\ref{lem:seam-repair-energy} gives a channel with repair error
\[
 \Delta=\Delta_{\rm seam}(\epsilon,\eta)=2\sqrt\eta+\sqrt{2\epsilon}+\tfrac32\epsilon,\qquad
 \epsilon=\|\omega_{EB}-(\tau_M)_{EB}\|_1,\qquad
 \eta=I(E:C\mid B)_\omega.
\]
Here $\epsilon$ measures the boundary agreement, and $\eta$ is the
conditional mutual information of the reference state $\omega$.
Let $\alpha>0$ and $m\ge1$ be two parameters such that
\[
 2R_0J\Delta_{\rm seam}\!\left(\alpha,\frac{4R_0\log \widehat D}{m}\right)
 <\varepsilon_{\rm cl}.
\]
Then, it suffices to choose widths $b,c,r$ so that the following inequalities hold simultaneously,
\begin{equation}\label{eq:choose_widths}
 \epsilon\le\alpha,\qquad \eta\le\frac{4R_0\log\widehat D}{m}.
\end{equation}
We now use the iterative construction and selection described in
Section~\ref{subsec:energy-repair} to get $b,c,r$ together with $M_{\max}$, with onsite dimension
$\widehat D$. Start with $b_1\ge R_0+1$ such that $b_1+R_0$
is even and $6b_1\ge M_{\min}$. Let
$(b_i,U_i)_{i=1}^m$ be the resulting candidate pairs, and write
\[
 t_i=2R_0+2b_i+2U_i,\qquad M_{\max}=2t_m.
\]
The construction makes $b_i$ and $t_i$ increasing. All these
numbers are fixed before specifying the circumference
$L$ or the state $\rho$, and depend only on the parameters $\widehat D, R_0, J, \varepsilon_{\rm cl}, M_{\min}$ in the lemma.

For every $L>M_{\max}$, all candidate intervals fit on the
source ring. The selection in Section~\ref{subsec:energy-repair}
then gives an index $i$ and widths
\[
 b=b_i,\qquad b_i\le c<r\le U_i,
\]
for which
\[
 \epsilon\le\alpha,\qquad
 \eta\le\frac{4R_0\log\widehat D}{m}.
\]
Lemma~\ref{lem:seam-repair-energy} therefore provides the repair
channel required in Step~1, with
$2R_0J\Delta_{\rm seam}(\epsilon,\eta)<\varepsilon_{\rm cl}$.

It remains to check the shorter-ring circumference
$M=2R_0+2b+2c+2r$. The choice of the initial width gives
\[
 M\ge6b_i\ge6b_1\ge M_{\min},
\]
while the candidate bounds give
\[
 M\le t_i+2U_i\le2t_i\le2t_m=M_{\max}<L.
\]
Thus the selected widths satisfy all requirements of Step~1.
Applying its energy comparison proves
Eq.~\eqref{eq:shorter-ring-variational}.
\end{proof}

We comment on the importance of translation invariance in Lemma~\ref{lem:shorter-ring-variational}.
For any Hamiltonian, one may define
$e_{y,L}=\Tr(\rho\widehat h_{y,L})$ and
$\bar h_\rho=L^{-1}\sum_y e_{y,L}$. The role of translation invariance
in Lemma~\ref{lem:shorter-ring-variational} is to justify Eq.
\eqref{eq:shorter-ring-energy-average}, which converts the selected
local energies in Eq. \eqref{eq:shorter-ring-selected-energies} into
$M/L$ times the source energy. This conversion is needed above to
reduce the source excitation energy $A_*$ to $A_*M/L$.

To isolate this issue for a position-dependent interaction, suppose
that the local interaction matrices still agree under the window
identifications $\beta_x$ used in that lemma. Even with this
agreement, marginal repair gives only
\[
 \Tr(\chi_M\widehat H_M)
 \le\frac ML\Tr(\rho\widehat H_L)
    +\mathcal D_{M,L}(\rho)+2R_0J\Delta,
 \qquad
 \mathcal D_{M,L}(\rho)
 :=\sum_{x\in\T_M}e_{j(x),L}
       -\frac ML\sum_{y\in\T_L}e_{y,L}.
\]
Translation invariance of both the interaction and $\rho$ makes
all $e_{y,L}$ equal, and thus $\mathcal D_{M,L}(\rho)=0$ exactly.
For a position-dependent interaction, translation invariance of
$\rho$ alone does not make these expectations equal. The repair
estimate controls $2R_0J\Delta$ but supplies no small upper bound
on $\mathcal D_{M,L}(\rho)$, so defining coordinate-dependent local
energies does not establish the required shorter-ring energy bound.
If the interaction matrices also fail to agree under $\beta_x$,
their differences produce an additional error.

\subsection{Proof of Theorem \ref{thm:energy}}
\label{subsec:proof-energy-defect-bound}

Based on the shorter-ring construction in Lemma~\ref{lem:shorter-ring-variational}, we now prove the energy-defect bound in Theorem \ref{thm:energy}.

\begin{proof}[Proof of Theorem \ref{thm:energy}]
Recall 
\[
 K_N=T_N(\widehat H_N-\widehat E_N\one)T_N^{\dagger},\qquad
 \mathsf{N}_N=\sum_{x\in\T_N}Q_x,\qquad
 \gamma=\gamma_0,
\]
where $Q_x=\one-\ket{e_x}\bra{e_x}$ and
$T_N\ket{\Psi_N}=\ket{\Omega_N}=\bigotimes_x \ket{e_x}$. For the case $R_0 =0$, $\widehat H_N$ consists of on-site terms, so we choose $T_N$ to be a product of on-site unitaries mapping the doubled onsite ground states to vacuum. In this case, it is obvious that
$K_N\ge\gamma\mathsf{N}_N$.

So in the following we focus on the case where $R_0\geq 1$. The proof is by contradiction. Suppose the claim in Theorem~\ref{thm:energy} fails. By Lemma~\ref{lem:failure-states-energy}, we get fixed constants $A_*<\infty$
and $\delta_*>0$, such that for every sufficiently large integer $n$, there are
$2n\le L_n\le3n$ and a translation-invariant physical-frame state $\rho_n$
on the enlarged $L_n$-ring satisfying
\begin{equation}\label{eq:Ln_energy_bound}
 \Tr[\rho_n(\widehat H_{L_n}-\widehat E_{L_n}\one)]\le A_*,\qquad
 \frac1{L_n}\Tr(T_{L_n}\rho_nT_{L_n}^{\dagger}\mathsf{N}_{L_n})\ge\delta_*.
\end{equation}
We will apply the shorter-ring construction to such a state and
obtain incompatible bounds on the output state's excitation energy.

\smallskip\noindent\emph{Step 1: fix the tolerances and the shorter-ring sizes.}
Let $c_f>0,C_b\ge0$ be the interval-defect constants from
Lemma~\ref{lem:translation-defects-energy}, and let $C_s\ge0$ be the
defect loss from Lemma~\ref{lem:retained-defects-energy}.
Set
\[
 d_*:=c_f\delta_*/6,\qquad g_*:=\gamma d_*,\qquad
 \varepsilon_{\rm cl}:=g_*/4.
\]
Here $d_*$ is the target lower bound on the defect density, $g_*$ is the corresponding
energy lower bound from the original gap, and $\varepsilon_{\rm cl}$ is the
allowed cost of closing the shorter ring.
Recall the constants $C_E,c_E>0$ in Eq.
\eqref{eq:energyconvergence}.  Fix an integer $L_{\rm valid}>L_0$
above which Eq.~\eqref{eq:energyconvergence}, Theorem~\ref{thm:prep-exact}, and
Lemmas~\ref{lem:translation-defects-energy} and
\ref{lem:retained-defects-energy} apply.
Here $L_0$ is the size threshold in the original gap assumption Eq.
\eqref{eq:assumption}.
Choose an integer $M_{\min}$ with
\[
 M_{\min}\ge\max\{L_{\rm valid},1/c_E\},\qquad
 C_b+C_s\le d_*M_{\min},\qquad
 \widehat C_E M_{\min}e^{-c_E M_{\min}}<g_*/4.
\]
with $\widehat C_E = 2C_{E}$ since we are discussing the enlarged system here.
For every $M\ge M_{\min}$, these conditions give
\begin{equation}
 C_b+C_s\le d_*M,\qquad \widehat C_E M e^{-c_E M}<g_*/4,
 \label{eq:variational-size-bounds}
\end{equation}
since {as $M$ increases,} $M e^{-c_E M}$ decreases for $M\ge1/c_E$.
Since $M_{\min}\ge L_{\rm valid}>L_0\ge2R_0$,
Lemma~\ref{lem:shorter-ring-variational} applies with these
$\varepsilon_{\rm cl},M_{\min}$ and fixes a finite $M_{\max}$ before we choose
the source state.

\smallskip\noindent\emph{Step 2: choose the source and construct the shorter-ring state.}
Select $(L,\rho)=(L_n,\rho_n)$ with
\[
 L>M_{\max},\qquad A_*M_{\max}/L<g_*/4.
\]
This is possible because the source circumferences are unbounded.
By Lemma~\ref{lem:shorter-ring-variational}, we get a constructed state which we now denote by $\zeta_M$. Thus
$M\in[M_{\min},M_{\max}]$ and $I=[a,M-a-1]$ with
$1\le a\le\lfloor M/3\rfloor$, and
\[
 (\zeta_M)_I=\rho_I,\qquad |I|\ge M/3,\qquad
 \Tr(\zeta_M\widehat H_M)
 \le\frac ML\Tr(\rho\widehat H_L)+\varepsilon_{\rm cl}.
\]
Note that $\zeta_M$ is not necessarily translation invariant.
{Now we have}
\begin{equation}
\begin{split}
 \Tr[\zeta_M(\widehat H_M-\widehat E_M\one)]
 &\le\frac ML\Tr[\rho(\widehat H_L-\widehat E_L\one)]
   +M\left|\frac{\widehat E_L}{L}-\frac{\widehat E_M}{M}\right|
   +\varepsilon_{\rm cl}\\
 &\le\frac{A_*M_{\max}}{L}+\widehat C_E M e^{-c_E M}+\varepsilon_{\rm cl}
 <\frac{3g_*}{4}.
\end{split}
\end{equation}
The second line uses $M<L$, the longer ring excitation energy density bound Eq.
\eqref{eq:Ln_energy_bound}, and the energy-density estimate in Proposition \ref{prop:circumference}.  The last inequality uses Eq.
\eqref{eq:variational-size-bounds} and $\varepsilon_{\rm cl}=g_*/4$.

\smallskip\noindent\emph{Step 3: the opposite bound.}
Define the source defect expectations
\[
 n_x=\Tr(\rho T_L^{\dagger}Q_xT_L),\qquad
 \bar n=L^{-1}\sum_x n_x\ge\delta_*.
\]
Lemma~\ref{lem:translation-defects-energy} gives
$\sum_{x\in I}n_x\ge c_f\delta_*|I|-C_b$.
Moreover, $L>M$ and $1\le a\le M/3<M/2$, so
$(\zeta_M)_I=\rho_I$ meets all the hypotheses of
Lemma~\ref{lem:retained-defects-energy}.
Consequently,
\begin{equation}
 \begin{aligned}
 \Tr(T_M\zeta_MT_M^{\dagger}\mathsf{N}_M)
 &\ge\sum_{x\in I}n_x-C_s\\
 &\ge c_f\delta_*|I|-C_b-C_s\\
 &\ge(c_f\delta_*/3)M-d_*M=d_*M.
 \end{aligned}
\end{equation}
Finally, $M\ge L_{\rm valid}>L_0$, so the original gap implies
\[
 K_M\ge\gamma(\one-P_{\T_M}^\Omega),\qquad
 \mathsf{N}_M\le M(\one-P_{\T_M}^\Omega),\qquad
 P_{\T_M}^\Omega=\ket{\Omega_M}\bra{\Omega_M}.
\]
Hence $K_M\ge(\gamma/M)\mathsf{N}_M$ and
\begin{equation}
 \Tr[\zeta_M(\widehat H_M-\widehat E_M\one)]
 =\Tr(T_M\zeta_MT_M^{\dagger}K_M)
 \ge\frac\gamma M\Tr(T_M\zeta_MT_M^{\dagger}\mathsf{N}_M)
 \ge\gamma d_*=g_*.
\end{equation}
This contradicts the upper bound $3g_*/4$ for the same state.
The assumed failure is therefore impossible, proving the existence of
uniform $\kappa>0$ and $L_1$ in Theorem~\ref{thm:energy}.

\end{proof}

\begin{remark}[Threshold $L_1$] We can show that the length threshold of the energy-defect inequality can be relaxed to $L_0$. For every $L>L_0$, the original gap assumption gives 
\[
\widehat H_L-\widehat E_L\mathbf1
\ge\gamma_0\bigl(\mathbf1-\ket{\Psi_L}\bra{\Psi_L}\bigr).
\]
So conjugating by $T_L$ we get $ K_L\geq \gamma_0(\one - P^\Omega_{\T_L})$. Denote the value of $ L_1$ given in the above proof as $L_{\rm large}$. For $L_0<L<L_{\rm large}$, we get $K_L\geq \frac{\gamma_0}{L}\mathsf N_L \geq \frac{\gamma_0}{L_{\rm large}}\mathsf N_L$. Hence if we define $\kappa' := \min\{\kappa, \frac{\gamma_0}{L_{\rm large}}\}>0 $, then the relation $K_L\geq \kappa' \mathsf N_L$ holds for all $L>L_0$.

\end{remark}

\section{Proof of Theorem \ref{thm:main} and Corollary \ref{corollary: continuity of gap}: The fermionic part}
\label{sec:fermions}

In this section we prove Theorem \ref{thm:main} and Corollary \ref{corollary: continuity of gap} for fermionic systems. This proof closely parallels the proof for the spin part in Sec. \ref{sec: proof of main theorem}, with modifications required by fermion parity and canonical anti-commutation relation (CAR). Table~\ref{tab:fermion-transfer} lists the steps of the proof in the fermionic case with their corresponding steps in the spin case.

\subsection{Fermionic setup and common tools}
\label{sec:fermion-algebra}

In the fermionic setup, most differences from its spin counterpart arise from the fermion parity requirement of states and operators. We first introduce the basic setup of fermion Hilbert spaces and operators, where fermion parity plays an important role in organizing them. Then we discuss the fermionic version of conditional expectation and channel that preserves parity covariantly.

\begin{table}[!ht]
\caption{Correspondence between results used in the spin and
fermionic proofs. A reference appearing in both columns denotes
a result stated for both settings. Bold entries require additional arguments
required in the fermionic case. }
\label{tab:fermion-transfer}
\label{tab:fermion-transfer-preparation-energy}
\centering\small
\renewcommand{\arraystretch}{1.12}
\begin{tabular}{@{}p{0.56\linewidth}ll@{}}
\toprule
Result & Spin & Fermion \\
\midrule
Filtering and transport locality
 & \ref{lem:appendix-locality}
 & {\bfseries\ref{lem:appendix-locality}} \\
Ground-state projection transport
 & \ref{lem:spectral-flow-transport}
 & \ref{lem:spectral-flow-transport} \\
Vacuum form bound
 & \ref{lem:vacuum}
 & \ref{lem:fermion-vacuum} \\
Normal-form shell representation
 & \ref{lem:normal-form-transport}
 & \ref{lem:fermion-normal-form} \\
Transported relative bound
 & \ref{lem:transported-gap}
 & \ref{lem:fermion-transported-gap} \\
Normal-form stability
 & \ref{thm:productstability}
 & \ref{thm:fermion-productstability} \\
Conjugated shell representation
 & \ref{prop:shelltransfer}
 & \ref{prop:fermion-shelltransfer} \\
Gap stability
 & \ref{thm:main}
 & \ref{thm:main} \\
Gap continuity
 & \ref{corollary: continuity of gap}
 & \ref{cor:fermion-gap-continuity} \\
\midrule
Agreement between rings
 & \ref{prop:circumference}
 & \ref{prop:fermion-circumference} \\
Decoupling
 & \ref{lem:decoupling}
 & {\bfseries\ref{lem:fermion-decoupling}} \\
Recovery after erasure
 & \ref{lem:erasure}
 & {\bfseries\ref{lem:fermion-erasure}} \\
Local comparison between rings
 & \ref{lem:local-ring-comparison}
 & \ref{lem:fermion-local-ring-comparison} \\
\midrule
Exact preparation
 & \ref{thm:prep-exact}
 & {\bfseries\ref{thm:fermion-prep-exact}} \\
Recovery unitary
 & \ref{lem:prep-insert}
 & {\ref{lem:fermion-prep-insert}} \\
Preparation at scale $r$
 & \ref{lem:prep-approx}
 & \ref{lem:fermion-prep-approx} \\
Circuit corrections
 & \ref{lem:prep-correct}
 & {\bfseries\ref{lem:fermion-prep-correct}} \\
Decoupling input for preparation
 & \ref{cor:prep-decoupling}
 & \ref{cor:fermion-prep-decoupling} \\
Nonuniform preparation
 & \ref{cor:prep-nonti}
 & \ref{cor:fermion-prep-nonti} \\
Local generation
 & \ref{thm:prep-lga}
 & \ref{thm:fermion-prep-lga} \\
\midrule
Energy-defect bound
 & \ref{thm:energy}
 & \ref{thm:fermion-energy} \\
Block compression
 & \ref{lem:block-compression-energy}
 & \ref{lem:fermion-block-compression} \\
Translation and defect estimates
 & \ref{lem:translation-defects-energy}
 & \ref{lem:fermion-translation-defects} \\
Comparison of compressed energies
 & \ref{lem:compressed_interaction_size_agreement}
 & \ref{lem:fermion-compressed-comparison} \\
States witnessing failure of the bound
 & \ref{lem:failure-states-energy}
 & \ref{lem:fermion-failure-states} \\
Exact marginal adjustment
 & \ref{lem:marginal-adjustment-energy}
 & \ref{lem:marginal-adjustment-energy} \\
One-seam repair
 & \ref{lem:seam-repair-energy}
 & \ref{lem:seam-repair-energy} \\
Entropy-band selection
 & \ref{lem:entropy-selection-energy}
 & \ref{lem:entropy-selection-energy} \\
Block exchange
 & \ref{lem:oriented-recurrence-energy}
 & {\bfseries\ref{lem:fermion-oriented-recurrence}} \\
Retained defects after seam repair
 & \ref{lem:retained-defects-energy}
 & \ref{lem:fermion-seam-defects} \\
Shorter-ring variational estimate
 & \ref{lem:shorter-ring-variational}
 & \ref{lem:fermion-shorter-ring} \\
\bottomrule
\end{tabular}
\end{table}

For a fermionic lattice system with $\nu\geqslant 1$ modes per site, let $\mathcal F_X$ be the Fock space of a region $X$, with an on-site Hilbert space dimension $D=2^\nu$. The local CAR algebra
$\A_X\cong\mathcal B(\mathcal F_X)$ is generated by the mode creation and annihilation operators $a_i^\dagger, a_i$,
which satisfy $\{a_i,a_j^{\dagger}\}=\delta_{ij}\one$ and $\{a_i,a_j\}=0$, where $i$ and $j$ label the modes in $X$ ($i\in X$ abbreviates all modes located at sites in $X$).
Define the fermion parity operator $\Gamma_X=(-1)^{\mathsf n_X}$ with $\mathsf n_X=\sum_{i\in X} a_i^\dagger a_i$. An operator $A\in\A_X$ is
homogeneous of $\mathbb Z_2$-grading degree $\deg(A)\in\{0,1\}$ when
$\Gamma_XA\Gamma_X=(-1)^{\deg(A)}A$. {Such an operator is called even if $\deg(A)=0$ and odd if $\deg(A)=1$.} For {such homogeneous operators $A$ and $B$ with} disjoint supports,
$AB=(-1)^{\deg(A)\deg(B)}BA$. All Hamiltonian and perturbation terms should be even, otherwise the Lieb-Robinson bound and thus locality will be violated. {We write the total fermion parity as $\Gamma:=\Gamma_{\T_L}:=(-1)^{\mathsf{n}}$, where $\mathsf{n}:=\sum_{i\in\T_L}a_i^\dag a_i$. A pure state is said to be even (odd) under the total fermion parity if it is an eigenstate of $\Gamma$ with eigenvalue 1 ($-1$).}
The unique ground state of our Hamiltonian, $\ket{\psi_L}$, can have either parity $p_L\in\{+1,-1\}$, {i.e., $\Gamma|\psi_L\rangle=p_L|\psi_L\rangle$}, while the density matrix $\rho_L=\ket{\psi_L}\bra{\psi_L}$ is always even for either
value of $p_L$.

We impose the fermion-parity superselection, i.e., all physical density matrices {$\rho$ (not only the unique ground state)} satisfy $[\rho,\Gamma]=0$. Thus $\rho$ has no coherence between the even and odd parity sectors, but may have components in both sectors\footnote{A density matrix cannot be odd, since $\Gamma\rho\Gamma=-\rho$ would imply
$\Tr\rho=-\Tr\rho$, contradicting with $\Tr\rho=1$.}. Our uniform gap condition with a unique ground state is still given by Eq. \eqref{eq:assumption}, which means that $|\psi_L\rangle$ has the lowest energy and is separated by an energy $\gamma_0$ from all orthogonal states, including states with either even or odd fermion parity.

Next, we consider joining a few Fock spaces. For two Fock spaces $\mathcal{F}_X$ and $\mathcal{F}_Y$ in regions $X$ and $Y$, respectively, we want a joint Fock space $\mathcal{F}_{XY}$ such that all operators preserve CAR. The Fock spaces are joined by ordinary tensor product, $\mathcal F_X \otimes \mathcal{F}_Y$, but the operators are obtained by the following graded tensor product.
We take a fixed ordering of modes from $X$ to $ Y$, and embed the operators $A\in \A_X$ as $\iota(A)=A\otimes \one_Y$, embed the operators $B\in \A_Y$ as $\iota(B)=\Gamma_X^{\deg(B)} \otimes B$. Then the embedded operators are still graded and preserving the CAR, namely, for all parity-definite operators $A,B$ the relation $\iota(A)\iota(B) = (-1)^{\deg(A)\deg(B)} \iota(B)\iota(A)$ holds. 
We use the symbol $\widehat\otimes$ to denote this graded tensor product,
\begin{equation}
    A\widehat \otimes B := A \Gamma_X^{\deg(B)} \otimes B, \quad A\in\A_X, B\in \A_Y.
\end{equation}
{In the above equation, $B$ is assumed to be a homogeneous operator with degree $\deg(B)$. For general operators in $\mathcal{A}_Y$, they can be written as a linear combination of homogeneous operators, and the above equation extends linearly.} For convenience, we will just use $A, B$ to denote $\iota(A),\iota(B)$ in the joint Hilbert space when there is no ambiguity. 

For embedding an operator in region $X\subset\T_L$ into the full system $\T_L$ with a fixed ordering in $\T_L$, we define an embedding map $\iota_X: \A_X\to \A_{\T_L}$ as follows. {First, we note that under the chosen fixed ordering in $\T_L$, $X$ may not be before or after $X^c$, which will be demonstrated by an example below.} Let $\mathcal U_X:\mathcal F_{\T_L}\to
\mathcal F_X\otimes\mathcal F_{X^c}$ be the unitary that regroups the modes
of $X$ before those of $X^c$, while preserving their order within each group.
On occupation-basis vectors, this regrouping includes the sign from exchanging occupied modes. The embedding map is concretely
\begin{equation}
 \iota_X(A)=\mathcal U_X^{\dagger}(A\otimes\one_{X^c})\mathcal U_X
\end{equation}

On the other hand, {starting from the entire system on $\T_L$ and restricting it to a subregion $X$,} the reduced density matrix is obtained by first regrouping the modes,
\begin{equation}\label{eq:partial_trace_formula}
 \rho_X=\Tr_{\mathcal F_{X^c}}(\mathcal U_X\rho\mathcal U_X^{\dagger})
\end{equation} 
If $X$ is an interval in the chosen mode ordering, its CAR reduced density matrix agrees with the ordinary partial trace over the complementary factors. But for a general disconnected region $X$, the fermionic regrouping signs must be included before taking the partial trace. {Under the above definitions, $\Tr( \iota_X(A)\rho)=\Tr_{\mathcal{F}_X}(A\rho_X)$ for any $A\in\mathcal{A}_X$.}

Let us demonstrate these concepts via a concrete example.
 
Consider a four-fermion system, with its modes ordered as 0,1,2,3.
 Suppose that each mode is located at a different site.
The creation operators are $a_i^\dagger,0\leq i\leq3$. The basis states of this Fock space are
\[
(a_0^{\dagger})^{n_0} (a_1^{\dagger})^{n_1} (a_2^{\dagger})^{n_2} (a_3^{\dagger})^{n_3}\ket{0},\quad n_i\in\{0,1\},
\]
where $|0\rangle$ is the vacuum state, i.e., the state without any fermion. Consider a simple two-particle state 
\[
\ket{\psi} = \frac{1}{\sqrt2} ( a_0^\dagger a_1^\dagger \ket{0} + a_1^\dagger a_2^\dagger \ket{0}),\quad \rho =\ket{\psi}\bra{\psi}.
\]
Let $X=\{0,2\}$ and $X^c=\{1, 3\}$, then $X$ is neither before nor after $X^c$ in the originally chosen ordering $0, 1, 2, 3$, and the regrouping unitary acts on the basis states as
\[
\mathcal U_X (a_0^{\dagger})^{n_0} (a_1^{\dagger})^{n_1} (a_2^{\dagger})^{n_2} (a_3^{\dagger})^{n_3}\ket{0} = (-1)^{n_1 n_2}(a_0^{\dagger})^{n_0} (a_2^{\dagger})^{n_2}\ket{0}_X \otimes (a_1^{\dagger})^{n_1} (a_3^{\dagger})^{n_3}\ket{0}_{X^c},
\]
where $|0\rangle_X$ and $|0\rangle_{X^c}$ are the vacuum states in $X$ and $X^c$, respectively. For $\ket{\psi}$, it gives 
\[
\mathcal U_X\ket{\psi} = \frac{1}{\sqrt2} \left(a_0^\dagger \ket{0}_X\otimes a_1^\dagger \ket{0}_{X^c} - a_2^\dagger\ket{0}_X \otimes a_1^\dagger\ket{0}_{X^c}\right) .
\]
Hence under the partial trace formula Eq. \eqref{eq:partial_trace_formula}, we get
\[
\rho_X = \frac{1}{2} (a_0^\dagger - a_2^\dagger)\ket{0}_X \bra{0}_X (a_0 - a_2).
\]
One can verify that for all operators $O$ in $\A_X$, it satisfies $\Tr[\rho_X O] = \Tr[\rho O]$. 

In the following subsection, we discuss the proof of fermionic counterparts of Theorem \ref{thm:main} and Corollary \ref{corollary: continuity of gap}. Most lemmas and theorems follow from the spin case, while a few need extra care about fermion parity. We present the conclusions in the fermionic case and stress the differences in their proofs if any.

\subsection{Normal-form stability: counterparts of Section~\ref{sec: proof of main theorem}}
\label{sec:fermion-normal-form}

In this subsection, we discuss the fermionic counterpart of Section \ref{sec: proof of main theorem}. The key difference is that we require that the Hamiltonians and shell interactions be even under fermion parity. 

\subsubsection{Locality and normal-form interactions} 

Following Section \ref{subsubsec:locality_and_filtering}, the definitions and properties of shell representations and filtering apply to fermions naturally without obstacles, where the vacuum state $|\Omega_L\rangle$ means the pure state where all fermionic modes are unoccupied in this context. Later we will also write $|\Omega_X\rangle$ for a region $X\subseteq\T_L$ to represent a state in $\mathcal{F}_{X}$, where all fermionic modes in $X$ are unoccupied. Under fermionic Hamiltonian dynamics, the locality of operators is preserved by the fermionic version of Lieb-Robinson bound. A different point is that we need to consider the locality of odd local operators, which we discuss in detail in the technical note (see \ref{lem:appendix-locality}). 

In the following, we discuss the counterparts of Lemma~\ref{lem:vacuum} to Theorem~\ref{thm:productstability}.

\begin{fermionlemma}[Product-vacuum form bound]
\label{lem:fermion-vacuum}
Let $\A_X$ be the CAR algebra and let $P_X^{\Omega} = \ket{\Omega_X}\bra{\Omega_X}$ be the vacuum projector in region $X$. Let $\mathsf N_X = \sum_{x\in X} (1-\ket{0_x}\bra{0_x})$. 
\begin{enumerate} 
\item If $a=a^\dagger \in \A_X$ and $a P_X^\Omega = 0$, then $-\norm{a}\mathsf N_X\leq a\leq \norm{a} \mathsf N_X$.
\item If $a=a^\dagger$ satisfies $a\ket{\Omega_L} = 0$, then $\E^\Omega_X(a) P^\Omega_X = 0$.
\end{enumerate}
\end{fermionlemma}
The proof of this lemma is exactly the same as Lemma~\ref{lem:vacuum}.

\begin{fermionlemma}\label{lem:fermion-normal-form}
Let $\{G_L\}$ be a family of fermionic Hamiltonians, and let $g_L(x,r)$ be shell interactions such that 
\begin{equation}
G_L=\sum_{x\in\T_L}\sum_{r=0}^{M_L}g_L(x,r),\qquad
g_L(x,r)=g_L(x,r)^\dagger\in\A_{B_L(x,r)},\qquad
 \|g_L(x,r)\|\le Je^{-ar^\alpha},\quad
 M_L=\lfloor L/2\rfloor,
\end{equation}
where $J,a>0,\,0<\alpha\leq 1$. 
Suppose $G_L\ket{\Omega_L} = 0$ and $G_L\geq \gamma(1-P_{\T_L}^{\Omega})$ for fixed $\gamma>0$. Then there are self-adjoint shell interactions $\theta_L(x,r)\in \A_{B_L(x,r)}$ and constants $J_\theta, a_\theta, C_0>0, 0<\alpha_\theta\leq 1$ such that
\begin{equation}
G_L=\sum_{x\in\T_L}\sum_{r=0}^{M_L}\theta_L(x,r),\qquad
 \norm{\theta_L(x,r)} \le J_\theta e^{-a_\theta r^{\alpha_\theta}},\qquad \theta_L(x,r)\ket{\Omega_{B_L(x,r)}}=0,
\end{equation}
and 
\begin{equation}
0\le G_L\le C_0\mathsf N_L,\qquad
 \mathsf N_L=\sum_{x\in\T_L}(\one-\ket{0_x}\bra{0_x}).
\end{equation}
\end{fermionlemma}

The proof is nearly the same as Lemma~\ref{lem:normal-form-transport}, as the even fermion parity of interactions like $G_L, F_{G_L,w_g}(g_{x,L})$ is preserved throughout the construction in the proof of Lemma~\ref{lem:normal-form-transport}. {In the proof of the current fermionic case, the maps $\E^{\Omega}_X$ and $\E^{\tr}_{X}$ in the spin case should be viewed as $\E^{\omega}_X$ in Eq. \eqref{eq:fermion-expectations}, with $\omega_{X^c}=|\Omega_{X^c}\rangle\langle\Omega_{X^c}|$ and $\omega=\one_{X^c}/\dim(\mathcal{F}_{X^c})$, respectively.}

\begin{fermionlemma}
\label{lem:fermion-transported-gap}
Let $\{G_L\}$ be a family of fermionic normal-form Hamiltonians satisfying
$G_L\ge\kappa\mathsf N_L$ for an $L$-independent $\kappa>0$. Suppose for an $L$-independent $0<\varepsilon\leq 1$,
\[
 V_L=\sum_{x,r}v_L(x,r),\qquad
 v_L(x,r)=v_L(x,r)^{\dagger}\in\A_{B_L(x,r)},\qquad
 \|v_L(x,r)\|\le\varepsilon f_0(r)
\]
where $f_0$ is an $L$-independent stretched-exponential function.
Fix $0<g<\kappa$, a filter function $w_g(t)$ and $0<s_*\le1$. Assume that
$G_L(s)=G_L+sV_L$ has a unique ground state and a gap greater than $g$
for $0\le s\le s_*$. Let $e_L(s)$ be the ground state energy of $G_L(s)$, and let
$U_L(s)$ be the spectral flow with this filter threshold, defined by
\[
 \dot U_L(s)=iU_L(s)\mathcal D_{G_L(s), w_g}(V_L),\qquad U_L(0)=\one,
\]
where $\mathcal{D}_{G_L(s), w_g}(V_L)$ is defined through Eq. \eqref{eq:filter}. Then there is a constant $C_*$, depending only on the uniform shell bounds (i.e., $J$, $a$ and $\alpha$ in Eq. \eqref{eq:shell}),
$f_0$, $g$ and the filter function $w_g$, such that
\begin{equation}\label{eq:transported-energy-defect-bound_fermion}
 U_L(s)(G_L(s)-e_L(s)\one)U_L(s)^{\dagger}
 \ge(\kappa-sC_*\varepsilon)\mathsf N_L,
 \qquad 0\le s\le s_*.
\end{equation}
Moreover, whenever $\kappa-sC_*\varepsilon>0$,
$\gap(G_L(s))\ge\kappa-sC_*\varepsilon$.
The constant $C_*$ is independent of $L$, $s_*$, and $\varepsilon$.
\end{fermionlemma}
The proof is the same as Lemma~\ref{lem:transported-gap}.

\begin{fermiontheorem}[Stability of even normal-form Hamiltonians]
\label{thm:fermion-productstability}
Let $\{G_L\}$ be a family of fermionic normal-form Hamiltonians, namely, each $G_L$ admits a shell representation as in Eq.~\eqref{eq:shell} and
$G_L\ket{\Omega_L}=0$. Suppose that
$G_L\ge\kappa \mathsf{N}_L$ for a fixed $\kappa>0$.  Let $\{V_L\}$ be a family of perturbations with a shell representation and satisfy
\begin{equation}
    V_L = \sum_{x,r} v_{L}(x,r),\qquad \norm{v_L(x,r)}\leq \varepsilon f_0(r)
\end{equation}
with $\varepsilon$ an $L$-independent constant and $f_0(r)$ an $L$-independent stretched-exponential function in $r$. 
Then there is an $L$-independent $\varepsilon_*>0$, such that
$\gap(G_L+V_L)\ge\kappa/2$ whenever $\varepsilon<\varepsilon_*$.  The ground state of $G_L+V_L$ remains unique.
\end{fermiontheorem}
This theorem is fermionic counterpart of Theorem~\ref{thm:productstability}. The proof is the same as the proof of Theorem~\ref{thm:productstability}.

\subsubsection{Conjugation and the main conclusions}

By analogy with the spin case, we need to enlarge the system to include its time reversal copy. 
For the physical ring and its conjugate copy, we define
\begin{equation}\label{eq:fermion-enlargement}
 H_L^c=\overline H_L,
 \qquad \widehat H_L=H_L\otimes\one+\one\otimes H_L^c.
\end{equation}
The ground state energy of $\widehat H_L$ is $2E_L$, and its energy gap is $\gap(\widehat H_L) \geq \gamma_0$.

Next we need to specify the ground state of $\widehat H_L$. In the occupation basis, we order all physical modes before
its conjugate-copy modes along the ring coordinates, schematically as $0, 1, \cdots, L-1,\bar 0,\bar 1,\cdots \overline{L-1}$. If the unique ground vector
$\ket{\psi_L}$ has parity $p_L\in\{+1,-1\}$, define
\[
 \ket{\psi_L^c}=\ket{\overline{\psi_L}},\qquad
 \ket{\Psi_L}=d_{p_L}\ket{\psi_L}\otimes\ket{\psi_L^c},
 \qquad d_+=1,\quad d_-=-i.
\]
Then $\ket{\Psi_L}$ is the normalized even ground state of
$\widehat H_L$, with ground energy $2E_L$.
The phase $d_{p_L}$ makes this definition {identical to} the
canonical purification {of $\ket{\psi_L}\bra{\psi_L}$} used in
Section~\ref{sec:fermion-preparation}.

\begin{fermionproposition}[Shell representation under even conjugation]
\label{prop:fermion-shelltransfer}
Suppose $T_L$ is a parity-even unitary such that $T_L\ket{\Psi_L}=\ket{\Omega_L}$, and $T_L$ is exponentially local
in the sense of Eq. \eqref{eq:uniform_locality}.
Then $K_L$ in Eq. \eqref{eq:mechanism} is a fermionic normal-form Hamiltonian and satisfies $0\le K_L\le C \mathsf{N}_L$.

If
$V_L=\sum_X\Xi_L(X)$ has range $R$, then its lift
$\widehat V_L:=V_L\otimes\one$ satisfies
$T_L\widehat V_LT_L^{\dagger}=\sum_{x,m}\widetilde v_L(x,m)$, where
\begin{equation}
    \tilde v_L(x, m)=\tilde v_L(x, m)^\dag\in\mathcal{A}_{B_L(x, m)},
    \quad
    \norm{\tilde v_L(x,m)} \leq C_R \norm{\Xi_L}_{\rm inc} e^{-bm},
\end{equation}
where $C_R>0$ and $b>0$ are $L$-independent constants.
\end{fermionproposition}
\begin{proof}
This proposition is fermionic counterpart of Proposition~\ref{prop:shelltransfer}.
Conjugation by $T_L$ and the normalized-trace CAR
expectations preserve evenness, so the differences
$\Delta_{x,m}$ in Proposition~\ref{prop:shelltransfer} are even.
The vacuum-annihilating shells are obtained separately from
Lemma~\ref{lem:fermion-normal-form}.
\end{proof}

\begin{fermionmain}
Let $H_L=\sum_{x\in\T_L}h_{x,L}$ be a family of fermionic
Hamiltonians with fixed onsite dimension $D=2^\nu$, where
$h_{x,L}$ is the cyclic CAR translate of a fixed even self-adjoint
term supported on $R_0+1$ consecutive sites. Suppose that for
every $L>L_0\ge2R_0$ there is a normalized unique ground vector
$\ket{\psi_L}$ with energy $E_L$ and
\[
 H_L-E_L\one\ge\gamma_0(\one-\ket{\psi_L}\bra{\psi_L}),
 \qquad \gamma_0>0.
\]
For every fixed perturbation range $R$, there exist
$L_*\ge L_0$ and $\varepsilon_R,\gamma_R>0$ such that the
following holds. If $L>L_*$ and
$V_L=\sum_X\Xi_L(X)$ has even self-adjoint terms, each supported
within a cyclic interval of at most $R+1$ sites, with
\[
 \|\Xi_L\|_{\inc}
 :=\sup_{y\in\T_L}\sum_{X\ni y}\|\Xi_L(X)\|
 <\varepsilon_R,
\]
then $H_L+V_L$ has a unique ground state and
\[
 \gap(H_L+V_L)\ge\gamma_R.
\]
The constants depend only on the
fixed unperturbed interaction, $D,L_0,\gamma_0$, and $R$, and not
on $L$ or the choice of perturbation.
\end{fermionmain}
\begin{proof}
The even preparation in Theorem~\ref{thm:fermion-prep-exact},
Proposition~\ref{prop:fermion-shelltransfer}, and
Theorem~\ref{thm:fermion-energy} give the four normal-form
hypotheses listed in Section~\ref{subsec:main_proof}. Set
$\varepsilon_R=\varepsilon_*/C_R$ and $\gamma_R=\kappa/2$.
Choose $L_*$ above the size thresholds in the preparation,
shell construction, and energy-defect bound.
Theorem~\ref{thm:fermion-productstability} gives
$\gap(\widehat H_L+V_L\otimes\one )\ge\kappa/2$.
Evenness makes the two-copy sum an ordinary tensor sum in the
grouped-copy representation, so
\[
 \gap(\widehat H_L+V_L\otimes\one )
 =\min\{\gap(H_L+V_L),\gap(H_L)\}.
\]
The gap bound and uniqueness follow as in the spin proof.
\end{proof}

\begin{fermioncorollary}[Continuity of gap of fermionic systems]
\label{cor:fermion-gap-continuity}
Under the assumptions of Theorem~\ref{thm:main}, fix a perturbation range
$R$. There are constants $L_*,\eta_R,c_R>0$ such that, for every $L>L_*$
and every interaction $\Xi_L$ with range at most $R$ and strength
$\|\Xi_L\|_{\inc}<\eta_R$,
\begin{equation}
 \bigl|\gap(H_L+V_L)-\gap(H_L)\bigr|
 \le c_R\|\Xi_L\|_{\inc},
 \label{eq:gap-continuity_fermion}
\end{equation}
where $V_L=\sum_X\Xi_L(X)$. The constants are independent of $L$ and whether the
perturbation is translation invariant. 
\end{fermioncorollary}
\begin{proof}
Perturb both copies by the even operator
$\widehat V_L=V_L\otimes\one+\one\otimes V_L^c$.
The two perturbed copies have identical spectra, so
$\gap(\widehat H_L+s\widehat V_L)=\gap(H_L+sV_L)$.
The conjugated perturbation has even shells bounded by
$2C_R\|\Xi_L\|_{\inc}e^{-br}$.
Lemma~\ref{lem:fermion-transported-gap} preserves evenness and gives
both sides of Eq.~\eqref{eq:transported-relative-bound}.
Lemma~\ref{lem:fermion-normal-form} supplies $K_L\le C_0\mathsf N_L$.
The min--max comparison in the proof of
Corollary~\ref{cor:gap-continuity_corollary} gives the asserted
bound with $c_R=2C_RC_*C_0/\kappa$ after decreasing $\eta_R$.
\end{proof}

\begin{remark}[Preservation of ground-state parity]
Within the perturbative regime of the fermionic stability theorem,
the ground state of $H_L+V_L$ has the same fermion parity as that
of $H_L$. That is, if
$
 \Gamma_{\T_L}\ket{\psi_L}=p_L\ket{\psi_L},
\,p_L\in\{+1,-1\},
$
then the normalized ground state $\ket{\psi_L^V}$ of $H_L+V_L$
satisfies
$
\Gamma_{\T_L}\ket{\psi_L^V}=p_L\ket{\psi_L^V}.
$
Indeed, the stability theorem applies to $H_L+sV_L$ for every
$s\in[0,1]$, since
$\|s\Xi_L\|_{\mathrm{inc}}\le\|\Xi_L\|_{\mathrm{inc}}$.
Its rank-one ground-state projection $P_L(s)$ is continuous in $s$.
Because $H_L+sV_L$ commutes with $\Gamma_{\T_L}$, the quantity
$\operatorname{Tr}[\Gamma_{\T_L}P_L(s)]$ belongs to $\{+1,-1\}$.
Continuity therefore makes it constant along the path.
\end{remark}

\subsection{Agreement across ring sizes: counterparts of Section~\ref{sec:across_sizes}}
\label{sec:fermion-information}

{Starting from this subsection until the end of this section, we generalize Sections \ref{sec:across_sizes}, \ref{sec:state_preparation} and \ref{sec:coercivity} to the fermionic case, which supply the necessary lemmas for the proof of the fermionic part of Theorem \ref{thm:main}.

We start with the generalization of Section \ref{sec:across_sizes}.}

In the following, the density matrix $\rho_L=\ket{\psi_L}\bra{\psi_L}$ is the unique
ground state density matrix of the even translation invariant Hamiltonian $H_L$. 
Denote $\langle O\rangle_\rho:=\Tr(\rho O)$. A tensor product
of even marginal density matrices is written in the displayed order of
regions, and changes of order need to use fermionic regrouping unitaries.

\begin{fermionproposition}[Agreement across ring sizes]
\label{prop:fermion-circumference}
Let $H_L$ be a local translation invariant fermionic Hamiltonian satisfying the uniform gap condition with a unique ground state on a ring of length $L$, as discussed around Eq. \eqref{eq:assumption}. For such a length-$L$ ring, denote by $E_L$ the ground state energy and by $\rho_L=|\psi_L\rangle\langle\psi_L|$ the density matrix of the ground state. There are constants $\mathcal E_\infty\in\mathbb R$, $C_E,c_E,C_{\rm marg}>0$, and $L_1$ such that
for $L,N\ge L_1$,
\begin{equation}\label{eq:energyconvergence_fermion}
 \left|\frac{E_L}{L}-\frac{E_N}{N}\right|\le C_E e^{-c_E \min(L,N)},
 \qquad |E_L-L\mathcal E_\infty|\le C_ELe^{-c_E L}.
\end{equation}
For $L\geq N\geq L_1$ and $1\le \ell\leq\lfloor N/12\rfloor$,
let $I=\{0,\ldots,\ell-1\}$ in both rings. Identifying the corresponding on-site Hilbert spaces, then the following inequality holds:
\begin{equation}\label{eq:marginal_agreement_fermion}
 \|\rho_{L,I}-\rho_{N,I}\|_1\le C_{\rm marg}e^{-c_E N/4},
\end{equation}
where $\rho_{L, I}$ and $\rho_{N, I}$ are the reduced density matrices of $\rho_L$ and $\rho_N$ on the region $I$.
\end{fermionproposition}
The proof strategy is similar to the Proposition~\ref{prop:circumference}, with only some difference in the following lemma.

\begin{fermionlemma}[Schmidt tails and trace-norm decoupling]
\label{lem:fermion-decoupling}
For the fermionic Hamiltonians $\{H_L\}$ satisfying the uniform gap condition with a unique ground state,
the Schmidt-tail and trace-norm estimates of ground state $\ket{\psi_L}$ as stated in
Lemma~\ref{lem:decoupling} hold: denote by $\rho_{L, X}$ the reduced density matrix of the ground state $\rho_L$ on a region $X\subseteq\mathbb{T}_L$, and by $\lambda_j(\rho_{L, X})$ the $j$-th eigenvalue of $\rho_{L, X}$, ordered as $\lambda_1\geqslant\lambda_2\geqslant\cdots$. There are $L$-independent constants $C,c,a>0$, such that for every interval $X$, the Schmidt tail satisfies \begin{equation}\label{eq:schmidttail_fermion}
\sum_{j>K}\lambda_j(\rho_{L,X})\le CK^{-a},\qquad \forall\ K\ge1.
\end{equation}
Moreover, for disjoint intervals $X$ and $Y$ with a distance at least $r$,
\begin{equation}\label{eq:decoupling_fermion}
 \|\rho_{L,XY}-\rho_{L,X}\otimes\rho_{L,Y}\|_1\le Ce^{-cr}.
\end{equation}

\end{fermionlemma}
\begin{proof}
The proof of the Schmidt tail in Eq. \eqref{eq:schmidttail_fermion} is the same as in the Lemma~\ref{lem:decoupling} by the folding trick. After the folding operation, the ring becomes an open chain, and a fermionic local Hamiltonian can be mapped to a local Hamiltonian of qubits under the Jordan-Wigner transformation. So Eq. \eqref{eq:schmidttail_fermion} can be proved by considering the transformed spin system of the original fermionic system. We now focus on the trace-norm decoupling property in Eq. \eqref{eq:decoupling_fermion}.

As done in Lemma~\ref{lem:decoupling}, we first bound the connected correlation functions $\left|\langle AB\rangle_{\rho_L}-\langle A\rangle_{\rho_L}\langle B\rangle_{\rho_L}\right|$ for $A\in\mathcal{A}_{X}$ and $B\in\mathcal{A}_{Y}$, and then we combine this bound and Eq. \eqref{eq:schmidttail_fermion} to prove Eq. \eqref{eq:decoupling_fermion}.

To do so, we decompose $A$ and $B$ into components with definite fermion parities, i.e., $A=A_++A_-$, $B=B_++B_-$, where $A_\pm=(A\pm\Gamma_XA\Gamma_X)/2$ and $B_\pm=(B\pm\Gamma_YB\Gamma_Y)/2$. Then
\begin{equation}
\begin{split}
    \left|\langle AB\rangle_{\rho_L}-\langle A\rangle_{\rho_L}\langle B\rangle_{\rho_L}\right|
    =&\left|\langle A_+B_+\rangle_{\rho_L}-\langle A_+\rangle_{\rho_L}\langle B_+\rangle_{\rho_L}+\langle A_-B_-\rangle_{\rho_L}\right|\\
    \leqslant&
    \left|\langle A_+B_+\rangle_{\rho_L}-\langle A_+\rangle_{\rho_L}\langle B_+\rangle_{\rho_L}\right|
    +
    \left|\langle A_-B_-\rangle_{\rho_L}\right|,
\end{split}
\end{equation}
where we have used the evenness of $\rho_L$ and the fact that the expectation value of any fermion-parity-odd operator vanishes in $\rho_L$.

To bound $\left|\langle AB\rangle_{\rho_L}-\langle A\rangle_{\rho_L}\langle B\rangle_{\rho_L}\right|$, we can separately bound the two terms above, i.e., $\left|\langle A_+B_+\rangle_{\rho_L}-\langle A_+\rangle_{\rho_L}\langle B_+\rangle_{\rho_L}\right|$ and $\left|\langle A_-B_-\rangle_{\rho_L}\right|$. To do so, we will invoke the fermionic version of Lieb-Robinson bound: $\|[\tau_t(A_\pm),B_\pm]_{\rm gr}\|\leqslant C\|A_\pm\|\|B_\pm\|(e^{v|t|}-1)e^{-\mu r}$ with constants $C, v, \mu>0$, where $[\tau_t(A_\pm),B_\pm]_{\rm gr}:=\tau_t(A_\pm) B_\pm-(-1)^{\deg(A_\pm)\deg(B_\pm)}B_\pm \tau_t(A_\pm)$. The constant $C$ and the $C$ in Eq. \eqref{eq:schmidttail_fermion} can be chosen to be the same. So we can bound $\left|\langle A_+B_+\rangle_{\rho_L}-\langle A_+\rangle_{\rho_L}\langle B_+\rangle_{\rho_L}\right|$ following exactly the same strategy as in Lemma~\ref{lem:decoupling}, as only fermion-parity-even operators are involved in every step of the derivation. So $\left|\langle A_+B_+\rangle_{\rho_L}-\langle A_+\rangle_{\rho_L}\langle B_+\rangle_{\rho_L}\right|$ satisfies an analog of Eq. \eqref{eq:covariance}.

The way to bound $\left|\langle A_-B_-\rangle_{\rho_L}\right|$ is similar: we first show that $\langle A_-B_-\rangle_{\rho_L}$ and $\langle \{A_\eta^-, B_-\}\rangle_{\rho_L}$ are close, and then use the Lieb-Robinson bound to bound $\langle \{A_\eta^-, B_-\}\rangle_{\rho_L}$, where
\[
 A^-_\eta=\tfrac12 A_-+\frac{i}{2\pi}\operatorname{pv}\!\int_{\mathbb R}
             e^{-\eta t^2}\frac{\tau_t(A_-)}{t}\,dt.
\]
To show that $\langle A_-B_-\rangle_{\rho_L}$ and $\langle \{A_\eta^-, B_-\}\rangle_{\rho_L}$ are close, we write the energy eigenbasis of $H_L-E_L\one$ as $\{|j\rangle\}$ with the ground state labeled as $|0\rangle$. Then
\[
 \langle A^-_\eta B_- \rangle_{\rho_L}
 =\sum_{j>0}f_\eta(-E_j)(A_-)_{0j}(B_-)_{j0},
 \qquad
 \langle B_-A^-_\eta\rangle_{\rho_L}
 =\sum_{j>0}f_\eta(E_j)(B_-)_{0j}(A_-)_{j0}
\]
with
\[
 f_\eta(E)=\frac{1}{2}{\rm erfc}\left(\frac{E}{2\sqrt{\eta}}\right),\quad {\rm erfc}(z) = \frac{2}{\sqrt{\pi}}\int_z^\infty e^{-u^2} du.
\]
Since $f_\eta(-E)=1-f_\eta(E)$ and
$f_\eta(E)\le\tfrac12e^{-E^2/(4\eta)}$ for $E\ge0$,
\begin{equation}
 |\langle A_-B_-\rangle_{\rho_L}
 -\langle \{A_\eta^-, B_-\}\rangle_{\rho_L}|
 =\left|\sum_{j>0}f_\eta(E_j)\big((A_-)_{0j}(B_-)_{j0}-(B_-)_{0j}(A_-)_{j0}\big)\right|
 \le C\|A_-\|\|B_-\|e^{-\gamma_0^2/(4\eta)}.
\end{equation}
In the last step we have used Cauchy-Schwarz inequality. Here the constant $C$ and the constant $C$ in the Lieb-Robinson bound can be chosen to be the same, and we have used the facts that $E_j\geqslant\gamma_0$ and that $f_\eta(E)$ decreases monotonically. Then $\langle \{A_\eta^-, B_-\}\rangle_{\rho_L}$ can be bounded using the fermionic version of Lieb-Robinson bound in the same way as its spin counterpart in Lemma \ref{lem:decoupling}.

So in the end we get the fermionic analog of Eq. \eqref{eq:covariance}:
\begin{equation}\label{eq:fermion-correlations}
 |\langle AB\rangle_{\rho_L}
 -\langle A\rangle_{\rho_L}\langle B\rangle_{\rho_L}|
 \le Ce^{-c_0r}\|A\|\|B\|.
\end{equation}

{Next, we combine Eqs. \eqref{eq:schmidttail_fermion} and \eqref{eq:fermion-correlations} to prove Eq. \eqref{eq:decoupling_fermion}, using a similar strategy as in the proof of Lemma \ref{lem:decoupling}.}

Since $\rho_{L,X}$ is even, its
spectral cutoff $P$ can be chosen to be even, including the case when the
cutoff eigenvalue is degenerate. Write
$q=\Tr(\rho_{L,X}P)$ and $\sigma=q^{-1}P\rho_{L,XY}P$. {Then $\|\sigma-\rho_{L, XY}\|_1\leqslant 2\sqrt{C}K^{-a/2}$.} The remainder of the trace-norm estimate follows from Step~3 \footnote{Choose the orthonormal operator basis to have definite parity. For odd \(A,B\), use $A\otimes B = (A\Gamma_X)B$ to turn it into fermion product for the use of Eq.~\eqref{eq:fermion-correlations}, so the same correlation bound applies. Even–even terms follow the spin argument, and mixed-parity terms vanish.} of
Lemma~\ref{lem:decoupling}, using
Eq.~\eqref{eq:fermion-correlations}.

\end{proof}

\begin{fermionlemma}[Local comparison between rings]
\label{lem:fermion-local-ring-comparison}
Let $r\geq 1$ be an integer, and let $\eta$ be a pure translation invariant fermionic state on a ring of length $L\geq 12r$. Suppose 
\[
\norm{\eta_{XY} - \eta_X\otimes \eta_Y }_1\leq \delta
\]
for all disjoint intervals $X,Y$ with
$d_L(X,Y)\ge r$. Then, for every $N\ge8r$, there is a translation invariant state
$\sigma_N$ on the ring of length $N$ such that
\begin{equation}\label{eq:loca_ring_construction_fermion}
 \|(\sigma_N)_I-\eta_I\|_1
 \le\delta+4\sqrt{\delta}
\end{equation}
for every interval $I$ with $|I|\le r$.
Here $\eta_I$ denotes the marginal on a matching interval
of the seed ring with length $L$, with sites identified in the same
forward order.
\end{fermionlemma}

To prove this lemma, the proof of Lemma~\ref{lem:local-ring-comparison} applies if we
use the parity-covariant recovery maps from
Lemma~\ref{lem:fermion-erasure}.
By Lemma~\ref{lem:fermion-channel}, their extensions preserve the evenness and exterior marginals, and commute on disjoint recovery regions. Thus the construction and error estimates are unchanged.

\begin{proof}[Proof of Proposition~\ref{prop:fermion-circumference}]
The proof of Proposition~\ref{prop:circumference} applies
using Lemma~\ref{lem:fermion-local-ring-comparison} with
the decoupling estimate from Lemma~\ref{lem:fermion-decoupling}.
\end{proof}

\subsection{Even state preparation: counterparts of Section~\ref{sec:state_preparation}}
\label{sec:fermion-preparation}

Now we move to prove the fermionic counterpart of Section~\ref{sec:state_preparation}.

\subsubsection{Canonical purification and exact preparation}

To discuss the preparation of $\ket{\psi_L}\otimes\ket{\overline\psi_L}$, we first discuss the canonical purification for fermions. The difference from the spin case is that it requires a careful treatment of fermion anticommutation relations.

Denote a fermionic canonical purification of a density matrix $\sigma$ by $\ket{\can_f(\sigma)}$, which is in a doubled Fock space. For a region $X$, we use $\overline X$ to denote its auxiliary copy in the doubled space. We want this purification to satisfy the following two properties for convenience:
\begin{enumerate}
    \item $\ket{\can_f(\ket{\phi}\bra{\phi})} = e^{i\alpha}\ket{\phi}\otimes\ket{\overline\phi}$ for pure density matrix $\ket{\phi}\bra{\phi}$, up to a global phase $e^{i\alpha}$.
    \item For a tensor product density matrix $\sigma_X\otimes\sigma_Y$, its canonical purification can be factorized,
    \begin{equation}\label{eq:fermion-canonical-product}
    \mathcal R_{X,Y}\ket{\operatorname{can}_f(\sigma_X\otimes\sigma_Y)}=\ket{\operatorname{can}_f(\sigma_X)}\otimes\ket{\operatorname{can}_f(\sigma_Y)},
    \end{equation}
    where $\mathcal{R}_{X,Y}: \mathcal{F}_{XY\overline{X}\overline{Y} }\to \mathcal F_{X\overline X}\otimes \mathcal F_{Y\overline Y}$ is fermion regrouping unitary operator.
\end{enumerate}

There are different ways of defining the canonical purifications, and we take the following convention. Consider a fermionic system with $m$ modes, and denote their creation operators by $a_j^\dagger, 1\leq j\leq m$, and their auxiliary counterpart by $b_j^\dagger$. Then define the canonical purification of density matrix $\sigma$ as
\begin{equation}
    \ket{\can_f(\sigma)} = (\sqrt{\sigma}\otimes \one)\ket{\mathcal I_{m}},\qquad \ket{\mathcal{I}_m} = \prod_{j=1}^m (1-i a_j^\dagger b_j^\dagger) \ket{0}.
\end{equation}
The above definition is fully specified by the operator algebra, independent of the fermion ordering.

Now we take an ordering for the computation. Order the $2m$ modes in $\ket{\mathcal{I}_m}$ as $1,2,\cdots m, \bar1,\bar2,\cdots \bar m$. Then it can be verified that $\ket{\mathcal I_m} = \sum_{S} f_{|S|} \ket{S}\otimes\ket{S}$ where $S$ is a bit string $n_1n_2\cdots n_m$ associated with the creation operators $(a_1^\dagger)^{n_1} (a_2^\dagger)^{n_2}\cdot (a_m^\dagger)^{n_m}$. Here $|S|$ denotes the sum $\sum_i n_i$, and $f_{|S|} = (-i)^{|S|}(-1)^{|S|(|S|-1)/2}$. Indeed, 
$f_{|S|} = 1$ for even $|S|$ and $f_{|S|}=-i$ for odd $|S|$. 

We now verify the two properties for this definition. For a pure state $\ket{\phi}\bra{\phi}$ of parity $p\in\{+,-\}$, it only contains components with $(-1)^{|S|}=p$, so we get $\ket{\can_f(\ket{\phi}\bra{\phi})} = \ket{\phi}\otimes\ket{\overline\phi}$ for even state $\ket\phi$, and $\ket{\can_f(\ket{\phi}\bra{\phi})} = -i\ket{\phi}\otimes\ket{\overline\phi}$ for odd $\ket{\phi}$. This verifies the first property. For the second property, we only need to prove $\mathcal R_{X,Y}\ket{\mathcal I_{XY}} = \ket{\mathcal{I}_{X}}\otimes \ket{\mathcal I_Y}$. This can be seen by considering any component $f_{|S|}\ket{S}_{XY}\otimes \ket{S}_{\overline X\overline Y}$. Suppose this component contains $k$ occupations in $\overline X$ and $l$ occupations in $Y$ (notice that $X$ and $\overline X$ have the same occupation in $\ket{S}\otimes\ket{S}$; so do $Y$ and $\overline Y$), then regrouping to $X\overline X Y\overline Y$ changes {this component into $(-1)^{kl}f_{|S|}(|S_1\rangle_{X}\otimes |S_1\rangle_{\overline X})\otimes(|S_2\rangle_Y\otimes|S_2\rangle_{\overline{Y}})=[f_{|S_1|}(|S_1\rangle_{X}\otimes |S_1\rangle_{\overline X})]\otimes [f_{|S_2|}|S_2\rangle_Y\otimes|S_2\rangle_{\overline{Y}}]$, where $S_1$ and $S_2$ stand for the bit strings in the regions $X$ and $Y$, respectively.} So we proved $\mathcal R_{X,Y}\ket{\mathcal I_{XY}} = \ket{\mathcal{I}_{X}}\otimes \ket{\mathcal I_Y}$ and thus the second property.

The canonical purification has two useful properties. First, if $V$ is an even isometry from a Fock space with $m$ modes to a Fock space with $n$ modes, by expanding in the basis $\ket{\mathcal I} = \sum_{S}f_{|S|}\ket{S}\otimes \ket{S}$, we get
\begin{equation}\label{eq:fermion-isometry}
 \ket{\operatorname{can}_f(V\sigma V^\dagger)}
 =(V\otimes \overline V)\ket{\operatorname{can}_f(\sigma)}.
\end{equation}
The other property is 
\begin{equation}\label{eq:fermion-canonical-continuity}
\norm{\ket{\operatorname{can}_f(\sigma)}
-\ket{\operatorname{can}_f(\tau)}}
\le\sqrt{\|\sigma-\tau\|_1}.
\end{equation}

Based on definitions above, we present the fermion counterpart of Theorem~\ref{thm:prep-exact}. 

\begin{fermiontheorem}[Exact preparation by even unitaries]
\label{thm:fermion-prep-exact}
Let $\{\varrho_L\}$ be pure fermionic density matrices satisfying the
translation-invariance, trace-norm decoupling, and inter-ring comparison
hypotheses of Theorem~\ref{thm:prep-exact}.
Replace the doubled state by
$\ket{\Phi_L}=\ket{\operatorname{can}_f(\varrho_L)}$.
Then the conclusions of Theorem~\ref{thm:prep-exact} hold with
$[T_L,\widehat\Gamma_{\T_L}]=0$ with $\widehat \Gamma_{T_L}$ the parity operator of doubled system, and every preparation and correction gate even.
The estimates Eqs. \eqref{eq:prep-locality}, \eqref{eq:prep-compatible},
and \eqref{eq:prep-truncation} apply to any local operators in $\mathcal{A}_{\T_L}$. The truncated
circuits have the bulk period $448s$ for the gates away from the seam region.
\end{fermiontheorem}
{The proof of this theorem will be given at the end of the next subsubsection, after proving a few lemmas there.}

\subsubsection{Recovery and correction gates}

\begin{fermionlemma}[Even local recovery unitary]
\label{lem:fermion-prep-insert}
Under the geometry of Lemma~\ref{lem:prep-insert}: $B$ an interval and $C=C_-\cup C_+$ two collars of $B$, and $F$ the remaining sites, $G$ an exterior region to $BC$.
Assume that the pure reference state $\sigma_{BCF}=\ket{\psi_\sigma}\bra{\psi_\sigma}$ satisfies $\norm{\sigma_{BF} - \sigma_B\otimes \sigma_F}_1\leq\delta$. Then
there is an even unitary $U_{\rm loc}\in\A_{BC\bar B\bar C}$, depending
only on $\sigma_{BCF}$, such that
\begin{equation}\label{eq:fermion-robust-insertion}
 \left\|U_{\rm loc}\left(\ket{0}_{B\bar B}\otimes
       \ket{\operatorname{can}_f(\tau_{CG})}\right)
       -\ket{\operatorname{can}_f(\tau_{BCG})}\right\|
 \le4(\delta+\epsilon/2)^{1/4}
\end{equation}
{for any state $\tau_{BCG}$ that satisfies 
\begin{equation}
\|\sigma_{BC}-\tau_{BC}\|_1\leqslant\epsilon.
\end{equation}
}
Here $U_{\rm loc}$ acts trivially on $G\bar G$.
The unitary $U_{\rm loc}$ depends only on $\sigma_{BCF}$,
and is independent of the finite exterior system $G$
and target state $\tau_{BCG}$.
\end{fermionlemma}

\begin{proof}
The proof follows that of Lemma~\ref{lem:prep-insert}.
By Corollary~\ref{cor:fermion-decoupled-purification}
we choose the comparison isometry $W$ to be even.
The canonical-purification identities Eqs.
\eqref{eq:fermion-isometry},
\eqref{eq:fermion-canonical-product}, and
\eqref{eq:fermion-canonical-continuity}
give the same purification estimates as in Lemma~\ref{lem:prep-insert}.
The two states connected by $G_0$ are even,
so $G_0$ can also be chosen to be even.
Consequently, the compressed operator $K$ is even, and
taking its polar decomposition separately in the two
local parity sectors gives an even unitary $U_{\rm loc}$.
The error bound and the dependence only on the reference
state follow exactly as in Lemma~\ref{lem:prep-insert}.

\end{proof}

\begin{fermionlemma}[Preparation at scale $r$]
\label{lem:fermion-prep-approx}
Under the hypotheses of Theorem~\ref{thm:fermion-prep-exact},
the conclusions of Lemma~\ref{lem:prep-approx} hold with
$\ket{\Phi_L}=\ket{\operatorname{can}_f(\varrho_L)}$ and every unitary gate even. More concretely,
let $\{\ket{\Phi_L}\}$ and $\{\varrho_L\}$ be the sequences of fermionic states satisfying the assumptions of Theorem \ref{thm:fermion-prep-exact}. For any integers $r\geq 1$ and $L\ge96r$, there are
depth-two unitary circuits $U_{L,r}$, with gate intervals of length at most $7r$, and constants $C,c>0$ such that
\begin{align}
 \|U_{L,r}\ket{\Omega_L}-\ket{\Phi_L}\|&\le C(L/r)e^{-cr},
 \label{eq:prep-global_fermion}\\
 \|(U_{L,r}\ket{\Omega_L}\bra{\Omega_L}U_{L,r}^{\dagger})_X
-( \ket{\Phi_L}\bra{\Phi_L})_X\|_1
 &\le C(1+|X|/r)e^{-cr}
 \label{eq:prep-local_fermion}
\end{align}
for every interval $X$, including the whole ring.
The bulk gates and their locations
are independent of $L$, and only a single gate near the closing seam
has size dependent on $(L\mod 4r)$.

\end{fermionlemma}
\begin{proof}
The only difference from the proof of Lemma~\ref{lem:prep-approx} is inspecting the fermion parity of unitary gates.
An even unitary prepares each even block vector
$\ket{\operatorname{can}_f(\nu_{|A_i|})}$ from the block vacuum.
Then, Lemma~\ref{lem:fermion-prep-insert} supplies even recovery gates.
The estimations in the proof hold with no changes.
The reference gates therefore have the same
support lengths.
\end{proof}

\begin{fermionlemma}[Even corrections of a nearly vacuum circuit]
\label{lem:fermion-prep-correct}
Let $\ket{\Omega_L}$ be the product-vacuum state on a ring with length $L$, and let $Q$ be a depth-$d$ even unitary circuit whose gates have length at most $a$, on a ring of circumference $L\geq 24da$. Consider the partition of ring as in Eqs. \eqref{eq:da_partiaion_Y} and \eqref{eq:da_partion_Z}. Suppose that, for every collar or core denoted as $I$,
\begin{equation}
\bra{\Omega_I}
\bigl(Q\ket{\Omega_L}\bra{\Omega_L}Q^{\dagger}\bigr)_I
\ket{\Omega_I}
\ge 1-\epsilon,
\qquad 0\le\epsilon<\frac16.
\end{equation}
Then there is a unitary circuit \(R\) such that
\begin{equation}
RQ\ket{\Omega_L}=e^{i\alpha}\ket{\Omega_L},\quad \alpha\in[0,2\pi).
\end{equation}
Its depth is at most four, and each gate \(G\) is supported on an interval of length at most $18da$ and satisfies
\begin{equation}
\norm{G-\mathbf1}\le \sqrt{6\epsilon}.
\end{equation}
The constants are independent of the onsite dimension and circumference. The gates in $R$ can be chosen using only the gates of \(Q\) in neighborhoods of size \(O(da)\). All gates in $R$ are even.
\end{fermionlemma}
\begin{proof}
The construction of $R$ is the same as Lemma~\ref{lem:prep-correct}, and we keep track of the parity of gates and states. The factorization $Q=VW$ in Eq.~\eqref{eq:QW_boundary} uses
products and conjugations of even gates, so every factor
$W_{X_j}$ and $V_{Y_j}$ is even. The block vectors
$\ket{\phi_j}=W_{X_j}\ket{\Omega_{X_j}}$ are hence even.
The leading eigenvalue of $\sigma_E$ satisfies
$\lambda_{\max}(\sigma_E)\ge1-\epsilon>1/2$, so the leading eigenvector $\ket{\zeta_E}$ is non-degenerate with definite parity.
The estimate in the spin proof gives
\[
 |\langle\phi_j|\zeta_{X_j}\rangle|^2\ge1-3\epsilon>0,
 \qquad \ket{\zeta_{X_j}}=\bigotimes_{E\subset X_j}\ket{\zeta_E}.
\]
The product vector $\ket{\zeta_{X_j}}$ has definite parity.
The nonzero overlap $\langle\phi_j|\zeta_{X_j}\rangle$ forces the
parity to be even. The first rank-two correction (rotating $\ket{\phi_j}$ to $\ket{\zeta_j}$) can therefore be
chosen inside the even sector.

For the second correction on a collar or core $I$, the spin proof
gives, with $p_I=\lambda_{\max}(\sigma_I)$,
\[
 p_I\ge1-\epsilon,\qquad
 p_I\bigl(1-|\langle\Omega_I|V_I|\zeta_I\rangle|^2\bigr)
 \le\epsilon.
\]
Because $\epsilon<1/6$, the overlap $\ev{\Omega_I|V_I|\zeta_I}$ is nonzero. The vector
$V_I^\dagger\ket{\Omega_I}$ is even, so $\ket{\zeta_I}$ is even.
The second rank-two correction also acts within the even sector.
Both corrections extend by the identity on the odd sector; their
norm estimates and region supports are unchanged.
\end{proof}

Now we can prove Theorem~\ref{thm:fermion-prep-exact}.

\begin{proof}[Proof of Theorem~\ref{thm:fermion-prep-exact}]
Apply Lemmas~\ref{lem:fermion-prep-approx} and
\ref{lem:fermion-prep-correct} at the dyadic scales of
Theorem~\ref{thm:prep-exact}. The initial preparation gates,
correction gates, and final rotation map even vectors to even
vectors and can all be chosen even. Fix occupation bases, reference
gates, and polar extensions independently of the target ring.
Lemma~\ref{lem:fermion-prep-insert} then transfers the reference
gates to all target marginals with the original error bounds.
Even gates commute with arbitrary disjoint CAR observables;
therefore the circuit support estimates prove
Eqs.~\eqref{eq:prep-locality}--\eqref{eq:prep-truncation} for both
parities of $A$. The common-reference construction retains the
bulk period and inter-ring compatibility.
\end{proof}

\subsubsection{Further properties of the preparation}

Similar to spin case, the following properties hold for fermions: (1) existence of preparation unitary $T_L$ independent of translation symmetry of Hamiltonian and (2) $T_L$ is the unitary dynamics generated by a quasi-local Hamiltonian.

\begin{fermioncorollary}[Preparation from decoupling]
\label{cor:fermion-prep-decoupling}
Let $\{\varrho_L=\ket{\phi_L}\bra{\phi_L}\}$ be pure fermion states satisfying the trace-norm decoupling inequality \eqref{eq:prep-input-decouple} for disjoint intervals, with constants independent of $L$. Translation invariance and Eq.
\eqref{eq:prep-input-compare} are not required. Then
there are even unitaries $\{T_L\}$ such that
\[
 T_L\ket{\Phi_L}=\ket{\Omega_L},\qquad
 \ket{\Phi_L}=\ket{\can_f(\ket{\phi_L}\bra{\phi_L})},
\]
satisfying Eq. \eqref{eq:prep-locality} and admitting
truncated circuits $\mathcal U_{L,<s}$ with the support bound and
error estimate Eq. \eqref{eq:prep-truncation} as in
Theorem~\ref{thm:prep-exact}.  All constants and the initial scale $r_0$
can be chosen independently of $L$, although the gates may depend on
position and $L$.
\end{fermioncorollary}
The proof is the same as Corollary~\ref{cor:prep-decoupling}.

\begin{fermioncorollary}[Preparation without translation invariance]
\label{cor:fermion-prep-nonti}
Let $H_L=\sum_{x\in\T_L}h_{x,L}$ be fermion Hamiltonians, where $h_{x,L}$ is supported on the cyclic interval $\{x,\ldots,x+R_0\}$ and $\|h_{x,L}\|\le J$, for fixed
$R_0,J$.  Suppose each $H_L$ has a unique ground state
$\ket{\phi_L}$ and a spectral gap at least $\gamma_0>0$, uniformly in
$L$. No translation invariance is assumed.
Then the doubled ground states $\ket{\phi_L}\otimes\ket{\overline{\phi}_L}$ satisfy the conclusions of
Corollary~\ref{cor:fermion-prep-decoupling}, with constants depending only on
$D,R_0,J,\gamma_0$.
\end{fermioncorollary}
The proof is the same as Corollary \ref{cor:prep-nonti}.

\begin{fermiontheorem}[Local generation by even interactions]
\label{thm:fermion-prep-lga}
The conclusion of Theorem~\ref{thm:prep-lga} holds for the preparations
of Theorem~\ref{thm:fermion-prep-exact} and
Corollaries~\ref{cor:fermion-prep-decoupling}--\ref{cor:fermion-prep-nonti},
with the generating interactions
$\Phi_{L,t}(Z)$ self-adjoint and even. The interaction and
time-derivative bounds are unchanged.
\end{fermiontheorem}
The proof is the same as spin case (Theorem~\ref{thm:prep-lga}).

\subsection{The energy-defect bound: counterparts of Section~\ref{sec:coercivity}}
\label{sec:fermion-energy}

Next, we turn to the fermionic counterpart of Section~\ref{sec:coercivity}.

Use Eq.~\eqref{eq:fermion-enlargement} and the compatible even
$T_L$ from Theorem~\ref{thm:fermion-prep-exact}. Write
$K_L=T_L(\widehat H_L-2E_L\one)T_L^\dagger$ and
$\mathsf N_L=\sum_xQ_x$. For every block compression and every
shell estimate below, use the even vacuum-annihilating shells
$\theta_L(x,r)$ from Lemma~\ref{lem:fermion-normal-form}.

\begin{fermiontheorem}[Energy-defect inequality]
\label{thm:fermion-energy}
Under the even fermionic hypotheses of Eq.~\eqref{eq:assumption},
the conclusion of Theorem~\ref{thm:energy} holds for the doubled
Hamiltonian \eqref{eq:fermion-enlargement} and the compatible even
preparation of Theorem~\ref{thm:fermion-prep-exact}:
\[
 K_L\ge\kappa\mathsf N_L\qquad(L\ge L_1),
\]
where $\kappa>0$ and $L_1$ depend only on the fixed family and
preparation construction.
\end{fermiontheorem}

The proof of this theorem is given at the end of this section, after proving a few lemmas.

\subsubsection{Compression, translations, and failure states}

\begin{fermionlemma}[Block compression]
\label{lem:fermion-block-compression}
The statement is the same as
Lemma~\ref{lem:block-compression-energy}. The isometry is
$J_I\ket{u}=\mathcal U_I^\dagger
(\ket{u}\otimes\ket{\Omega_{I^c}})$, and the extension of
$G_{L,I}=J_I^\dagger K_LJ_I$ is $\iota_I(G_{L,I})$.
The shell constant $C_\partial$ is computed from the
vacuum-annihilating shells of Lemma~\ref{lem:fermion-normal-form}.
\end{fermionlemma}

\begin{fermionlemma}[Defect non-concentration]
\label{lem:fermion-translation-defects}
Under the hypotheses of Lemma~\ref{lem:translation-defects-energy},
assume that $W$ and the preparation $T_L$ are even. 
Then Eq.\eqref{eq:vacuum-unitary-comparison}--\eqref{eq:defect-density-energy} hold for fermions.
\end{fermionlemma}

\begin{fermionlemma}[Comparison of compressed energies]
\label{lem:fermion-compressed-comparison}
The Lemma~\ref{lem:compressed_interaction_size_agreement} holds for fermionic compressed Hamiltonians $G_{L,I}$.
\end{fermionlemma}

The proof of the above three lemmas is the same as the proof of Lemma~\ref{lem:block-compression-energy}--\ref{lem:compressed_interaction_size_agreement}.

\begin{fermionlemma}[Bounded energy and positive defect density]
\label{lem:fermion-failure-states}
Let $\{T_L\}$ be the family of unitaries constructed in
Theorem~\ref{thm:fermion-prep-exact} for different $L$, with $T_L\ket{\Psi_L}=\ket{\Omega_L}$.
If Theorem~\ref{thm:fermion-energy} fails, then there are $A_*<\infty$ and $\delta_*>0$
such that, for every sufficiently large $n$, there is a circumference
$2n\le L_n\le3n$ and a translation-invariant physical-frame state $\rho_n$ on the whole $L_n$-ring
satisfying
\begin{equation}
 \operatorname{Tr}[\rho_n(\widehat H_{L_n}-\widehat E_{L_n})]\le A_*,
 \qquad
 \operatorname{Tr}(T_{L_n}\rho_nT_{L_n}^{\dagger}\mathsf{N}_{L_n})\ge\delta_*L_n.
 \label{eq:failure-states-energy_fermion}
\end{equation}
\end{fermionlemma}
\begin{proof}
For a density $\sigma$ violating a candidate bound, replace $\sigma$
by $\sigma^{\rm ev}=\tfrac12(\sigma+\Gamma\sigma\Gamma)$.
Evenness of $K_L$ and $\mathsf N_L$ gives
\[
 \Tr(\sigma^{\rm ev}K_L)=\Tr(\sigma K_L),\qquad
 \Tr(\sigma^{\rm ev}\mathsf N_L)=\Tr(\sigma\mathsf N_L).
\]
Compression by $\one-P_I^\Omega$, insertion of the even exterior
vacuum, conjugation by $T_L$, and CAR translation averaging all
preserve evenness. Use
Lemmas~\ref{lem:fermion-block-compression},
\ref{lem:fermion-translation-defects}, and
\ref{lem:fermion-compressed-comparison} in the block selection
of Lemma~\ref{lem:failure-states-energy}.
\end{proof}

\subsubsection{Closing the shorter ring and the energy contradiction}

\begin{fermionlemma}[Defects through the seam]
\label{lem:fermion-seam-defects}
The statement is the same as
Lemma~\ref{lem:retained-defects-energy}, using the compatible
preparations of Theorem~\ref{thm:fermion-prep-exact}.
\end{fermionlemma}

\begin{fermionlemma}[fermionic shorter ring state]
\label{lem:fermion-shorter-ring}
Assume $R_0\ge1$ and let $H_L$ be the translation invariant fermionic Hamiltonian. For every energy $\varepsilon_{\rm cl}>0$ and integer
$M_{\min}>2R_0$, there is a finite integer
$M_{\max}\ge M_{\min}$, depending only on $\widehat D$ (on-site Hilbert space dimension of enlarged system), $R_0$ (interaction range of $H_L$), $J$ (interaction strength $\norm{\widehat h_{x,L}} \leq J$), and
$\varepsilon_{\rm cl},M_{\min}$, with the following property.
For every $L>M_{\max}$ and every translation invariant density matrix
$\rho$ on the enlarged $L$-ring, there exist integers $M,a$ and a density
matrix $\chi_M$ on the enlarged $M$-ring, such that
\[
 M_{\min}\le M\le M_{\max},\qquad
 1\le a\le\lfloor M/3\rfloor,\qquad I=[a,M-a-1],
\]
and
\begin{equation}
 (\chi_M)_I=\rho_I,\qquad
 \Tr(\chi_M\widehat H_M)
 \le\frac ML\Tr(\rho\widehat H_L)+\varepsilon_{\rm cl}.
 \label{eq:shorter-ring-variational_fermion}
\end{equation}
In particular, $|I|=M-2a\ge M/3$.  The bound $M_{\max}$ is independent
of the source circumference $L$ and the source state $\rho$.
\end{fermionlemma}
\begin{proof}

The proof follows that of
Lemma~\ref{lem:shorter-ring-variational}, using
Lemmas~\ref{lem:entropy-selection-energy},
\ref{lem:fermion-oriented-recurrence}, and
\ref{lem:seam-repair-energy} for entropy selection,
fermionic boundary exchange, and parity-covariant seam
repair, respectively. 
\end{proof}

\begin{proof}[Proof of Theorem~\ref{thm:fermion-energy}]
$R_0=0$ case is obvious.
For $R_0\ge1$, repeat its contradiction
argument using Proposition~\ref{prop:fermion-circumference}
and Lemmas~\ref{lem:fermion-failure-states},
\ref{lem:fermion-translation-defects},
\ref{lem:fermion-seam-defects}, and
\ref{lem:fermion-shorter-ring}.

\end{proof}

\section{Two-dimensional example of gap instability}
\label{sec:2d-instability}

In this section, we give more details of the two dimensional translation invariant example in the main text that has an unstable gap. We first introduce some basics of tiling in Sec. \ref{subsec:tiling-definitions} and \ref{subsec:wang-aperiodicity}, and then discuss infinite-volume generalization of this example in Sec. \ref{subsec:2d-infinite-ground-states}.

\subsection{Wang tiles}
\label{subsec:tiling-definitions}

We start by reviewing the basic notion of Wang tiles \cite{Wang1961}. In a square lattice system, a \emph{Wang tile} is a type of site with a color on each of its four links. Fix a finite set of tile labels $\mathcal T$ and a finite color set $\mathcal C$. The data that characterizes the tiles is captured by four color functions 
\begin{equation}\label{eq:wang-edge-functions}
    E, W, N, S:\quad \mathcal T\to \mathcal{C}, 
\end{equation}
giving the east, west, north and south edge colors. These functions specify the abstract tile label, which does not depend on the position of the site. In the following, we will take each site of the lattice to be a specific tile, so that each link gets colors from the two sites it connects, according to color functions in Eq. \eqref{eq:wang-edge-functions}.

Let $\vec e_1=(1,0)$ and $\vec e_2=(0,1)$ be the primitive lattice vectors. A tile configuration on a region
$\Lambda\subseteq\Z^2$ is a function $a:\Lambda\to\mathcal T$ with $x\mapsto a_x$. It is a \emph{valid tiling} if
\begin{equation}
 E(a_{\vec x})=W(a_{\vec x+\vec e_1}),\qquad
 N(a_{\vec x})=S(a_{\vec x+\vec e_2})
 \label{eq:wang-matching}
\end{equation}
whenever $\{\vec x, {\vec x+\vec e_1}\}$ and $\{\vec x, {\vec x+\vec e_2}\}$ belong to $\Lambda$. In other words, in a valid tiling pattern all links receive consistent colors from the two sites they connect. On an open square,
no condition is imposed across its outer boundary. The rules in Eq. \eqref{eq:wang-matching} are
translation invariant in the sense that they are independent of $\vec x$, so checking whether a tiling configuration is valid
requires only checking the labels at neighboring sites.

The set of all valid tilings on the infinite plane will be denoted by
\begin{equation}
 X_{\mathcal T}
 =\{a\in\mathcal T^{\Z^2}:a\text{ satisfies Eq. \eqref{eq:wang-matching}}\}.
 \label{eq:wang-configuration-space}
\end{equation}
Note that $X_{\mathcal T}$ is purely determined by $\mathcal{T}$, $\mathcal{C}$ and the color functions in Eq. \eqref{eq:wang-edge-functions}. A plane tiling $a$ has a \emph{period} $\vec p\in\Z^2\setminus\{0\}$ if
\begin{equation}
 a_{\vec x+\vec p}=a_{\vec x}\qquad\text{for every }\vec x\in\Z^2.
 \label{eq:wang-period}
\end{equation}
We call $a$ periodic if it has such a period, and aperiodic if it has none.
A \emph{doubly periodic} tiling has two linearly independent periods.
An \emph{aperiodic tile set} satisfies
\begin{equation}
 X_{\mathcal T}\ne\varnothing,
 \qquad
 \text{every }a\in X_{\mathcal T}\text{ is aperiodic}.
 \label{eq:wang-aperiodic-set}
\end{equation}
Again, whether a tile set is aperiodic is purely determined by $\mathcal{T}$, $\mathcal{C}$ and the color functions in Eq. \eqref{eq:wang-edge-functions}.

According to the above definition, to prove that a tile set is aperiodic, it is insufficient to construct an aperiodic tiling pattern, because this tile set may also allow periodic tiling patterns. It is also insufficient to rule out all periodic tiling patterns, because we also need to show that a valid tiling pattern exists.

\subsection{The eleven-tile set and its  aperiodicity}
\label{subsec:wang-aperiodicity}

The example of local translation invariant Hamiltonians with a unique ground state and an unstable gap discussed in the main text is based on a tile set constructed by Jeandel and Rao, which is the smallest aperiodic tile set \cite{JeandelRao2021}. In this subsection, we briefly explain this tile set, and more details can be found in Ref. \cite{JeandelRao2021}.

In this tile set, $\mathcal T=\{1,\ldots,11\}$,
$\mathcal C=\{\rm R, G, B, Y\}$, with color functions given in
Table~\ref{tab:sm-wang-tiles}. This is the four-color set of Jeandel and
Rao~\cite[Fig.~4 and Sec.~7.2]{JeandelRao2021}.

\begin{table}[h]
\centering
\caption{The eleven Wang tiles used in the two dimensional construction.
The entries specify the four functions in Eq. \eqref{eq:wang-edge-functions}.}
\label{tab:sm-wang-tiles}
\begin{tabular}{c rrrr @{\qquad} c rrrr}
\toprule
$a$&$W(a)$&$E(a)$&$S(a)$&$N(a)$&
$a$&$W(a)$&$E(a)$&$S(a)$&$N(a)$\\
\midrule
1 & R & R & G & R & 7 & Y & G & G & G\\
2 & R & Y & B & G & 8 & Y & G & B & B\\
3 & G & R & B & B & 9 & Y & Y & Y & G\\
4 & G & G & R & B & 10 & B & B & G & R\\
5 & G & Y & B & Y & 11 & B & B & R &B\\
6 & Y & R & G & G & & & & &\\
\bottomrule
\end{tabular}
\end{table}

\begin{theorem}[Jeandel--Rao]\label{thm:sm-wang-aperiodicity}
The tile set in Table~\ref{tab:sm-wang-tiles} admits a valid tiling of
$\Z^2$, and no valid tiling has a nonzero period.
\end{theorem}

We briefly explain its proof. Details can be found in
Secs.~5-7 of Ref.~\cite{JeandelRao2021}.

First, we consider a row of sites in the lattice. For each site attached with a tile label, we color its legs according to the 
functions $E,W,N,S$.
A horizontally valid row requires the left and right legs match with their neighbors according to first equation in Eq. \eqref{eq:wang-matching}. 
The labels $S(a)$ and $N(a)$ record its lower and upper leg colors, and stacking
rows requires the entire upper color sequence of one row to agree with
the lower sequence of the next. 

The horizontal colors of all tiles in Table~\ref{tab:sm-wang-tiles} are divided into two types, one with only colors $\{\rm R,G,Y\}$ (i.e., $a=1, 2, \cdots, 9$) and the other with only color $\{\rm B\}$ (i.e., $a=10, 11$). Consequently, each
row consists entirely of tiles $1,\ldots,9$ or entirely of tiles $10,11$.
We denote these two types of row by $r_y=\alpha$ and $r_y=\beta$, respectively. A valid row coloring alone does not give a plane tiling. For example,
tile $1$ can be repeated along a row, but two such rows cannot be stacked together,
because $N(1)=R\ne G=S(1)$. Vertical compatibility restricts which rows can coexist.

To simplify the analysis of plane tilings, Jeandel and Rao first analyzed a five-color version
$\widetilde{\mathcal T}$. It differs from the Table \ref{tab:sm-wang-tiles} only by replacing
$S(4)={\rm R}$ and $N(10)={\rm R}$ with $S(4)={\rm P}$ and $N(10)={\rm P}$.
For the valid plane tiling of this five-color version, their analysis shows that the row-type sequence
decomposes into strips (rows from top to bottom)
\begin{equation}
 A=\beta\alpha\alpha\alpha\alpha=\beta\alpha^4,\qquad B=\beta\alpha\alpha\alpha=\beta\alpha^3.
 \label{eq:wang-strip-blocks}
\end{equation}
Note that both types of stripes start with a $\beta$-type row, and the entire configuration is made up of such strips. As $A,B$ consist of building blocks of all valid tilings, we refer to $A$ and $B$ as {\it letters}, a string of symbols from the alphabet $\{A,B\}$ as a \emph{word}, and a contiguous part of
such a string is called a \emph{subword}. Here each symbol represents an entire strip as in Eq.
\eqref{eq:wang-strip-blocks}. 

Define the substitution
\begin{equation}
 \sigma(A)=AB,\qquad \sigma(B)=A.
 \label{eq:wang-fibonacci-substitution}
\end{equation}
At each step, we can replace every letter of the current word simultaneously
and concatenate the replacements in their original order. Starting
from $w_0=A$ and setting $w_{n+1}=\sigma(w_n)$ give
\begin{equation}
 A\longmapsto AB\longmapsto ABA\longmapsto ABAAB
 \longmapsto ABAABABA\longmapsto\cdots.
 \label{eq:wang-fibonacci-words}
\end{equation}
For example, the step following $ABAAB$ is
\begin{equation}
 \sigma(ABAAB)=(AB)(A)(AB)(AB)(A)=ABAABABA,
 \label{eq:wang-fibonacci-example}
\end{equation}
where the parentheses separate the replacements of the five input
letters. The lengths of words $w_n$ form a Fibonacci sequence as follows,
\begin{equation}
 w_{n+1}=w_nw_{n-1}\quad(n\ge1),\qquad
 |w_{n+1}|=|w_n|+|w_{n-1}|.
 \label{eq:wang-fibonacci-recursion}
\end{equation}
The first identity holds for $w_2=(AB)(A)$ and follows inductively
because the substitution respects concatenation. Thus the word lengths are $1,2,3,5,8,\ldots$. Each word is a prefix of the next, so their limit
defines a one-sided infinite \emph{Fibonacci word}, indexed by
$0,1,2,\ldots$. Jeandel and Rao proved that, for every valid tiling
by $\widetilde{\mathcal T}$, a finite word occurs in its $A,B$
strip sequence if and only if that word occurs in some
$w_n=\sigma^n(A)$.

This constraint excludes the periodicity of all valid tiling patterns through the frequencies
of the two types of strip. Let $F_0=0$, $F_1=1$, and
$F_{n+2}=F_{n+1}+F_n$. Counting letters in the substitution gives
\begin{equation}
 \#_A(w_n)=F_{n+1},\qquad \#_B(w_n)=F_n,\qquad |w_n|=F_{n+2}.
 \label{eq:wang-fibonacci-counts}
\end{equation}
Consequently, the fractions of letters of each type approach
\begin{equation}
 f_A=\lim_{n\to\infty}\frac{F_{n+1}}{F_{n+2}}=\varphi^{-1},
 \qquad
 f_B=\lim_{n\to\infty}\frac{F_n}{F_{n+2}}=\varphi^{-2},
 \qquad
 \varphi=\frac{1+\sqrt5}{2}.
 \label{eq:wang-fibonacci-frequencies}
\end{equation}
Indeed, for a fixed $n\ge1$, any long subword of the infinite Fibonacci
word can be partitioned, apart from two end fragments, into copies of
$\sigma^n(A)=w_n$ and $\sigma^n(B)=w_{n-1}$. For large $n$, both block
types have letter fractions close to the limits above, and as the subword
length grows, the end fragments contribute a vanishing fraction. Since
every finite subword of an allowed two-sided sequence occurs in the
infinite Fibonacci word, the same limits hold for that sequence.
Both frequencies are irrational. By contrast, in a periodic sequence
with period $p$, the frequency of $A$ is $m/p$, where $m$ is its number
of occurrences in one period, and is therefore rational. A vertical
period of a tiling by
$\widetilde{\mathcal T}$ would preserve the $\beta$-type rows and hence induce
a period of its $A,B$ sequence, giving a contradiction. This contradiction proves that a valid tiling by $\widetilde{\mathcal T}$, if it exists, cannot be doubly periodic. Indeed, any doubly periodic tiling will have a period along the vertical direction by simple linear combinations of integers.

The above arguments exclude the periodicity along the $y$ direction (and hence doubly periodic tiling), and we still need to exclude possible periodicity $(x,y)\to (x+p_1, y+p_2)$ with $p_1,p_2$ both nonzero. For finite Wang rules, excluding doubly periodic tilings also excludes
every nonzero period~\cite[Lemma~1]{JeandelRao2021}. 

On the other hand, the same strip
analysis does give valid strips of unbounded height. There exist valid plane tilings for $\widetilde T$. 

The four-color set is obtained by identifying colors P and R.
This identification preserves the existence of a tiling, but a separate
argument is needed to show that it does not introduce a periodic one.
The analysis in Sec.~7.2 of Ref.~\cite{JeandelRao2021} shows that, the tile set $\mathcal{T}$ allows, in addition
to $A$ and $B$, two more types of strips with row words
\begin{equation}
 C=\beta\alpha^8,\qquad D=\beta\alpha^3\beta^2\alpha^3.
 \label{eq:wang-extra-strips}
\end{equation}
Jeandel and Rao showed that each $C$ or $D$ strip occurring
in a valid plane tiling can be replaced by an $AB$ strip,
preserving the matching and periodicity
\cite[Sec.~7.2, Corollary~3]{JeandelRao2021}.
Consequently, a doubly periodic four-color tiling would yield
a doubly periodic tiling by $\widetilde{\mathcal T}$,
contradicting the Fibonacci constraint on its $A,B$ strip sequence. Hence by Ref. \cite[Lemma~1]{JeandelRao2021}, this four-color tiling is aperiodic.

The same finite-state
reasoning applies to an arbitrary period vector $\vec v\ne0$. To see it, we identify sites
related by $\vec v$.
This gives a cylinder of finite circumference. After choosing an integer
lattice basis with one direction parallel to $\vec v$, its successive cross
sections form a one-dimensional sequence over a finite alphabet, with
finite-range matching constraints. Such constraints can be encoded by a
finite directed graph whose vertices remember a fixed number of cross
sections. An infinite valid path contains a directed cycle. Repeating
that cycle gives a second period transverse to $\vec v$, while preserving the
first. Thus the existence of any periodic tiling would imply the
existence of a doubly periodic one, which the strip argument excludes.

\subsection{Infinite-volume version of the Hamiltonian}
\label{subsec:2d-infinite-ground-states}

In this subsection, we generalize the two dimensional example to infinite plane. Our goal is to show that the infinite-volume version of this example has many ground states.

On the infinite plane, the Hamiltonian without the perturbation is a formal sum:
\begin{equation}
    H^{(2)}=J
    \left[
    \sum_{\langle\vec x, \vec y\rangle}
    (P_{\vec x}-P_{\vec y})^2+\sum_{\vec x}(C^{(1)}_{\vec x}+C^{(2)}_{\vec x})\right],
\end{equation}
where the symbols have the same meaning as in the main text. This formal sum defines a derivation $\delta_{H^{(2)}}(A)=[H^{(2)}, A]$ for any local operator $A$. Recall that a ground state in such a setting is defined by Eq. \eqref{eq:infinite-ground-state-condition}.

\begin{proposition}[Many infinite-plane ground states]
\label{prop:infinite-plane-ground-states}

The ground states of $\delta_{H^{(2)}}$ include both the vacuum and every product state obtained from a valid plane tiling with arbitrary on-site spin states.

\end{proposition}
\begin{proof}
Choose a valid plane tiling $a:\mathbb Z^2\to\mathcal T$, supplied by
Theorem~\ref{thm:sm-wang-aperiodicity}, and normalized spin state $\ket{s_x}$.
Their product state is specified on each finite region by
\[
 (\omega_{a,\boldsymbol s})_\Lambda
 =\bigotimes_{x\in\Lambda}
   \ket{a_x,s_x}\bra{a_x,s_x}.
\]
At $\varepsilon=0$, all sites are occupied and all colors match, so
every local term annihilates this state. The same holds for the vacuum state.
For a positive local term $h$ annihilating a state,
$\omega(A^{\dagger}[h,A])=\omega(A^{\dagger}hA)\ge0$.
Summing over the finitely many terms meeting the support of $A$ proves Eq.
\eqref{eq:infinite-ground-state-condition} for both types of states. So both types of states are ground states.

\end{proof}

\subsection{Ground state of perturbed Hamiltonian}\label{subsec:ground_state_2D}

In the main text, we have mentioned that when the perturbation strength satisfies $\varepsilon L_1L_2>J(L_1+L_2)$, certain states where all sites host defects become the degenerate ground state when $J_s=0$. Now we explain the reason.

Recall in the main text,
\begin{equation}
    V_{\vec L}^{(2)}=\varepsilon\sum_{
    \vec x
    }P_{\vec x},
\end{equation}
where we have already set $J_s=0$. Suppose we have an eigenstate with some vacant sites, we show that by adding defects to these sites we can always construct a state with lower energy. 

Let the number of vacant sites in this state be $h>0$, the number of links that connect vacant sites and occupied sites be $b$, and left the number of links that connect two occupied sites giving inconsistent colors be $m$. Such an eigenstate has an eigenenergy under $H^{(2)}_{\vec L}+V^{(2)}_{\vec L}$:
\begin{equation}
    E(h, b, m)=J(m+b)+\varepsilon h.
\end{equation}

The new state with all sites being occupied by a defect and a lower energy can be constructed as follows. We choose a horizontal cut and a vertical cut on the torus such that each of them crosses as few vacant-vacant bonds as possible. Then these two cuts cross at most $h/L_1$ and $h/L_2$ vacant-vacant bonds, respectively. Next, we consider a valid tiling pattern on the infinite plane $\mathbb{Z}^2$ and wrap it on our torus $\T^2_{\vec L}$. Then we use this tiling pattern on $\T^2_{\vec L}$ to assign the tiles to the vacant sites (and let the spin of these defects be any pure state). Note that after this process, among vacant-vacant bonds in original state, the newly added defects can give at most $h/L_1 +h/L_2$ mismatches. After this is done, the energy of the new state is at most
\begin{equation}
    E'(h, b, m) = J\left(m+b+\frac{h}{L_1}+\frac{h}{L_2}\right).
\end{equation}
When $\varepsilon L_1L_2>J(L_1+L_2)$, the new state has an energy lower than $E(h, b, m)$. 

Therefore, if $\varepsilon L_1L_2>J(L_1+L_2)$, the new ground states must have all sites being occupied by a defect. Note that the above argument does not prove (or disprove) that such ground states are obtained from a tiling pattern that results from wrapping a valid tiling pattern on $\Z^2$ onto $\T^2_{\vec L}$. What we have shown is that the states with any vacant site cannot be the ground state, so the ground states must have all sites being occupied by defects, and such states should correspond to tiling patterns on $\T^{2}_{\vec L}$ that have the smallest number of links with inconsistent colors.

\section{Two dimensional example with a non-topological 1-form symmetry} \label{sec: 1-form}

In this section, we discuss an application of our results to higher-form symmetries.

The notion of higher-form symmetries was originally developed in the context of quantum field theories \cite{Gaiotto2014}, and it was later applied to lattice systems relevant to condensed matter physics (see, for example, Refs. \cite{McGreevy2022, Luo2023}, for reviews). However, Ref. \cite{Liu2026} pointed out some important differences between the higher-form symmetries often discussed in condensed matter systems and the topological higher-form symmetries in field theories. Concretely, consider systems in two spatial dimensions, then 1-form symmetries are generated by operators supported on closed loops. For a 1-form symmetry to be topological in the sense of quantum field theory, all states in the theory should be eigenstates of such operators that are supported on contractible loops, and they should have the {\it same} eigenvalue. However, this condition is generally not satisfied by condensed matter systems, where the total Hilbert space is often a tensor product of local Hilbert spaces. In order for the 1-form symmetries of the condensed matter system to flow to a topological 1-form symmetry at low energies, the low-energy states of this system must be eigenstates of such loop-like operators with the same eigenvalue. Let us denote this eigenvalue by $\lambda_0$, and other eigenvalues of the loop-like operators by $\lambda_i$ with $i=1,2, \cdots$. The above discussion means that in order for a two dimensional condensed matter system to have a topological 1-form symmetry that results from the microscopic 1-form symmetry, the eigenstates of the loop-like operators with eigenvalues $\lambda_i\neq\lambda_0$ must be gapped, and the energy scale below which the topological 1-form symmetry emerges is such a gap.

On the other hand, topological higher-form symmetries in quantum field theory cannot be broken by any local perturbation, because such symmetries do not act on local operators,  i.e., local perturbations are always invariant under such symmetries. So if a two dimensional condensed matter system has an emergent topological 1-form symmetry, the excitations that are eigenstates of the loop-like operators with eigenvalues $\lambda_i\neq \lambda_0$ should have a {\it stable} gap under local perturbations that are symmetric under the 1-form symmetry.

In this section, we give an example of two dimensional Hamiltonian with the following properties.

\begin{enumerate}
    
    \item The system is gapped, local and translation invariant.

    \item Ground states of the system can be viewed as the ground states of the toric code Hamiltonian \cite{Kitaev1997} stacked with a product state.

    \item The Hamiltonians have a lattice version of 1-form symmetry as defined in the Introduction of Ref. \cite{Liu2026}.

    \item There is certain arbitrarily weak translation invariant and {\it 1-form symmetric} local perturbation, under which the system becomes gapless and the low-energy states are eigenstates of the 1-form symmetry generators with {\it different} eigenvalues. This phenomenon reflects the non-topological nature of the 1-form symmetry in this example.
    
\end{enumerate}
This new example is interesting, and no example in Ref. \cite{Liu2026} satisfies all the conditions above. For instance, the examples in Sec. II and Sec. III. C of Ref. \cite{Liu2026} are not gapped, and the example in Sec. III. D of Ref. \cite{Liu2026} is not translation invariant and there does not appear any arbitrarily weak local perturbation that can make it gapless.

This new example can be obtained from the example of Sec. \ref{sec:2d-instability} by adding one qubit at each link of the square lattice. We denote the Pauli operators of the qubit on the link $l$ by $X_l$, $Y_l$ and $Z_l$. The Hamiltonian of the system is
\begin{equation}
    H'_{\vec L}= H_{\vec L}^{(2)} + J_1\sum_p (\one-B_p)+ J_2\sum_{\vec x} P_{\vec x}(\one-A_{\vec x}),
\end{equation}
where $H_{\vec L}^{(2)}$ is the Hamiltonian for the 23-dimensional qudits in the example of Sec. \ref{sec:2d-instability}, $B_p=\prod_{l\in\partial p}Z_l$, $P_{\vec x}=|0_{\vec x}\rangle\langle 0_{\vec x}|$ is the projector onto the on-site vacuum for the original 23-dimensional qudit at site $\vec x$, $A_{\vec x}=\prod_{\partial l\ni\vec x}X_l$, and $J_{1, 2}>0$. In other words, the $B_p$ and $A_{\vec x}$ operators are precisely the plaquette operators and vertex operators in the standard toric code model \cite{Kitaev1997}. Just like the standard toric code model, $H'_{\vec L}$ has a lattice version of 1-form symmetry, which is generated by the operators $A_{\vec x}$ for all $\vec x\in\T^2_{\vec L}$.

All terms in $H'_{\vec L}$ can be simultaneously minimized. It is straightforward to check that $H'_{\vec L}$ is gapped, and its ground states have the form $\otimes_{\vec x}|0_{\vec x}\rangle \otimes|{\rm TC}\rangle$, where $|{\rm TC}\rangle$ stands for any ground state of the standard toric code model, which is an eigenstate of all $B_p$ and $A_{\vec x}$ operators with eigenvalue 1. So such a ground state is precisely the ground state of the toric code stacked on a product state $\otimes_{\vec x}|0_{\vec x}\rangle$. Note that here the ground states are 4-fold degenerate, just as the toric code model, and the gap here refers to the energy difference that separates this degenerate subspace and the rest of the states.

At this point, the first three properties enumerated above have been verified. To verify the last property, we consider the following perturbation: 
\begin{equation}
    V'_{\vec L}=\frac{1}{2}\varepsilon\left[\sum_{\vec x} \frac{1+A_{\vec x}}{2} (\one- \sigma_{\vec x}^3 - \sigma_{\vec x- \vec e_1}^{1} \cdot \sigma_{\vec x}^{1} )\right],
\end{equation}
where $\sigma^3_{\vec x}=(\sum_{a_{\vec x}\in\mathcal{T}}|a_{\vec x}\rangle\langle a_{\vec x}|)\otimes(|\uparrow_{\vec x}\rangle\langle\uparrow_{\vec x}|-|\downarrow_{\vec x}\rangle\langle\downarrow_{\vec x}|)$ and $\sigma_{\vec x}^{1} = (\sum_{a_{\vec x}\in\mathcal{T}}|a_{\vec x}\rangle\langle a_{\vec x}|)\otimes(|\downarrow_{\vec x}\rangle\langle\uparrow_{\vec x}|+|\uparrow_{\vec x}\rangle\langle\downarrow_{\vec x}|)$.
This local translation invariant perturbation preserves the 1-form symmetry generated by $A_{\vec x}$. Intuitively, this perturbation consists of a bunch of Ising chains at their quantum critical point, and each Ising term is turned on or off by an onsite $A_{\vec x}$ charge, i.e., it is on for states with eigenvalue 1 under $A_{\vec x}$ and off for states with eigenvalue $-1$ under $A_{\vec x}$. In the following, we call states with $A_{\vec x}=1$ charge neutral and states with $A_{\vec x}=-1$ charged under the 1-form symmetry.

We now analyze the spectrum and eigenstates of the $H_{\vec L}'+ V_{\vec L}'$. For simplicity, we consider the case $L_1=L_2=L$, and it easily generalizes to other aspect ratios. The perturbation $V_{\vec L}'$ consists of $L$ rows as Ising chains coupled to charge-free projectors. First, for a periodic chain of length $n$ without $A_{\vec x}$ charge, its contribution to $V_{\vec L}'$ is
\[
\frac{n}{2} - \frac{1}{2}\sum_{j=1}^n (\sigma_j^3 + \sigma_{j}^1 \sigma_{j+1}^1), \quad \sigma_{n+1}^1 = \sigma_{1}^1.
\]
The lowest energy of such an Ising chain is $e_{\rm p}(n)=\frac{n}{2} - \csc\frac{\pi}{2n}$ for $n\geq 3$ \cite{Grimm_2002}. Let $\mu = \frac{2}{\pi}-\frac{1}{2}>0$, then we get $e_{\rm p}(L) = -\mu L -\frac{\pi}{12L} +O(L^{-3})$. Hence, we see that the fully occupied states are favored compared to the vacuum state by $V'_{\vec L}$ when $L$ is sufficiently large, because the mismatch from tiling introduces energy penalty at most $\sim JL$, while a fully occupied neutral (no $A_{\vec x}$ charge) state has $-\varepsilon\mu L^2$ energy awards from the perturbation.

We still need to consider charged states as competitors of ground states. A site with an $A_{\vec x}$ charge will break its row into an open segment of length $n$ where the $\sigma_{\vec x}^3$ on this site is removed. Then the contribution of this open segment to $V_L'$ is
\[
\frac{n-1}{2} - \frac{1}{2}\sum_{j=2}^n \sigma_j^3 - \frac{1}{2}\sum_{j=1}^{n-1} \sigma_j^1\sigma_{j+1}^1.
\]
Its lowest energy is $e_{\rm o}(n) = \frac{n}{2} - \frac{1}{2}\cot\frac{\pi}{4n} $ for $ n\geq 1$ \cite{Campostrini_2015}. We rewrite $e_{\rm o}(n) = -\mu n+ f(n)$ with $f(n) = \frac{2n}{\pi} - \frac{1}{2}\cot\frac{\pi}{4n}>0$. Suppose that charges divide a full row into segments of lengths $n_1, n_2, \cdots, n_r$ with $\sum_j n_j = L$, then its lowest energy is
\[
\sum_j e_{\rm o}(n_j)  =  -\mu L +\sum_j f(n_j) > -\mu L > e_{\rm p}(L).
\]
We find that $A_{\vec x}$ charges increase the energy from the perturbation $V'_{\vec L}$ relative to the fully neutral state. Moreover, $A_{\vec x}$ charges also increase the energy from $H'_{\vec L}$ due to the term proportional to $J_2$, if $P_{\vec x}=1$, and it keeps the energy invariant for states with $P_{\vec x}=0$. So in all cases within the fully occupied sector, introducing $A_{\vec x}$ charges never decreases the energy relative to the charge neutral states.

We may also consider states with vacancies, and they cannot be the new ground state, according to an argument similar to that in Sec. \ref{subsec:ground_state_2D}. Starting from a state with fixed occupation, tile types and charge sector, we show that it is always possible to construct a state with lower energy by completely filling these vacancies. Let us count the number of mismatched occupied-occupied bonds as $m$, and number of occupied-vacant bonds as $b$. Then energy from $H^{(2)}_{\vec L}$ is $Jm+Jb$. Now we choose a horizontal cut and a vertical cut, such that each of them crosses minimal number of vacant-vacant bonds. Then we fix an arbitrary tiling pattern (called $\mathcal T_{\T^2_{\vec L}}$) on this torus whose mismatches only appear across the chosen cuts. We fill the holes of original state with corresponding tiles of $\mathcal T_{\T_{\vec L}^2}$ together with $\ket{\uparrow}$ on each hole. This newly constructed state has energy from $H_{\vec L}^{(2)}$ being at most $Jm+Jb+Jk$ with $k\leq 2h/L$, where $h$ is the number of vacant sites. Next we count the energy lowered by this filling: if a site has $A_{\vec x}=+1$, then filling it lowers energy by $\varepsilon/2$; if it has $A_{\vec x} = -1$, then filling it lowers energy by $2J_2$. Notice that the filled spin part is $\ket{\uparrow_{\vec x}}$, which makes the contribution of $\sigma_{\vec x-\vec e_1}^1\sigma_{\vec x}^1$ vanish. So the total energy variation after filling is $E_{\rm filled} - E_{\rm original}\leq 2Jh/L - \varepsilon h_+ /2  - 2J_2h_-$ with $h=h_++h_-,\,h_{\pm}$ the number of vacant sites with $A_{\vec x} = \pm1$. For sufficiently large $L$, $E_{\rm filled} - E_{\rm original}<0$, hence we get a state with lower energy than original state. This excludes the possibility of vacant states being the ground state.

In conclusion, the ground state space of $H_{\vec L}'+ V'_{\vec L}$ is spanned by states of the form $\ket{a_{\rm tile}}\otimes\ket{\Omega_{\rm Ising}}\otimes\ket{\rm TC}$, where $|a_{\rm tile}\rangle$ is any tiling pattern that has the smallest number of links with mismatched colors, and $|\Omega_{\rm Ising}\rangle$ is the ground state of all length-$L$ Ising chains. Notice that the energy difference of two charge-one rows and two neutral rows is $\delta_L = 2\varepsilon(e_{\rm o}(L) - e_{\rm p}(L)) = \varepsilon\tan \frac{\pi}{4L} \sim \frac{\pi\varepsilon}{4L}$, so we can find many low-lying eigenstates with fluctuating $A_{\vec x}$ charges. The quasi-continuous spectrum above the ground state energy is clear from the critical Ising chains. This discussion verifies the last property mentioned at the beginning of this section.

\section{Stabilization of gap} \label{sec: stabilization}

In this section, we prove that gapped local Hamiltonians for one dimensional spin and fermionic systems with a unique ground state, no matter whether they have a stable or unstable gap, can make their gap stable under certain arbitrarily weak local perturbations. In higher dimensions, this result also holds if the ground states are product states. The one dimensional case is discussed in Sec. \ref{subsec: stabilization 1d} and the higher dimensional case is discussed in Sec. \ref{subsec: stabilization higher dimensions}.

\subsection{Stabilization of gap in one dimension} \label{subsec: stabilization 1d}

The following theorem shows that the gap of any gapped local Hamiltonians in one dimension with a unique ground state can be stabilized.

\begin{theorem}[stabilization of gap on finite-ring]\label{thm:gap-stabilization} Consider a ring with circumference $L$. Let $H_L = \sum_{X}\Phi_L(X)$ be a spin or fermionic Hamiltonian on this ring, with fixed on-site Hilbert space dimension, range at most $R_0$, and $\norm{\Phi_L}_{\rm inc}<J$. Suppose for all $L\geq L_0$, $H_L$ has a unique ground state $\ket{\psi_L}$, ground state energy $E_L$ and a uniform gap:
\[
H_L -E_L\one \geq \gamma_0(\one - P_L),\qquad P_L = \ket{\psi_L}\bra{\psi_L}, \qquad \gamma_0>0.
\]
Then, there are positive interactions $\mathcal{K}_L$, constants $R_K, L_*, t_0,\nu>0$, such that the following holds: for each finite perturbation range $r$, there are constants $c_r, D_r>0$ such that
\begin{equation}\label{eq:stab-statement}
\begin{gathered}
 K_L=\sum_X\mathcal K_L(X),\qquad
 \mathcal K_L(X)\ge0,\qquad
 \operatorname{range}(\mathcal K_L)\le R_K,\qquad
 \|\mathcal K_L\|_{\inc}\le1,\\
 \gap(H_L+tK_L+V_L)\ge\gamma_0+\nu t,\\
 \big|\gap(H_L+tK_L+V_L)-\gap(H_L+tK_L)\big|
       \le \frac{D_r}{t}\|\Xi_L\|_{\inc}
\end{gathered}
\end{equation}
for all $L\geq L_*\geq\max\{L_0,R_K+2\}$, $0<t\leq t_0$ and $V_L=\sum_X \Xi_L(X)$ of range at most $r$  with $\norm{\Xi}_{\rm inc}<c_r t$. Each $H_L+tK_L+V_L$ has a unique ground state. 
The constants $c_r,D_r$ are independent of $L,t$.
\end{theorem}

The last inequality in Eq. \eqref{eq:stab-statement} shows the continuity of $\gap(H+tK_L+V_L)$ in $\norm{\Xi}_{\rm inc}$ for each finite $t$. Notice that this theorem does not claim the compatibility of $K_L$ as described in Eq.~\eqref{eq:local-dynamics-compatible}, and this compatibility is proved in Corollary \ref{cor:stab-compatible} under further assumptions.

\begin{proof}
The family of interactions $\mathcal{K}_L$ can be constructed by defect operators in the product-vacuum frame.
In the following, we first construct a quasi-local relative of $\mathcal{K}_L$, and show that it stabilizes the gap of the Hamiltonian $H_L$. Next, we truncate this quasi-local relative to get a finite-range interaction $\mathcal{K}_L$ while preserving the gap-stabilization property{, i.e., $H_L+tK_L$ has a stable gap where $K_L=\sum_X\mathcal{K}_L(X)$}. Finally, we prove the continuity of the gap of the Hamiltonian $H_L + t K_L$.

{\it Step 1: quasi-local relative of $\mathcal{K}_L$.} {We first construct a quasi-local relative of $\mathcal{K}_L$.}

By Corollaries~\ref{cor:prep-nonti} and
\ref{cor:fermion-prep-nonti}, there exists an exponentially local unitary $T_L$ such that $T_L(\ket{\psi_L}\otimes\ket{\psi_L^c}) = \ket{\Omega_L}$. Let $Q_x$ be the doubled on-site defect projection from
Eq.~\eqref{eq:defects}, and define
\begin{equation}\label{eq:stab-compression}
 J_L|u\rangle=|u\rangle\otimes|\psi_L^c\rangle,\qquad
 m_{x,L}=J_L^\dagger T_L^\dagger Q_xT_LJ_L,\qquad \mathsf M_L=\sum_xm_{x,L}.
\end{equation}
It is straightforward to verify that 
\begin{equation}\label{eq:stab-penalty}
\begin{gathered}
 0\le m_{x,L}\le\one,\quad m_{x,L}|\psi_L\rangle=0,\quad
 \mathsf M_L\ge\one-P_L, \quad
 \|m_{x,L}-\E^{\tr}_{B_L(x,n)}m_{x,L}\|\le C e^{- c n},
\end{gathered}
\end{equation}
where $C,c$ are constants {obtained from Corollaries~\ref{cor:prep-nonti} and
\ref{cor:fermion-prep-nonti}, which are} independent of $x,L$. The last inequality can be proved by using the triangle inequality with $b_{x,L}^{[n]} = J_L^\dagger \E_{\widehat B_L(x,n)}^{\rm tr}[T_L^\dagger Q_x T_L]J_L$, where $\widehat B_L(x,n)$ contains both copies of $B_L(x,n)$ and $\E_{\widehat B_L(x,n)}^{\rm tr}$ traces out its complement. Note that $\E^{\rm tr}_{B_L(x,n)}b_{x, L}^{[n]}= b_{x,L}^{[n]}$ because $b_{x,L}^{[n]}$ is supported on $B_L(x,n)$ already and the extra conditional expectation does nothing. So we have
\[
\begin{split}
\norm{m_{x,L} - \E^{\tr}_{B_L(x,n)} m_{x,L} } &\leq \norm{m_{x,L} - b_{x,L}^{[n]}} + \norm{ \E^{\tr}_{B_L(x,n)}\left( b_{x,L}^{[n]} - m_{x,L}\right)} \leq C e^{-c n}.
\end{split}
\]
where Corollaries~\ref{cor:prep-nonti} and
\ref{cor:fermion-prep-nonti} have been used in the last step.

{The interaction $m_{x, L}$ constructed above is a quasi-local relative of $\mathcal{K}_L$, from which we can construct $\mathcal{K}_L$ in Step 3.}

{\it Step 2: stability.} By Eq. \eqref{eq:stab-penalty}, $H_L + \tau \mathsf M_L$ is gapped for arbitrary $\tau>0$, and we now use the spectral flow to show the stability of its gap under weak perturbations.

Fix an input shell profile $f(n)=e^{-\alpha n^\beta}$ with $\alpha>0$ and $0<\beta<1$,
and write a perturbation as
$W_L=\sum_{x,n}w_L(x,n)$, with
$\|w_L(x,n)\|\le\varepsilon f(n)$. {Note that this perturbation is allowed to have a stretched-exponential tail and does not have to be strictly local.} Also fix a constant $t_0>0$. Take $B_L=H_L+\tau \mathsf M_L$, $0<\tau\le t_0$, then $B_L$ has the same unique gapped ground state $\ket{\psi_L}$ as $H_L$. For a fixed length $L$, we use the spectral flow of
$B_L+sW_L$ with the threshold $g=\gamma_0/2$ for filtering and the unitary defined as Eq.~\eqref{eq: transport unitary}. On every interval of $s$ with the
gap greater than $g$, Eqs.~\eqref{eq:anchor},
\eqref{eq:transported-derivative} and \eqref{eq:transported_perturbation_bound} give
\begin{equation}
 \frac{d}{ds}\big[U_s(B_L+sW_L-e_s\one)U_s^\dagger\big]
       =\sum_xa_{x,L}(s),
\end{equation}
where $e_s$ is the ground energy of $B_L+sW_L$, and
$a_{x,L}(s)$
is a self-adjoint operator, annihilates $\ket{\psi_L}$, and has norm and
stretched-exponential locality bounds with a prefactor proportional to $\varepsilon$ (denoted as $C_{{\rm l},a}\varepsilon$). Hence by Lemma~\ref{lem:bounded_axL} that will be proved later in this subsection, we can get
\begin{equation}\label{eq:stab-bootstrap}
 U_s(B_L+sW_L-e_s\one)U_s^\dagger
 \ge H_L-E_L\one+(\tau-s C_a\varepsilon)\mathsf M_L .
\end{equation}
with $C_a\ge1$ a constant which includes the prefactor $C_{{\rm l},a}$ and depends on the fixed locality bounds (i.e., $f$, $t_0$ and $g$), but is independent of $L,\tau,s$ and $\varepsilon$. Taking a constant $c_{\rm p} = \frac{1}{4C_a}$, then $\varepsilon\leq c_{\rm p}\tau$ implies $C_a \varepsilon\leq \tau/4$. The right side of Eq. \eqref{eq:stab-bootstrap} has a gap at least $\gamma_0+3\tau/4 > g$. Similar to the proof of Theorem \ref{thm:productstability}, we can extend this estimation to $s=1$ via the bootstrap argument. Hence we conclude that $H_L+\tau \mathsf M_L$ has a stable gap for $\tau\leq t_0$ and perturbation strength $\varepsilon\leq c_{\rm p}\tau$.

{\it Step 3: finite-range truncation.} We now consider a truncation of $m_{x,L}$ to get the finite-range interaction $\mathcal{K}_L$ and show that the gap of the Hamiltonian $H_L+tK_L$ is still stable. 

Since $\norm{m_{x,L} - \E^{\rm tr}_{B_L(x,n)}m_{x,L}}\leq C e^{-c n}$, and the exponent $\beta$ in $f(n)$ is smaller than 1, we consider the shell representation of the truncated terms of radius $R\geq 0$:
\[
D_{L}^{(R)} = \mathsf M_L - \sum_{x} \E^{\tr}_{B_L(x,R)} m_{x,L} = \sum_x\sum_{ n=R+1}^{\lfloor L/2\rfloor} d_{x,L}^{[n]},\qquad d_{x,L}^{[n]}:=(\E_{B_L(x,n)}^\tr - \E_{B_L(x,n-1)}^\tr)[m_{x,L}].
\]
They satisfy
\begin{equation}
\frac{\norm{d_{x,L}^{[n]}}}{f(n)}
\le C(1+e^{c})\sup_{\substack{k\in\mathbb N\\k>R}}
e^{-c k+\alpha k^\beta}
=: \delta_R \to 0 \text{ as } R\to \infty,
\qquad R<n\le\lfloor L/2\rfloor.
\end{equation}
Therefore, we can choose $R$ such that $\delta_R< c_{\rm p}/4$ and take
\begin{equation}\label{eq:stab-finite-range}
    {\mathcal{K}_L(X=B_L(x, R))=\frac{1}{2R+1} \E^{\rm tr}_{B_{L}(x,R)} m_{x,L},
    \qquad K_L=\sum_{X}\mathcal{K}_L(X)}.
\end{equation}
{The above sum over $X$ is the same as a sum over $x\in\T_L$.} {Each $\mathcal{K}_L(X)$} has range at most $R_K:=2R$ and incidence norm at most one by the normalization $1/(2R+1)$. {This $K_L$ will be our local perturbation that stabilizes the gap of $H_L$.}

Fix an integer $L_*\geq\max\{L_0,2R+2\}$. We want to show for $0<t\leq t_0$ that
\begin{equation}
    H_L +tK_L = H_L + \frac{t}{2R+1} \left(\mathsf M_L - \sum_{x}\sum_{n>R} d_{x,L}^{[n]}\right)
\end{equation}
has a stable gap. For convenience, denote $\tau = t/(2R+1)$, so $H_L+t K_L = H_L +\tau\mathsf M_L - \tau D_L^{(R)}$. 
We regard $-\tau D_L^{(R)}$ as a perturbation to $H_L+\tau \mathsf M_L$, and combine it with an arbitrary range-$r$ perturbation $V_L$ of incidence strength \(\eta:=\|\Xi_L\|_{\rm inc}\). Choose a shell representation for $V_L$ that satisfies
$\norm{v_L(x,n)}\le C_r\eta f(n)$ with $C_r=e^{\alpha r^\beta}$ \footnote{For instance, such a shell representation can be constructed as follows: attach each $X$ a special site denoted as $s(X)$, and denote $l(X) = \max_{y} \{d_L(s(X), y), y\in X\}$. These are used to ensure each $X$ is used exactly once. Then let $v_L(x,n) = \sum_{\substack{ X: s(X) = x\\ l(X)=n }} \Xi_L(X)$. Then its shell has norm $\norm{v_{L}(x,n)}\leq \eta$. For $n\leq r$, $C_r f(n) = e^{\alpha (r^\beta - n^\beta)}\geq1$ with equality holding at $n=r$, so the shell norm $\norm{v_L(x,n)}\leq C_r\eta f(n)$ holds. For $n>r$, the shell terms vanish.}.
Consequently, the combined perturbation has shell terms
\[
w_L(x,n)=
\begin{cases}
v_L(x,n),&n\le R,\\
-\tau d_{x,L}^{[n]}+v_L(x,n),&n>R,
\end{cases}
\]
with
$\norm{w_L(x,n)}
\le (\tau\delta_R+C_r\eta) f(n).$ If
$\eta<\frac{c_{\mathrm{p}}}{4C_r}\tau,$
then
{$\|w_L(x, n)\|<(\frac{c_{\rm p} \tau}{4}+\frac{c_{\rm p}\tau}{4})f(n)<c_{\rm p}\tau f(n)$}.
Applying the stability of gap of $H_L+\tau\mathsf M_L$ below Eq. \eqref{eq:stab-bootstrap} gives
\begin{equation}\label{eq:HL_and_KL_gap}
\gap(H_L+t K_L+V_L)
\ge\gamma_0+\frac{3\tau}{4} = \gamma_0 + \nu t,
\qquad \nu:=\frac{3}{4(2R+1)}.
\end{equation}
with a unique ground state, uniformly for $L $ greater than some length scale $ L_*$. Hence we conclude that for $0<t\leq t_0$ and $c_r\leq c_{\rm p}/[4(2R+1)C_r]$, the Hamiltonian $H_L+tK_L+V_L$ has a unique ground state and a uniform gap for $L\geq L_*$.

{\it Step 4: continuity of energy gap.} 
{In the path $B_L+sW_L$ in Step 2, we take $W_L=-\tau D_L^{(R)}$, so that this path becomes a path of $B_L-s\tau D_L^{(R)}$, with an end point $B_L-\tau D_L^{(R)}=H_L+tK_L$. Let $U^{(0)}_{L, t}$ be the endpoint unitary for this path.} Let $U_{L,t}(s)$ be the spectral flow
along $A_L(s):=H_L+tK_L+sV_L$, with $U_{L,t}(0)=\one$.
Both use the threshold $g=\gamma_0/2$.
We write $e_{L,t}(s)$ for the ground state energy of $A_L(s)$, and define
\[
 \mathcal G_{L,t}(s):=
 U_{L,t}^{(0)}U_{L,t}(s)(A_L(s)-e_{L,t}(s)\one)
 U_{L,t}(s)^\dagger(U_{L,t}^{(0)})^\dagger,\qquad
 G_{L,t}:=\mathcal G_{L,t}(0).
\]
By construction, $\mathcal{G}_{L,t}(s)$ annihilates $\ket{\psi_L}$.
Then we use Eq. \eqref{eq:stab-bootstrap} to get a bound $G_{L,t}\geq \nu t \mathsf M_L$. Concretely, $G_{L,t}$ is equal to left hand side of Eq. \eqref{eq:stab-bootstrap} with $W_L = -\tau D_L^{(R)}$. The $\varepsilon$ in Eq.  \eqref{eq:stab-bootstrap}, in turn, is determined by shell terms of $W_L$. Since we have $\norm{-\tau d_{x,L}^{[n]}} \leq \tau\delta_R f(n)$, taking $\varepsilon = \tau \delta_R$ is valid. So Eq.~\eqref{eq:stab-bootstrap} gives
$G_{L,t}\ge H_L-E_L\one+
       (\tau-C_a\tau\delta_R)\mathsf M_L.$
Next, since $H_L-E_L\geq0$, and we have chosen $\delta_R<c_{\rm p}/4$ and $c_p\leq1/(4C_a)$, so $C_a\delta_R \leq 1/16$, so $1-C_a\delta_R>3/4$, then combine with \eqref{eq:HL_and_KL_gap} we get this second inequality.

\begin{equation}\label{eq:stab-reference-bound}
 G_{L,t}\ge H_L-E_L\one+
       (\tau-C_a\tau\delta_R)\mathsf M_L
 \ge\nu t\,\mathsf M_L.
\end{equation}
As in Eq. \eqref{eq:transported-derivative}, the derivative of
$\mathcal G_{L,t}(s)$ is a sum of self-adjoint terms annihilating
$\ket{\psi_L}$. The locality bounds following Eq. \eqref{eq:anchor},
followed by conjugation with $U_{L,t}^{(0)}$ and
Lemma~\ref{lem:bounded_axL}, give a constant $C_{{\rm cont},r}$ such that
\begin{equation}\label{eq:bound_of_GLt}
 -C_{{\rm cont},r}\eta\mathsf M_L
 \le\frac{d}{ds}\mathcal G_{L,t}(s)
 \le C_{{\rm cont},r}\eta\mathsf M_L.
\end{equation}
Here $C_{{\rm cont},r}\ge1$ is independent of $L,t,s,\eta$.
Integrating and using Eq. \eqref{eq:stab-reference-bound} give
\begin{equation}\label{eq:stab-relative-continuity}
 \left(1-\frac{C_{{\rm cont},r}\eta}{\nu t}\right)G_{L,t}
 \le\mathcal G_{L,t}(1)
 \le\left(1+\frac{C_{{\rm cont},r}\eta}{\nu t}\right)G_{L,t}.
\end{equation}
Choose
\[
 c_r:=\min\left\{\frac{c_{\rm p}}{4(2R+1)C_r},
                    \frac{\nu}{2C_{{\rm cont},r}}\right\}.
\]
For $\eta<c_rt$, the left side coefficient in Eq. \eqref{eq:stab-relative-continuity} is positive.
Minimizing the quadratic forms on $\ket{\psi_L}^{\perp}$ (orthogonal space of $\ket{\psi_L}$) yields
\[
 \big|\gap(H_L+tK_L+V_L)-\gap(H_L+tK_L)\big|
 \le\frac{C_{{\rm cont},r}\eta}{\nu t}\,\Delta_{L,t},
 \qquad \Delta_{L,t}:=\gap(H_L+tK_L).
\]

To bound $\Delta_{L,t}$, let $\ket{\chi_{L,t}}$ be the normalized
ground state of $H_L+tK_L$ and choose an on-site unitary $u_x$
with $\bra{\chi_{L,t}}u_x\ket{\chi_{L,t}}=0$.
For spins, take a cyclic permutation in an eigenbasis of the one-site
marginal; for fermions, take a Majorana operator, whose expectation
vanishes because the unique ground state has definite parity.
The variational principle and the incidence bounds give
\[
 \Delta_{L,t}
 \le\bra{\chi_{L,t}}u_x^\dagger[H_L+tK_L,u_x]\ket{\chi_{L,t}}\leq \norm{[H_L+tK_L, u_x]}
 \le2(J+t_0)=:\Delta_*.
\]
For fermions, interaction terms away from $x$ commute with $u_x$
because they are even. Thus the last inequality in Eq.
\eqref{eq:stab-statement} holds with
$D_r:=C_{{\rm cont},r}\Delta_*/\nu$, independent of $L,t$.
The preparation, penalties and spectral flows are even in the
fermionic case, so the same argument uses the even vacuum bound
from Section~\ref{sec:fermions}.
\end{proof}

{\bf Remark.}
The above proof works for any interaction $\mathcal{M}_L = \sum_{x} p_{x,L}$ which satisfies the properties in Eq. \eqref{eq:stab-penalty} and satisfies $\lambda \mathsf M_L\leq \mathcal{M}_L\leq \Lambda \mathsf M_L$ with $\lambda, \Lambda>0$. Replacing $\mathsf M_L$ by $\mathcal{M}_L$ from Step 2 to the end above is valid, up to some changes of constants in these bounds. 

To finish the above proof, we need to prove the following lemma. This lemma extends the relative bound of Lemma~\ref{lem:vacuum} in the product-vacuum frame to a relative bound of sums of quasi-local operators annihilating $\ket{\psi_L}$, which concern the physical frame, with $\mathsf M_L$ replacing the defect operator $\mathsf N_L$ in the product-vacuum frame.

\begin{lemma}\label{lem:bounded_axL}
    Under the assumptions of Theorem~\ref{thm:gap-stabilization}, let
    $\{a_{x,L}:x\in\T_L\}$ be self-adjoint operators (which are further assumed to be even for fermionic systems)
    satisfying
    \[
    a_{x,L}\ket{\psi_L} = 0,\qquad \|a_{x,L}\|\le\varepsilon,\qquad
    \norm{a_{x,L} - \E^{\tr}_{B_L(x,n)} a_{x,L}}\leq \varepsilon e^{-\alpha n^\beta}
    \]
    for $0\le n\le\lfloor L/2\rfloor$, with constants $\alpha>0$
    and $0<\beta<1$. Then
    \begin{equation}\label{eq:axL-inequality}
        -C_a \varepsilon \mathsf M_L \leq \sum_x a_{x,L} \leq C_a\varepsilon \mathsf M_L,
    \end{equation}
    where $C_a$ depends only on $\alpha,\beta$ and the uniform locality
    bounds of $T_L$, and is independent of $L,\varepsilon$.
\end{lemma}

\begin{proof}
Set $b_x:=T_L(a_{x,L}\otimes\one)T_L^\dagger$.
Then $\|b_x\|\le\varepsilon$ and $b_x\ket{\Omega_L}=0$. {In the following, we will prove Eq. \eqref{eq:axL-inequality} by bounding $\sum_xb_x$.}

{We first prove the locality of $b_x$ by controlling $\|b_x-\E^\Omega_{\widehat B_L(x,n)}b_x\|$, where $\widehat B_L(x,n)$ contains both the physical and auxiliary copies of $B_L(x,n)$.} Let $C_T,\mu_T>0$ be the uniform constants that control the exponential locality for $T_L$.
Taking $k=\lfloor n/2\rfloor$, we first approximate $a_{x,L}$ by
$\E^{\rm tr}_{B_L(x,k)}a_{x,L}$ and then apply the locality bound
for $T_L$ with enlargement radius $n-k$.
Contractivity of the conditional expectations then gives
\begin{equation}
\begin{split}
\|b_x-\E^\Omega_{\widehat B_L(x,n)}b_x\|
=&\|T_L\E^{\tr}_{B_L(x, k)}(a_{x, L})\otimes\one T_L^\dagger
-
\E^\Omega_{\widehat{B}_{L}(x, n)}(T_L\E^{\tr}_{B_L(x, k)}(a_{x, L})\otimes\one T_L^\dagger)\\
+&
T_L[(a_{x, L}-\E^\tr_{B_L(x, k)}(a_{x, L}))\otimes\one]T_L^\dagger
-\E^\Omega_{\widehat{B}_L(x, n)}[T_L[(a_{x, L}-\E^\tr_{B_L(x, k)}(a_{x, L}))\otimes\one]T_L^\dagger]
\|\\
\leqslant&\|T_L\E^{\tr}_{B_L(x, k)}(a_{x, L})\otimes\one T_L^\dagger
-
\E^\Omega_{\widehat{B}_{L}(x, n)}(T_L\E^{\tr}_{B_L(x, k)}(a_{x, L})\otimes\one T_L^\dagger)\|\\
+&
\|T_L[(a_{x, L}-\E^\tr_{B_L(x, k)}(a_{x, L}))\otimes\one]T_L^\dagger
-\E^\Omega_{\widehat{B}_L(x, n)}[T_L[(a_{x, L}-\E^\tr_{B_L(x, k)}(a_{x, L}))\otimes\one]T_L^\dagger]
\|\\
\le&\varepsilon F(n)
\end{split}
\end{equation}
where
\[
F(n):=2C_T(2\lfloor n/2\rfloor+1)e^{-\mu_T(n-\lfloor n/2\rfloor)} + 2e^{-\alpha\lfloor n/2\rfloor^\beta}.
\]

Put $m:=\lfloor L/2\rfloor$ and define
\[
 b_{x,0}:=\E^\Omega_{\widehat B_L(x,0)}b_x,\qquad
 b_{x,n}:=(\E^\Omega_{\widehat B_L(x,n)}
          -\E^\Omega_{\widehat B_L(x,n-1)})b_x
 \quad(1\le n\le m).
\]
These shells sum to $b_x$, are self-adjoint, and annihilate their
local vacuum by Lemma~\ref{lem:vacuum}. They satisfy
$\|b_{x,n}\|\le\varepsilon g(n)$ with
$g(0):=1$ and $g(n):=F(n)+F(n-1)$ for $n\ge1$.
The vacuum form bound {in Lemma~\ref{lem:vacuum}} therefore gives
\[
 \pm\sum_xb_x
 \le\varepsilon\sum_{x,n}g(n)\mathsf N_{B_L(x,n)}
 \le C_a\varepsilon\mathsf N_L,\qquad
 C_a:=\sum_{n\ge0}(2n+1)g(n)<\infty.
\]
Conjugating by $T_L^\dagger$ and compressing with the isometry $J_L$
from Eq. \eqref{eq:stab-compression} yields
\[
 \pm\sum_xa_{x,L}
 =J_L^\dagger T_L^\dagger\left(\pm\sum_xb_x\right)T_LJ_L
 \le C_a\varepsilon J_L^\dagger T_L^\dagger\mathsf N_LT_LJ_L
 =C_a\varepsilon\mathsf M_L.
\]
This proves Eq. \eqref{eq:axL-inequality}.
\end{proof}

Next, for those Hamiltonians with thermodynamically convergent ground states, the following corollary constructs the compatible interactions (i.e., interactions with the same interaction terms across different system sizes) that can stabilize the gap. 
\begin{corollary}[Compatible stabilization]
\label{cor:stab-compatible}
Under the hypotheses of Theorem~\ref{thm:gap-stabilization}, suppose that
$[H_L,A]$ is exactly independent of $L$ for every fixed
local observable $A$, and that $\rho_{L,I}$ converges in trace norm for every fixed interval $I$. Then the $\mathcal{K}_L$ can be chosen so that
$[K_L,A]$ is exactly independent of $L$ as well.
All conclusions of Theorem~\ref{thm:gap-stabilization}, including gap continuity, remain valid.
\end{corollary}

The convergence of $\rho_{L,I}$ can be guaranteed, for instance, under the condition that the interaction in $H_L$ has a unique, gapped ground state functional if put on the infinite chain. {This convergence also holds if $H_L$ is translation invariant, as shown in Proposition \ref{prop:circumference}. Of course, by Theorem \ref{thm:main}, in this case the original Hamiltonian already has a stable gap.}

\begin{proof}
The key difference of the construction here is to choose an interaction $\mathcal{M}_L=\sum_{x} p_{x,L}$ that is independent of $T_L$, using the remark below the proof of Theorem \ref{thm:gap-stabilization}. We first use a filter to get such $p_{x,L}$, and then show its truncation gives the compatible dynamics. We mainly discuss the spin case, and briefly mention the fermionic case at the end.

{\it Step 1: convergent quasi-local operators.} Let us construct $p_{x,L}$ from local operators. Consider a site $x$, we fix a ball of radius $\ell$, $B_L(x, \ell)$. Choose a basis of Hermitian operators in $B_L(x, \ell)$, denoted by $\{O_{x,j}\}_{j=1}^{n_\ell}$, by translating a fixed basis in the on-site coordinates (for fermion case, choose a basis of the even Hermitian
operators). Each $O_{x,j}$ is assumed to have a norm at most one. Then define \footnote{Indeed, the equal weight summation of $p_{x,L}$ is not necessary. Taking bounded numbers $w_{x,j}\in[w_-, w_+]$ with $w_+,w_->0$ as weights of $|a_{x,j,L}|$ is also valid, which only changes the constant $\lambda$ in the bound $\mathcal M_L\geq \lambda \mathsf M_L$.}
\begin{equation}
    a_{x,j,L} = F_{H_L, w}(O_{x,j}) - \ev{\psi_L|O_{x,j}|\psi_L}\one, \qquad p_{x,L} = \frac{1}{2n_\ell} \sum_j |a_{x,j,L}|,\qquad \mathcal{M}_L =\sum_{x} p_{x,L}
\end{equation}
Here the filter $F_{H_L,w}(\cdot)$ is as described in Section~\ref{subsubsec:locality_and_filtering}, and we choose $g=\gamma_0/2$ and a nonnegative kernel $w$ with $\int w(u)du=1$. The locality bounds for filtering give stretched-exponential tails for $a_{x,j,L}$, as discussed in Lemma~\ref{lem:appendix-locality}. These bounds also hold for $|a_{x,j,L}|$, since for self-adjoint $A,B$ with norms at most two,
\[
 \big\||A|-|B|\big\|
 \le\|A^2-B^2\|^{1/2}\le2\|A-B\|^{1/2}.
\]
Applying this to $A=a_{x,j,L}$ and $B=\E^{\tr}_{B_L(x,n)}a_{x,j,L}$, and using contractivity once more, gives constants $C_p,c_p>0$ and $0<\beta_p<1$ such that
\begin{equation}\label{eq:stab-compatible-tails}
 0\le p_{x,L}\le\one,\qquad
 \|p_{x,L}-\E^{\tr}_{B_L(x,n)}p_{x,L}\|
 \le C_p e^{-c_p n^{\beta_p}},
\end{equation}
uniformly in $x,L$.

We can verify the properties of $p_{x,L}$ described in the remark below the Theorem \ref{thm:gap-stabilization}:
\begin{enumerate}

\item By the filter construction, {$p_{x, L}$} annihilates $\ket{\psi_L}$ {because $a_{x, j, L}|\psi_L\rangle=0$}.

\item By Lemma \ref{lem:bounded_axL}, $\mathcal{M}_L\leq \Lambda \mathsf M_L$ for some constant $\Lambda>0$.

\item To see $\mathcal M_L\geq \lambda \mathsf M_L$ for some $\lambda$, we use the truncation of $m_{x,L}$, denoted as $B_{x,L} = \E^{\rm tr}_{B_L(x, \ell)} m_{x,L}$, as an intermediate step. For convenience, define $\mathcal F_L(O):=F_{H_L,w}(O) - \ev{O}_{\psi_L}\one$. Then $\sum_x\mathcal F_L(B_{x,L}) = F_{H_L, w}(\mathsf M_L) - \sum_x \mathcal F_L(m_{x,L}-B_{x,L})$. We want to show $2n_\ell c_\ell\mathcal M_L\geq \sum_x \mathcal F_L(B_{x,L})\geq \frac{\kappa}{2}\mathsf M_L$ for some constants $n_\ell, c_\ell, \kappa$.

The exponential tails of $m_{x,L}$, the filtering bounds and Lemma~\ref{lem:bounded_axL} give
\begin{equation}\label{eq:fml_bound}
 -\epsilon_\ell\mathsf M_L
 \le\sum_x\mathcal F_L(m_{x,L}-B_{x,L}) = \mathcal{F}_L(\mathsf M_L) - \sum_x \mathcal{F}_L(B_{x,L})
 \le\epsilon_\ell\mathsf M_L,\qquad \epsilon_\ell\to0 {\text{ as } \ell\to \infty}.
\end{equation}
Indeed, we can check $\mathcal F_L(m_{x,L}-B_{x,L})$ satisfies the assumption of Lemma~\ref{lem:bounded_axL}:  $m_{x,L}-B_{x,L}$ has norm at most $Ce^{-c\ell}$ and tails bounded by 
\[
\norm{D_{x,L}^{(\ell)} - \E^{\tr}_{B_L(x,n)}(D_{x,L}^{(\ell)}) }\leq C e^{-c \ell/2}e^{-c n/2},\qquad D_{x,L}^{(\ell)} = m_{x,L} - B_{x,L}.
\]
This is because for $n<\ell$, we have $\norm{D_{x,L}^{(\ell)} - \E^{\tr}_{B_L(x,n)}(D_{x,L}^{(\ell)}) }\leq Ce^{-c \ell}$ and for $n\geq \ell$ we have $\norm{D_{x,L}^{(\ell)} - \E^{\tr}_{B_L(x,n)}(D_{x,L}^{(\ell)}) }\leq Ce^{-c n}$, taking $e^{-c(n+\ell)/2}$ is valid for both $n< \ell$ and $n\geq \ell$.

Moreover, filtering preserves the small prefactor, with a fixed stretched-exponential profile, so Lemma~\ref{lem:bounded_axL} applies.
Therefore, we can choose sufficiently large $\ell$ such that $\sum_x\mathcal{F}_L(B_{x,L}) \geq \frac{\kappa}{2} \mathsf M_L$ with some constant $\kappa>0$ \footnote{
This is because Lemma~\ref{lem:bounded_axL} applied to $\tau_u^{H_L}(m_{x,L})$ gives a constant $C_{\mathsf M}$ such that $\tau_u^{H_L}(\mathsf M_L)\leq C_{\mathsf M}\mathsf M_L$ for $|u|\le1$,
and the same $C_{\mathsf M}$ applies to $\tau_{-u}^{H_L}(\mathsf M_L)$, so we get $\mathsf M_L = \tau_{u}^{H_L} (\tau^{H_L}_{-u}(\mathsf M_L)) \leq C_{\mathsf M} \tau_{u}^{H_L} (\mathsf M_L) $.
Hence $F_{H_L,w}(\mathsf M_L) = \int w_g(u) \tau_u^{H_L}(\mathsf M_L) du \geq C^{-1}_{\mathsf M}\int_{-1}^1 w_g(u)  \mathsf M_L du  \geq \kappa \mathsf M_L$, where $\kappa$ can be explicitly taken as $C_{\mathsf M}^{-1} \int_{-1}^{1} w(u) du>0$. Then Eq. \eqref{eq:fml_bound} gives $\sum_{x} \mathcal F_L(B_{x,L})\geq (\kappa -\epsilon_\ell)\mathsf M_L$. We just need to take $\ell$ sufficiently large such that $\epsilon_\ell<\kappa/2$.
}.

Next, we decompose $B_{x,L}$ in the basis $O_{x,j}$,
\begin{equation}
B_{x,L} = \sum_{j=1}^{n_\ell} b_{x,j,L} O_{x,j}, \qquad \mathcal F_{L} (B_{x,L}) = \sum_{j=1}^{n_\ell} b_{x,j,L} a_{x,j,L}
\end{equation}
Take a uniform upper bound of $|b_{x,j,L}|$ as $c_\ell$. Then we get $2n_\ell c_\ell p_{x,L} \geq \mathcal F_L(B_{x,L})$. So summing up $x$ we get $2n_\ell c_\ell \mathcal{M}_L\geq \sum_x \mathcal F_L(B_{x,L})$, which completes the proof of $\mathcal M_L \geq\lambda \mathsf M_L$ with $\lambda=\kappa/(4n_\ell c_\ell)$.
\end{enumerate}

Next, we prove that the $\mathcal{M}_L$ converges as $L\to \infty$. Two assumptions are used here: 1) compatibility \footnote{
The exact compatibility here can be relaxed by the norm convergence
of $[H_L,A]$: $\lim_{M\to\infty}\sup_{L,L'>M}\norm{[H_{L},A]-[H_{L'},A]}=0$ for every fixed local observable $A$, with all
other assumptions remaining unchanged. The same proof applies.} of $H_L$ as in Eq. \eqref{eq:local-dynamics-compatible}, 2) the local marginal convergence which gives the convergence of $\ev{O_{x,j}}_{\psi_L}$. 
Here convergence of $\mathcal{M}_L$ means norm convergence of each $p_{x,L}$ with fixed $x$, and with the rings embedded in the infinite chain using the centered coordinates of Section~\ref{sec: setup and summary}. 
By compatibility Eq. \eqref{eq:local-dynamics-compatible}
and the standard Lieb-Robinson construction of the
thermodynamic-limit dynamics
\cite[Sec.~3.3]{NSYspin}, there exists a dynamics $\tau_u$
such that, for every local operator $O$ and every $T<\infty$,
\[
 \lim_{L\to\infty}\sup_{|u|\le T}
 \|\tau_u^{H_L}(O)-\tau_u(O)\|=0.
\]

Let $\omega(O):=\lim_{L\to\infty}\Tr(\rho_LO)$ for
each local operator $O$, and define
$$
 a_{x,j}:=\int_{\mathbb R}w(u)\tau_u(O_{x,j})\,du
          -\omega(O_{x,j})\one,
 \qquad
 p_x:=\frac{1}{2n_\ell}\sum_{j=1}^{n_\ell}|a_{x,j}|.
$$
For fixed $x,j$, convergence of the local dynamics,
integrability of $w$, and convergence of the ground-state
marginals imply $a_{x,j,L}\to a_{x,j}$ in norm.
Continuity of the absolute-value map therefore gives
$p_{x,L}\to p_x$ in norm, with $0\le p_x\le\one$.

{\it Step 2: finite-range truncations.} In this step, we first show that $H_L + \tau \mathcal{M}_L$ has a stable gap, and then apply truncations to the terms $p_{x,L}$ to get a finite-range interaction. 

To prove the stability of the gap of $H_L+\tau\mathcal M_L$,
we repeat Step~2 of the proof of
Theorem~\ref{thm:gap-stabilization}, using
$\mathcal M_L\ge\lambda\mathsf M_L$ obtained from Step~1.
Fix $f(n)=e^{-\alpha n^\beta}$ with $\alpha>0$ and
$0<\beta<\beta_p$.
The Step~2 gives a constant $c_{\rm p}>0$,
independent of $L$ and $\tau$, such that, for
$0<\tau\le t_0$, $H_L+\tau\mathcal M_L+W_L$ has
a unique ground state and gap at least
$\gamma_0+3\lambda\tau/4$ whenever the shell terms
of $W_L$ satisfy
$\|w_L(x,n)\|\le c_{\rm p}\tau f(n)$.

Now we move to the truncation of $q_{x,L}^{(R)}$. Define the positive contractions
\[
 q_{x,L}^{(R)}:=\E^{\tr}_{B_L(x,R)}p_{x,L},\qquad
 q_x^{(R)}:=\E^{\tr}_{B_\infty(x,R)}p_x,\quad B_{\infty}(x,R) = \{ y\in\mathbb Z,: |y-x|\leq R \}.
\]
For every fixed $x,R$, $q_{x,L}^{(R)}\to q_x^{(R)}$ in norm. Also,
\begin{equation}\label{eq:stab-compatible-shells}
 \mathcal M_L-\sum_xq_{x,L}^{(R)}
 =\sum_x\sum_{n>R}d_{x,L}^{[n]},\qquad
 \|d_{x,L}^{[n]}\|\le\delta_R f(n),\qquad \delta_R\to 0,
\end{equation}
where $d_{x,L}^{[n]}=(\E^{\tr}_{B_L(x,n)}-\E^{\tr}_{B_L(x,n-1)})p_{x,L}$ and the shell sums end at $\lfloor L/2\rfloor$. The last bound follows from Eq. \eqref{eq:stab-compatible-tails} and $\beta<\beta_p$. Fix $R\ge\ell$ with $\delta_R<c_{\rm p}/4$, and choose $\epsilon_R>0$ with $\epsilon_R/f(R)<c_{\rm p}/4$.

Notice that convergence of $q_{x,L}^{(R)}\to q_x^{(R)}$ for each fixed $x$ need not be uniform over all sites of a ring. So to construct $K_L$, we only keep those $q_{x}^{(R)}$ for $x$ far from the seam of the centered coordinate Eq.~\ref{eq:origin_coordinate} as follows.
For every integer $j\ge1$, choose increasing integers $\ell_{(j)}\ge\max\{L_0,2(j+R)+3\}$ such that
\[
 \max_{|x|\le j}\|q_{x,L}^{(R)}-q_x^{(R)}\|\le\epsilon_R
 \qquad\text{for every }L\ge 
 \ell_{(j)}.
\]
For each $L\ge \ell_{(1)}$, let $j_L:=\max\{j:\ell_{(j)}\le L\}$. Then $j_L\to\infty$ as $L\to\infty$, and $[-j_L-R,j_L+R]$ does not cross this seam in centered coordinate Eq.~\eqref{eq:origin_coordinate} by definition. Set
\begin{equation}\label{eq:stab-compatible-interaction}
 \widetilde q_{x,L}^{(R)}:=
 \begin{cases}
 q_x^{(R)},&|x|\le j_L,\\
 q_{x,L}^{(R)},&|x|>j_L,
 \end{cases}
 \qquad
 K_L:=\frac{1}{2R+1}\sum_x\widetilde q_{x,L}^{(R)}.
\end{equation}
These summands define a positive interaction $\mathcal K_L$ with range at most $2R$ and $\|\mathcal K_L\|_{\inc}\le1$. For $A\in\mathcal A_I$ with fixed finite $I\subset\mathbb Z$, all sites at distance at most $R$ from $I$ satisfy $|x|\le j_L$ for sufficiently large $L$. Therefore
\begin{equation}\label{eq:stab-compatible-commutator}
 [K_L,A]=\frac{1}{2R+1}
 \sum_{x:\,d(x,I)\le R}[q_x^{(R)},A],
\end{equation}
which is exactly independent of $L$.

{\it Step 3: stability and continuity of gap.} It remains to check the gap estimates for this choice of $K_L$. Put $\tau=t/(2R+1)$ and $\eta=\|\Xi_L\|_{\inc}$. Relative to $H_L+\tau\mathcal M_L$, the perturbation is
\begin{equation}
 H_L+tK_L+V_L-(H_L+\tau\mathcal M_L)
 =-\tau\sum_{x,n>R}d_{x,L}^{[n]}
  +\tau\sum_{|x|\le j_L}(q_x^{(R)}-q_{x,L}^{(R)})+V_L.
\end{equation}
The second sum has radius-$R$ terms of norm at most $\tau\epsilon_R$. A range-$r$ perturbation has a shell representation bounded by $C_r\eta f(n)$, with $C_r=e^{\alpha r^\beta}$. Hence the combined shell bound of $H_L+tK_L+V_L - (H_L +\tau \mathcal M_L)$ is bounded by $\varepsilon_{\rm total} f(n)$, with
\begin{equation}
 \varepsilon_{\rm total}
 :=\tau\delta_R+\frac{\tau\epsilon_R}{f(R)}+C_r\eta
 <\frac34c_{\rm p}\tau
 \quad\text{if}\quad
 \eta<\frac{c_{\rm p}\tau}{4C_r}.
\end{equation}
The preceding stability estimate proves uniqueness and
\[
 \gap(H_L+tK_L+V_L)\ge\gamma_0+\nu t,
 \qquad \nu:=\frac{3\lambda}{4(2R+1)},
\]
for $L\ge L_*=\ell_{(1)}$ and $0<t\le t_0$.

For the continuity of gap, use Step~4 with the reference perturbation
\[
 W_L^{(0)}=-\tau\sum_{x,n>R}d_{x,L}^{[n]}
       +\tau\sum_{|x|\le j_L}(q_x^{(R)}-q_{x,L}^{(R)}).
\]
In its notation, the transported operators satisfy
\[
 G_{L,t}\ge\frac{3\tau}{4}\mathcal M_L,\qquad
 \pm\frac{d}{ds}\mathcal G_{L,t}(s)
 \le C_{{\rm cont},r}\eta\mathcal M_L.
\]
The constant $C_{{\rm cont},r}$ is independent of $L,t,s,\eta$, since both paths have the uniform shell bounds just established. Integrating and minimizing on $\ket{\psi_L}^{\perp}$, as in Eq. \eqref{eq:stab-relative-continuity}, give
\[
 \big|\gap(H_L+tK_L+V_L)-\gap(H_L+tK_L)\big|
 \le\frac{4(2R+1)C_{{\rm cont},r}}{3t}\,
          \eta\,\gap(H_L+tK_L)
 \le\frac{D_r}{t}\eta,
\]
with the fixed choices
\[
 c_r:=\min\left\{\frac{c_{\rm p}}{4(2R+1)C_r},
                  \frac{3}{8(2R+1)C_{{\rm cont},r}}\right\},
 \qquad
 D_r:=\frac{8(2R+1)C_{{\rm cont},r}(J+t_0)}{3}.
\]
Here the upper bound $\gap(H_L+tK_L)\le2(J+t_0)$ is the same as in Step~4. All constructed terms are even in the fermionic case. In particular, Eq. \eqref{eq:stab-compatible-commutator} also holds for odd $A$.

\end{proof}

The compatible stabilizing direction is robust under small changes of the coefficients in the space of Hamiltonians. Namely, if we get a compatible stabilizing perturbation $\mathcal K_L^{(0)}$, as we will see, changing $\mathcal K_L^{(0)}$ to $\mathcal K_L'$ as follows gives another compatible stabilizing direction. Indeed, following the proof of Corollary \ref{cor:stab-compatible}, we can construct another stabilizing perturbation $K_L' = K_L + \Delta K_L$ such that $H_L+ t K_L'$ is also stable. Keep the profile $f(n) = e^{-\alpha n^\beta}$, the radius $R$ and the constant $c_{\rm  p}$, from the proof.
Let $\Delta\mathcal K_L$ be any self-adjoint interaction of range at
most $s$, compatible across sizes, with
\[
 \sup_L\|\Delta\mathcal K_L\|_{\mathrm{inc}}< \frac{c_{\rm p}}{4(2R+1)}f(s),
 \qquad
 \mathcal K_L'=\mathcal K_L^{(0)}+\Delta\mathcal K_L,
 \qquad K_L'=\sum_X\mathcal K_L'(X).
\]
Consider a further range-$r$ perturbation $V_L=\sum_X\Xi_L(X)$, and suppose $\eta=\|\Xi_L\|_{\mathrm{inc}}<c_r't$ with $c_r'=\frac{c_{\rm p}}{4 C_r (2R+1)}$. Then we can write $H_L + t K_L' + V_L = H_L + \tau \mathcal M_L + (tK_L' + V_L - \tau \mathcal M_L)$. Let $ tW'_L = tK_L' + V_L - \tau \mathcal M_L$, then Step~3 bounds the
shells of \(t\Delta K_L\) by
\(t\|\Delta\mathcal K_L\|_{\rm inc}f(n)/f(s)\).
Then the shell norm of $t W'_L$ is bounded by $\varepsilon'_{\rm total}$ with 
\begin{equation}
 \varepsilon_{\rm total}' =  \tau \delta_R + \frac{\tau \epsilon_R}{f(R)} + C_r\eta + \frac{t}{f(s)}\norm{\Delta \mathcal{K}_L}_{\rm inc} < c_{\rm p}\tau.
\end{equation}
Hence the total error remains inside the reference critical perturbation strength of $H_L + \tau \mathcal M_L$.
Consequently, for $0<t\le t_0$ and all sufficiently large $L$,
$H_L+tK_L'+V_L$ has a unique ground state and gap at least
$\gamma_0+\nu t$, with $\nu>0$ uniform over these choices.
Thus, for every fixed \(s\), all compatible variations
\(\Delta\mathcal K_L\) of interaction range at most \(s\)
preserve gap stabilization when
\(\sup_L\|\Delta\mathcal K_L\|_{\rm inc}\) is sufficiently small,
with the allowed size depending on \(s\). No condition \(\Delta K_L\ket{\psi_L}=0\) is required.
In other words, for every fixed range bound \(s\ge R_K\), the stabilizing directions contain an open neighborhood of the constructed interaction family \(\{\mathcal K_L\}\) in the uniform interaction incidence norm, within the space of compatible interactions of range at most \(s\).

\subsection{Stabilization of gap in higher dimensions} \label{subsec: stabilization higher dimensions}

The gap stabilization discussion can be generalized to higher dimensions, under the condition of unique ground state being vacuum. We present two conclusions in the following theorem.

\begin{theorem}\label{thm:stabilization-product-higher-dim-MZ}
Let $H_{\vec L}=\sum_X\Phi_{\vec L}(X)$ be spin Hamiltonians on the $d$-dimensional tori
$\T_{\vec L}$, with fixed local dimension, range at most $R_0$ and
$\|\Phi_{\vec L}\|_{\inc}\le J$, uniformly in ${\vec L}$. Denote $L_{\min}({\vec L}) = \min_{1\leq i\leq d} L_i$. 
Suppose their unique ground state is the product vacuum
$\ket{\Omega_{\vec L}}$. Let $Q_x=\one-\ket{0_x}\bra{0_x}$ and $\mathsf N_{\vec L} = \sum_x Q_x$. Then the following two statements hold.
\begin{enumerate}
\item[(i)] For every fixed $t>0$ and interaction range $r$, there are $\varepsilon(t,r)>0$ and $L_*(t,r)$ satisfying the following property: for every $V_{\vec L} =\sum_{X} \Xi_{{\vec L}}(X)$ with range $r$ and $\norm{\Xi_{\vec L}}_{\rm inc}<\varepsilon(t,r)$, 
$H_{\vec L}+t\mathsf N_{\vec L} +V_{\vec L}$ has a unique ground state and energy gap at least $t/2$ for $L_{\min}({\vec L})\geq L_*(t,r)$. No uniform initial gap is needed for $H_{\vec L}$.

\item[(ii)] Suppose, in addition, that
$\inf_{\vec L:\,L_{\min}(\vec L)\ge L_0}
    \gap(H_{\vec L})\ge\gamma_0>0$.
Then there exist $L_*\ge L_0$ and $t_0>0$ such that,
for every fixed perturbation range $r$, there are constants
$c_r,D_r>0$, independent of ${\vec L}$, with the following
property. For all $L_{\min}(\vec L)\ge L_*$, $0<t\le t_0$, and self-adjoint
perturbations $V_{\vec L}=\sum_X\Xi_{\vec L}(X)$ of range at most $r$
with $\|\Xi_{\vec L}\|_{\inc}<c_rt$, the Hamiltonian
$H_{\vec L}+t\mathsf N_{\vec L}+V_{\vec L}$ has a unique ground state and
\[
 \gap(H_{\vec L}+t\mathsf N_{\vec L}+V_{\vec L})
 \ge \gamma_0+\frac34t,
\]
\[
 \left|\gap(H_{\vec L}+t\mathsf N_{\vec L}+V_{\vec L})
       -\gap(H_{\vec L}+t\mathsf N_{\vec L})\right|
 \le \frac{D_r}{t}\|\Xi_{\vec L}\|_{\inc}.
\]
This conclusion also holds for fermionic $H_{\vec L}$ whose unique ground state is a fermionic product vacuum, i.e., the ground state satisfies $b_i|\Omega_{\vec L}\rangle=0$ for all $i$, where $i$ labels a fermionic mode and $b$ is either the creation or annihilation operator of the mode $i$. 
\end{enumerate}
\end{theorem}
We remark that if the ground state of $H_{\vec L}$ is a general product state, this theorem also holds, since any product state can be rotated back to vacuum via product of onsite unitaries, and the gap-stabilizing interaction $K_{\vec L}$ follows by a rotation of $\mathsf N_{\vec L}$.

\begin{proof}
\emph{(i)}
Fix $t>0$ and a perturbation range $r$. Let $E_{\vec L}$ be the
ground energy of $H_{\vec L}$, and define
\[
 G_{\vec L}:=H_{\vec L}-E_{\vec L}\one+t\mathsf N_{\vec L},
 \qquad
 P_{\vec L}:=\ket{\Omega_{\vec L}}\bra{\Omega_{\vec L}}.
\]
Since $H_{\vec L}-E_{\vec L}\one\ge0$ and
$\mathsf N_{\vec L}\ge\one-P_{\vec L}$, we have
\begin{equation}\label{eq:stab-product-initial-bound}
 G_{\vec L}\ket{\Omega_{\vec L}}=0,
 \qquad
 G_{\vec L}\ge t\mathsf N_{\vec L}\ge t(\one-P_{\vec L}).
\end{equation}
Thus $G_{\vec L}$ has the unique ground state $\ket{\Omega_{\vec L}}$
and gap at least $t$. In the following, we apply the same procedures {to prove the stability of gap of normal-form Hamiltonians} in Theorem \ref{thm:productstability} to $G_{\vec L}$.

Write $\eta:=\|\Xi_{\vec L}\|_{\inc}$ and initially assume
$\eta\le1$. Consider the path
\[
 G_{\vec L}(s):=G_{\vec L}+sV_{\vec L},\qquad 0\le s\le1.
\]
On any interval starting at $s=0$ where $G_{\vec L}(s)$ has a
unique ground state and gap greater than $t/2$, denote
its ground energy and ground projection by $e_{\vec L}(s)$
and $P_{\vec L}(s)$, respectively. Use the spectral-flow construction from
Theorem~\ref{thm:productstability}, with filter threshold
$g=t/2$, to obtain unitaries satisfying
\[
 U_{\vec L}(0)=\one,\qquad
 U_{\vec L}(s)P_{\vec L}(s)U_{\vec L}(s)^\dagger=P_{\vec L}.
\]
Define the transported excitation Hamiltonian by
\[
 \mathcal G_{\vec L}(s):=
 U_{\vec L}(s)\bigl(G_{\vec L}(s)-e_{\vec L}(s)\one\bigr)U_{\vec L}(s)^\dagger.
\]
It has the same excitation spectrum as $G_{\vec L}(s)$,
annihilates $\ket{\Omega_{\vec L}}$, and satisfies
$\mathcal G_{\vec L}(0)=G_{\vec L}$.

The derivative decomposition used in
Theorem~\ref{thm:productstability} gives self-adjoint
operators $a_{x,n,\vec L}(s)$ such that
\begin{equation}\label{eq:stab-product-shells}
 \begin{gathered}
 \dot{\mathcal G}_{\vec L}(s)
   =\sum_{x\in\T_{\vec L}}\sum_{n=0}^{m_{\vec L}}a_{x,n,\vec L}(s),
 \qquad m_{\vec L}:=\operatorname{diam}(\T_{\vec L}^d)=\sum_{i=1}^d\lfloor L_i/2\rfloor,\\
 a_{x,n,\vec L}(s)\in\mathcal A_{B_{\vec L}(x,n)},\qquad
 a_{x,n,\vec L}(s)\ket{\Omega_{B_{\vec L}(x,n)}}=0,\\
 \|a_{x,n,\vec L}(s)\|\le\eta f_{t,r}(n).
 \end{gathered}
\end{equation}

Here \(B_{\vec L}(x,n)\) is the radius-\(n\) ball about \(x\) (in the sense of graph distance). The bound
\[
|B_{\vec L}(x,n)|\le(2n+1)^d
\]
holds uniformly in the side lengths. It implies uniform
summability and convolution bounds for the decay functions
\((1+u)^{-p}e^{-bu^\beta}\), with \(p>d\), \(b>0\), and
\(0<\beta\le1\).
Then Lieb--Robinson, filtering, and spectral-flow locality
estimates apply uniformly on these rectangular tori.
They yield constants \(A_{t,r},b_{t,r}>0\) and
\(0<\beta_{t,r}<1\) for which we may take
\[
 f_{t,r}(n):=A_{t,r}e^{-b_{t,r}n^{\beta_{t,r}}}.
\]
These constants may depend on the fixed locality data,
\(d\), \(t\), and \(r\), but are independent of \(\vec L\),
\(s\), and the particular perturbation with \(\eta\le1\).
No restriction on the aspect ratios is needed.
With these uniform bounds, the construction of the shells
and the derivation of Eq.~\eqref{eq:stab-product-shells}
follow the proof of Theorem~\ref{thm:productstability}.

We now bound $\dot{\mathcal G}_{\vec L}(s)$ using Eq.
\eqref{eq:stab-product-shells}.
Since each $a_{x,n,\vec L}(s)$ annihilates its local vacuum,
Lemma~\ref{lem:vacuum} gives
\begin{equation}
 -\|a_{x,n,\vec L}(s)\|\mathsf N_{B_{\vec L}(x,n)}
 \le a_{x,n,\vec L}(s)
 \le \|a_{x,n,\vec L}(s)\|\mathsf N_{B_{\vec L}(x,n)}.
\end{equation}
For every radius \(n\),
\[
 |B_{\vec L}(x,n)|
 \le\prod_{i=1}^d\min\{L_i,2n+1\}
 \le(2n+1)^d. 
\]
By symmetry of the distance, each site belongs to at most
\((2n+1)^d\) such balls. Hence
\[
 \sum_{x\in\T_{\vec L}}
    \mathsf N_{B_{\vec L}(x,n)}
 \le(2n+1)^d\mathsf N_{\vec L}.
\]
Summing the preceding operator inequalities yields
\begin{equation}\label{eq:stab-product-relative-bound}
 -C(t,r)\eta\mathsf N_{\vec L}
 \le\dot{\mathcal G}_{\vec L}(s)
 \le C(t,r)\eta\mathsf N_{\vec L},
\end{equation}
where
\begin{equation}\label{eq:stab-product-constant}
 C(t,r):=\max\left\{
 1,\sum_{n\ge0}(2n+1)^d f_{t,r}(n)\right\}<\infty.
\end{equation}
The series converges because $f_{t,r}$ decays
stretched exponentially. In particular, $C(t,r)$
is independent of $\vec L$.

Integrating Eq. \eqref{eq:stab-product-relative-bound}
and using Eq. \eqref{eq:stab-product-initial-bound}, we obtain
\begin{equation}\label{eq:stab-product-bootstrap}
 \mathcal G_{\vec L}(s)
 \ge G_{\vec L}-sC(t,r)\eta\mathsf N_{\vec L}
 \ge\bigl(t-sC(t,r)\eta\bigr)\mathsf N_{\vec L}.
\end{equation}
Choose
$\varepsilon(t,r):=
 \min\left\{1,\frac{t}{4C(t,r)}\right\},$
with $C(t,r)$ defined in Eq. \eqref{eq:stab-product-constant}.
For $\eta<\varepsilon(t,r)$, Eq.
\eqref{eq:stab-product-bootstrap} implies
\[
 \mathcal G_{\vec L}(s)\ge\frac{3t}{4}\mathsf N_{\vec L}
 \ge\frac{3t}{4}(\one-P_{\vec L}).
\]
Hence the gap of $G_{\vec L}(s)$ is at least $3t/4$
wherever the spectral-flow argument applies.

Then we can run the same gap-continuity or bootstrap argument as in Theorem \ref{thm:productstability}. The ground state remains unique and the gap
exceeds $t/2$ throughout $0\le s\le1$.

All bounds are uniform in $\vec L$, giving the claimed
perturbation threshold and size cutoff.
Finally, subtracting $E_{\vec L}\one$ does not change the gap,
so $H_{\vec L}+t\mathsf N_{\vec L}+V_{\vec L}$ has a unique ground state
and gap at least $t/2$.
No uniform lower bound on $\gap(H_{\vec L})$ was used.

\medskip
{\it (ii)} For (ii), the proof follows from the proof of Theorem \ref{thm:gap-stabilization}. Use the spectral-flow proof of
Theorem~\ref{thm:gap-stabilization} directly with the $\mathsf M_{\vec L}=\mathsf N_{\vec L}$. The required relative bound is exactly the
shell argument from Lemma~\ref{lem:vacuum}. In dimension $d$ its only
change is
\[
 \sum_x \mathsf N_{B_{\vec L}(x,n)}\le (2n+1)^d \mathsf N_{\vec L}.
\]
The locality estimates in Eq.~\eqref{eq:anchor} also extend:
the functions $(1+u)^{-p}e^{-bu^\beta}$ with $p>d$ are uniformly
summable and have bounded convolution constants on these tori.
The Lieb--Robinson and filter estimates therefore retain
stretched-exponential decay; the extra ball-volume factors are
polynomial and are absorbed by weakening its coefficient.
Thus the same derivative bound gives
\[
 U_s(H_{\vec L}+t\mathsf N_{\vec L}+sV_{\vec L}-e_s\one)U_s^\dagger
       \ge H_{\vec L}-E_{\vec L}\one+(t-sC_r\|\Xi_{\vec L}\|_{\inc})\mathsf N_{\vec L} .
\]
With the fixed filter threshold $\gamma_0/2$, continuation yields
the asserted $\gamma_0+3t/4$ gap for
$\|\Xi_{\vec L}\|_{\inc}<c_rt$. The continuity proof above applies with
$U_0=\one$ and $\tau=t$, and gives its $D_r/t$ modulus.
Even fermionic terms and the vacuum expectations preserve parity;
the same estimates test both parity sectors. This proves (ii).
\end{proof}
The compatibility property of $\mathsf N_{\vec L}$ in (ii) across different sizes is obvious since it consists of the same onsite terms for all sizes. The robustness property of gap stabilization below Corollary~\ref{cor:stab-compatible} also applies to item (ii) in higher dimension cases.

\section{Some technical notes}
\label{app:technical}

In this final section, we collect some useful general results that are frequently used in the other parts of this SM.

\subsection{Uniform Lieb-Robinson bounds on rings}
\label{subsec:uniform-lr}

We use the following form of the Lieb--Robinson bound,
which follows from Ref.~\cite[Theorem~3.1]{NSYspin}
for spins and Ref.~\cite[Theorem~3.1]{NSYfermion}
for fermions.

\begin{lemma}[Uniform Lieb--Robinson bound]
\label{lem:uniform-lr}
Let $G_L(t)$ be Hamiltonians on the rings $\T_L$,
with norm-continuous shell interactions
\[
 G_L(t)=\sum_{x\in\T_L}\sum_{n=0}^{\lfloor L/2\rfloor}
 g_L(x,n;t),
 \qquad
 g_L(x,n;t)=g_L(x,n;t)^\dagger
 \in\A_{B_L(x,n)},
\]
satisfying
\[
 \|g_L(x,n;t)\|\le J e^{-a n^\alpha},
 \qquad J,a>0,\quad 0<\alpha\le1,
\]
uniformly in $L,x,n$ and $t$ in a time interval $I$.
In the fermionic case, assume that every interaction
term is even.

Let
\[
 i\partial_t\mathcal U_L(t,s)
 =G_L(t)\mathcal U_L(t,s),
 \qquad \mathcal U_L(s,s)=\one,
\]
and define
\[
 \tau^L_{t,s}(A)
 =\mathcal U_L(t,s)^\dagger A\mathcal U_L(t,s).
\]
For spins, set $[A,B]_{\rm gr}:=[A,B]$.
For fermionic operators of definite parities
$p(A),p(B)\in\{0,1\}$, set
\[
 [A,B]_{\rm gr}
 :=AB-(-1)^{p(A)p(B)}BA.
\]

There are constants $C,\mu,v>0$, depending only on
$J,a,\alpha$, such that for disjoint regions
$X,Y\subset\T_L$, operators $A\in\A_X$, $B\in\A_Y$,
and $s,t\in I$,
\begin{equation}\label{eq:uniform-lr}
 \bigl\|[\tau^L_{t,s}(A),B]_{\rm gr}\bigr\|
 \le
 C\|A\|\|B\|
 \bigl(e^{v|t-s|}-1\bigr)
 \sum_{x\in X}\sum_{y\in Y}
 e^{-\mu d_L(x,y)^\alpha}.
\end{equation}
In the fermionic case, $A$ and $B$ are assumed to
have definite parities. Thus the bracket is a
commutator if either operator is even and an
anticommutator if both are odd.

The constants are independent of $L,X,Y,A,B,s,t$.
No spectral-gap assumption is required.
\end{lemma}

Here $d_L$ is the periodic distance on the ring.
The uniformity in $L$ follows from the uniform shell
bounds and the uniform summability and convolution
bounds for the spatial decay functions on the rings.
Finite-range interactions are included by taking
$\alpha=1$ and adjusting the shell-bound constants.

\subsection{Locality of spectral flow}
\label{subsec:anchored-locality}

This section introduces some technical details of spectral flow.

We first verify the transport property of ground-state projection stated in
Section~\ref{subsubsec:locality_and_filtering}.
\begin{lemma}[Transport of the ground-state projection]\label{lem:spectral-flow-transport}
Let $G_L(s)=G_L+sV_L$, $0\le s\le s_*$, be a path of self-adjoint
Hamiltonians on a finite ring. Suppose that its ground state is unique
and its gap is greater than a fixed filter threshold $g>0$ throughout
this interval. Let $e_L(s)$ and $P_L(s)$ denote its ground-state energy
and ground projection. Using the filters in Eq.~\eqref{eq:filter}, define
\[
 \mathcal D_L(s)=\mathcal D_{G_L(s)}(V_L),\qquad
 \dot U_L(s)=iU_L(s)\mathcal D_L(s),\qquad U_L(0)=\one.
\]
Then
\begin{equation}\label{eq:spectral-flow-transport}
 \dot P_L(s)=-i[\mathcal D_L(s),P_L(s)],\qquad
 U_L(s)P_L(s)U_L(s)^{\dagger}=P_L(0),
 \qquad 0\le s\le s_*.
\end{equation}
The same statements also hold for fermionic Hamiltonians.
\end{lemma}

\begin{proof}
We will take the derivative of the ground state equation $G_L(s) P_L(s) = e_L(s) P_L(s)$ and use the definition of $U_L(s)$ to show that
$U_L(s)P_L(s)U_L(s)^\dagger$ does not change with $s$, therefore giving Eq.~\eqref{eq:spectral-flow-transport}.
Since $G_L(s)=G_L+sV_L$ depends analytically on $s$ and its ground state remains unique and separated from the rest of the spectrum, both $e_L(s)$ and $P_L(s)$ are analytic, and hence differentiable \cite[Chapter~II, Theorem~6.1]{Kato1995}.

We first show
$[F_{G_L(s),w_g}(V_L),P_L(s)]=0$. Denote the eigenstate basis as $\ket{0},\ket{1},\cdots$. Then $\ev{n|F_{G_L(s),w_g}(V_L)|m} = \widehat w_g(E_n-E_m)\ev{n|V_L|m}$, where $\widehat w_g(\omega)$ is the Fourier transformation of $w_g(t)$. By the gap assumption and defining property of $w_g$,  we have $\widehat w_g(\pm(E_n-E_0) )=0$ for any $n>0$. Hence $P_L(s)F_{G_L(s),w_g}(V_L)(1-P_L(s)) = (1-P_L(s))F_{G_L(s),w_g}(V_L) P_L(s) = 0 $, which gives $[F_{G_L(s),w_g}(V_L),P_L(s)]=0$. 

Next, we derive the differential equation in Eq.~\eqref{eq:spectral-flow-transport}. Without ambiguity, we suppress $L$ and $s$ for convenience. Differentiating $GP=eP$, with
$\dot G=V$, gives
\[
 VP+G\dot P=\dot e\,P+e\dot P.
\]
Multiplying on the left by $\one-P$ and on the right by $P$, and using
$P^2=P$ and $(\one-P)P=0$, we obtain
\begin{equation}\label{eq:eq_Ge}
 (\one-P)(G-e\one)\dot P P=-(\one-P)VP.
\end{equation}
On the other hand, the filtering identity Eq.~\eqref{eq:exactfilter} and $[F_{G, w_g}(V), P] = 0$ give
\begin{equation}\label{eq:eq_Ge2}
 0=(\one-P)F_{G,w_g}(V)P
   =(\one-P)VP-i(\one-P)(G-e\one)\mathcal D P.
\end{equation}
Since $G-e\one$ is invertible on the orthogonal complement of the
ground state, Eqs. \eqref{eq:eq_Ge} and \eqref{eq:eq_Ge2} imply
$(\one-P)\dot P P=-i(\one-P)\mathcal D P$.
Notice that differentiating $P^2=P$ gives 
$P\dot P P=(\one-P)\dot P(\one-P)=0$.
Together with the adjoint of the preceding relation, this yields
$\dot P=-i[\mathcal D,P]$. Hence
\[
 \frac{d}{ds}(U_LP_LU_L^{\dagger})
 =U_L(\dot P_L+i[\mathcal D_L,P_L])U_L^{\dagger}=0.
\]
Since $U_L(0)=\one$, integration gives
$U_L(s)P_L(s)U_L(s)^{\dagger}=P_L(0)$.

The proof of fermionic cases is the same.
\end{proof}

Now we establish the locality statements used in
Section~\ref{subsubsec:locality_and_filtering}.
We use $\E_{x,m}:=\E^{\tr}_{B_L(x,m)}$ for convenience.

\begin{lemma}[Locality of filtering and transport]
\label{lem:appendix-locality}
Let $G_L(s)$ and $\mathcal D_L(s)$ be self-adjoint
Hamiltonians satisfying the shell bounds Eq. \eqref{eq:shell}
uniformly in $L$ and $0\le s\le1$, and suppose
$\mathcal D_L(s)$ is continuous in $s$.
Let
\[
 \dot U_L(s)=iU_L(s)\mathcal D_L(s),
 \qquad U_L(0)=\one.
\]
Suppose an operator $A_{x,L}$ satisfies
\[
 \|A_{x,L}\|\le J_A,\qquad
 \|A_{x,L}-\E_{x,m}A_{x,L}\|
 \le J_Ae^{-bm^\beta},
 \qquad 0<\beta\le1.
\]
For fixed $g>0$, each of the operators
\[
 \begin{gathered}
 F_{G_L(s),w_g}(A_{x,L}),\quad
 \mathcal D_{G_L(s),w_g}(A_{x,L}),\quad
 U_L(s)A_{x,L}U_L(s)^\dagger
 \end{gathered}
\]
satisfies
\begin{equation}\label{eq:appendix-locality-output}
 \|B_{x,L}(s)\|\le CJ_A,\qquad
 \|B_{x,L}(s)-\E_{x,m}B_{x,L}(s)\|
 \le CJ_Ae^{-cm^\eta}
\end{equation}
for some $C,c,\eta>0$. Here $B_{x, L}(s)$ stands for any of the above three operators.
The same conclusion holds for
$U_L(s)F_{G_L(s),w_g}(A_{x,L})U_L(s)^\dagger$.
The constants depend only on the stated locality
parameters and $g$, and are independent of $x,L,s,m$
and $J_A$. No spectral gap is required.

The same conclusion holds for fermionic Hamiltonians $G_L(s),\, \mathcal D_L(s)$, and all fermion operators $A_{x,L}$ as described.
\end{lemma}

\begin{proof}
We first prove the estimates for an operator supported
inside a ball. Then we approximate $A_{x,L}$ by such an operator and bound the error from this approximation.

We begin with bounds on the norms of the resulting operators.
Unitarity of the evolutions gives, for every operator $A$,
\[
 \begin{split}
 \|F_{G_L(s),w_g}(A)\|
 &\le \|w_g\|_1\|A\|,\\
 \|\mathcal D_{G_L(s),w_g}(A)\|
 &\le \|W_g\|_1\|A\|,\\
 \|U_L(s)AU_L(s)^\dagger\|&=\|A\|.
 \end{split}
 \qquad
 \|k\|_1:=\int_{\mathbb R}|k(u)|\,du.
\]
Thus, with $F_{G_L,w_g}(\cdot), \mathcal D_{G_L(s),w_g}(\cdot), U_L(s)(\cdot) U_L(s)^\dagger$ denoted by symbol $\mathcal M$, and with
\[
 K_g:=\max\{1,\|w_g\|_1,\|W_g\|_1\},
\]
each of the three maps $\mathcal M$ in the statement satisfies
$\|\mathcal M(A)\|\le K_g\|A\|$.

\medskip
\noindent
\emph{Step 1: evolution of an operator supported inside a ball.}
Write $g_L(x,n;s)$ and $d_L(x,n;s)$ for the shell terms
of $G_L(s)$ and $\mathcal D_L(s)$, respectively.
Choose common constants $J,a>0$ and $0<\alpha\le1$ such that
\begin{equation}\label{eq:appendix-common-shells}
 \|g_L(x,n;s)\|+\|d_L(x,n;s)\|
 \le J e^{-an^\alpha}.
\end{equation}
Both terms are supported on $B_L(x,n)$.

Let $O\in\A_{B_L(x,q)}$, let $m\ge q$, and put
$\ell=m-q$. Denote by $\tau_{t,r}$ the evolution
generated either by $G_L(s)$ with $s$ fixed, or by
the time-dependent generator $\mathcal D_L(t)$.
The Lieb--Robinson bound Eq.~\eqref{eq:uniform-lr},
together with the conditional-expectation estimate
of Ref.~\cite[Lemma~4.2]{NSYspin}, gives
\begin{equation}\label{eq:appendix-dynamics-locality}
 \|\tau_{t,r}(O)-\E_{x,m}\tau_{t,r}(O)\|
 \le C(2q+1)\|O\|
       e^{v|t-r|-\mu\ell^\alpha}.
\end{equation}
Here $C,\mu,v>0$ depend only on $J,a,\alpha$.
The geometric estimates used to obtain this bound are
\[
 |B_L(x,q)|\le2q+1,\qquad
 \#\{y:d_L(z,y)=n\}\le2,
\]
and, for $z\in B_L(x,q)$,
\[
 \sum_{y\notin B_L(x,m)}
 e^{-\mu_0d_L(z,y)^\alpha}
 \le 2\sum_{n\ge\ell}e^{-\mu_0n^\alpha}
 \le C_0e^{-\mu\ell^\alpha},
 \qquad 0<\mu<\mu_0.
\]
These estimates are independent of $L$.
The fermionic version of Eq.
\eqref{eq:appendix-dynamics-locality} is justified below.

Applying Eq. \eqref{eq:appendix-dynamics-locality} to
the evolution generated by $\mathcal D_L(s)$ over
$0\le s\le1$ gives $\tau_{s,0}^{\mathcal D_{L}(s)} (\cdot) = U_L(s)(\cdot) U_L(s)^\dagger$, and
\[
 \|U_L(s)OU_L(s)^\dagger
   -\E_{x,m}[U_L(s)OU_L(s)^\dagger]\|
 \le Ce^v(2q+1)\|O\|e^{-\mu\ell^\alpha}.
\]
This proves the required estimate for conjugation by
$U_L(s)$ when the input is supported on $B_L(x,q)$.

\medskip
\noindent
\emph{Step 2: filtering an operator supported inside a ball.}
The pointwise bounds on $w_g$ and $W_g$ stated in
Section~\ref{subsubsec:locality_and_filtering}
allow us to choose $A_g,b_g>0$ such that
\[
 |w_g(u)|+|W_g(u)|
 \le A_g e^{-b_g\sqrt{|u|}}.
\]
Consequently, for either $k=w_g$ or $k=W_g$,
integration with the substitution $z=\sqrt u$ gives
\[
 \begin{split}
 \int_{|u|>T}|k(u)|\,du
 &\le 2A_g\int_T^\infty e^{-b_g\sqrt u}\,du\\
 &=\frac{4A_g}{b_g^2}
   (1+b_g\sqrt T)e^{-b_g\sqrt T}\\
 &\le C_g e^{-c_g\sqrt T},
 \qquad T\ge0,
 \end{split}
\]
where, for example, one may take
$C_g=8A_g/b_g^2$ and $c_g=b_g/2$.

To treat the two filters together, write
\[
 \mathcal F_{k,s}(O)
 :=\int_{\mathbb R}k(u)\tau_u^{G_L(s)}(O)\,du.
\]
This is $F_{G_L(s),w_g}(O)$ when $k=w_g$, and
$\mathcal D_{G_L(s),w_g}(O)$ when $k=W_g$.
Split the integral into $|u|\le T$ and $|u|>T$.
For $|u|\le T$, use Eq.
\eqref{eq:appendix-dynamics-locality}.
For $|u|>T$, use
\[
 \|\tau_u^{G_L(s)}(O)
   -\E_{x,m}\tau_u^{G_L(s)}(O)\|
 \le2\|O\|.
\]
It follows that
\[
 \begin{split}
 &\|\mathcal F_{k,s}(O)
       -\E_{x,m}\mathcal F_{k,s}(O)\|\\
 &\quad\le
 C(2q+1)\|O\|\|k\|_1
 e^{vT-\mu\ell^\alpha}
 +2C_g\|O\|e^{-c_g\sqrt T}.
 \end{split}
\]
Choose $T=\frac{\mu}{2v}\ell^\alpha$, then the two exponential factors become
\[
 e^{vT-\mu\ell^\alpha}
 =e^{-\mu\ell^\alpha/2},
 \qquad
 e^{-c_g\sqrt T}
 =e^{-c_g\sqrt{\mu/(2v)}\,\ell^{\alpha/2}}.
\]
Combining these estimates with Step 1, we obtain,
for each of the three maps $\mathcal M$,
\begin{equation}\label{eq:appendix-local-input}
 \|\mathcal M(O)-\E_{x,m}\mathcal M(O)\|
 \le C(2q+1)\|O\|
 e^{-c(m-q)^{\alpha/2}},
 \qquad m\ge q.
\end{equation}

\medskip
\noindent
\emph{Step 3: approximation of the given $A_{x,L}$.}
Set
$
 q=\lfloor m/2\rfloor,\,
 A^{[q]}_{x,L}:=\E_{x,q}A_{x,L}.$
Then $A^{[q]}_{x,L}$ is supported on $B_L(x,q)$,
and the assumptions and contractivity give
\[
 \|A^{[q]}_{x,L}\|\le J_A,\qquad
 \|A_{x,L}-A^{[q]}_{x,L}\|
 \le J_Ae^{-bq^\beta}.
\]
Using $\|\mathcal M(A)\|\le K_g\|A\|$ and Eq.
\eqref{eq:appendix-local-input}, we obtain
\[
 \begin{split}
 &\|\mathcal M(A_{x,L})
       -\E_{x,m}\mathcal M(A_{x,L})\|\\
 &\quad\le
 2K_gJ_Ae^{-bq^\beta}
 +C(2q+1)J_Ae^{-c(m-q)^{\alpha/2}}\\
 &\quad\le C'J_Ae^{-c'm^\eta},
 \qquad
 \eta=\min\{\beta,\alpha/2\}.
 \end{split}
\]
Together with the norm bounds at the beginning,
this proves Eq. \eqref{eq:appendix-locality-output}.

The filtered operator $F_{G_L(s),w_g}(A_{x,L})$ itself satisfies these input bounds, so applying the estimate for conjugation by
$U_L(s)$ once more proves the claim of locality of $U_L(s)F_{G_L(s),w_g}(A_{x,L})U_L(s)^\dagger$.

\medskip
\noindent
\emph{Fermionic operators.}
It remains to justify Eq.
\eqref{eq:appendix-dynamics-locality} for fermions.
Let $H_L(u)$ denote either of the even generators above.
For $Y=B_L(x,m)$, let $H_{L,Y}(u)$ contain only its
interaction terms supported inside $Y$, and write
$\tau^Y_{t,r}$ for the resulting evolution.
For $r\le t$, the two evolutions satisfy
\[
 \begin{split}
 \tau_{t,r}(O)-\tau^Y_{t,r}(O)
 =i\int_r^t \tau_{u,r}\!\left(
 [H_L(u)-H_{L,Y}(u),\tau^Y_{t,u}(O)]
 \right)\,du.
 \end{split}
\]
Terms supported entirely outside $Y$ commute with
$\tau^Y_{t,u}(O)$. Every remaining interaction term
is even, so the commutator estimate
Eq. \eqref{eq:uniform-lr} applies for either parity of $O$.
Summing these estimates using
Eq. \eqref{eq:appendix-common-shells} gives
\[
 \|\tau_{t,r}(O)-\tau^Y_{t,r}(O)\|
 \le C(2q+1)\|O\|e^{v|t-r|-\mu(m-q)^\alpha},
\]
with possibly different constants.
Since $\tau^Y_{t,r}(O)\in\A_Y$ and the CAR expectation
$\E_Y^{\rm tr}$ is contractive and fixes $\A_Y$,
\[
 \|\tau_{t,r}(O)-\E_Y^{\rm tr}\tau_{t,r}(O)\|
 \le2\|\tau_{t,r}(O)-\tau^Y_{t,r}(O)\|.
\]
This proves Eq. \eqref{eq:appendix-dynamics-locality};
reversing the time interval covers $t<r$.
An arbitrary $O$ is treated by decomposing it into
its even and odd parts.
Steps 2 and 3 then apply without further changes.
\end{proof}

To apply this lemma to Eq.~\eqref{eq:anchor}, set
$v_{x,L}:=\sum_r v_L(x,r)$.
The assumed shell bounds give
\[
 \|v_{x,L}\|\le CJ_v,\qquad
 \|v_{x,L}-\E_{x,m}v_{x,L}\|
 \le2J_v\sum_{r>m}e^{-ar^\alpha}
 \le C'J_ve^{-c'm^\alpha}.
\]
Thus Lemma~\ref{lem:appendix-locality} gives precisely
the claimed norm and locality bounds for
\[
 A_{x,L}(s)
 =U_L(s)F_{G_L(s),w_g}(v_{x,L})U_L(s)^\dagger.
\]

The locality assumption on the spectral-flow generator
also follows from the same lemma.
For $V_L=\sum_xv_{x,L}$, define
\[
 D_{x,L}(s):=\mathcal D_{G_L(s),w_g}(v_{x,L}).
\]
These operators satisfy Eq.
\eqref{eq:appendix-locality-output}.
With $\E_{x,-1}=0$, the terms
\[
 d_L(x,n;s):=
 (\E_{x,n}-\E_{x,n-1})D_{x,L}(s)
\]
therefore give an exact stretched-exponential shell
representation of
$\mathcal D_{G_L(s),w_g}(V_L)$.
For a fixed upper bound on $J_v$, the resulting
transport estimates are uniform in the perturbation
strength.

On cubic tori in fixed dimension $d$, the same proof
uses $(2q+1)^d$ in Eq. \eqref{eq:appendix-local-input};
the additional polynomial factor does not change
the conclusion.

\subsection{Uhlmann's theorem and purification comparison}
\label{app:purification}

First, we briefly recall Uhlmann's purification theorem~\cite{Uhlmann1976} and the
Fuchs--van de Graaf inequalities~\cite{FuchsVanDeGraaf1999}.

For density matrices $\rho,\omega$ on a finite-dimensional Hilbert space,
we use the root fidelity and trace distance
\[
 F(\rho,\omega):=\|\sqrt\rho\sqrt\omega\|_1,\qquad
 D_{\rm tr}(\rho,\omega):=\tfrac12\|\rho-\omega\|_1.
\]
Thus $F$ is the unsquared fidelity.  A purification of $\rho$ is a unit
vector $\ket{u}\in\mathcal H\otimes\mathcal K$ with
$\Tr_{\mathcal K}\ket{u}\bra{u}=\rho$.
The following two standard results are stated with these conventions
(see \cite[Theorems 3.22 and 3.33]{Watrous}).

\begin{theorem}[Uhlmann's theorem, isometric form]
\label{thm:appendix-uhlmann}
Let $\ket{u}\in\mathcal H\otimes\mathcal K$ and
$\ket{v}\in\mathcal H\otimes\mathcal K'$ be fixed purifications of
$\rho$ and $\omega$, respectively, and suppose
$\dim\mathcal K'\ge\dim\mathcal K$.  Then
\[
 \max_{W^{\dagger}W=\one_{\mathcal K}}
 \bigl|\bra{v}(\one_{\mathcal H}\otimes W)\ket{u}\bigr|
 =F(\rho,\omega),
 \qquad W:\mathcal K\longrightarrow\mathcal K'.
\]
The overall phase of a maximizing isometry can be chosen so that the
overlap itself equals the nonnegative number $F(\rho,\omega)$.
\end{theorem}
This form follows from the usual unitary form by embedding $\mathcal K$
isometrically into $\mathcal K'$ and then optimizing over unitaries on
$\mathcal K'$.  The physical system $\mathcal H$ is unchanged, and only the
purifying system is aligned.

\begin{theorem}[Fuchs--van de Graaf inequalities]
\label{thm:appendix-fvdg}
For density matrices $\sigma,\tau$,
\begin{equation}\label{eq:matrixfacts}
1-F(\sigma,\tau) \leq \frac{1}{2}\norm{\sigma-\tau}_1\le\sqrt{1-F(\sigma,\tau)^2}.
\end{equation}
In particular, $\|\sigma-\tau\|_1\le\delta$ implies
$F(\sigma,\tau)\ge1-\delta/2$.
\end{theorem}

\begin{corollary}[Purification of an approximately decoupled state]
\label{cor:appendix-decoupled-purification}
Let $\ket{\psi_\sigma}\in\mathcal H_B\otimes\mathcal H_C\otimes\mathcal H_F$
be a unit vector, with density matrix $\sigma$, such that
\[
 \|\sigma_{BF}-\sigma_B\otimes\sigma_F\|_1\le\delta.
\]
Fix purifications $\ket{\beta_{BC_1}}$ of $\sigma_B$ and $\ket{\xi_{C_2F}}$ of
$\sigma_F$, and assume
$\dim(\mathcal H_{C_1}\otimes\mathcal H_{C_2})\ge\dim\mathcal H_C$.
There is an isometry
$W:\mathcal H_C\to\mathcal H_{C_1}\otimes\mathcal H_{C_2}$ such that
\begin{equation}\label{eq:appendix-purification-distance}
 \ket{\widetilde\psi_\sigma}:=(\one_B\otimes W\otimes\one_F)\ket{\psi_\sigma},
 \qquad
 \|\ket{\widetilde\psi_\sigma}-\ket{\beta}\otimes\ket{\xi}\|\le\sqrt\delta.
\end{equation}
\end{corollary}
\begin{proof}
Set $\rho=\sigma_{BF}$ and $\omega=\sigma_B\otimes\sigma_F$.
Regrouping tensor factors, $\ket{\psi_\sigma}$ purifies $\rho$ with auxiliary
space $\mathcal H_C$, while $\ket{\beta}\otimes\ket{\xi}$ purifies $\omega$ with
auxiliary space $\mathcal H_{C_1}\otimes\mathcal H_{C_2}$.
Theorem~\ref{thm:appendix-uhlmann} therefore gives an isometry $W$,
including a choice of its overall phase, for which
\[
 \bigl(\bra{\beta}\otimes\bra{\xi}\bigr)\ket{\widetilde\psi_\sigma}
 =F(\rho,\omega).
\]
Both vectors have norm one.  Theorem~\ref{thm:appendix-fvdg} now yields
\[
 \begin{aligned}
 \|\ket{\widetilde\psi_\sigma}-\ket{\beta}\otimes\ket{\xi}\|^2
 &=2-2\operatorname{Re}
       \bigl(\bra{\beta}\otimes\bra{\xi}\bigr)\ket{\widetilde\psi_\sigma}\\
 &=2\bigl(1-F(\rho,\omega)\bigr)\\
 &\le\|\rho-\omega\|_1\le\delta.
 \end{aligned}
\]
Taking square roots proves Eq. \eqref{eq:appendix-purification-distance}.
\end{proof}
In Lemma~\ref{lem:prep-insert}, the abstract purifying spaces have dimensions
$\dim\mathcal H_{C_1}=d_B$ and $\dim\mathcal H_{C_2}=d_Bd_C$.
Their product is $d_B^2d_C\ge d_C$, so the isometry has enough room.
Moreover, purity gives $\operatorname{rank}\sigma_F\le d_Bd_C$, ensuring
that the prescribed purification $\ket{\xi_{C_2F}}$ exists.

Another useful inequality that will be used is the Powers--St{\o}rmer inequality (see Ref.~\cite[Lemma~3.34]{Watrous}):
\begin{equation}\label{eq:powers-stormer}
    \norm{\sqrt\sigma - \sqrt\tau}_2^2 \leq \norm{\sigma-\tau}_1,\qquad \norm{X}_2 = \sqrt{\Tr(X^\dagger X) }.
\end{equation}

\begin{lemma}[Recovery after erasing one region]\label{lem:erasure}
Let \(B,C,F\) be pairwise disjoint finite-dimensional subsystems, and let \(\eta=\ket{u}\bra{u}\) be a pure density matrix on \(B\cup C\cup F\), where \(\ket{u}\in\mathcal H_B\otimes\mathcal H_C\otimes\mathcal H_F\). Let $\eta_{B},\eta_{F},\eta_{BF}$ be its reduced density matrices on regions $B,F,BF$. If
$\|\eta_{BF}-\eta_B\otimes\eta_F\|_1\le\delta$, then there is a channel $\mathcal R:\A_C\to \A_{BC}$ such that
\begin{equation}\label{eq:erasure}
 \|(\mathcal R\otimes\id_F)(\eta_{CF})-\eta\|_1\le4\sqrt\delta.
\end{equation}
\end{lemma}

Because the full state is pure and \(B\) is nearly uncorrelated with \(F\), Uhlmann’s theorem allows us to reorganize \(C\) into two auxiliary parts, approximately separating its entanglement with \(B\) from its entanglement with \(F\). We can then restore the erased region \(B\) by preparing a fresh entangled pair, while leaving the part correlated with \(F\) untouched, and finally map the auxiliary systems back to \(C\).

\begin{proof}
Choose purifications $\ket{\beta_{BB'}}$ of $\eta_B$ and $\ket{\xi_{F'F}}$ of
$\eta_F$.  Thus $\ket{q}=\ket{\beta_{BB'}}\otimes\ket{\xi_{F'F}}$ is a purification of
$\eta_B\otimes\eta_F$.  By Uhlmann's theorem, after padding $F'$ if needed, purification
comparison gives an isometry $V:\mathcal H_C\to \mathcal{H}_{B'F'}$ and a choice of overall phase
such that, for
\[
 \ket{u'}=(\one_B\otimes V\otimes\one_F)\ket{u},
\]
one has
\begin{equation}
 \braket{q}{u'}
 =F(\eta_{BF},\eta_B\otimes\eta_F)
 \ge1-\tfrac12\|\eta_{BF}-\eta_B\otimes\eta_F\|_1
 \ge1-\delta/2.
\end{equation}
Here the overlap is real and nonnegative after the phase choice.  Since
$\ket{u'}$ and $\ket{q}$ are unit vectors,
\begin{equation}
 \|\ket{u'}-\ket{q}\|^2=2-2\braket{q}{u'}\le\delta,
 \qquad
 \bigl\|\ket{u'}\bra{u'}-\ket{q}\bra{q}\bigr\|_1
 \le2\|\ket{u'}-\ket{q}\|\le2\sqrt\delta.
 \label{eq:erasure-purifications}
\end{equation}
The fidelity bound and the purification comparison are the finite matrix
facts recorded in Eq. \eqref{eq:matrixfacts}.

We next describe the recovery channel.  Let $\widetilde B'$ be a copy
of $B'$, and let $\ket{\widetilde\beta_{B\widetilde B'}}$ be the corresponding
copy of $\ket{\beta_{BB'}}$.  Define the reset channel
\[
 \mathcal S:\A_{B'F'}\longrightarrow \A_{B\widetilde B'F'},\qquad
 \mathcal S(Y_{B'F'})
 =\ket{\widetilde\beta}\bra{\widetilde\beta}_{B\widetilde B'}
   \otimes\Tr_{B'}Y_{B'F'}.
\]
Thus $\mathcal S$ discards the $B'$ produced by $V$ and tensors the
reduced state on $F'$ with a newly prepared purification of
$\eta_B$.  After applying $\mathcal S$, identify $\widetilde B'$ with
$B'$ and $\ket{\widetilde\beta}$ with $\ket{\beta}$.  On the ideal marginal this gives
\[
 (\mathcal S\otimes\id_F)
 \bigl(\Tr_B\ket{q}\bra{q}\bigr)
 =\ket{\beta}\bra{\beta}_{BB'}\otimes\ket{\xi}\bra{\xi}_{F'F}
 =\ket{q}\bra{q}.
\]
Partial trace and channels contract the trace norm.  Hence
\begin{equation}
 \left\|(\mathcal S\otimes\id_F)
 \bigl(\Tr_B\ket{u'}\bra{u'}\bigr)-\ket{q}\bra{q}\right\|_1
 \le2\sqrt\delta.
 \label{eq:erasure-reset}
\end{equation}

Finally, map $B'F'$ back to $C$.  For any density matrix $\omega_C$,
define the channel $\mathcal D: \mathcal{A}_{B'F'} \to \mathcal{A}_{C}$
\[
 \mathcal D(Y)=V^{\dagger}YV+\Tr[(\one-VV^{\dagger})Y]\,\omega_C.
\]
The second term makes the map trace preserving outside the range of $V$,
and $\mathcal D(VXV^{\dagger})=X$ for every matrix $X$ on $C$.  Thus
$\mathcal D$ is a left inverse to the isometric channel induced by $V$.
Writing $\mathcal V(X)=VXV^{\dagger}$ for that isometric channel, define
\[
 \mathcal R=(\id_B\otimes\mathcal D)\circ\mathcal S\circ\mathcal V:
 \A_C\longrightarrow \A_{BC}.
\]

Apply $\id_B\otimes\mathcal D\otimes\id_F$ to Eq.
\eqref{eq:erasure-reset}.  By the left-inverse property, the decoder maps
$\ket{u'}\bra{u'}$ exactly to $\eta$.  The triangle inequality,
contractivity, Eq. \eqref{eq:erasure-reset}, and Eq.
\eqref{eq:erasure-purifications} therefore give
\[
 \begin{aligned}
 \| (\mathcal R\otimes\id_F)(\eta_{CF})-\eta\|_1
 &\le
 \left\|(\mathcal S\otimes\id_F)
 \bigl(\Tr_B\ket{u'}\bra{u'}\bigr)-\ket{q}\bra{q}\right\|_1
 +\bigl\|\ket{q}\bra{q}-\ket{u'}\bra{u'}\bigr\|_1 \\
 &\le4\sqrt\delta.
 \end{aligned}
\]

\end{proof}

\begin{fermionlemma}[Parity-covariant erasure recovery]
\label{lem:fermion-erasure}
The conclusion of Lemma~\ref{lem:erasure} holds when the pure
reference vector has definite total parity. The recovery channel
$\mathcal R:\A_C\to\A_{BC}$ can be chosen parity-covariant, and
its exterior extension is the extension of
Lemma~\ref{lem:fermion-channel}. The error remains $4\sqrt\delta$.
\end{fermionlemma}
\begin{proof}
Corollary~\ref{cor:fermion-decoupled-purification} supplies an even
comparison isometry $V:C\to C_1C_2$, an even purification
$\ket{\beta_{BC_1}}$, and a purification $\ket{\xi_{C_2F}}$ with
the parity of the reference vector. Encoding by $V$, resetting
$C_1$ to $\ket{\beta}\bra{\beta}$, and decoding by
\[
 \mathcal D(Y)=V^\dagger YV+
 \Tr[(\one-VV^\dagger)Y]\omega_C
\]
are parity-covariant when $\omega_C$ is even.
Lemma~\ref{lem:fermion-channel} extends the three maps to the
exterior. The two trace-distance comparisons in
Lemma~\ref{lem:erasure} each contribute $2\sqrt\delta$.
\end{proof}

Finally, we present a short lemma for construction of a minimal unitary for rotations between two vectors.

\begin{lemma}[Rotation between two vectors].\label{lem:rotation} Let \(\ket u,\ket v\) be normalized vectors in a finite-dimensional Hilbert space. There exist a unitary \(R\) and a phase \(\alpha\) such that
\[
R\ket u=e^{i\alpha}\ket v,
\qquad
\|R-\mathbf1\|
=\sqrt{2\bigl(1-|\langle u|v\rangle|\bigr)}
\le\sqrt{2\bigl(1-|\langle u|v\rangle|^2\bigr)}.
\]
Moreover, \(R\) can act as the identity on
\(\operatorname{span}\{\ket u,\ket v\}^{\perp}\).
\end{lemma}

\begin{proof} Choose the phase of \(\ket v\) so that
\(\langle u|v\rangle=\cos\theta\ge0\), with \(0\le\theta\le\pi/2\).
If \(\theta=0\), take \(R=\mathbf1\). Otherwise, take the rotation through angle \(\theta\) sending \(\ket u\) to \(\ket v\) within their two-dimensional span, extended by the identity on its orthogonal complement. More concretely, write
\[
\ket v=\cos\theta\,\ket u+\sin\theta\,\ket w,
\qquad \langle u|w\rangle=0.
\]
In the orthonormal basis \((\ket u,\ket w)\), we choose
\[
R=
\begin{pmatrix}
\cos\theta&-\sin\theta\\
\sin\theta&\cos\theta
\end{pmatrix}.
\]

Its eigenvalues on that span are \(e^{\pm i\theta}\), so
\[
\|R-\mathbf1\|
=2\sin(\theta/2)
=\sqrt{2(1-\cos\theta)}
\le\sqrt{2(1-\cos^2\theta)}.
\]
\end{proof}

\subsection{Marginal adjustment and local repair}
\label{app:marginal-repair}

The following finite-dimensional channel estimates apply to arbitrary
density matrices and require no Hamiltonian or gap assumption.
Section~\ref{subsec:energy-repair} uses them to repair a closing seam (and Section~\ref{sec:fermion-energy} for fermions).

Let \(E,B,C\) be pairwise disjoint finite-dimensional subsystems and let \(\rho=\rho_{EBC}\) be a density matrix on their joint Hilbert space. Write $S(\rho) = -\Tr(\rho\ln\rho)$,
$I(E:C)_\rho=S(\rho_E)+S(\rho_C)-S(\rho_{EC})$, and
$I(E:C\mid B)_\rho=S(\rho_{EB})+S(\rho_{BC})-S(\rho_B)-S(\rho_{EBC})$.
There is a recovery channel $\mathcal R:B\to BC$ such that
\begin{equation}
 \mathcal R(\rho_B)=\rho_{BC},\qquad
 \|(\operatorname{id}_E\otimes\mathcal R)(\rho_{EB})-\rho_{EBC}\|_1
 \le2\sqrt{I(E:C\mid B)_\rho}.
 \label{eq:recovery-energy}
\end{equation}
Estimate Eq. \eqref{eq:recovery-energy} follows from
\cite[Corollary 4.1, Eq.~(101), and Remark 2.4(2)]{Junge_2018}:
the averaged rotated Petz map recovers $\rho_{BC}$ from $\rho_B$ exactly
and has root fidelity at least $e^{-I(E:C\mid B)/2}$ with $\rho_{EBC}$.
The Fuchs-van de Graaf inequalities in
Theorem~\ref{thm:appendix-fvdg}, together with $1-e^{-u}\le u$,
give the trace bound in Eq. \eqref{eq:recovery-energy}.
For the case that $\rho_B$ is not full-rank, the recovery map $\mathcal R$ is only trace preserving on the support of $\rho_B$ 
, and we can extend
the map to
a CPTP map in the whole Hilbert space $\mathcal H_B$ by $X\mapsto\mathcal R(PXP)+\Tr[(\one-P)X]\omega_{BC}$, $P$ the projection to $\supp\rho_B$,
with any fixed density $\omega_{BC}$ {\cite[discussion preceding Eq.~(5),(9) and Remark~2.2]{Junge_2018}}. Positivity implies that $\rho_{EB}$
is supported on $\one_E\otimes P$. 
Both \(\rho_B\) and \(\rho_{EB}\) are supported on \(P\) in the \(B\) factor, so the extension leaves their recovered states unchanged.

The recovery channel needs exactly the reference marginal $\rho_B$ as input.
The following elementary channel corrects a nearby input marginal before
recovery.
\begin{lemma}[Exact marginal adjustment]
\label{lem:marginal-adjustment-energy}
Let $\sigma_B,\tau_B$ be density matrices and
$\epsilon=\|\sigma_B-\tau_B\|_1$.  There is a channel
$\mathcal L_B:\mathcal{B}(\mathcal{H}_B)\to \mathcal{B}(\mathcal{H}_B)$ with $\mathcal L_B(\sigma_B)=\tau_B$ such that every
extension $\sigma_{EB}$ satisfies
\begin{equation}
 \|(\operatorname{id}_E\otimes\mathcal L_B)(\sigma_{EB})-\sigma_{EB}\|_1
 \le\sqrt{2\epsilon}+\epsilon/2.
 \label{eq:marginal-adjustment-energy}
\end{equation}
For fermionic states $\sigma_B,\tau_B$ under the same assumption, the channel
$\mathcal L_B$ can be chosen to satisfy parity-covariant condition in Eq.~\eqref{eq:parity_covariant_channel}, with its action on the exterior region $E$ understood through
the CAR extension of Lemma~\ref{lem:fermion-channel}. 
\end{lemma}
\begin{proof}
To construct $\mathcal L_B$, we define $D_+=(\sigma_B-\tau_B)_+$,
and $t=\operatorname{Tr}D_+=\epsilon/2$ \footnote{For a Hermitian operator
$O=\sum_j\lambda_j\ket{j}\bra{j}$, its positive part is
$O_+=\sum_{\lambda_j>0}\lambda_j\ket{j}\bra{j}$, so
$O=O_+-(-O)_+$.  If $\Tr O=0$, then
$\Tr O_+=\tfrac12\|O\|_1$.  Here $O=\sigma_B-\tau_B$ is traceless.}. 
And let {$\omega=\tau_B+D_+$, which satisfies $\omega\geqslant \sigma_B$} and $\omega\ge\tau_B$,
and $F=\sqrt{\tau_B}\,\omega^{-1/2}$, where the inverse is taken on the
support of $\omega$.
Then $F^{\dagger}F\le1$, $F\sigma_BF^{\dagger}\le\tau_B$, and
$p=1-\operatorname{Tr}F\sigma_BF^{\dagger}\le t$.
To see the last inequality, using, on the support of $\omega$,
$1-F^{\dagger}F=\omega^{-1/2}D_+\omega^{-1/2}$ and
$\omega^{-1/2}\sigma_B\omega^{-1/2}\le1$, then cyclicity gives
\[
 p=1-\Tr(\sigma_BF^\dagger F)=\Tr(\sigma_B(1-F^\dagger F))=\Tr\!\left(D_+\omega^{-1/2}\sigma_B\omega^{-1/2}\right)
 \le\Tr D_+=t.
\]
For $p>0$ put $\nu=(\tau_B-F\sigma_BF^{\dagger})/p$ and set
\[
 \mathcal L_B(X)=FXF^{\dagger}+\operatorname{Tr}[(1-F^{\dagger}F)X]\nu.
\]
For $p=0$, choose any density matrix $\nu$.
The two terms define a completely positive trace-preserving map, and
$\mathcal L_B(\sigma_B)=F\sigma_BF^{\dagger}+p\nu=\tau_B$.
Moreover
\[
 \operatorname{Tr}[\sigma_B(F-1)^{\dagger}(F-1)]
 \le\|\sqrt{\tau_B}-\sqrt\omega\|_2^2\le t,
\]
For the first inequality, the matrix
$C=\omega^{-1/2}\sqrt{\sigma_B}$, defined on the support of $\omega$, is a
contraction and satisfies $\sqrt{\sigma_B}=\sqrt\omega C$.  Thus
$(F-1)\sqrt{\sigma_B}=(\sqrt{\tau_B}-\sqrt\omega)C$.
The second inequality \footnote{Recall that $\norm{X}_2 = \sqrt{\Tr(X^{\dagger} X)}$ for an operator $X$.} follows because $\sqrt\omega\ge\sqrt{\tau_B}$ and
$\operatorname{Tr}\sqrt\omega\sqrt{\tau_B}\ge1$.

Let $\ket{v}\in\mathcal H_E\otimes\mathcal H_B\otimes\mathcal H_R$
purify $\sigma_{EB}$ and put $\ket{w}=(1_E\otimes F\otimes1_R)\ket{v}$.
The preceding estimate gives $\|\ket{w}-\ket{v}\|\le\sqrt t$, while $F^{\dagger}F\le1$ gives
$\|\ket{w}\|\le\|\ket{v}\|=1$.  Using
$\ket{w}\bra{w}-\ket{v}\bra{v}
=(\ket{w}-\ket{v})\bra{w}+\ket{v}(\bra{w}-\bra{v})$
and $\|\ket{a}\bra{b}\|_1=\|\ket{a}\|\,\|\ket{b}\|$, we obtain
\[
 \bigl\|\ket{w}\bra{w}-\ket{v}\bra{v}\bigr\|_1
 \le(\|\ket{w}\|+\|\ket{v}\|)\|\ket{w}-\ket{v}\|\le2\sqrt t.
\]
The positive output of the replacement branch is
\[
 R_{EB}:=\Tr_B\!\left[(1_E\otimes(1-F^{\dagger}F))\sigma_{EB}\right]\otimes\nu
 \ge0,\qquad
 \|R_{EB}\|_1=\Tr R_{EB}=p\le t.
\]
Thus $(\operatorname{id}_E\otimes\mathcal L_B)(\sigma_{EB})
=\Tr_R\ket{w}\bra{w}+R_{EB}$.  Contractivity under partial trace
and the triangle inequality now give
\[
 \begin{aligned}
 &\|(\operatorname{id}_E\otimes\mathcal L_B)(\sigma_{EB})-\sigma_{EB}\|_1\\
 &\quad\le
 \bigl\|\Tr_R(\ket{w}\bra{w}-\ket{v}\bra{v})\bigr\|_1
 +\|R_{EB}\|_1
 \le2\sqrt t+t=\sqrt{2\epsilon}+\epsilon/2,
 \end{aligned}
\]
which proves Eq. \eqref{eq:marginal-adjustment-energy}.

For fermionic case, the above construction carries over, and the obtained channel is parity-covariant.
\end{proof}

\begin{lemma}[One-seam repair]
\label{lem:seam-repair-energy}
Let $E,B,C$ be disjoint finite-dimensional systems and set
$A=BC$. Let $\rho_{EBC}$ be the desired reference marginal and let
$\sigma_{EBC}$ be the input marginal.  Put
\[
 \epsilon=\|\rho_{EB}-\sigma_{EB}\|_1,
 \qquad \eta=I(E:C\mid B)_\rho.
\]
There is a channel $\Lambda_A$ acting only on $A=BC$ such that
$\zeta_{EBC}=(\operatorname{id}_E\otimes\Lambda_A)(\sigma_{EBC})$ satisfies
\begin{equation}
 \zeta_A=\rho_A,\qquad
 \|\zeta_{EBC}-\rho_{EBC}\|_1\le \Delta_{\rm seam}(\epsilon,\eta)
 :=2\sqrt\eta+\sqrt{2\epsilon}+\tfrac32\epsilon.
 \label{eq:seam-repair-energy}
\end{equation}

For the fermionic states $\rho_{EBC},\sigma_{EBC}$, the repair channel $\Lambda_A$ can be chosen
parity-covariant as in Eq.~\eqref{eq:parity_covariant_channel} with its action on exterior region $E$ understood through
the extension of Lemma~\ref{lem:fermion-channel}. 
\end{lemma}
\begin{proof}
Use Lemma~\ref{lem:marginal-adjustment-energy} with
$\tau_B=\rho_B$ and define
\[
 \chi_{EB}=(\operatorname{id}_E\otimes\mathcal L_B)(\sigma_{EB}),
 \qquad
 \Lambda_A=\mathcal R_{B\to BC}\circ\mathcal L_B\circ\Tr_C,
\]
with $\mathcal{R}_{B\to BC}$ the recovery channel mentioned in Eq.~\eqref{eq:recovery-energy}.
The adjusted marginal obeys $\chi_B=\rho_B$, and therefore
\[
 \zeta_A=\mathcal R_{B\to BC}(\chi_B)=\mathcal R_{B\to BC}(\rho_B)=\rho_{BC}=\rho_A.
\]
Moreover, Lemma~\ref{lem:marginal-adjustment-energy} implies that
\[
 \|\chi_{EB}-\rho_{EB}\|_1
 \le\|\chi_{EB}-\sigma_{EB}\|_1+\epsilon
 \le\sqrt{2\epsilon}+\tfrac32\epsilon.
\]
Contractivity of trace norm under $\operatorname{id}_E\otimes\mathcal R_{B\to BC}$
and Eq. \eqref{eq:recovery-energy} now give
\begin{align*}
 \|\zeta_{EBC}-\rho_{EBC}\|_1
 &\le
 \|(\operatorname{id}_E\otimes\mathcal R_{B\to BC})
       (\chi_{EB}-\rho_{EB})\|_1\\
 &\quad+
 \|(\operatorname{id}_E\otimes\mathcal R_{B\to BC})(\rho_{EB})
       -\rho_{EBC}\|_1
 \le \Delta_{\rm seam}(\epsilon,\eta).
\end{align*}
The final assertion follows because a trace-preserving channel on $A$
preserves every marginal on the complementary subsystem.

For fermions, replace $\mathcal R_{B\to BC}$ by
\[
 \mathcal R_{\rm cov}(X)
 =\tfrac12\bigl[
 \mathcal R_{B\to BC}(X)+
 \Gamma_{BC}\mathcal R_{B\to BC}(\Gamma_BX\Gamma_B)\Gamma_{BC}
 \bigr].
\]
This makes $\mathcal R_{\rm cov}$ parity-covariant. Using the parity-covariant adjustment from
Lemma~\ref{lem:marginal-adjustment-energy}, the same
proof then applies with CAR extensions.
\end{proof}

\subsection{Entropy and exchange of boundary blocks}
\label{app:boundary-selection}

The following two lemmas apply to arbitrary translation invariant states
on finite rings. They provide the sufficient conditions for repairing the boundary part of shorter-ring state in
Section~\ref{subsec:energy-repair} (and Section~\ref{sec:fermion-energy} for fermions).

For a translation invariant state on a ring with an on-site dimension $d$, let
$S(j)$ denote the entropy of any length-$j$ interval and put
$s_j=S(j)-S(j-1)$, with $S(0)=0$.  Strong subadditivity gives
$s_{j+1}\le s_j$.  Subadditivity and the Araki--Lieb inequality give
$-\log d\le s_j\le\log d$.

\begin{lemma}[Entropy-band selection]
\label{lem:entropy-selection-energy}
Let $\rho$ be a translation invariant state on a ring of length $L$
with on-site Hilbert space dimension $d$. Consider six consecutive blocks
$E_-B_-C_-C_+B_+E_+$ with positive integer lengths
$|E_\pm|=R_0$, $|B_\pm|=b$, and $|C_\pm|=c$.
Set $E=E_-E_+$, $B=B_-B_+$, $C=C_-C_+$, and suppose
$t=2b+2c+2R_0\le L$. Then
\begin{equation}
 I(E:C\mid B)_\rho\le2R_0(s_{b+1}-s_t).
 \label{eq:entropy-selection-energy}
\end{equation}
Let $m\ge1$ be an integer, and let $(b_i,c_i)_{i=1}^m$ be predetermined
pairs of positive integer block widths.  Set $t_i=2b_i+2c_i+2R_0$.  If
\[
b_{i+1}+1\ge t_i\quad(1\le i<m),\qquad t_m\le L,
\]
then some index $i$ satisfies
\begin{equation}\label{eq:s_difference_bound}
 s_{b_i+1}-s_{t_i}\le\frac{2\log d}{m}.
\end{equation}
For this index, the six-block partition with $b=b_i$ and $c=c_i$ satisfies
\begin{equation}\label{eq:cmi_bound}
 I(E:C\mid B)_\rho\le\frac{4R_0\log d}{m}.
\end{equation}
The conclusion also holds for a translation invariant fermion density matrix $\rho$. 
\end{lemma}

\begin{proof}
For two disjoint systems $F,G$, we use \emph{quantum conditional entropy} $S(F\mid G)_\rho=S(\rho_{FG})-S(\rho_G)$.
The conditional-entropy chain rule and strong subadditivity \cite[Theorem~5.36]{Watrous} give
\[
 \begin{aligned}
 S(E\mid B)_\rho
 &=S(E_-\mid B_-B_+)_\rho
   +S(E_+\mid E_-B_-B_+)_\rho\\
 &\le S(E_-\mid B_-)_\rho+S(E_+\mid B_+)_\rho\\
 &=2[S(b+R_0)-S(b)].
 \end{aligned}
\]
The last equality uses translation invariance of $\rho$.  Since $BC$ and $EBC$
are contiguous intervals, $S(E\mid BC)_\rho=S(t)-S(t-2R_0)$.
Thus monotonicity of the entropy increments (i.e., $s_{j+1}\leqslant s_j$) gives
\[
 \begin{aligned}
 I(E:C\mid B)_\rho
 &=S(E\mid B)_\rho-S(E\mid BC)_\rho\\
 &\le2[S(b+R_0)-S(b)]-[S(t)-S(t-2R_0)]\\
 &\le2R_0(s_{b+1}-s_t),
 \end{aligned}
\]
which proves Eq. \eqref{eq:entropy-selection-energy}.  

For the increment ranges $[b_i+1,t_i]$, the monotonicity gives
\[
 \sum_{i=1}^m(s_{b_i+1}-s_{t_i})
 \le s_{b_1+1}-s_{t_m}\le2\log d.
\]
Hence, we see that at least one summand is not greater than $2\log d/m$, which proves Eq. \eqref{eq:s_difference_bound}. Then applying Eq.
\eqref{eq:entropy-selection-energy} proves Eq. \eqref{eq:cmi_bound}.

For fermions, the strong subadditivity holds for the CAR marginals under our even parity superselection rule, and
the same entropy argument applies.
\end{proof}

The {next lemma supplies a small $\epsilon=\|\rho_{EB}-\sigma_{EB}\|_1$ required in Lemma \ref{lem:seam-repair-energy}. In the context of this lemma, the relevant marginals are $\rho_{EB}=q_c$ and
$\sigma_{EB}=\mathsf S q_r\mathsf S^\dagger$.} 

Fix integers $b,R_0\ge1$ and put $\ell=b+R_0$.
For an integer $u\ge b$ with $2\ell+2u\le L$, let $q_u$ be
the joint marginal of the two forward-ordered blocks
\[
 [0,\ell-1]\quad\hbox{and}\quad
 [\ell+2u,2\ell+2u-1],
\]
so the intervening region contains $2u$ sites. We call such a $q_u$ a \emph{pair marginal}.  Translation invariance makes $q_u$
independent of the starting site.  Using the fixed onsite bases to identify both block spaces with
$\mathbb C^{d^\ell}$, let $\mathsf S$ exchange the two factors:
\[
 \mathsf S(v\otimes w)=w\otimes v.
\]
This preserves the internal site order within each block.

\begin{lemma}[Block exchange]
\label{lem:oriented-recurrence-energy}
Suppose $\ell=b+R_0$ is even.  For every $\alpha>0$ there is a finite number
$U=U(d,b,R_0,\alpha)\ge b$ such that every translation invariant state on a
ring of length $L\ge2\ell+2U$ admits integers $c,r$, such that
\begin{equation}
 b\le c<r\le U,\qquad \|q_c-\mathsf S q_r\mathsf S^{\dagger}\|_1\le\alpha.
 \label{eq:oriented-recurrence-energy}
\end{equation}
\end{lemma}

\begin{proof}
Put $d_{\rm blk}=d^\ell$. Choose an integer $k\ge4$ divisible by four
and $\delta>0$ such that
\[
 \frac{32d_{\rm blk}^3}{k}\le\frac\alpha2,
 \qquad (16d_{\rm blk}^3+2)\delta\le\frac\alpha2.
\]
We reduce from the desired comparison to a finite selection of blocks. The choices above ensure that the errors in these reductions
sum to at most $\alpha$.

\smallskip\noindent\emph{Step 1. Reduce the comparison to one nearly symmetric pair marginal.}
To prove Eq. \eqref{eq:oriented-recurrence-energy}, it suffices to find
$b\le c<r\le U$ and a two-block density matrix $\omega$ such that
\begin{equation}
 \|q_c-\omega\|_1,\ \|q_r-\omega\|_1\le\delta,
\end{equation}
and
\begin{equation}\label{eq:omega_swap_estimate}
    \norm{\omega-\mathsf S\omega\mathsf S^{\dagger}}_1
 \le d_{\rm blk}^3\left(\frac{32}{k}+16\delta\right)
\end{equation}
Indeed, the triangle inequality and unitary invariance of the trace norm give
\[
 \begin{aligned}
 \|q_c-\mathsf S q_r\mathsf S^{\dagger}\|_1
 &\le\|q_c-\omega\|_1
   +\|\omega-\mathsf S\omega\mathsf S^{\dagger}\|_1
   +\|\mathsf S(\omega-q_r)\mathsf S^{\dagger}\|_1\\
 &\le2\delta+d_{\rm blk}^3\left(\frac{32}{k}+16\delta\right)
 \le\alpha.
 \end{aligned}
\]
In Step~4 we will select $k$ disjoint $\ell$-site
blocks $J_1,\ldots,J_k$ ordered by their starting positions along the ring coordinate,
such that for any pairs $1\leq i<j\leq k$ and $1\leq p<q\leq k$,
$\|\rho_{J_iJ_j}-\rho_{J_pJ_q}\|_1\le\delta$. We then set $\omega=\rho_{J_1J_2}$,
so that
\[
 \|\rho_{J_iJ_j}-\omega\|_1\le\delta
 \qquad (1\le i<j\le k).
\]
Assuming such blocks have been selected, Steps~2--3
prove the required exchange estimate Eq. \eqref{eq:omega_swap_estimate} for $\omega$.
The pairs $(J_1,J_2)$ and $(J_1,J_3)$ then determine
$c$ and $r$, respectively.

\smallskip\noindent\emph{Step 2. Reduce exchange symmetry to scalar estimates.}
We now assume the existence of the $k$ blocks described in Step 1, and reduce inequality \eqref{eq:omega_swap_estimate} into Eq. \eqref{eq:omega_swap_reduce}.
To prove the required bound on
$Y=\omega-\mathsf S\omega\mathsf S^{\dagger}$, it suffices to show that
\begin{equation}\label{eq:omega_swap_reduce}
 |\Tr[\omega(A\otimes B-B\otimes A)]|
 \le\frac{32}{k}+16\delta
\end{equation}
for all Hermitian operators $A,B$ on one block with norm at most 1. To see this, choose
a Hermitian Hilbert--Schmidt orthonormal basis
$\{H_a\}_{a=1}^{d_{\rm blk}^2}$ {and decompose $Y$ into this basis}. Each $H_a$ has operator norm at most one,
and the coefficient of $H_a\otimes H_b$ in $Y$ is
\[
 \Tr[Y(H_a\otimes H_b)]
 =\Tr[\omega(H_a\otimes H_b-H_b\otimes H_a)].
\]
The scalar estimate Eq. \eqref{eq:omega_swap_reduce} therefore bounds all $d_{\rm blk}^4$ coefficients.
By the Cauchy-Schwarz inequality,
\[
 \|Y\|_1\le d_{\rm blk}\|Y\|_2
 \le d_{\rm blk}^3\left(\frac{32}{k}+16\delta\right).
\]

\smallskip\noindent\emph{Step 3. Reduce the scalar estimates to positivity on $k$ blocks.}
Suppose the $k$ blocks described in Step~1 have been found, and label
them $1,\ldots,k$ in spatial order. {Now we will prove Eq. \eqref{eq:omega_swap_reduce}.} 

For fixed $A,B$, set
\[
 X=A+iB,\qquad a_0=\Tr[\omega(X^{\dagger}\otimes X)],\qquad
 G_{jl}=\Tr(\rho X_j^{\dagger}X_l).
\]
Here $X_j$ is the same operator as $X$ which acts on the $j$th block. Since
\[
 \operatorname{Im}a_0=\Tr[\omega(A\otimes B-B\otimes A)],
\]
it suffices to bound $|\operatorname{Im}a_0|$. The common origin of
the pair marginals gives the needed constraint: for every $v\in\mathbb C^k$,
\[
 v^{\dagger}Gv=\Tr\!\left[\rho
    \left(\sum_jv_jX_j\right)^{\dagger}\left(\sum_lv_lX_l\right)\right]\ge0.
\]
Moreover, $\|X\|\le2$ gives $0\le G_{jj}=\Tr(\rho X_j^\dagger X_j)\leq \norm{X}^2\le4$, so $|a_0|=|\Tr[\omega X^\dagger\otimes X]|\le4$, and
\[
|G_{jl}-a_0| = |\Tr[(\rho_{J_j J_l} - \omega) X^\dagger\otimes X ]| \leq \norm{\rho_{J_jJ_l} - \omega}_1\cdot\norm{X^\dagger\otimes X}\leq 4\delta
\]
for $j<l$.

To extract the imaginary part, we only need the positivity of the matrix $G$:
take $h=k/4$ and let $w$ be the following vector
\[
 w=\frac1{\sqrt k}
 (\underbrace{1,\ldots,1}_{h},
  \underbrace{i,\ldots,i}_{h},
  \underbrace{-1,\ldots,-1}_{h},
  \underbrace{-i,\ldots,-i}_{h})^T.
\]
and $\bar w$ its complex conjugate.
Then we use the positivity condition $w^\dagger G w \geq 0$ and $(\bar w)^\dagger G \bar w \geq 0$: let $G_{jl} = a_{0} + e_{jl}$ with $|e_{jl}|\leq 4\delta$, and use
\begin{equation}
    v^\dagger G v = \sum_{j} |v_j|^2G_{jj} + 2\operatorname{Re}[\sum_{j<l} \bar v_j G_{jl} v_l ],
\end{equation}
we get 
\begin{equation}
    w^\dagger G w = \frac{1}{k}\sum_{j} G_{jj} + 2\operatorname{Re} [\sum_{j<l} \bar w_j  a_0 w_l] + 2\operatorname{Re}[\sum_{j<l} \bar w_j e_{jl} w_l].
\end{equation}
Then let $R_{+} = 2\operatorname{Re}[\sum_{j<l} \bar w_j e_{jl} w_l],\, R_-=2\operatorname{Re}[\sum_{j<l} w_j e_{jl} \bar w_l]$ and using
\[
 \sum_{j<l}\bar w_jw_l=-\frac12+\frac{ik}{8},\quad R_{\pm} 
\le2\sum_{j<l}|w_j||w_l||e_{jl}|
\le2\cdot\frac{k(k-1)}2\cdot\frac1k\cdot4\delta
=4\delta(k-1).
\]
and writing $D=k^{-1}\sum_jG_{jj}$, we get
\[
 0\le D-\operatorname{Re}a_0\mp\frac k4\operatorname{Im}a_0+R_\pm,
 \qquad 0\le D\le4.
\]
Consequently,
\[
 \frac k4|\operatorname{Im}a_0|
 \le D-\operatorname{Re}a_0+4\delta(k-1)
 \le8+4\delta(k-1),
\]
which proves the estimate in Eq.~\eqref{eq:omega_swap_reduce}. Therefore, to complete the proof, it remains only to construct
the $k$ blocks and check their separations.

\smallskip\noindent\emph{Step 4. Obtain the blocks from a finite cover.}
We now show that the $k$ blocks can be found if the ring is sufficiently long.

Each two-block marginal acts on
$\mathbb C^{d_{\rm blk}}\otimes\mathbb C^{d_{\rm blk}}$, of dimension
$d^{2\ell}$, and all these marginals form a compact set. We can therefore cover it by finitely many
sets $\mathcal C_1,\ldots,\mathcal C_{N_{\rm col}}$
such that
\[
 \|\sigma-\tau\|_1\le\delta
 \qquad\text{whenever }\sigma,\tau\in\mathcal C_a.
\]
Now for a collection of candidate blocks, and for each pair of candidate blocks $J_j,J_l$, we give it a label \(\chi(j,l)\in\{1,\ldots,N_{\rm col}\}\) indicating which set $\mathcal C_a$ contains \(\rho_{J_jJ_l}\). If several sets contain it, choose the smallest label. We take the following selection strategy. Choose the leftmost block, say $J_{i_1}$, and find its most frequent label with other blocks, say $\gamma_1$, and pick up all $J_j$ satisfying $\chi(i_1, j)=\gamma_1$ with $j>i_1$. Then select iteratively in these kept $J_j$ blocks: take the leftmost block $J_{i_2}$, and find its most frequent label with other blocks $J_l,\, l>i_2$, denoted as $\gamma_2$, and then pick up the blocks $\chi(i_2, l)=\gamma_2$. Continue this procedure, we can get a sequence of blocks $J_{i_s}$ such that $\chi(i_s, i_t)=\gamma_s,\, i_t>i_s$. Keep doing this until we select $s_0 = N_{\rm col} (k-1)+1$ blocks, which ensures that there are at least $k$ blocks in the same set $\mathcal C_a$. Then the defining property of these $\mathcal C_a$ shows these $k$ blocks satisfy the requirement of Step 1.

In above strategy, we need to select $s_0$ blocks, which requires the collection of candidate blocks contains sufficiently many blocks. More concretely, let $T=\sum_{j=0}^{s_0} N_{\rm col}^j$, then as long as the collection of candidate blocks contains more than $T$ blocks, we can get $s_0$ blocks from the above strategy. For instance, after first selection, the remaining blocks are at least $(T-1)/N_{\rm col} = \sum_{j=0}^{s_0-1} N_{\rm col}^j$. After $m$-th selection, we are left with at least $\sum_{j=0}^{s_0-m} N_{\rm col}^j$ blocks.

Based on above analysis, the number of spatial
blocks needed for the selection is 
$ T=\sum_{j=0}^{s_0}N_{\rm col}^{j}. $ So we set $p=\ell+2b$ and take $U=[p(T-1)-\ell]/2$. Then $U\geq b$ is an integer, and $L\ge2\ell+2U$ allows the blocks
\[
 J_j=\{(j-1)p,\ldots,(j-1)p+\ell-1\},\qquad 1\le j\le T,
\]
to fit in the ring without wrapping around the ring.

Finally, in the $k$ blocks, we pick up the first three blocks, and take their left endpoints as $z_1<z_2<z_3$. Then define
\[
 2c=z_2-z_1-\ell,\qquad 2r=z_3-z_1-\ell.
\]
The $z_1, z_2, z_3$ are multiples of the even integer $p$, so $c,r$ are
integers; their spacing and total span give $b\le c<r\le U$.
Translation invariance identifies the first--second and first--third
pair marginals with $q_c,q_r$. Both are within $\delta$ of $\omega$,
so step~1 proves the claimed comparison. The construction of $U$
depends only on $d,b,R_0,\alpha$.
\end{proof}

Now we extend Lemma~\ref{lem:oriented-recurrence-energy} to the fermionic case. For two identical block spaces define the fermionic exchange
\begin{equation}\label{eq:fermion-exchange}
 \mathsf S_f(\ket{v}\otimes \ket{w})=(-1)^{ab}\ket{w}\otimes \ket{v} \quad \text{ for }
 \quad\Gamma \ket v=(-1)^a\ket{v},\quad\Gamma \ket w=(-1)^b\ket w.
\end{equation}

\begin{fermionlemma}[Fermionic block exchange]
\label{lem:fermion-oriented-recurrence}
Let $\rho$ be translation-invariant fermion state. Under the size and
width conditions of Lemma~\ref{lem:oriented-recurrence-energy},
the conclusion holds with $\mathsf S$ replaced by $\mathsf S_f$:
\[
 b\le c<r\le U,\qquad
 \|q_c-\mathsf S_fq_r\mathsf S_f^\dagger\|_1\le\alpha.
\]
The density matrices $q_u$ are the fermionic restrictions to the two blocks, in the order of first block before the second. The bound
$U=U(d,b,R_0,\alpha)$ remains independent of $L$ and $\rho$.
\end{fermionlemma}

\begin{proof}
Use the choices of $k,\delta,d_{\rm blk}$ and the
block-selection argument in
Lemma~\ref{lem:oriented-recurrence-energy}.
This gives $k$ ordered blocks whose pair marginals
are within $\delta$ of a common density $\omega$.
We only modify the exchange estimate.

For any homogeneous block operator $X$, let $X_j$
be its CAR copy on block $j$.
The same positivity argument for
$G_{ij}=\langle X_i^\dagger X_j\rangle_\rho$ gives
\[
 \bigl|\operatorname{Im}
 \langle X_1^\dagger X_2\rangle_\omega\bigr|
 \le \|X\|^2(8/k+4\delta).
\]
For even Hermitian operators $A,B$ with norm smaller than 1, the spin
argument applies unchanged. For odd $A,B$, fermionic
exchange instead gives
\[
 \mathsf S_f^\dagger A_1B_2\mathsf S_f
 =A_2B_1=-B_1A_2.
\]
Thus we must bound
$\langle A_1B_2+B_1A_2\rangle_\omega$.
Set $X_\pm=A\pm B$. Since
$X_{\pm,1}X_{\pm,2}$ is anti-Hermitian and
$\|X_\pm\|\le2$, the preceding estimate and
\[
 A_1B_2+B_1A_2
 =\tfrac12\bigl(
 X_{+,1}X_{+,2}-X_{-,1}X_{-,2}\bigr)
\]
give
\[
 |\langle A_1B_2+B_1A_2\rangle_\omega|
 \le32/k+16\delta.
\]
Mixed-parity expectations vanish.

For the trace-norm estimate, choose a homogeneous
Hermitian Hilbert--Schmidt orthonormal block basis
$\{E_\mu\}$. The even--even products
$E_{\mu,1}E_{\nu,2}$ and odd--odd products
$iE_{\mu,1}E_{\nu,2}$ form an orthonormal basis
of the joint even algebra. The same coefficient
estimate as in the spin proof therefore yields
\[
 \|\omega-\mathsf S_f\omega\mathsf S_f^\dagger\|_1
 \le d_{\rm blk}^3(32/k+16\delta).
\]
The choice of $c,r$ and the final triangle inequality
are unchanged.
\end{proof}

\subsection{fermionic conditional expectations and channels}

\begin{fermionlemma}[Conditional expectations of fermions]
\label{lem:fermion-expectations}
Let $\omega_{X^c}$ be a density matrix \footnote{By superselection rule, $\omega_{X^c}$ is always parity even.}. In the order $X,X^c$, define
\begin{equation}
 \E_X^\omega(O)
 =\iota_X\!\left(
   \Tr_{\mathcal F_{X^c}}\!\left[
    (\one_X\otimes\omega_{X^c})
    \mathcal U_X O\mathcal U_X^\dagger
   \right]\right).
 \label{eq:fermion-expectations}
\end{equation}
The map $\E_X^\omega$ is a parity-preserving unital completely positive contraction fixing $\A_X$, and satisfies
\[
 \E_X^\omega(ab)=a\,\Tr(\omega_{X^c} b),
 \qquad a\in\A_X,\quad b\in\A_{X^c};\qquad
 \E_X^\omega(\A_Y)\subset\A_{X\cap Y}
 \]
\end{fermionlemma}
\begin{proof}
We first show that $\E^\omega_X$ is a parity-preserving unital completely positive contraction. {It is parity-preserving because the actions of $\mathcal{U}_X$ and $\iota_X$, which commute with the total fermion parity $\Gamma$, and partial traces are both parity-preserving. It is unital because $\E^{\omega}_{X}(\one)=\one$.} To show its complete positivity, consider the spectral decomposition $\omega_{X^c} = \sum_{j} p_j \ket{j}_{X^c}\bra{j}_{X^c}$ with $p_j\geq 0$ and $\sum_j p_j=1$. Then $\Tr_{\mathcal F_{X^c}}( (\one\otimes\omega_{X^c}) O ) = \sum_j p_j (\one_X\otimes \bra{j}_{X^c}) O (\one_X\otimes\ket{j}_{X^c})$, which is a Kraus representation. So $\Tr_{\mathcal{F}_{X^c}}(\cdot)$ is completely positive. Since unitary conjugation is completely positive and compositions of completely positive maps are completely positive, $\Tr_{\mathcal F_{X^c}}\left[(\one_X\otimes\omega)\mathcal{U}_X(\cdot)\mathcal{U}_X^\dag\right]$ and $\E^{\omega}_{X}$ are completely positive. The Schwarz inequality \cite{choi1974schwarz} for a unital completely positive map gives
\[
\E_{X}^\omega (O)^\dagger \E_X^\omega(O) \leq \E_{X}^\omega (O^\dagger O) \leq \norm{O}^2\one,
\]
where the last inequality is due to the fact that $O^\dagger O\leq \norm{O^\dagger O}\one$ and $\E^\omega_X(\cdot)$ is a positive and unital channel. Therefore, $\norm{\E^\omega_X(O)}\leq \norm{O}$, which proves the contractivity of $\E^\omega_X$. 

Next, we prove $\E^\omega_X(ab) = a\Tr(\omega_{X^c} b)$. This is clearly true for $b=\one_{X^c}$. For more general operator $b\in\mathcal{A}_{X^c}$, let us first consider the case where $b$ has a fixed parity $\deg(b)$. Its embedding is $\iota(b)=\Gamma_X^{\deg(b)}\otimes b$. Consequently, $\mathcal U_X (ab) \mathcal U_X^\dagger = a \Gamma_X^{\deg(b)}\otimes b$, and hence we get 
\[
\E^\omega_X(ab) = a\Gamma_X^{\deg(b)} \Tr(\omega_{X^c} b ).
\]
If $b$ is even, then $\deg(b)=0$ and $\E^\omega_X(ab) = a\Tr(\omega_{X^c} b)$ holds. If $b$ is odd, then the evenness of $\omega$ implies that $\Tr(\omega_{X^c} b)=0$ and $\E^\omega_X(ab) = a\Tr(\omega_{X^c} b)$ also holds. For a general $b$ as a sum of even and odd terms, linearity of $\E^\omega_X$ implies that $\E^\omega_X(ab) = a\Tr(\omega_{X^c} b)$ holds.

The claim that $\E^\omega_{X}(\A_Y)\subset\A_{X\cap Y}$ is obvious.

\end{proof}

\begin{fermionlemma}[Extension of a parity-covariant channel]
\label{lem:fermion-channel}
Let $A$ and $A'$ be two regions. Let $\Lambda:\A_A\to\A_{A'}$ be a channel satisfying parity covariance:
\begin{equation}\label{eq:parity_covariant_channel}
 \Lambda(\Gamma_A O \Gamma_A)
 =\Gamma_{A'}\Lambda(O)\Gamma_{A'}, \quad \text{for every }O\in \A_A.
\end{equation}
For an exterior $F$ disjoint from the input region $A$ and output region  $A'$, define the channel $\widetilde\Lambda^{F}:\A_{FA}\to\A_{FA'}$ by
\begin{equation}
 \widetilde\Lambda^{F}=\operatorname{id}_F\otimes\Lambda
 \quad\text{in the ordered Fock representations }(F,A)\text{ and }(F,A').
 \label{eq:fermion-channel-tensor}
\end{equation}
The map $\widetilde\Lambda^{F}$ is completely positive and
trace-preserving. Its adjoint satisfies the CAR product identity
\begin{equation}
 (\widetilde\Lambda^{F})^{\dagger}(b_Fa_{A'})
 =b_F\Lambda^{\dagger}(a_{A'}),
 \qquad b_F\in\A_F,\quad a_{A'}\in\A_{A'},
 \label{eq:fermion-extension}
\end{equation}
where the local operators are embedded in the joint CAR algebras.
In particular,
\begin{equation}
 (\widetilde\Lambda^{F})^{\dagger}(b_F)=b_F,
 \qquad (\widetilde\Lambda^{F}(\rho))_F=\rho_F,
 \label{eq:fermion-channel-exterior}
\end{equation}
where $\rho$ is a density matrix on $FA$. For every $F_0\subseteq F$,
\begin{equation}
 (\widetilde\Lambda^{F}(\rho))_{F_0A'}
 =\widetilde\Lambda^{F_0}(\rho_{F_0A}).
 \label{eq:fermion-channel-restriction}
\end{equation}
If $\Xi:\A_B\to\A_{B'}$ is another parity-covariant channel on a
disjoint region, the extensions $\tilde\Xi^{FA}={\rm id}_{FA}\otimes\Xi$ and $\tilde\Xi^{FA'}={\rm id}_{FA'}\otimes\Xi$ obey
\begin{equation}
 \widetilde\Lambda^{FB'}\circ\widetilde\Xi^{FA}
 =\widetilde\Xi^{FA'}\circ\widetilde\Lambda^{FB}
 :\A_{FAB}\longrightarrow\A_{FA'B'}.
 \label{eq:fermion-channel-commutation}
\end{equation}
\end{fermionlemma}

\begin{proof}
The extension $\widetilde\Lambda^F$ is completely positive and trace-preserving because $\Lambda$ is. We now prove Eq. \eqref{eq:fermion-extension}. Taking the adjoint of Eq. \eqref{eq:parity_covariant_channel}, we can see that the adjoint channel $\Lambda^\dagger$ is also parity covariant. This implies that $\Lambda^\dagger(a)$ and $a$ have the same parity for a parity-definite operator $a$. Let \(b\in\mathcal A_F\) be arbitrary and \(a\in\mathcal A_{A'}\) be homogeneous of degree \(q\). In the output ordering \((F,A')\), $ba=b\Gamma_F^q\otimes a$, so
\[
(\widetilde\Lambda^{F})^{\dagger}(ba)=(\operatorname{id}_F\otimes \Lambda^\dagger)
(b\Gamma_F^q\otimes a)
=b\Gamma_F^q\otimes \Lambda^\dagger(a).
\]
Since \(\Lambda^\dag(a)\) has parity \(q\), the last expression is precisely the representation of \(b\Lambda^\dag(a)\) in the input ordering \((F,A)\). Hence
\[
(\widetilde\Lambda^F)^\dagger(ba)
=b\Lambda^\dagger(a).
\]
Every operator is a sum of its even and odd parts, so the identity holds for arbitrary \(a\) by linearity. Then, taking $a=\one$ we get the first equation in Eq.~\eqref{eq:fermion-channel-exterior}. The second equation in Eq.~\eqref{eq:fermion-channel-exterior} can be seen by the following relation for any operator $b_F\in \A_{F}$,
\begin{equation}
\Tr( (\widetilde \Lambda^F(\rho))_F b_F ) = \Tr\left(\rho (\tilde\Lambda^F)^{\dagger}(b_F)\right) = \Tr(\rho_F b_F) = \Tr(\rho b_F).
\end{equation}
For Eq. \eqref{eq:fermion-channel-restriction}, similarly, it can be proved by taking arbitrary $b_{F_0}$ and $a_{A'}$,
\begin{equation}
\operatorname{Tr}\!\left[
(\widetilde\Lambda^{F}(\rho))_{F_0A'}
\,b_{F_0}a_{A'}
\right]=\operatorname{Tr}\!\left[
\rho\,b_{F_0}\Lambda^\dagger(a_{A'})
\right]=\operatorname{Tr}\!\left[
\rho_{F_0A}\,b_{F_0}\Lambda^\dagger(a_{A'})
\right]=\operatorname{Tr}\!\left[
\widetilde\Lambda^{F_0}(\rho_{F_0A})
\,b_{F_0}a_{A'}
\right].
\end{equation}
For Eq. \eqref{eq:fermion-channel-commutation}, it suffices to prove its version in terms of their adjoints on region $FA'B'$, namely $(\widetilde \Xi^{FA})^\dagger\circ(\widetilde\Lambda^{FB'})^\dagger = (\widetilde\Lambda^{FB})^\dagger \circ (\widetilde\Xi^{FA'})^\dagger$. Take $f_{F} a_{A'} b_{B'}$ with $a_{A'}, b_{B'}$ parity $p,q\in\{0,1\}$,
\[
\begin{aligned}
(\widetilde\Lambda^{FB'})^\dagger(f_Fa_{A'}b_{B'})
&=(-1)^{pq}
(\widetilde\Lambda^{FB'})^\dagger(f_Fb_{B'}a_{A'})\\
&=(-1)^{pq}f_Fb_{B'}\Lambda^\dagger(a_{A'})\\
&=f_F\Lambda^\dagger(a_{A'})b_{B'}.
\end{aligned}
\]
Applying the $B$-channel next therefore yields
\[
(\widetilde\Xi^{FA})^\dagger
\circ(\widetilde\Lambda^{FB'})^\dagger(f_Fa_{A'}b_{B'}) = f_F\Lambda^\dagger(a_{A'})\Xi^\dagger(b_{B'}).
\]
On the other hand, we have
\[
(\widetilde\Xi^{FA'})^\dagger(f_Fa_{A'}b_{B'})
=f_Fa_{A'}\Xi^\dagger(b_{B'}),
\]
and 
\[
(\widetilde\Lambda^{FB})^\dagger
(\widetilde\Xi^{FA'})^\dagger(f_Fa_{A'}b_{B'})
=
f_F\Lambda^\dagger(a_{A'})\Xi^\dagger(b_{B'}).
\]
Hence we get Eq. \eqref{eq:fermion-channel-commutation} proved.

\end{proof}

\begin{fermionlemma}[Parity-preserving purification comparison]
\label{lem:fermion-uhlmann}
Let $\ket{u}\in\mathcal H_R \otimes\mathcal K$ and
$\ket{v}\in\mathcal H_R \otimes\mathcal K'$ be normalized pure states with the same definite total fermion parity $p$.
Write $\rho_R^u$ and $\rho_R^v$ for their restrictions to $R$.
Suppose $\dim\mathcal K'_t\ge\dim\mathcal K_t$ in each parity
sector $t=\pm1$. There is an even isometry
$V:\mathcal K\to\mathcal K'$ satisfying
\[
 \bra{v}(\one_R\otimes V)\ket{u}
 =F(\rho_R^u,\rho_R^v),\qquad
 \|(\one_R\widehat\otimes V)\ket{u}-\ket{v}\|
 \le\sqrt{\|\rho_R^u-\rho_R^v\|_1}.
\]
\end{fermionlemma}

\begin{proof}
This Lemma follows by applying the ordinary Uhlmann's theorem separately in the two parity sectors. Project $\ket{u}$ and $\ket{v} $ to parity-definite subspaces, $\ket{u_s}\in \mathcal{H}_{R,s} \otimes \mathcal K_{ps}$ and $\ket{v_s}\in \mathcal{H}_{R,s} \otimes \mathcal K'_{ps}$ with $s=\pm1$ and $\mathcal H_{R,s}$ is the subspace of parity $s$ and $\mathcal K_{ps}$ is the subspace of parity $ps$. We further define 
\[
\rho_s=\operatorname{Tr}_{\mathcal K}|u_s\rangle\langle u_s|,
\qquad
\sigma_s=\operatorname{Tr}_{\mathcal K'}|v_s\rangle\langle v_s|,
\]
so that $\rho_R^u=\rho_+\oplus\rho_-,\,
\rho_R^v=\sigma_+\oplus\sigma_-.$ Then we apply Uhlmann's theorem within each sector. Ordinary Uhlmann's theorem gives $V_{ps}: \mathcal K_{ps} \to \mathcal{ K'}_{ps}$ such that 
\[
\ev{v_s|\one\otimes V_{ps}|u_s} = \norm{\sqrt{\rho_s}\sqrt{\sigma_s}}_1.
\]
Now we assemble the two sector maps, $V= V_+\oplus V_-$, which satisfies $V^\dagger V = \one,\, \Gamma_{\mathcal K'} V = V\Gamma_{\mathcal K}$. Hence we get
\begin{equation}
\begin{split}
\langle v|(\mathbf1_R\otimes V)|u\rangle
&=\sum_{s=\pm1}
\langle v_s|(\mathbf1\otimes V_{ps})|u_s\rangle\ =\sum_{s=\pm1}\|\sqrt{\rho_s}\sqrt{\sigma_s}\|_1\\
&=\|\sqrt{\rho_R^u}\sqrt{\rho_R^v}\|_1 =F(\rho_R^u,\rho_R^v).
\end{split}
\end{equation}
The first equality of second line holds because the matrices are block diagonal. Then, we get
\[
\begin{aligned}
\|(\mathbf1_R\otimes V)|u\rangle-|v\rangle\|^2
&=2-2F(\rho_R^u,\rho_R^v)\\
&\le\|\rho_R^u-\rho_R^v\|_1.
\end{aligned}
\]
The last line is the Fuchs–van de Graaf inequality
\[
1-F(\rho,\sigma)\le\tfrac12\|\rho-\sigma\|_1.
\]
This leads to the claim
\[
\|(\mathbf1_R\otimes V)|u\rangle-|v\rangle\|
\le\sqrt{\|\rho_R^u-\rho_R^v\|_1}.
\]

\end{proof}

\begin{fermioncorollary}[Even isometry for an approximately decoupled state]
\label{cor:fermion-decoupled-purification}
Let $\ket{\psi}\in\mathcal F_{BCF}$ be normalized and have parity $p\in\{+1,-1\}$. Take
$\sigma=\ket{\psi}\bra{\psi}$, and suppose
$\|\sigma_{BF}-\sigma_B\otimes\sigma_F\|_1\le\delta$.
Choose an even purification $\ket{\beta_{BC_1}}$ of $\sigma_B$ and a
parity-$p$ purification $\ket{\xi_{C_2F}}$ of $\sigma_F$. If
\[
 \dim(\mathcal F_{C_1}\otimes\mathcal F_{C_2})_t
 \ge\dim(\mathcal F_C)_t,\qquad t=\pm1,
\]
there is an even isometry $W:\mathcal F_C\to
\mathcal F_{C_1}\otimes\mathcal F_{C_2}$ such that
\[
 \|(\one_B\otimes W\otimes\one_F)\ket{\psi}
       -\ket{\beta}\ket{\xi}\|\le\sqrt\delta.
\]
\end{fermioncorollary}
\begin{proof}
Regroup the modes so that $BF$ is before $C$. The vectors  $\ket{\psi}$ and $\ket\beta\otimes\ket\xi$ then purify \(\sigma_{BF}\) and \(\sigma_B\otimes\sigma_F\), respectively, and both have total parity \(p\). The assumed sector dimensions allow us to apply Lemma~\ref{lem:fermion-uhlmann}, yielding an even isometry $W:C\to C_1C_2$ with vector error at most $\sqrt\delta$. Returning to the original mode ordering gives the stated inequality. Note that it is the evenness of $W$ that ensures compatibility with the fermionic regrouping signs.
\end{proof}

\bibliography{reference}